%% file: 3LoopMGFs.tex
\documentclass[11pt,a4paper]{article}
\pdfoutput=1

\usepackage{jheppub}

\usepackage{amsmath, amssymb, amsthm} 
\usepackage{mathtools}
\usepackage{thmtools}
\usepackage{bm}
\usepackage{dsfont}
\usepackage{braket}

\newtheorem{thm}{Theorem}[section]
\newtheorem{lem}[thm]{Lemma}
\newtheorem{prop}[thm]{Proposition}
\newtheorem{cor}[thm]{Corollary}

\numberwithin{equation}{section}

\usepackage{enumitem}
\setlist[itemize]{noitemsep}
\setlist[itemize]{leftmargin=*}
\setlist[description]{noitemsep}

\usepackage{xcolor}
\usepackage{tikz}

\usepackage[numbers, sort&compress]{natbib}
\usepackage{hyperref}
\hypersetup{colorlinks=true, linktoc=page, linkcolor=purple, citecolor=blue}

\renewcommand{\a}{\alpha}
\renewcommand{\b}{\beta}
\newcommand{\g}{\gamma}
\newcommand{\ep}{\varepsilon}

\newcommand{\sm}{\smallskip}
\newcommand{\no}{\nonumber}

\input{preambledefs}

\newcommand{\intM}{ \int_\cM \! \frac{d^2\tau}{\tau_2^2} \, }
\newcommand{\intML}{ \int_{\cM_L} \! \frac{d^2\tau}{\tau_2^2} \, }
\newcommand{\intMR}{ \int_{\cM_R} \! \frac{d^2\tau}{\tau_2^2} \, }

\title{Integrating three-loop modular graph functions and transcendentality of string amplitudes}
    
\date{October 12, 2021}

\author[a]{Eric D'Hoker}
\author[a]{and Nicholas Geiser}

\affiliation[a]{
    Mani L. Bhaumik Institute for Theoretical Physics\\
    Department of Physics and Astronomy\\
    University of California, Los Angeles, CA 90095, USA}

\emailAdd{dhoker@physics.ucla.edu}
\emailAdd{ngeiser@physics.ucla.edu}

\arxivnumber{2110.06237}


\abstract{
Modular graph functions (MGFs) are $\mathrm{SL}(2,\mathbb{Z})$-invariant functions on the Poincar\'e upper half-plane associated with Feynman graphs of a conformal scalar field on a torus. The low-energy expansion of genus-one superstring amplitudes involves suitably regularized integrals of MGFs over the fundamental domain for $\mathrm{SL}(2,\mathbb{Z})$. In earlier work, these integrals were evaluated for all MGFs up to two loops and for higher loops up to weight six. These results led to the conjectured uniform transcendentality of the genus-one four-graviton amplitude in Type~II superstring theory. In this paper, we explicitly evaluate the integrals of several infinite families of three-loop MGFs and investigate their transcendental structure. Up to weight seven, the structure of the integral of each individual MGF is consistent with the uniform transcendentality of string amplitudes. Starting at weight eight, the transcendental weights obtained for the integrals of individual MGFs are no longer consistent with the uniform transcendentality of string amplitudes. However, in all the cases we examine, the violations of uniform transcendentality take on a special form given by the integrals of triple products of non-holomorphic Eisenstein series. If Type~II superstring amplitudes do exhibit uniform transcendentality, then the special combinations of MGFs which enter the amplitudes must be such that these integrals of triple products of Eisenstein series precisely cancel one another. Whether this indeed is the case poses a novel challenge to the conjectured uniform transcendentality of genus-one string amplitudes.
}

\begin{document}

\maketitle

\newpage

\section{Introduction}
\label{sec:intro}

\emph{Modular graph functions} (MGFs) are $\mathrm{SL}(2,\bbZ)$-invariant functions on the Poincar\'e upper half-plane associated with Feynman graphs for a conformal scalar field theory on a torus. MGFs may be partially organized by their loop order and their transcendental weight, which are the number of loops and the number of scalar Green function edges, respectively, in their corresponding Feynman graphs. MGFs may be added and multiplied together so that the space of MGFs forms a graded ring, where the grading is given by the transcendental weight. The modular weight of a MGF vanishes and is not to be confused with its transcendental weight, which is often simply referred to as the \emph{weight}.

\sm

MGFs of small weight were encountered in the low-energy expansion of the genus-one four-graviton amplitude in Type~II superstring theory~\cite{Green:1999pv, Green:2008uj}. The coefficients of the effective interactions in this expansion involve suitably regularized integrals of MGFs over the moduli space of compact genus-one Riemann surfaces. The weights of the MGFs which enter this expansion are intimately related with the space-time structure of these effective interactions. For example, the coefficients of the effective interactions of the form $D^{2w} \cR^4$ involve integrals of MGFs of weight $w$.

\sm

MGFs of arbitrary weight and loop order were defined in terms of multiple Kronecker-Eisenstein sums in~\cite{DHoker:2015gmr, Zerbini:2015rss, DHoker:2015wxz}, where their systematic study was initiated. \emph{Modular graph functions} were generalized in~\cite{DHoker:2016mwo} to \emph{modular graph forms}, which are modular forms of non-vanishing modular weight associated with decorated Feynman graphs. Modular graph forms encompass holomorphic, anti-holomorphic, and non-holomorphic modular forms and may be assigned a generalized weight. Modular graph functions and forms obey a hierarchy of algebraic identities and differential equations which preserve their weight but generally mix different loop orders. A systematic method to derive these identities at two-loop order and arbitrary weight was developed in~\cite{DHoker:2015gmr} and was generalized to arbitrary loop orders and weight in~\cite{DHoker:2016mwo, DHoker:2016quv, Gerken:2018zcy,Gerken:2019cxz}. A Mathematica package containing the systematic implementation of all identities among MGFs up to weight six was introduced in~\cite{Gerken:2020aju, Gerken:2020xte}. Various special cases of these identities were proven in~\cite{DHoker:2015sve, Basu:2015ayg, Basu:2016xrt, Basu:2016kli, Basu:2019idd, Kleinschmidt:2017ege}.

\sm

Modular graph functions are closely related to other mathematical objects. For instance, the Kronecker-Eisenstein series representation of MGFs shows that they may be viewed as natural generalizations of multiple zeta-values. The loop order and weight of MGFs may be identified with the depth and transcendental weight of multiple zeta-values, respectively. Weight provides a grading on the ring of MGFs generalizing the grading by transcendental weight on the space of (motivic) multiple zeta-values~\cite{Brown:2013gia}. Identities between MGFs generalize identities between multiple zeta-values, such as those collected in~\cite{Blumlein:2009cf}. Additionally, MGFs may be viewed as generalizations of the non-holomorphic Eisenstein series, which themselves provide a one-dimensional basis for all one-loop MGFs. Moreover, MGFs may be obtained as special values of elliptic modular graph functions which are closely related to single-valued elliptic polylogarithms~\cite{DHoker:2015wxz, Zerbini:2018hgs, Zerbini:2018sox, Broedel:2018izr} and iterated modular integrals~\cite{Broedel:2019vjc, Gerken:2020yii}.

\sm

Especially important for the present work are the Fourier and Poincar\'e series for arbitrary connected two-loop MGFs obtained in~\cite{DHoker:2017zhq, Ahlen:2018wng, DHoker:2019txf,Dorigoni:2019yoq, Basu:2020kka}. These series were used in~\cite{DHoker:2019mib} to evaluate the integrals of two-loop MGFs using the unfolding trick familiar from the Rankin-Selberg-Zagier method~\cite{Rankin:1939, Selberg:1940, Zagier:1982}. These integrals may be expressed in terms of zeta-values and assigned a definite transcendental weight, thereby providing the starting point for a systematic investigation of the transcendentality properties of the genus-one four-graviton amplitude in Type~II superstring theory in~\cite{DHoker:2019blr}.

\sm

In particular, the low-energy expansion of the genus-one four-graviton amplitude in Type~II superstring theory was computed explicitly and shown to exhibit uniform transcendentality up to the order of $D^{12} \cR^4$ using the integrals of MGFs up to weight six. Although the corresponding integrands involve MGFs with as many as five loops, the identities of~\cite{DHoker:2016quv} may be used to re-express the integrands in terms of one-loop and two-loop MGFs plus a single three-loop~MGF. This three-loop MGF may then be integrated using the differential equation it satisfies, which was obtained in~\cite{Broedel:2018izr}. Thus, up to the order of~$D^{12} \cR^4$, only the integrals of one-loop and two-loop~MGFs are effectively needed.

\sm

The low-energy expansion of the amplitude beyond the order of $D^{12} \cR^4$ requires integrals of MGFs of weight seven and higher as well as integrals of MGFs with three or more loops which cannot be reduced to two-loop order. Few such integrals have been calculated prior to this work.

\sm

Significant partial results, to be explained below, support the validity of uniform transcendentality to arbitrary order in the low-energy expansion of the genus-one four graviton amplitude in Type~II superstring theory, as conjectured in~\cite{DHoker:2019blr}. However, the results of the present paper raise the possibility of violations of uniform transcendentality starting at weight eight, which corresponds to the order of $D^{16} \cR^4$ in the low-energy expansion. Before turning to the detailed calculations involved, we shall provide a brief overview of the questions pursued and the results obtained in the sequel of the paper.

\subsection{Transcendental weight assignments}
\label{sec:trans}

To explain the goal and results of the present paper, we shall begin with a brief summary of the assignments of transcendental weight to superstring amplitudes and to the various mathematical ingredients out of which these amplitudes are constructed.

\sm

The Riemann zeta function $\zeta(z)$ is defined for $\Re(z)>1$ by the series,\footnote{In \autoref{apdx:zeta}, we review the salient properties needed in this paper of zeta functions, zeta-values, finite harmonic sums, and multiple zeta-values, including their transcendental weight assignments.}
\begin{align}
    \zeta(z)
    =
    \sum_{n=1}^\infty \frac{1}{n^z}
\end{align}
In number theory, it is standard to assign (transcendental) weight zero to rational numbers, weight one to $\pi$ and to the natural logarithm of rational numbers, and weight $n$ to the zeta-value $\zeta(n)$ for integer $n \geq 2$. This standard assignment of transcendental weight is consistent with the fact that the even zeta-values $\zeta(2n)$ may be written as follows for $n \geq 1$, 
\begin{align}
    \zeta(2n)
    &=
    \half (-)^{n+1} \,
    (2\pi)^{2n} \,
    \frac{ B_{2n} }{ (2n)! }
\end{align}
where the Bernoulli numbers $B_n$ are rational.

\sm

In physics, the tree-level four-graviton amplitude in Type~II superstring theory is given by the following expression, up to a kinematic multiplicative factor,
\begin{align}
\label{eq:A0}
    \frac{1}{stu}
    \frac{ \Gamma(1-s) \Gamma(1-t) \Gamma(1-u) }
         { \Gamma(1+s) \Gamma(1+t) \Gamma(1+u) }
    =
    \frac{1}{stu}
    \exp
    \bigg\{
    2
    \sum_{m=1}^\infty
    \frac{ \zeta(2m+1) }{ 2m+1 } \, 
    \sigma_{2m+1}(s,t,u)
    \bigg\}
\end{align}
where $s,t,u$ are dimensionless kinematic variables and ${\sigma_k(s,t,u) = s^k + t^k + u^k}$ are the corresponding symmetric polynomials subject to momentum conservation~${\sigma_1 = 0}$. Taylor expanding the right-hand side of~\eqref{eq:A0} in powers of $s,t,u$ produces the low-energy expansion of this tree-level amplitude.

\sm

\emph{Uniform transcendentality} of an amplitude refers to the property that all terms in its low-energy expansion may be assigned the same transcendental weight. For the tree-level amplitude~\eqref{eq:A0}, this is achieved by adopting the standard number theoretic transcendental weight assignments for $\zeta(n)$ and choosing the transcendental weight of $s,t,u$ to be~$-1$. With these assignments, each term in the low-energy expansion of~\eqref{eq:A0} has transcendental weight three. Thus, the tree-level four-graviton amplitude in Type~II superstring theory exhibits uniform transcendentality in the sense defined above. Remarkably, string amplitudes have non-trivial transcendental structure already at tree-level while in quantum field theory, non-trivial transcendental structure only arises from loop integrals~\cite{Kotikov:2002ab, Beccaria:2009vt, Arkani-Hamed:2010pyv, Broedel:2018qkq}.

\sm

At loop-level in Type~II superstring theory, the transcendental structure of the genus-one four-graviton amplitude was investigated in~\cite{DHoker:2019blr}. There it was shown that the low-energy expansion of this amplitude exhibits uniform transcendentality up to the order of~$D^{12} \cR^4$ provided we make the following additional assignments of transcendental weight.
\begin{itemize}
\item The logarithmic derivative of the Riemann zeta function $\zeta'(n)/\zeta(n)$ with integer $n \geq 2$ and the Euler-Mascheroni constant $\gamma_E$ have transcendental weight one.\footnote{This assignment is a slightly stronger version of Assumption 3 of~\cite{DHoker:2019blr}.}
\sm
\item The finite harmonic sums $H_1(m) = \sum_{k=1}^{m-1} \frac{1}{k}$ have transcendental weight one.
\end{itemize}
This last assumption is delicate. One should think of $H_1(m)$ not as its value for a single~$m$ (which would give a rational number whose natural transcendental weight assignment is zero) but instead as a function of $m$ to be inserted into an infinite series in $m$. For instance,~$H_1(m)$ occurs in this manner in the double zeta-value $\zeta(n,1)$,
\begin{align}
    \zeta(n,1)
    =
    \sum_{m > k > 0}
    \frac{1}{m^n \, k}
    =
    \sum_{m = 2}^\infty
    \frac{H_1(m)}{m^n}
\end{align}
The standard transcendental weight assignments of $\zeta(n)$ and $\zeta(n,1)$ are $n$ and $n+1$, respectively, which justifies assigning transcendental weight one to the function $H_1(m)$. In fact, this assignment of non-zero transcendental weight to finite harmonic sums is familiar to $\cN = 4$ supersymmetric quantum field theory amplitudes~\cite{Kotikov:2002ab, Beccaria:2009vt}.

\subsection{Overview of goals and results}
\label{sec:overv}

To describe our goals and results requires some additional set-up. We shall parametrize the Poincar\'e upper half-plane $\cH$ by the variable ${\tau = \tau_1 + i\tau_2 \in \bbC}$ with $\tau_1, \tau_2 \in \bbR$ and ${\tau_2 > 0}$. The modular group $\mathrm{SL}(2,\bbZ)$ acts on $\cH$ by its normal subgroup $\mathrm{PSL}(2,\bbZ) = \mathrm{SL}(2,\bbZ) / \bbZ_2$. The moduli space of compact genus-one Riemann surfaces is given by ${\cM = \mathrm{PSL}(2,\bbZ) \backslash \cH}$ and may be represented by the standard fundamental domain,
\begin{align}
    \cM
    =
    \left\{
    \tau \in \cH
    \,:\,
    |\Re(\tau)| \leq \half, \,
    |\tau|\geq 1
    \right\}
\end{align}
which contains a single cusp at $\tau = i \infty$. The modular-invariant Poincar\'e metric~\smash{$ d^2\tau / \tau_2^2$} provides a volume form on $\cM$ to integrate modular-invariant functions, such as modular graph functions. We shall review the properties of MGFs in \autoref{sec:MGFs}.

\sm

Generically, a modular graph function $\cC$ will have polynomial growth at the cusp so that its integral over $\cM$ diverges. To associate a finite integral to $\cC$, we partition the fundamental domain~${\cM = \cM_L \cup \cM_R}$ into a neighborhood of the cusp $\cM_R$ and the truncated fundamental domain~$\cM_L$, defined by,
\begin{align}
    \cM_L &= \cM \cap \{ \Im(\tau) \leq L \}
    &
    \cM_R &= \cM \cap \{ \Im(\tau) > L \}
\end{align}
for some cut-off $L > 1$. The integral of $\cC$ over $\cM_L$ is convergent but $L$-dependent. Mathematically, a finite integral of $\cC$ over $\cM$ may be defined by adding to the integral of $\cC$ over~$\cM_L$ the integral over $\cM_R$ of a truncated version of $\cC$ in which the polynomial growth is subtracted~\cite{Zagier:1982}. We shall review the technical details of these integrals in \autoref{sec:ints}. 

\sm

Physically, the divergence of the integral of $\cC$ over $\cM_R$ must be considered from the vantage point of the full genus-one Type~II superstring amplitude. The integral formula for the full amplitude is absolutely convergent only when the dimensionless kinematic variables~$s,t,u$ are purely imaginary. The existence and uniqueness of the analytic continuation of the full amplitude in $s,t,u$ was established long ago in~\cite{DHoker:1994gnm}. Therefore, the source of the divergence of the integral over $\cM$ of the modular graph function $\cC$ stems from the non-uniformity across~$\cM$ of the Taylor expansion in powers of $s,t,u$ of the amplitude's integrand. Analytically continuing the integral over~$\cM_R$ in $s,t,u$ produces the non-analyticities expected on physical grounds from the presence of massless strings. These non-analyticities include the logarithms $\ln s, \ln t,$ and $ \ln u$ which give rise to the physical branch cuts produced by massless string pairs required by the unitarity of the string amplitude, as described in~\cite{DHoker:2019blr}. 

\sm

The genus-one contributions to the low-energy effective interactions of the string amplitude are computed by an integral over the full fundamental domain $\cM$, which may be partitioned into the regions~$\cM_L$ and~$\cM_R$ defined above. The cut-off~$L > 1$ is clearly arbitrary and necessarily cancels out of the full amplitude. The contribution from $\cM_L$ to the full amplitude is analytic in $s,t,u$ so that the contributions from~$\cM_L$ to any order in the low-energy expansion are given by a sum of integrals over~$\cM_L$ of MGFs. When the modular graph function $\cC$ appears in this context, the $L$-dependence in its integral over $\cM_L$ will be cancelled by the integral of the full string integrand over $\cM_R$. See for instance~\cite{Green:2008uj, DHoker:2019blr}.

\sm

The partial results regarding the transcendentality of the genus-one four-graviton amplitude in Type~II superstring theory beyond the order of $D^{12} \cR^4$ in the low-energy expansion, alluded to earlier, consist of two parts~\cite{DHoker:2019blr}. First, the coefficients of the $\ln s, \ln t, \ln u$ non-analyticities exhibit uniform transcendentality. This result is fully expected from the factorization of the genus-one amplitude and follows from the uniform transcendentality of the tree-level amplitude~\eqref{eq:A0}. Second, the contribution to the analytically continued integral which arises from the integral over~$\cM_R$ and which is analytic in $s,t,u$ exhibits uniform transcendentality to all orders in the low-energy expansion. This second result is not known to follow from factorization and constitutes a non-trivial motivation for the uniform transcendentality conjecture of the full genus-one amplitude. Thus, to prove the conjecture up to an arbitrary order in the low-energy expansion, it remains to study the transcendental structure of the integrals over~$\cM_L$ of the MGFs that appear in this low-energy expansion.

\sm

Before considering three-loop MGFs, we shall first review the results at two-loops. The integrals over $\cM_L$ of arbitrary connected two-loop MGFs were evaluated in~\cite{DHoker:2019mib}. In this paper, we shall also evaluate the integrals of arbitrary disconnected two-loop MGFs. From these explicit calculations, we find that the transcendental structure of the integrals of all two-loop MGFs is consistent with the uniform transcendentality of superstring amplitudes. Specifically, we prove the following proposition in \autoref{sec:I2loops}.

\begin{prop}
\label{thm:2looptrans}
The integral over $\cM_L$ of an arbitrary two-loop modular graph function~$\cC$ of weight $w$, has the following structure,
\begin{align}
    \intML
    \cC
    =
    A_{w+1} + B_w \, H_1 + \cO(L^\pm)
\end{align}
where $\cO(L^\pm)$ indicates the omission of terms with non-trivial $L$-dependence; $A_{w+1}$ and $B_w$ have transcendental weight $w+1$ and $w$, respectively; and $H_1$ is a sum of rational numbers which may be interpreted as a sum of finite harmonic sums of transcendental weight one.
\end{prop}

Prior to the present work, it was an open question whether this transcendental structure persisted in the integrals of three-loop MGFs. In this paper, we shall evaluate the integrals over $\cM_L$ of an infinite number of three-loop MGFs belonging to several special infinite families. We shall also investigate the transcendental structure of these integrals. We restrict to these families because the evaluation of the integrals of the most general three-loop MGFs appears prohibitively involved with the methods presently available. Our results from the study of these integrals are as follows. 
\begin{itemize}
\item Up to weight seven included, the transcendental structure of the integral of each individual three-loop MGF is consistent with the uniform transcendentality of genus-one superstring amplitudes.
\sm
\item At weight eight and higher, the transcendental structure of the integrals of certain individual three-loop MGFs is not consistent with the uniform transcendentality of superstring amplitudes.
\sm
\item If the full genus-one superstring amplitude is to exhibit uniform transcendentality, then special cancellations between the integrals of individual three-loop MGFs must occur.
\end{itemize}

The nature of the violations of uniform transcendentality that occur in the integrals of individual three-loop MGFs may be illustrated by considering the simplest such case, the integral of a triple product of non-holomorphic Eisenstein series. The non-holomorphic Eisenstein series $E_s^*(\tau,\bar{\tau}) = \half \Gamma(s) \, E_s(\tau,\bar{\tau})$ of weight~$s$ will be defined in \autoref{sec:1loopMGF} and is a one-loop modular graph function. Zagier evaluated the integral of the triple product $E_s^* E_t^* E_u^*$ in terms of a quadruple product of zeta functions in~\cite{Zagier:1982}. For integer weights~${s,t,u \geq 2}$,
\begin{align}
    \intML E_s^* E_t^* E_u^*
    = 
    \frac{ c_{s,t,u} }{ \pi^{w-1} } \,
    \zeta(w-1) \,
    \zeta(w-2s) \,
    \zeta(w-2t) \,
    \zeta(w-2u)
    + \cO(L^\pm)
\end{align}
where $w=s+t+u$ and $\cO(L^\pm)$ indicates the omission of terms with non-trivial $L$-dependence. The proportionality factor $c_{s,t,u}$ will be given in \autoref{thm:IEEE} and has transcendental weight zero or one when $w$ is even or odd, respectively. Naively, the transcendental weight of the integral is $w$ or $w+1$. This result indeed holds as long as the three integers~$s,t,u \geq 2$ satisfy,
\begin{align}
   s+t-u, \,
   t+u-s, \,
   u+s-t
   \geq 2
   \qquad \iff \qquad
   \max(s,t,u) \leq \floor{w/2} - 1
\end{align}
However, if say $s+t-u < 0$, then the corresponding zeta-value $\zeta(w-2u) = \zeta(s+t-u)$ is a rational number with transcendental weight zero. In other words, when the zeta functions have a non-positive argument, they violate the expected counting of transcendental weight. As a result, the transcendental weight of the integral is now \emph{larger than $w+1$}, and no harmonic sum with positive weight can make up for the violation.

\sm

The main result of this paper is that the violations of uniform transcendentality occurring in the integrals of individual three-loop MGFs are all of the same form as the violations endemic to Zagier's integrals of triple products of Eisenstein series. Specifically, we prove the following proposition for the various infinite families of three-loop MGFs whose integrals we calculate in \autoref{sec:Eisen}, \autoref{sec:ECabc}, \autoref{sec:vk3}, and \autoref{sec:Ck111}.
\begin{prop}
\label{thm:3looptrans}
For an arbitrary three-loop modular graph function $\cC$ of weight $w$, there exist rational numbers $Q_{\ell_1,\ell_2,\ell_3}$ such that,
\begin{align}
    \intML
    \bigg(
    \cC
    -
    \sum_{\substack{ \ell_{1,2,3} \geq 2 \\
        \mathclap{\ell_1 + \ell_2 + \ell_3 = w} }}
    Q_{\ell_1,\ell_2,\ell_3} \,
    E_{\ell_1}^*
    E_{\ell_2}^*
    E_{\ell_3}^*
    \bigg)
    =
    A_{w+1} + B_w \, H_1 + \cO(L^\pm)
\end{align}
where $\cO(L^\pm)$ indicates the omission of terms with non-trivial $L$-dependence; $A_{w+1}$ and $B_w$ have transcendental weight $w+1$ and $w$, respectively; and $H_1$ is a sum of rational numbers which may be interpreted as a sum of finite harmonic sums of transcendental weight one.
\end{prop}

It is an important open question whether this proposition holds for arbitrary three-loop MGFs beyond the infinite families studied in this paper.

\sm

In any case, the necessity of subtracting these triple products of Eisenstein series constitutes a novel source of violations of uniform transcendentality and raises challenging questions for the transcendental structure of genus-one superstring amplitudes. Is uniform transcendentality in the physical genus-one superstring amplitudes for four gravitons violated at sufficiently high weight? Or do the violations found in the integrals of individual MGFs conspire to cancel in the full superstring amplitude? The present work constitutes a first step towards investigating and answering these questions.

\subsection{A useful byproduct}
\label{sec:byproduct}

Finally, an important byproduct emerges from our investigations concerning the two-loop modular graph functions $C_{a,b,c}$ for integer $a,b,c \geq 1$ which were introduced in~\cite{DHoker:2015gmr} and will be reviewed in \autoref{sec:2loopMGF}. It was shown in~\cite{DHoker:2015gmr} that these functions obey a system of inhomogeneous Laplace eigenvalue equations at each weight $w=a+b+c$ whose inhomogeneous part is a linear combination of the non-holomorphic Eisenstein series $E_w$ and the double products $E_k E_{w-k}$ for $2 \leq k \leq w-2$. The eigenvalues of the corresponding homogeneous system were shown to be of the form $s(s-1)$, where $s = w-2, w-4, \cdots \geq 0$. The explicit diagonalization of this homogeneous system was achieved only for small weights in~\cite{DHoker:2015gmr} but is obtained in \autoref{apdx:CabcCwmp} of this paper for arbitrary weights. 

\sm

In particular, the formulas of \autoref{apdx:CabcCwmp} make it possible to express the modular graph functions $C_{a,b,c}$ as a linear combination with rational coefficients of the modular functions \smash{$F^{+(s)}_{m,k}$} introduced in~\cite{Dorigoni:2021jfr, Dorigoni:2021ngn}. These functions obey the inhomogeneous Laplace equation \smash{$(\Delta-s(s-1)) F^{+(s)}_{m,k} = E_m E_k$} where the inhomogeneous term is a double product of non-holomorphic Eisenstein series.

\subsection{Organization}

The remainder of this paper is organized as follows.  In \autoref{sec:MGFs}, we review modular graph functions and forms. In \autoref{sec:ints}, we discuss several methods to evaluate the integrals of MGFs over the truncated fundamental domain $\cM_L$. In \autoref{sec:I2loops}, we evaluate the integrals of arbitrary two-loop MGFs using these methods. In the subsequent four sections, we evaluate the integrals of various special infinite families of three-loop MGFs and analyze their transcendental structures. Specifically we evaluate the integrals for the following integrands:
\begin{description}
\item[\autoref{sec:Eisen},] triple products of Eisenstein series and their derivatives using the results of~\cite{Zagier:1982};
\item[\autoref{sec:ECabc},] the disconnected three-loop MGFs $E_k^* C_{a,b,c}$ using their differential equations;
\item[\autoref{sec:vk3},] the connected three-loop MGFs $v_{k,3}$ using their differential equations;
\item[\autoref{sec:Ck111},] the three-loop MGFs $C_{k,1,1,1}$ using their Poincar\'e series and the unfolding trick.
\end{description}
We conclude in \autoref{sec:conc} and discuss open problems including the transcendental structure of physical superstring amplitudes. In \autoref{apdx:zeta}, we review zeta functions, zeta-values, multiple zeta-values, and finite harmonic sums. In \autoref{apdx:CabcCwmp}, we discuss the relation between the two-loop modular graph functions $C_{a,b,c}$ and $\mfC_{w;m;p}$ announced in \autoref{sec:byproduct}. In \autoref{apdx:Laurent} we review the Laurent polynomials of several two-loop modular graph functions. The final two appendices, \autoref{apdx:ILk111} and \autoref{apdx:Ik111exp}, contain several technical details used to evaluate the integral of~$C_{k,1,1,1}$.

\subsection*{Acknowledgements}

We are grateful to Michael Green and Oliver Schlotterer for discussions of various topics related to this work. The research of ED is supported in part by the National Science Foundation under research grant PHY-19-14412. The research of NG is supported by the Mani L. Bhaumik Institute for Theoretical Physics, by a Eugene V. Cota-Robles Fellowship, and by an NSF grant supplement from the Alliances for Graduate Education and the Professoriate Graduate Research Supplements (AGEP-GRS).

\newpage

\section{Modular graph functions and forms}
\label{sec:MGFs}

In this section, we shall review modular graph functions and forms, and we shall introduce several infinite families of one-loop modular graph forms and two-loop modular graph functions.

\subsection{Conventions and definitions}

The modular group $\mathrm{SL}(2,\bbZ)$ will be parametrized as follows,
\begin{align}
    \mathrm{SL}(2,\bbZ)
    =
    \left\{
    \(
    \begin{smallmatrix}
    \alpha & \beta \\
    \gamma & \delta
    \end{smallmatrix}
    \)
    \,:\,
    \alpha, \beta, \gamma, \delta
    \in \bbZ, \,
    \alpha \delta - \beta \gamma = 1
    \right\}
\end{align}
A modular transformation 
$
    \lambda
    =
    \( \begin{smallmatrix}
    \alpha & \beta \\
    \gamma & \delta
    \end{smallmatrix} \)
    \in
    \mathrm{SL}(2,\bbZ)
$ 
acts on the variable $\tau \in \bbC$ by a M\"obius transformation,
\begin{align}
    \lambda \tau
    =
    \frac{\alpha \tau + \beta}{\gamma \tau + \delta}
\end{align}
The modular group has several important subgroups. For instance, ${\bbZ_2 = \{ \pm \mathds{1} \} \subset \mathrm{SL}(2,\bbZ)}$ leaves $\tau$ invariant, and the Borel subgroup $\Gamma_\infty$ defined by,
\begin{align}
\label{eq:Borel}
    \Gamma_\infty
    =
    \left\{
    \pm
    \(
    \begin{smallmatrix}
    1 & n \\ 0 & 1
    \end{smallmatrix}
    \)
    \, : \,
    n \in \bbZ
    \right\}
    \subset
    \mathrm{SL}(2,\bbZ)
\end{align}
implements integer translations of $\tau$. The modular group is itself a subgroup of the group $\mathrm{SL}(2,\bbR)$ of isometries of the Poincar\'e upper half-plane $\cH$, which is defined by,
\begin{align}
\label{eq:Hplane}
    \cH
    =
    \mathrm{SL}(2,\bbR) / \mathrm{U}(1)
    =
    \{ \tau \in \bbC \,:\, \Im(\tau) > 0 \}
\end{align}
We shall parametrize $\cH$ by the complex variable ${\tau = \tau_1 + i\tau_2}$ with $\tau_1, \tau_2 \in \bbR$ and $\tau_2 > 0$. The modular group acts on $\cH$ by its normal subgroup $\mathrm{PSL}(2,\bbZ) = \mathrm{SL}(2,\bbZ) / \bbZ_2$.

\sm

The moduli space of compact genus-one Riemann surfaces is given by the quotient ${\cM = \mathrm{PSL}(2,\bbZ) \backslash \cH}$ and may be represented by the standard fundamental domain,
\begin{align}
\label{eq:domainM}
    \cM
    =
    \left\{
    \tau \in \cH
    \,:\,
    |\Re(\tau)| \leq \half
    \, , \,
    |\tau|\geq 1
    \right\}
\end{align}
which contains a single cusp at $\tau = i \infty$. As discussed in \autoref{sec:overv}, we shall partition the fundamental domain $\cM = \cM_L \cup \cM_R$ into a neighborhood of the cusp $\cM_R$ and the truncated fundamental domain $\cM_L$, which are defined as follows,
\begin{align}
\label{eq:ML}
    \cM_L &= \cM \cap \{ \Im(\tau) \leq L \}
    &
    \cM_R &= \cM \cap \{ \Im(\tau) > L \}
\end{align}
for some cut-off $L > 1$. In practice it will often be convenient to work with a small neighborhood $\cM_R$ with large $L \gg 1$.

\sm

We may also consider quotients of $\cH$ by subgroups of the full modular group. For example, the upper half-strip $\Gamma_\infty \backslash \cH$ is given by the quotient of $\cH$ by the Borel subgroup of translations defined in~\eqref{eq:Borel} and may be represented as follows,
\begin{align}
\label{eq:strip}
    \Gamma_\infty \backslash \cH
    =
    \{
    \tau \in \cH
    \,:\,
    0 \leq \Re(\tau) \leq 1
    \}
\end{align}
Elements of $\Gamma_\infty$ stabilize the cusp at $\tau = i \infty$.

\subsubsection{Modular forms}

A complex-valued function $f$ of $\tau \in \cH$ is said to have modular weight $(u,v)$, with $u,v \in \bbR$ and $u-v \in \bbZ$, if it has the following transformation law under $\mathrm{SL}(2,\bbZ)$,
\begin{align}
    f(\lambda \tau, \lambda \bar{\tau})
    =
    (\gamma \tau + \delta)^u \,
    (\gamma \bar\tau + \delta)^v \,
    f(\tau, \bar{\tau})
\end{align}
and may be referred to as a modular $(u,v)$-form. If $u=v=0$, then $f$ is modular invariant and called a modular function. If $f$ is holomorphic in $\tau$ then we must have $v=0$ and $u \in \bbZ$, and $f$ is referred to as a holomorphic modular form. Similarly when $f$ is anti-holomorphic, we must have $u=0$ and $v \in \bbZ$, and $f$ is referred to as an anti-holomorphic modular form. In all other cases, $f$ is referred to as non-holomorphic modular forms.

\sm

The variable $\tau_2$ is itself a modular form of weight $(-1,-1)$. Multiplication by $\tau_2^c$ provides a canonical equivalence relation  between modular forms of weight $(u,v)$ and modular forms of weight $(u-c,v-c)$. The equivalence class of modular forms of vanishing weight $(u,v) \approx (0,0)$ is referred to as the class of non-holomorphic modular functions.

\sm

In view of the transformation law of the differential, \smash{$d(\lambda \tau) = (\gamma \tau + \delta)^{-2} \, d\tau$}, there is a one-to-one correspondence between functions of modular weight $(u,v)$ and modular-invariant differential forms on $\cH$ given by \smash{$f(\tau, \bar{\tau}) \, (d\tau)^{u/2} \, (d \bar{\tau})^{v/2}$}. In particular, the upper half-plane supports the \smash{$\mathrm{SL}(2,\bbZ) \subset \mathrm{SL}(2,\bbR)$}-invariant Poincar\'e metric whose volume form is given by,
\begin{align}
    \frac{d^2\tau}{\tau_2^2}
    &=
    \frac{i d\tau \wedge d\bar\tau}{2 \tau_2^2}
    &
    \int_\cM \frac{d^2\tau}{\tau_2^2}
    &=
    \frac{\pi}{3}
\end{align}
and thereby provides a well-defined volume form on $\cM$ to integrate modular functions. 

\subsection{Modular graph forms}

Modular graph forms are modular forms associated to a decorated graph $(\Gamma, A, B)$ with $V$~vertices and $R$~edges. The $V \times R$ connectivity matrix $\Gamma$ has components~$\Gamma_{vr}$, where the index $v = 1, \dots, V$ labels the vertices and $r = 1, \dots, R$ labels the edges. No edge is allowed to begin and end on the same vertex. When edge $r$ contains vertex $v$ we have $\Gamma_{vr} = \pm 1$ while otherwise $\Gamma_{vr} = 0$. The decoration $(A, B)$ consists of two $R$-dimensional arrays,
\begin{align}
    A
    &=
    [ a_1 , \dots , a_R ]
    &
    B
    &=
    [ b_1 , \dots , b_R ]
\end{align}
whose entries satisfy $a_r, b_r \in \bbC$ with $a_r - b_r \in \bbZ$. The pair $(a_r, b_r)$ is associated with the edge~$r$. We refer to $a_r$ and $b_r$ as the holomorphic and anti-holomorphic exponents, respectively, and define the total exponents \smash{$a = \sum_{r=1}^R a_r$} and \smash{$b = \sum_{r=1}^R b_r$}.

\sm

To the decorated graph $(\Gamma,A,B)$ we associate a complex-valued function of $\tau \in \cH$ called a modular graph form, defined by the following Kronecker-Eisenstein series,
\begin{align}
\label{eq:MGFdef}
    \cC_\Gamma
    \! \begin{bmatrix}
    A \\ B
    \end{bmatrix}
    (\tau, \bar{\tau})
    =
    \( \frac{\tau_2}{\pi} \)
        ^{(a+b)/2}
    \!\!\!
    \sum_{p_1, \dots, p_R \in \Lambda'}
    \,\,
    \prod_{r=1}^R
    \frac{1}{(p_r)^{a_r}
        (\bar p_r)^{b_r}}
    \,\,
    \prod_{v=1}^V
    \delta
    \bigg(
    \sum_{s=1}^R
    \Gamma_{vs} \, p_s
    \bigg)
\end{align}
whenever the sum is absolutely convergent. Throughout the sequel of this paper, we shall often suppress the dependence on $\tau$ and $\bar{\tau}$ when no confusion is expected to arise.

\sm

The variables $p_r$ are summed over the lattice $\Lambda' = \Lambda \setminus \{0\}$ where $\Lambda = \bbZ\tau + \bbZ$, corresponding to the momenta on a torus with periods $\tau$ and $1$. We shall often use the parametrization $p_r = m_r \tau + n_r$ and $\bar p_r = m_r \bar \tau +n_r$ with $(m_r,n_r) \in \bbZ^2 \setminus \{(0,0)\}$. The Kronecker delta symbols enforce conservation of momentum at each vertex since $\delta( x = 0 ) = 1$ and $\delta( x \neq 0 ) = 0$. The number of loops in the graph, $R - V + 1$, is equal to the number of independent momenta.

\sm

The Kronecker-Eisenstein series~\eqref{eq:MGFdef} is absolutely convergent if, after solving all the delta symbols, the powers of each loop momentum in the denominator are greater than two. In string theory, the exponents $a_r$ and $b_r$ will always be non-negative integers satisfying $a_r + b_r \geq 2$ for all~$r$ so that absolute convergence is guaranteed. The theory of modular graph forms with non-integer exponents is not well-developed beyond one loop.

\sm

A modular graph form $\cC_\Gamma$ vanishes whenever the graph $\Gamma$ becomes disconnected upon severing a single edge or whenever the integer $a-b$ is odd since the summand is odd and the domain $\Lambda'$ is invariant under the reversal of the signs of all momenta $p_r$. For a connected graph $\Gamma$ which is the union of two graphs $\Gamma = \Gamma_1 \cup \, \Gamma_2$ whose intersection $\Gamma_1 \cap \, \Gamma_2$ consists of a single vertex, the modular graph form factorizes as $\cC_\Gamma = \cC_{\Gamma_1} \! \times \cC_{\Gamma_2}$ with the corresponding partition of exponents.

\subsubsection{Feynman rules for modular graph forms}

Modular graph forms are associated with vacuum Feynman graphs of a conformal scalar field on a torus. A decorated edge $r$ with exponents $(a_r$, $b_r)$ and momentum $p_r$ is drawn as follows and contributes the following factor to the Kronecker-Eisenstein summand,
\begin{align}
\label{eq:MGFedge}
    \begin{tikzpicture}[baseline=-0.5ex, scale=.8]
    \node[label={left:$\cdots$}] (a) at (-3,0) {};
    \node[label={above:$p_{r} \rightarrow$}] at (-1.8,0) {};
    \node[style=draw, fill=white] (b) at (0,0) 
        {$a_r, b_r$};
    \node[label={right:$\cdots$}] (c) at (3,0) {};
    \draw (a) to (b) to (c);
    \end{tikzpicture}
    ~=~
    \( \frac{\tau_2}{\pi} \)
        ^{(a_r+b_r)/2}
    \frac{1}{(p_r)^{a_r}
        (\bar p_r)^{b_r}}
\end{align}
Each vertex $v$ contributes a momentum-conserving factor \smash{$\delta ( \sum_{s=1}^R \Gamma_{vs} \, p_s )$} to the summand. These Feynman rules, along with the instructions to sum each momentum over the lattice~$\Lambda'$ and to connect edges to vertices according to the connectivity matrix $\Gamma$, reproduce the definition~\eqref{eq:MGFdef}.

\subsubsection{Modular transformations}

The modular graph form $\cC_\Gamma$ owes part of its designation to the fact that it transforms as a modular form under the modular group acting on $\tau$ with the arrays $A$ and $B$ invariant. Under a modular transformation
$
    \lambda
    =
    \( \begin{smallmatrix}
    \alpha & \beta \\
    \gamma & \delta
    \end{smallmatrix} \)
    \in
    \mathrm{SL}(2,\bbZ)
$,
\begin{align}
    \cC_\Gamma
    \! \begin{bmatrix}
    A \\ B
    \end{bmatrix}
    ( \lambda  \tau, \lambda \bar{\tau})
    =
    \( \frac{\gamma \tau + \delta}
        {\gamma \bar \tau + \delta} \)
        ^{(a-b)/2}
    \cC_\Gamma
    \! \begin{bmatrix}
    A \\ B
    \end{bmatrix}
    (\tau, \bar{\tau})
\end{align}
The modular weight $(\frac{a-b}{2}, \frac{b-a}{2})$ of any non-vanishing $\cC_\Gamma$ has integer entries in view of the fact that the integer $a-b$ must be even.

\sm

For the special case where $a=b$, the modular graph form $\cC_\Gamma$ is modular invariant and referred to as a modular graph function (MGF) with weight $w=a=b$. The weight of a MGF is not to be confused with its modular weight which necessarily vanishes.

\sm

The choice made for the exponent of the $\tau_2$ prefactor in the definition~\eqref{eq:MGFdef} of $\cC_\Gamma$ ensures that the modular weight vanishes for MGFs, for which $a=b$. However, when $a \neq b$ there is no canonical normalization. Two alternative normalizations \smash{$\cC_\Gamma^\pm$} are,
\begin{align}
\label{eq:C+-}
    \cC_\Gamma^\pm
    \! \begin{bmatrix}
    A \\ B
    \end{bmatrix}
    =
    (\tau_2)^{\pm(a-b)/2}
    \,
    \cC_\Gamma
    \! \begin{bmatrix}
    A \\ B
    \end{bmatrix}
\end{align}
The normalizations \smash{$\cC_\Gamma^+$} and \smash{$\cC_\Gamma^-$} have modular weights $(0, b-a)$ and $(a-b, 0)$, respectively. For MGFs, $a=b$ so that \smash{$\cC_\Gamma^{\mathstrut} = \cC_\Gamma^+ = \cC_\Gamma^-$}.

\subsubsection{Dihedral modular graph forms}

One-loop decorated graphs have only bivalent vertices. Connected modular graph forms with two or more loops may be distinguished by the number of non-bivalent vertices in their corresponding graphs. Following the nomenclature introduced in~\cite{DHoker:2016mwo}, this distinction is into dihedral graphs, trihedral graphs, tetrahedral graphs, and so on. All connected two-loop graphs are dihedral, but connected three-loop graphs come in dihedral, trihedral, and tetrahedral varieties. We shall also consider one-loop graphs as dihedral. 

\sm

In this paper, we shall study only dihedral modular graph forms. A generic dihedral modular graph form with $R \geq 2$ edges (and $R-1$ loops) has the following decorated graph, matrix of exponents, and Kronecker-Eisenstein series representation,
\begin{align}
\label{eq:dihedral}
    \begin{tikzpicture}[baseline=-0.5ex, scale=.8, yscale=.8, xscale=.9]
    \draw (-1,0) node[circle,fill,inner sep=1.5](a){};
    \draw (1,0) node[circle,fill,inner sep=1.5](b){};
    \draw (a) to [out=-30,in=-150]
        node[midway,yshift=3,fill=white,scale=1] 
        {$\vdots$} (b);
    \draw (-1,0) arc(180:0:1)
        node[style=draw,midway,fill=white,scale=.5]
        {$a_1,b_1$};
    \draw (a) to [out=30,in=150]
        node[style=draw,midway,fill=white,scale=.5] 
        {$a_2,b_2$} (b);
    \draw (-1,0) arc(-180:0:1)
        node[style=draw,midway,fill=white,scale=.5]
        {$a_R,b_R$};
    \end{tikzpicture}
    ~=~
    \cC
    \! \begin{bmatrix}
    a_1 & a_2 & \cdots & a_R \\
    b_1 & b_2 & \cdots & b_R
    \end{bmatrix}
    =
    \( \frac{\tau_2}{\pi} \)^{(a+b)/2}
    \!\!\!
    \sum_{\substack{p_r \in \Lambda' \\
        r = 1,\dots,R}}
    \frac{ \delta(p_1+\dots+p_R) }
         { p_1^{a_1} \bar p_1 ^{b_1} \cdots p_R^{a_R} \bar p_R ^{b_R} }
\end{align}
This Kronecker-Eisenstein series is absolutely convergent if $ \Re( a_r + b_r + a_s + b_s ) > 2$ for all pairs of $r,s = 1, 2, \dots, R$ with $r \neq s$.

\subsubsection{Poincar\'e series}

Modular graph functions may be written as Poincar\'e series with respect to $\Gamma_\infty \backslash \mathrm{PSL}(2,\bbZ)$, the coset of the modular group by the Borel subgroup of translations $\Gamma_\infty$, defined in~\eqref{eq:Borel}. A Poincar\'e series representation for the modular graph function $\cC_\Gamma$ is written as,
\begin{align}
\label{eq:Poincare}
    \cC_\Gamma
    \! \begin{bmatrix}
    A \\ B
    \end{bmatrix}
    (\tau, \bar{\tau})
    =
    \sum_{\lambda \in \Gamma_\infty \backslash \mathrm{PSL}(2,\bbZ)}
    \Lambda_\Gamma
    \! \begin{bmatrix}
    A \\ B
    \end{bmatrix}
    ( \lambda  \tau, \lambda \bar{\tau})
\end{align}
where $\Lambda_\Gamma$ is referred to as a Poincar\'e seed function for $\cC_\Gamma$. The Poincar\'e seed is not unique, and clever choices may simplify subsequent calculations. 

\sm

It was shown in~\cite{DHoker:2019txf} that a Poincar\'e seed for a generic dihedral modular graph function may be constructed by replacing the sum over any one lattice momentum $p_r$ in~\eqref{eq:dihedral} by a sum over $p_r = N \in \bbZ \setminus \{0\}$. Thus, a Poincar\'e seed for a generic dihedral modular graph function with weight $w=a=b$, is given by the following Kronecker-Eisenstein series,
\begin{align}
    \Lambda
    \! \begin{bmatrix}
    w-a' & a_2 & \cdots & a_R \\
    w-b' & b_2 & \cdots & b_R
    \end{bmatrix}
    =
    \( \frac{\tau_2}{\pi} \)^w
    \sum_{N \neq 0}
    \sum_{\substack{p_r \in \Lambda' \\
        r = 2,\dots,R}}
    \frac{ N^{ a'+b' } }{ |N|^{2w} }
    \frac{ \delta( N+p_2+\dots+p_{R} ) }
         { p_2^{a_2} \bar{p}_2^{b_2}
           \cdots
           p_{R}^{a_{R}} \bar{p}_{R}^{b_{R}} }
\end{align}
where $a' = \sum_{r=2}^R a_r$ and $b' = \sum_{r=2}^R b_r$. This Kronecker-Eisenstein sum is absolutely convergent for sufficiently large $\Re(w)$ and may be defined by analytic continuation in the weight~$w$ elsewhere, while keeping integer exponents $a_r, b_r \in \bbZ$ for $r = 2, \dots R$. The two-loop case of this analytic continuation was studied in~\cite{DHoker:2019txf}.

\subsubsection{Asymptotic expansions}

The asymptotics at the cusp $\tau = i \infty$ of a modular graph function with integer exponents and weight $w \geq 2$ are given by a Laurent polynomial in $\tau_2$ of degree $(w, 1-w)$ plus exponentially suppressed terms~\cite{DHoker:2015wxz},
\begin{align}
\label{eq:MGFAsy}
    \cC_\Gamma
    \! \begin{bmatrix}
    A \\ B
    \end{bmatrix}
    =
    \sum_{\ell = 1-w}^w
    \mfc_\cC^{(\ell)} \,
    \tau_2^\ell
    + \cO(e^{-2\pi\tau_2})
\end{align}
where the Laurent coefficients \smash{$\mfc_\cC^{(\ell)}$} are constants with transcendental weight $w$. It was shown in~\cite{DHoker:2015wxz} that \smash{$\pi^{-w} \, \mfc_\cC^{(w)}$} is a rational number and that each \smash{$\pi^{-\ell} \, \mfc_\cC^{(\ell)}$} for $1-w \leq \ell \leq w-1$ is a linear combination of single-valued multiple zeta-values with rational coefficients. Modular graph functions or forms with non-integer exponents do not have a Laurent polynomial in~$\tau_2$ at the cusp.

\sm

More generally, modular graph functions with integer exponents have an asymptotic expansion in powers of the nome $q = e^{2\pi i \tau}$ and its complex conjugate $\bar q$,
\begin{align}
\label{eq:MGFqexp}
    \cC_\Gamma
    \! \begin{bmatrix}
    A \\ B
    \end{bmatrix}
    =
    \sum_{M,N \geq 0}
    \mfc_\cC^{(M,N)}(\tau_2) \,
    q^M \bar q^N
\end{align}
where \smash{$\mfc_\cC^{(M,N)}(\tau_2)$} are Laurent polynomials in $\tau_2$. The expansion~\eqref{eq:MGFqexp} contains the Fourier expansion of $\cC_\Gamma$ in the variable $\tau_1$. Terms with $M=N$ constitute the constant Fourier mode which consists of the Laurent polynomial in $\tau_2$ (from $M = N = 0$) plus infinitely many exponential terms (from ${M = N > 0}$) that are suppressed at the cusp.

\subsection{Identities between modular graph forms}

For a systematic discussion of the identities obeyed by modular graph forms, we refer the reader to~\cite{DHoker:2016quv}. Here we shall summarize the essential identities needed for this paper.

\subsubsection{Algebraic identities}

\paragraph{Momentum conservation:}

Modular graph forms obey the following momentum conservation identity at each vertex $v = 1, \dots, V$,
\begin{align}
\label{eq:momcon}
    \sum_{r=1}^R
    \Gamma_{vr}
    \,
    \cC_\Gamma
    \! \begin{bmatrix}
    A - S_r \\ B
    \end{bmatrix}
    &=
    \sum_{r=1}^R
    \Gamma_{vr}
    \,
    \cC_\Gamma
    \! \begin{bmatrix}
    A \\ B - S_r
    \end{bmatrix}
    =
    0
    &
    S_r
    &=
    [\,
    \underbrace{0,\dots,0}_{r-1},
    1,
    \underbrace{0,\dots,0}_{R-r}
    \,]
\end{align}
where the $R$-dimensional row-vector $S_r$ has zeroes in all slots except for the $r^\text{th}$ which instead equals one. These momentum conservation identities yield linear algebraic relations between modular graph forms with the same modular weight and loop order.

\paragraph{Factorization:}

Modular graph forms with a vanishing pair of holomorphic and anti-holomorphic exponents $a_r=b_r=0$ obey factorization identities. In particular, dihedral graphs with $R \geq 3$ edges obey,
\begin{align}
\label{eq:MGFfact}
   \cC
    \! \begin{bmatrix}
    a_1 & \cdots & a_{R-1} & 0 \\
    b_1 & \cdots & b_{R-1} & 0
    \end{bmatrix}
    &=
    \prod_{r=1}^{R-1}
    \cC
    \! \begin{bmatrix}
    a_r & 0 \\
    b_r & 0
    \end{bmatrix}
    -
    \cC
    \! \begin{bmatrix}
    a_1 & \cdots & a_{R-1} \\
    b_1 & \cdots & b_{R-1}
    \end{bmatrix}
\end{align}
These factorization identities yield non-linear algebraic relations between modular graph forms with the same modular weight but different loop order.

\paragraph{Holomorphic subgraph reduction:}

Modular graph forms with two vanishing holomorphic or anti-holomorphic exponents may be simplified into modular graph forms with fewer loops using a procedure called holomorphic subgraph reduction~\cite{DHoker:2016mwo, DHoker:2016quv, Gerken:2018zcy}.

\sm

These simpler functions include the holomorphic and anti-holomorphic Eisenstein series~$G_\ell$ and $\bar{G}_\ell$, which are defined by the following Kronecker-Eisenstein series,
\begin{align}
\label{eq:Gdef}
    G_\ell(\tau)
    &=
    \frac{1}{\pi^{\ell/2}}
    \sum_{p \in \Lambda '}
    \frac{ 1 }{p^\ell}
    &
    \bar{G}_\ell(\bar{\tau})
    &=
    \frac{1}{\pi^{\ell/2}}
    \sum_{p \in \Lambda '}
    \frac{ 1 }{\bar p ^\ell}
\end{align}
which are absolutely convergent for integer $\ell > 2$ and vanish for odd $\ell \geq 3$. For $\ell=2$, the series~\eqref{eq:Gdef} are conditionally convergent. Siegel's prescription defines a non-holomorphic but modular-covariant regularization \smash{$\hat{G}_2$} as follows,
\begin{align}
    \hat G_2(\tau, \bar{\tau})
    &=
    \frac{1}{\pi} \,
    \lim_{s \to 0} \,
    \sum_{p \in \Lambda' }
    \frac{1}{ p^2 |p|^s }
\end{align}
Alternatively, one may preserve holomorphicity at the cost of modular invariance by defining the function $G_2(\tau) = -4\pi i \, \p_\tau \ln \eta (\tau)$ where $\eta (\tau)$ is the Dedekind $\eta$ function.

\sm

For dihedral graphs, the holomorphic subgraph reduction procedure is expressed by the following equation~\cite{Gerken:2018zcy},
\begin{align}
\label{eq:HSR}
    \cC
    \! \begin{bmatrix}
    a_+ & a_- & A \\ 0 & 0 & B
    \end{bmatrix}
    &=
    (-)^{a_-}_{\phantom{2}} \,
    \tau_2^{a_0/2} \,
    G_{a_0} \,
    \cC
    \! \begin{bmatrix}
    A \\  B
    \end{bmatrix}
    - \binom{a_0}{a_+} \,
    \cC
    \! \begin{bmatrix}
    a_0 & A \\ 0 & B
    \end{bmatrix}
    \no \\[1ex]
    & \quad
    +
    \sum_{\ell=4}^{\max(a_\pm)}
    \bigg\{
    \binom{a_0-1-\ell}{a_+-\ell}
    +
    \binom{a_0-1-\ell}{a_--\ell}
    \bigg\}
    \,
    \tau_2^{\ell/2} \,
    G_\ell \,
    \cC
    \! \begin{bmatrix}
    a_0-\ell & A \\ 0 & B
    \end{bmatrix}
    \no \\[1ex]
    & \quad
    + \binom{a_0-2}{a_+-1}
    \left\{
    \tau_2 \,
    \hat{G}_2 \, 
    \cC
    \! \begin{bmatrix}
    a_0-2 & A \\ 0 & B
    \end{bmatrix}
    +
    \cC
    \! \begin{bmatrix}
    a_0-1 & A \\ -1 & B
    \end{bmatrix}
    \,
    \right\}
\end{align}
where $A$ and $B$ are row vectors of exponents and $a_0 = a_+ + a_-$. When $A$ and $B$ are one-dimensional, the first term on the right-hand side of~\eqref{eq:HSR} is absent. In combinations appearing from the application of derivatives to modular graph forms, the two terms on the last line with abnormal modular weight always cancel. The reduction formula for modular graph forms with two vanishing anti-holomorphic exponents is given by the complex conjugate of~\eqref{eq:HSR}.

\subsubsection{Differential identities}

Modular graph functions and forms also obey a number of identities involving the differential operators $\nabla$, $\bar{\nabla}$, and $\Delta$.

\sm

The first-order Maass operators $\nabla$ and $\bar \nabla$ map the space of modular graph forms into itself. The Christoffel connection in the covariant derivatives vanishes in $\nabla$ when acting on a modular graph ${(0,b-a)}$-form $\cC^+_\Gamma$ and in $\bar \nabla$ when acting on a modular graph $(a-b,0)$-form~$\cC^-_\Gamma$, giving the following simple representations, 
\begin{align}
\label{eq:nabla}
    &
    \text{on a $(0,b-a)$-form}
    \quad
    \cC^+_\Gamma
    &
    \nabla
    &=
    +2 i \tau_2^2 \p_\tau
    =
    \tau_2^2 (\p_{\tau_2} + i \p_{\tau_1}) 
\no \\
    &
    \text{on a $(a-b,0)$-form}
    \quad
    \cC^-_\Gamma
    &
    \bar{\nabla}
    &=
    -2 i \tau_2^2 \p_{\bar{\tau}}
    =
    \tau_2^2 (\p_{\tau_2} - i \p_{\tau_1}) 
\end{align}
The action of these operators results in an algebraic action on the exponents,
\begin{align}
    \label{eq:nablaMGF}
    \nabla
    \cC_\Gamma^+
    \! \begin{bmatrix}
    A \\ B
    \end{bmatrix}
    &=
    \sum_{r=1}^R
    a_r \,
    \cC_\Gamma ^+
    \! \begin{bmatrix}
    A + S_r \\ B - S_r
    \end{bmatrix}
    &
    \bar \nabla
    \cC_\Gamma^-
    \! \begin{bmatrix}
    A \\ B
    \end{bmatrix}
    &=
    \sum_{r=1}^R
    b_r \,
    \cC_\Gamma^-
    \! \begin{bmatrix}
    A - S_r \\ B + S_r
    \end{bmatrix}
\end{align}
with $S_r$ defined in~\eqref{eq:momcon}. Thus, $\nabla$ maps modular graph $(0, b-a)$-forms to $(0, b-a-2)$-forms while $\bar{\nabla}$ maps modular graph $(a-b,0)$-forms to $(a-b-2,0)$-forms. 

\sm

The second-order Laplace-Beltrami operator $\Delta$ on the upper half-plane $\cH$ maps the space of modular graph functions, for which $a=b$, into itself. We may write several equivalent expressions for $\Delta$, 
\begin{align}
\label{eq:Delta}
    \Delta
    =
    \nabla \, \tau_2^{-2} \, \bar{\nabla}
    =
    \bar{\nabla} \, \tau_2^{-2} \, \nabla
    =
    4 \tau_2^2 \p_{\bar\tau} \p_\tau
    =
    \tau_2^2 ( \p_{\tau_1}^2 + \p_{\tau_2}^2 )
\end{align}
The action of $\Delta$ on a MGF with weight $w = a = b$ may be expressed in terms of an algebraic action on the exponents,
\begin{align}
\label{eq:DMGF}
    (\Delta + w) \,
    \cC_\Gamma
    \! \begin{bmatrix}
    A \\ B
    \end{bmatrix}
    =
    \sum_{r,s=1}^R
    a_r b_s \,
    \cC_\Gamma
    \! \begin{bmatrix}
     A + S_r - S_s \\ B - S_r + S_s
    \end{bmatrix}
\end{align}
with $S_r$ defined in~\eqref{eq:momcon}

\subsection{One-loop modular graph forms}
\label{sec:1loopMGF}

We may completely characterize the space of one-loop modular graph forms. Decorated one-loop graphs contain only bivalent vertices and are specified by a single holomorphic exponent~$a$ and a single anti-holomorphic exponent~$b$ since,
\begin{align}
    \cC
    \! \begin{bmatrix}
    a & 0 \\ b & 0
    \end{bmatrix}
    =
    (-)^{a_2 - b_2} \,
    \cC
    \! \begin{bmatrix}
    a_1 & a_2 \\ b_1 & b_2
    \end{bmatrix}
\end{align}
for $a = a_1 + a_2$ and $b = b_1 + b_2$. All one-loop modular graph forms, including those with complex non-integer exponents, may be written in terms of various Eisenstein series. 

\subsubsection{Non-holomorphic Eisenstein series}

The unique one-loop modular graph function with weight $w = a = b$ is the non-holomorphic (also called real-analytic) Eisenstein series $E_w$. For $\Re(w)>1$, $E_w$ has the following decorated graph, matrix of exponents, and Kronecker-Eisenstein series representation,
\begin{align}
\label{eq:Edef}
    E_w
    ~=~
    \begin{tikzpicture}[baseline=-0.5ex, scale=.7]
    \draw (-1,0) node[circle,fill,inner sep=1.5](a){};
    \draw (-1,0) arc(180:-180:1)
        node[style=draw,midway,fill=white,scale=.8]
        {$w,w$};
    \end{tikzpicture}
    ~=~
    \cC
    \! \begin{bmatrix}
    w & 0 \\ w & 0
    \end{bmatrix}
    =~
    \frac{\tau_2^w}{\pi^w}
    \sum_{p \in \Lambda '}
    \frac{1}{|p|^{2w}}
\end{align}
The Eisenstein series is an eigenfunction of the Laplace-Beltrami operator,
\begin{align}
\label{eq:DE}
    \Delta E_w
    =
    w(w-1) E_w
\end{align}
In fact, $E_w$ is the unique modular-invariant solution to this differential equation with polynomial growth at the cusp.

\sm

Although the Eisenstein series $E_w$ arises naturally in string theory, it will be convenient to work with the starred Eisenstein series, an alternative normalization defined as follows,
\begin{align}
\label{eq:Estar}
    E^*_w
    &=
    \half \Gamma(w) \, E_w
\end{align}
The starred Eisenstein series obeys the functional relation $E^*_{1-w} = E^*_{w}$ analogous to the functional relation $\zeta^*(1-w) = \zeta^*(w)$ obeyed by the starred zeta function,
\begin{align}
    \zeta^*(w) = \pi^{-w/2} \, \Gamma(w/2) \, \zeta(w)    
\end{align}
Other properties of the starred zeta function are discussed in \autoref{apdx:zeta}. The asymptotic expansion of~$E_w^*$ manifestly exhibits its functional relation,
\begin{align}
\label{eq:EAsy}
    E^*_w
    &=
    \zeta^*(2w) \, \tau_2^w
    + \zeta^*(2w-1) \, \tau_2^{1-w}
    + \cO(e^{-2\pi\tau_2})
\end{align}
A Poincar\'e series for $E_w^*$ is given by,
\begin{align}
\label{eq:Epoin}
    E_w^*(\tau, \bar \tau)
    =
    \zeta^*(2w)
    \sum_{\lambda \in \Gamma_\infty \backslash \mathrm{PSL}(2,\bbZ)}
    \big[ \Im (\lambda \tau) \big]^w
\end{align}
so that $\zeta^*(2w) \, \tau_2^w$ is a Poincar\'e seed for $E_w^*$.

\sm

The derivatives of $E_w^*$ span the space of all one-loop modular graph forms. Using~\eqref{eq:nablaMGF} to compute the action of $\nabla$ and $\bar{\nabla}$, we find,
\begin{align}
\label{eq:NablaE}
    \cC^+ \! \begin{bmatrix}
    w+\ell & 0 \\
    w-\ell & 0
    \end{bmatrix}
    &=
    2 \, \Gamma(w+\ell)^{-1} \,
    \nabla^\ell E_w^*
\no \\
    \cC^- \! \begin{bmatrix}
    w-\ell & 0 \\
    w+\ell & 0
    \end{bmatrix}
    &= 
    2 \, \Gamma(w+\ell)^{-1} \,
    \bar{\nabla}^\ell E_w^*
\end{align}
for $\Re(w) > 1$ and integer $\ell \geq 0$. 

\subsubsection{Holomorphic and anti-holomorphic Eisenstein series}

The derivatives of $E_w^*$ include the holomorphic and anti-holomorphic Eisenstein series $G_\ell$ and $\bar G_\ell$ defined in~\eqref{eq:Gdef}. Our normalization is standard in the literature on modular graph forms and is chosen so that,
\begin{align}
    \tau_2^{2\ell} \, 
    G_{2\ell}
    &=
    \cC^+
    \! \begin{bmatrix}
    2\ell & 0 \\
    0 & 0
    \end{bmatrix}
    =
    2 \, \Gamma(2\ell)^{-1} \,
    \nabla^\ell E_\ell^*
 \no \\
    \tau_2^{2\ell} \, 
    \bar{G}_{2\ell}
    &=
    \cC^-
    \! \begin{bmatrix}
    0 & 0 \\
    2\ell & 0
    \end{bmatrix}
    =
    2 \, \Gamma(2\ell)^{-1} \,
    \bar{\nabla}^\ell E_\ell^*
\end{align}
The Fourier series of $G_{2\ell}$ for integer $\ell \geq 2$ is given by,
\begin{align}
\label{eq:GAsy}
    G_{2\ell}
    &=
    2 \, 
    \frac{ \zeta(2\ell) }{ \pi^\ell}
    +
    2 \, 
    \frac{ (- 4 \pi)^\ell }{ \Gamma(2\ell) }
    \sum_{n=1}^\infty
    \sigma_{2\ell-1}(n) \, q^n
\end{align}
where $q = e^{2\pi i \tau}$ and \smash{$\sigma_p(n) = \sum_{d | n} d^p$} is the sum of $p^\text{th}$ powers of the positive divisors of~$n$. 

\subsection{Two-loop modular graph functions}
\label{sec:2loopMGF}

The space of two-loop modular graph forms is much more complicated than the space at one loop. Two-loop graphs may be connected or disconnected. Disconnected two-loop graphs factorize into one-loop graphs, and connected two-loop graphs are always dihedral. To simplify our discussion, we shall restrict to two-loop modular graph functions, and for the connected two-loop functions, we shall restrict to non-negative integer exponents.

\subsubsection{The disconnected functions $\cV_{s,t}^{(n)}$}

Disconnected two-loop modular graph functions are equal to the product of two one-loop modular graph forms whose total modular weight vanishes. For these disconnected functions we shall consider complex non-integer exponents.

\sm

We define the infinite family of functions \smash{$\cV_{s,t}^{(n)}$} with $\Re(s), \Re(t) > 1$ and integer $n \geq 0$ by the following decorated graph and matrix of exponents,
\begin{align}
\label{eq:V}
    \cV_{s,t}^{(n)}
    ~&=~
    \begin{tikzpicture}[baseline=-0.5ex, scale=.7]
    \draw (-1,0) node[circle,fill,inner sep=1.5](a){};
    \draw (-1,0) arc(180:-180:1)
        node[style=draw,midway,fill=white,scale=.6]
        {$s+n,s-n$};
    \end{tikzpicture}
    ~~
    \begin{tikzpicture}[baseline=-0.5ex, scale=.7]
    \draw (-1,0) node[circle,fill,inner sep=1.5](a){};
    \draw (-1,0) arc(180:-180:1)
        node[style=draw,midway,fill=white,scale=.6]
        {$t-n,t+n$};
    \end{tikzpicture}
    ~=~
    \cC
    \! \begin{bmatrix}
    s+n & 0 \\ s-n & 0
    \end{bmatrix}
    \cC
    \! \begin{bmatrix}
    t-n & 0 \\ t+n & 0
    \end{bmatrix}
\end{align}
These functions have weight $s+t$ and may be written in terms of derivatives acting on two Eisenstein series as follows,
\begin{align}
\label{eq:VE}
    \cV_{s,t}^{(n)}
    &=
    4 \,
    \Gamma(s+n)^{-1} \,
    \Gamma(t+n)^{-1} \,
    \tau_2^{-2n} \,
    \nabla^n E_s^* \,
    \bar{\nabla}^n E_t^* 
\end{align}
Many properties of these functions, such as their asymptotic expansion or Poincar\'e series, follow from the respective properties of the Eisenstein series.
 
\sm
 
This infinite family of functions spans the space of all disconnected two-loop MGFs. For instance, the case $n = 0$ reduces to the double product \smash{$\cV_{s,t}^{(0)} = 4 \, \Gamma(s)^{-1} \, \Gamma(t)^{-1} \, E_s^* E_t^*$} of two non-holomorphic Eisenstein series while the case $n=s=t \geq 2$ reduces to the product~\smash{${\cV_{n,n}^{(n)} = \tau_2^{2n} \, G_{2n} \, \bar{G}_{2n}}$} of a holomorphic and an anti-holomorphic Eisenstein series.

\subsubsection{The connected functions $C_{a,b,c}$}

We shall now consider connected two-loop modular graph functions with non-negative integer exponents. To simplify things further, we shall first consider the case where each pair of holomorphic and anti-holomorphic exponents is equal.

\sm

The infinite family of connected two-loop modular graph functions $C_{a,b,c}$ was introduced in~\cite{DHoker:2015gmr}. These functions have the following decorated dihedral graph, matrix of exponents, and Kronecker-Eisenstein series representation,
\begin{align}
\label{eq:Cabc}
    C_{a,b,c}
    ~&=~
    \begin{tikzpicture}[baseline=-0.5ex, scale=.8, yscale=.8, xscale=.9]
    \draw (-1,0) node[circle,fill,inner sep=1.5](a){};
    \draw (1,0) node[circle,fill,inner sep=1.5](b){};
    \draw (-1,0) arc(180:0:1)
        node[style=draw,midway,fill=white,scale=.5]
        {$a,a$};
    \draw (-1,0) arc(-180:0:1)
        node[style=draw,midway,fill=white,scale=.5]
        {$b,b$};
    \draw (a) to [out=0,in=180] 
        node[style=draw,midway,fill=white,scale=.5] 
        {$c,c$} (b);
    \end{tikzpicture}
    ~=~
    \cC
    \! \begin{bmatrix}
    a & b & c \\
    a & b & c
    \end{bmatrix}
    =
    \frac{\tau_2^{a+b+c}}{\pi^{a+b+c}}
    \sum_{p_1, p_2, p_3 \in \Lambda '}
    \frac{ \delta(p_1+p_2+p_3)}
         { |p_1|^{2a} \, |p_2|^{2b} \, |p_3|^{2c} }
\end{align}
The functions $C_{a,b,c}$ have weight $a + b + c$ and are invariant under permutations of $a,b,c$. For integer $a, b, c \geq 1$, the Kronecker-Eisenstein series is absolutely convergent and $C_{a,b,c}$ is real. The Laurent polynomial of $C_{a,b,c}$ for integer $a,b,c$ was calculated in~\cite{DHoker:2017zhq} and is reviewed in \autoref{apdx:Laurent}.

\sm

The space of functions $C_{a,b,c}$ obeys a system of inhomogeneous Laplace eigenvalue equations. The action of the Laplace-Beltrami operator $\Delta$ on $C_{a,b,c}$ may be computed using~\eqref{eq:DMGF} and is given by,
\begin{align}
\label{eq:DCabc}
    \Delta C_{a,b,c}
    & =
    \big( a(a-1) + b(b-1) + c(c-1) \big)
    C_{a,b,c}
    \no \\
    & \quad
    + ab \,
    \big( 
    C_{a-1,b+1,c} + C_{a+1,b-1,c} + C_{a+1,b+1,c-2} 
    - 2 C_{a,b+1,c-1} - 2 C_{a+1,b,c-1} 
    \big)
    \no \\
    & \quad
    + bc \,
    \big( 
    C_{a,b-1,c+1} + C_{a,b+1,c-1} + C_{a-2,b+1,c+1}
    - 2 C_{a-1,b,c+1} - 2 C_{a-1,b+1,c} 
    \big)
    \no \\
    & \quad
    + ca \,
    \big( 
    C_{a+1,b,c-1} + C_{a-1,b,c+1} + C_{a+1,b-2,c+1}
    - 2 C_{a+1,b-1,c} - 2 C_{a,b-1,c+1} 
    \big)
\end{align}
When the right-hand side of this equation involves a lower index $a'$, $b'$, or $c'$ which equals either~$0$ or~$-1$, the corresponding MGFs reduce to Eisenstein series as follows,
\begin{align}
\label{eq:inhom}
    C_{w-\ell,\ell,0}
    &=
    E_{\ell} E_{w-\ell} - E_{w}
    &
    C_{w+1-\ell,\ell,-1}
    &=
    E_{\ell} E_{w-\ell}
    + E_{\ell-1} E_{w-\ell+1} 
\end{align}
The right-hand side of either equation in~\eqref{eq:inhom} may involve the symbol $E_1$, formally corresponding to a divergent series, but its contribution will always cancel out of the right-hand side of~\eqref{eq:DCabc}. Thus, the Laplacian maps the space of $C_{a,b,c}$ with weight~${w = a + b + c}$ and integer ${a, b, c \geq 1}$ into itself (the homogeneous part) plus a linear combination of the Eisenstein series $E_w$ and double products $E_\ell E_{w-\ell}$ (the inhomogeneous part).

\subsubsection{The eigenfunctions \texorpdfstring{$\mfC_{w;m;p}$}{}}

It was shown in~\cite{DHoker:2015gmr} that the action~\eqref{eq:DCabc} of $\Delta$ on the space of functions $C_{a,b,c}$ may be diagonalized, resulting in eigenfunctions $\mfC_{w;m;p}$ which are linear combinations of the functions~$C_{a,b,c}$ with weight $w=a+b+c$ and integer $a,b,c \geq 1$.\footnote{Our notation for the second subscript differs from~\cite{DHoker:2015gmr}. Our $\mfC_{w;m;p} = \mfC_{w;w-2m;p}$ of~\cite{DHoker:2015gmr}.} Each eigenfunction obeys an inhomogeneous Laplace eigenvalue equation of the following form,
\begin{align}
\label{eq:DCwmp}
    \big(
    \Delta - (w-2m)(w-2m-1)
    \big)
    \mfC_{w;m;p}
    =
    \mfh^{(0)}_{w;m;p} \, E_w^*
    +
    \sum_{\ell = 2}^{\floor{w/2}}
    \mfh^{(\ell)}_{w;m;p} \,
    E_\ell^* E_{w-\ell}^*
\end{align}
where the $\mfh$ coefficients are constants and the labels $m$ and $p$ are integers which run over the following ranges,
\begin{align}
    {1 \leq m \leq \floor{\tfrac{w-1}{2}}}
    &&
    {0 \leq p \leq \floor{\tfrac{w-2m-1}{3}}}
\end{align}
The label $m$ specifies the eigenvalue $(w-2m)(w-2m-1)$ while $p$ labels the degeneracy in the spectrum. We have written the inhomogeneous part of the Laplace equation~\eqref{eq:DCwmp} in terms of starred Eisenstein series for later convenience.

\sm

At any fixed weight $w \geq 3$, the number of eigenfunctions $\mfC_{w;m;p}$ is equal to the number of functions $C_{a,b,c}$ modulo permutations of $a,b,c$. In other words,
\begin{align}
    \sum_{m=1}^{\floor{\frac{w-1}{2}}} \!
    \sum_{p=0}^{\floor{\frac{w-2m-1}{3}}}
    =
    \sum_{\substack{ a \geq b \geq c \geq 1 \\
                     a+b+c=w }}
\end{align}
Thus, we may systematically relate the two bases of functions. Without loss of generality, we choose representatives of the $C_{a,b,c}$ with $a \geq b \geq c$. We then expand each basis in terms of the other as follows,
\begin{align}
\label{eq:CabcCwmp}
    C_{a,b,c}
    &=
    \sum_{m=1}^{\floor{\frac{w-1}{2}}} \!
    \sum_{p=0}^{\floor{\frac{w-2m-1}{3}}}
    d_{a,b,c}^{w;m;p} \,
    \mfC_{w;m;p}
    &
    \mfC_{w;m;p}
    &=
    \sum_{\substack{ a \geq b \geq c \geq 1 \\
                     a+b+c=w }}
    \mfd_{w;m;p}^{a,b,c} \,
    C_{a,b,c}
\end{align}
In \autoref{apdx:CabcCwmp}, we derive explicit expressions for the coefficients \smash{$d_{a,b,c}^{w;m;p}$} and \smash{$\mfd_{w;m;p}^{a,b,c}$} as well as for the $\mfh$ coefficients which appear in the inhomogeneous Laplace eigenvalue equation~\eqref{eq:DCwmp}, thereby proving the following proposition. 

\begin{prop}
\label{thm:dhprop}
The coefficients \smash{$d_{a,b,c}^{w;m;p}$}, \smash{$\mfd_{w;m;p}^{a,b,c}$}, and \smash{$\mfh^{(\ell)}_{w;m;p}$} are all rational numbers. Moreover, we have \smash{$\mfh^{(\ell)}_{w;m;p} = 0$} for $2 \leq \ell \leq m$.
\end{prop}

In \autoref{apdx:Laurent}, we compare the Laurent polynomials of $C_{a,b,c}$ and $\mfC_{w;m;p}$. We also write the Laurent coefficients of $\mfC_{w;m;p}$ in terms of the $\mfh$ coefficients.

\sm

Similar inhomogeneous Laplace equations of the form \smash{$(\Delta-s(s-1)) F^{+(s)}_{m,k} = E_m E_k$} were recently studied in~\cite{Dorigoni:2021jfr, Dorigoni:2021ngn}. There it was shown that the solutions to these equations include the modular graph functions $\mfC_{w;m;p}$ as well as modular functions which are not modular graph function. The formulas of \autoref{apdx:CabcCwmp} make it possible to express the modular graph functions~$C_{a,b,c}$ as a linear combination of the modular functions \smash{$F^{+(s)}_{m,k}$}.

\subsubsection{The connected functions $\cC_{u,v;w}$}

We shall now consider the larger space of all connected two-loop modular graph functions with non-negative integer coefficients. A generic connected two-loop MGF has the following decorated graph, matrix of exponents, and Kronecker-Eisenstein series representation,
\begin{align}
\label{eq:C2loop}
    \begin{tikzpicture}[baseline=-0.5ex, scale=.8, yscale=.8, xscale=.9]
    \draw (-1,0) node[circle,fill,inner sep=1.5](a){};
    \draw (1,0) node[circle,fill,inner sep=1.5](b){};
    \draw (-1,0) arc(180:0:1)
        node[style=draw,midway,fill=white,scale=.5]
        {$a_1,b_1$};
    \draw (-1,0) arc(-180:0:1)
        node[style=draw,midway,fill=white,scale=.5]
        {$a_2,b_2$};
    \draw (a) to [out=0,in=180] 
        node[style=draw,midway,fill=white,scale=.5] 
        {$a_3,b_3$} (b);
    \end{tikzpicture}
    ~=~
    \cC
    \! \begin{bmatrix}
    a_1 & a_2 & a_3 \\
    b_1 & b_2 & b_3
    \end{bmatrix}
    =
    \frac{\tau_2^{w}}{\pi^{w}}
    \sum_{p_1, p_2, p_3 \in \Lambda '}
    \frac{ \delta(p_1+p_2+p_3) }
         { p_1^{a_1} \bar{p}_1^{b_1}
           p_2^{a_2} \bar{p}_2^{b_2}
           p_3^{a_3} \bar{p}_3^{b_3} }
\end{align}
where $w = a_1 + a_2 + a_3 = b_1 + b_2 + b_3$. This Kronecker-Eisenstein series is absolutely convergent if $a_r + b_r + a_s + b_s \geq 3$ for all pairs of $r,s = 1, 2, 3$ with $r \neq s$ which implies that~$w \geq 3$. This larger space of two-loop MGFs contains the subspace of functions $C_{a,b,c}$ (as well as their linear combinations, the functions $\mfC_{w;m;p}$).

\sm

The generic connected two-loop MGF~\eqref{eq:C2loop} simplifies when any two of its exponents vanish. When $a_r = b_r = 0$ for some $r$, when $a_r = a_s = 0$ for some $r \neq s$, or when $b_r = b_s = 0$ for some~$r \neq s$, the function may be written in terms of the disconnected two-loop MGFs~\smash{$\cV_{s,t}^{(n)}$} and the Eisenstein series $E_w^*$ using the factorization identity~\eqref{eq:MGFfact} and holomorphic subgraph reduction~\eqref{eq:HSR}. When $a_r = b_s = 0$ for some $r \neq s$ with all other exponents positive, the function belongs to the infinite family of functions $\cC_{u,v;w}$.

\sm

The infinite family of connected two-loop modular graph functions $\cC_{u,v;w}$ was introduced in~\cite{DHoker:2019txf}. These functions have unequal pairs of holomorphic and anti-holomorphic exponents and the following decorated dihedral graph, matrix of exponents, and Kronecker-Eisenstein series representation,
\begin{align}
\label{eq:Cuvw}
    \cC_{u,v;w}
    ~&=~
    \begin{tikzpicture}[baseline=-0.5ex, scale=.8, yscale=.8, xscale=1.1]
    \draw (-1,0) node[circle,fill,inner sep=1.5](a){};
    \draw (1,0) node[circle,fill,inner sep=1.5](b){};
    \draw (-1,0) arc(180:0:1)
        node[style=draw,midway,fill=white,scale=.5]
        {$u,0$};
    \draw (-1,0) arc(-180:0:1)
        node[style=draw,midway,fill=white,scale=.5]
        {$0,v$};
    \draw (a) to [out=0,in=180] 
        node[style=draw,midway,fill=white,scale=.5] 
        {$w-u,w-v$} (b);
    \end{tikzpicture}
    ~=~
    \cC
    \! \begin{bmatrix}
    u & 0 & w-u \\
    0 & v & w-v
    \end{bmatrix}
    =
    \frac{\tau_2^{w}}{\pi^{w}}
    \sum_{p_1, p_2, p_3 \in \Lambda '}
    \frac{ \delta(p_1+p_2+p_3)}
         { p_1^{u\vphantom{w-u}} \, \bar{p}_2^{v\vphantom{w-v}} \, p_3^{w-u} \, \bar{p}_3^{w-v} }
\end{align}
where $u$ and $v$ are integers satisfying $1 \leq u,v \leq w-1$ and $u+v \geq 3$. The Laurent polynomial, Fourier series, and Poincar\'e series of $\cC_{u,v;w}$ were calculated in~\cite{DHoker:2019txf}.

\sm

Like the space of functions $C_{a,b,c}$, the space of functions $\cC_{u,v;w}$ obeys a system of inhomogeneous Laplace-eigenvalue equations. The action of the Laplace-Beltrami operator~$\Delta$ on~$\cC_{u,v;w}$ may be computed using~\eqref{eq:DMGF} and is given by,
\begin{align}
\label{eq:DCuvw}
    \Delta \cC_{u,v;w}
    & =
    \big( w(w-1) + 2uv - w(u+v) \big) \,
    \cC_{u,v;w}
    + uv \, \cC_{u+1,v+1;w}
\no \\
    & \quad
    + u (2v-w) \, \cC_{u+1,v;w}
    + u (v-w) \, \cC_{u+1,v-1;w}
\no \\
    & \quad
    + v (2u-w) \, \cC_{u,v+1;w}
    + v (u-w) \, \cC_{u-1,v+1;w}
\end{align}
When the right-hand side of this equation involves an index $u'$ or $v'$ which equals $0$ or~$w$, the corresponding MGFs can be reduced to Eisenstein series or double products of their derivatives using the factorization identity~\eqref{eq:MGFfact} and holomorphic subgraph reduction~\eqref{eq:HSR}. Thus, the Laplacian maps the space of $\cC_{u,v;w}$ with integer $u,v,w$ into itself (the homogeneous part) plus a linear combination of the Eisenstein series $E_w^*$ and disconnected two-loop MGFs (the inhomogeneous part).

\sm

In principle, one may diagonalize the action of the Laplacian on the space of~$\cC_{u,v;w}$, thereby constructing eigenfunctions which are linear combinations of the~$\cC_{u,v;w}$ analogous to the eigenfunctions $\mfC_{w;m;p}$ which are linear combinations of the $C_{a,b,c}$. It may be possible to construct these eigenfunctions using a generalization of the construction of the~$\mfC_{w;m;p}$ in~\cite{DHoker:2015gmr}, which we review in \autoref{apdx:CabcCwmp}. Unfortunately, this problem is more challenging than the diagonalization of the $C_{a,b,c}$, and no analytic solution is known. It is, however, a simple matter to construct these eigenfunctions with a computer algebra system. Numerical studies at small weights indicate that the eigenfunctions constructed from~$\cC_{u,v;w}$ have eigenvalues~${(w-m)(w-m-1)}$ for integers~$m$ satisfying ${1 \leq m \leq w-1}$. This is to be compared with the eigenfunctions~$\mfC_{w;m;p}$ whose eigenvalues are~${(w-2m)(w-2m-1)}$ for integers~$m$ satisfying ${1 \leq m \leq \floor{\frac{w-1}{2}}}$.

\sm

The functions $\cC_{u,v;w}$ are also special because all connected two-loop MGFs with non-negative integer exponents and at most one vanishing exponent may be written as linear combinations of the $\cC_{u,v;w}$.

\begin{prop}
\label{thm:CuvwProp}
An arbitrary connected two-loop modular graph function with non-negative integer exponents, weight $w = \sum_{r=1}^3 a_r = \sum_{r=1}^3 b_r \geq 3$, and no more than one vanishing exponent $a_r=0$ or $b_r=0$ admits the following decomposition,
\begin{align}
\label{eq:CuvwProp}
    \cC
    \! \begin{bmatrix}
    a_1 & a_2 & a_3 \\
    b_1 & b_2 & b_3
    \end{bmatrix}
    =
    \sum_{\substack{ 1 \leq u,v \leq w-1 \\ u+v \geq 3 }}
    \cK_{u,v;w} \!
    \[
    \begin{smallmatrix}
    a_1 & a_2 & a_3 \\
    b_1 & b_2 & b_3
    \end{smallmatrix}
    \]
    \,
    \cC_{u,v;w}
\end{align}
where the coefficients are integers defined by,
\begin{alignat}{4}
    \cK_{u,v;w} \!
    \[
    \begin{smallmatrix}
    a_1 & a_2 & a_3 \\
    b_1 & b_2 & b_3
    \end{smallmatrix}
    \]
    &=
    (-)^{u+v+a_3}
    \span \span \span
\no \\
    & \quad
    \times
    &
    \smash[b]{\Big\{} \,
    &
    (-)^{b_2+w} \,
    \tbinom{ w-a_3-u-1 }{ a_2-1 }
    \tbinom{ v-b_2-1 }{ b_1-1 } \,
    \hfill
    \Theta(a_1-u)
    &&
    \Theta(b_3-w+v)
\no \\
    &
    &{}+{}&
    (-)^{b_1+w} \,
    \tbinom{ w-a_3-u-1 }{ a_1-1 }
    \tbinom{ v-b_1-1 }{ b_2-1 } \,
    \hfill
    \Theta(a_2-u)
    &&
    \Theta(b_3-w+v)
\no \\
    &
    &{}+{}&
    (-)^{b_2} \,\,
    \tbinom{ u-a_3-1 }{ a_2-1 }
    \tbinom{ v-b_2-1 }{ b_3-1 } \,
    \hfill
    \Theta(a_1-w+u) 
    &&
    \Theta(b_1-w+v)
\no \\
    &
    &{}+{}&
    (-)^{b_1} \,\,
    \tbinom{ u-a_3-1 }{ a_1-1 }
    \tbinom{ v-b_1-1 }{ b_3-1 } \,
    \hfill
    \Theta(a_2-w+u)
    &&
    \Theta(b_2-w+v)
    \, \smash[t]{\Big\}}
\end{alignat}
and the step function is defined by $\Theta(x \geq 0) = 1$ and $\Theta(x < 0) = 0$.
\end{prop} 

A similar decomposition formula was proved in~\cite{DHoker:2019txf}. To prove our proposition, we first perform a partial fraction decomposition on the holomorphic momenta in the summand of~\eqref{eq:C2loop} using $p_1 = - p_2 - p_3$. We write,
\begin{align}
\label{eq:ppp1}
    \frac{1}{ p_1^{a_1} p_2^{a_2} p_3^{a_3} }
    &=
    \sum_{u=1}^{a_1}
    \frac{ \tbinom{ a_1+a_2-u-1 }{ a_2-1 } }
         { p_1^{u\vphantom{w-u}} p_3^{w-u} }
    \, (-)^{a_1+a_2+u} 
    +
    \sum_{u=1}^{a_2}
    \frac{ \tbinom{ a_1+a_2-u-1 }{ a_1-1 } }
         { p_2^{u\vphantom{w-u}} p_3^{w-u} }
    \, (-)^{a_1+a_2+u} 
\no \\
    &=
    \sum_{u=1}^{w-1}
    (-)^{w+a_3+u}
    \bigg\{
    \frac{ \tbinom{ w-u-a_3-1 }{ a_2-1 } }
         { p_1^{u\vphantom{w-u}} p_3^{w-u} }
    \, \Theta(a_1-u) 
    +
    \frac{ \tbinom{ w-u-a_3-1 }{ a_1-1 } }
         { p_2^{u\vphantom{w-u}} p_3^{w-u} }
    \, \Theta(a_2-u) 
    \bigg\}
\end{align}
where in the second line we have introduced step functions to extend the upper range of the finite sums. We have also used the fact that no more than one $a_r$ vanishes which implies that ${w-1 \geq a_1, a_2}$. We similarly decompose the anti-holomorphic momenta in the summand of~\eqref{eq:C2loop} in two different ways,
\begin{align}
\label{eq:ppp2}
    \frac{1}{ \bar{p}_1^{b_1} \bar{p}_2^{b_2} \bar{p}_3^{b_3} }
    &=
    \sum_{v=1}^{w-1}
    (-)^{w+b_2+u}
    \bigg\{
    \frac{ \tbinom{ w-v-b_2-1 }{ b_1-1 } }
         { \bar{p}_3^{v\vphantom{v-w}} \bar{p}_2^{w-v} }
    \, \Theta(b_3-v) 
    +
    \frac{ \tbinom{ w-v-b_2-1 }{ b_3-1 } }
         { \bar{p}_1^{v\vphantom{v-w}} \bar{p}_2^{w-v} }
    \, \Theta(b_1-v) 
    \bigg\}
\no \\
    &=
    \sum_{v=1}^{w-1}
    (-)^{w+b_1+u}
    \bigg\{
    \frac{ \tbinom{ w-v-b_1-1 }{ b_3-1 } }
         { \bar{p}_2^{v\vphantom{v-w}} \bar{p}_1^{w-v} }
    \, \Theta(b_2-v) 
    +
    \frac{ \tbinom{ w-v-b_1-1 }{ b_2-1 } }
         { \bar{p}_3^{v\vphantom{v-w}} \bar{p}_1^{w-v} }
    \, \Theta(b_3-v) 
    \bigg\}
\end{align}
We use the first line of~\eqref{eq:ppp2} for the anti-holomorphic momenta which multiply the first term on the second line of~\eqref{eq:ppp1}, and we use the second line of~\eqref{eq:ppp2} for the second term in~\eqref{eq:ppp1}. After some straightforward manipulations, we arrive at~\eqref{eq:CuvwProp}. The coefficient~$\cK_{1,1;w}$ necessarily vanishes, so the decomposition includes only convergent $\cC_{u,v;w}$ with $u+v \geq 3$. This completes our proof of \autoref{thm:CuvwProp}. 

\sm

Our decomposition formula is not unique since we could have performed several different partial fraction decompositions. Moreover, the functions $\cC_{u,v;w}$ are themselves not algebraically independent. For instance, the decomposition formula~\eqref{eq:CuvwProp} for the function~$C_{2,1,1}$ defined in~\eqref{eq:Cabc} yields,
\begin{align}
    C_{2,1,1} = \cC_{2,2;4} - 2 \, \cC_{2,3;4} - \cC_{3,2;4}
\end{align}
Since $C_{2,1,1}$ is real, this expression must be equal to its complex conjugate. Acting on~$\cC_{u,v;w}$, complex conjugation interchanges $u$ and $v$. Thus, we conclude that $\cC_{2,3;4} = \cC_{3,2;4}$. There are many other linear algebraic identities between the $\cC_{u,v;w}$. Despite this ambiguity, the functions $\cC_{u,v;w}$ have many nice properties and have been extensively studied.

\subsubsection{Two loops and beyond}

We have shown that two infinite families of functions \smash{$\cV_{s,t}^{(n)}$} and $\cC_{u,v;w}$ span the space of all two-loop MGFs with non-negative integer exponents. This is to be compared with the case at one loop, where there is a unique MGF at each weight.

\begin{cor}
\label{thm:2loop}
An arbitrary two-loop modular graph function with non-negative integer coefficients and weight $w$ can be written as a linear combination with rational coefficients of the connected two-loop functions $\cC_{u,v;w}$, the disconnected two-loop functions \smash{$\cV_{s,t}^{(n)}$}, and the Eisenstein series $E_w^*$, where $u,v$ are integers satisfying $1 \leq u,v \leq w-1$ and $u+v \geq 3$ and~$s,t,n$ are integers satisfying $s, t \geq 2$, $s+t=w$, and $\min(s,t) \geq n \geq 0$.
\end{cor}

The space of three-loop MGFs is considerably more complicated, and consists of connected dihedral, trihedral, and tetrahedral graphs as well as disconnected graphs. Instead of tackling the general three-loop case, we shall restrict our attention in this paper to several special infinite families of connected and disconnected three-loop MGFs, to be introduced in the sequel.

\newpage

\section{Integrating modular graph functions over \texorpdfstring{$\cM_L$}{}}
\label{sec:ints}

In this section, we shall discuss the general structure of the integrals of modular graph functions over the truncated fundamental domain $\cM_L$ defined in~\eqref{eq:ML}. We shall also discuss several methods to calculate these integrals as well as their transcendental weights. These methods include the integration of certain exact differentials using Stokes' theorem and the integration of regularized Poincar\'e series using the unfolding trick.

\subsection{Integrals of modular graph functions}

Modular graph functions with integer exponents, such as those which arise in string theory, have an asymptotic expansion of the form~\eqref{eq:MGFAsy} near the cusp. Because they have polynomial growth at the cusp, their integrals over the full fundamental domain $\cM$ generally diverge, as we discussed in \autoref{sec:overv}. Their integrals over the truncated fundamental domain~$\cM_L$ are, however, finite functions of the cut-off $L$.

\sm

The general structure of the integral over $\cM_L$ of a modular graph function \smash{$\cC = \cC_\Gamma \[ \begin{smallmatrix} A \\ B \end{smallmatrix} \]$} with integer exponents and weight $w$ may be inferred from its Laurent polynomial~\eqref{eq:MGFAsy}. This integral is given by,\footnote{\label{foot:exp}Throughout the sequel of this paper we shall omit all terms that are exponentially suppressed in $L$ for large $L$ and suppress the corresponding symbol $\cO(e^{-2\pi L}$).}
\begin{align}
\label{eq:IC}
    \intML \cC
    =
    \cI_\cC
    + \mfc_\cC ^{(1)} \ln L
    + \sum_{\substack{ \ell=1-w \\ \ell \neq 1}}^w
    \mfc_\cC ^{(\ell)} \, \frac{ L^{\ell-1} }{ \ell-1 }
    + \cO(e^{- 2 \pi L})
\end{align}
where \smash{$\mfc_{\cC}^{(\ell)}$} are the Laurent coefficients of $\cC$ introduced in~\eqref{eq:MGFAsy}. The term $\cI_\cC$ is independent of $L$ and cannot be inferred from the Laurent polynomial of~$\cC$ since it receives contributions from the integral of terms exponentially suppressed in $\tau_2$, namely the terms with ${M=N \neq 0}$ in~\eqref{eq:MGFqexp}. Alternatively, $\cI_\cC$ may be defined as the following limit,
\begin{align}
    \cI_\cC
    =
    \lim _{L \to \infty}
    \bigg(
    \intML \cC - \mfc_\cC ^{(1)} \ln L
    - \sum _{\ell=2}^w
    \mfc_\cC^{(\ell)} \, \frac{ L^{\ell-1} }{ \ell-1 }
    \bigg)
\end{align}
Mathematically, one may then assign $\cI_\cC$ as the value of the renormalized integral of $\cC$ over~$\cM$, following Zagier~\cite{Zagier:1982}.

\sm

Physically, genus-one contributions to the low-energy effective interactions of string theory are computed by an integral over $\cM$. This integral may be partitioned into a contribution from $\cM_L$ and a contribution from its complement $\cM_R$. The cut-off $L > 1$ is arbitrary and necessarily cancels in the complete integral. The  contribution from $\cM_L$ is analytic in the dimensional kinematic variables $s,t,u$. At any order in the expansion in powers of these variables, this analytic contribution is given by a sum of integrals over~$\cM_L$ of MGFs with non-negative integer exponents times rational coefficients. When the modular graph function~$\cC$ appears in this context, the $L$-dependence in its integral over~$\cM_L$~\eqref{eq:IC} will be cancelled by the integral of the full string integrand over $\cM_R$. Thus, in the context of string theory we may drop the $L$-dependent terms in~\eqref{eq:IC}. However, a special role is played by the $\ln L$ term, whose cancellation against the integral over $\cM_R$ produces logarithmic terms in~$s,t,u$ which in turn give rise to the physical branch cuts required by unitarity of the string amplitude, as described in~\cite{DHoker:2019blr}. 

\sm

Therefore, in the sequel of this paper we shall systematically write the integrals over $\cM_L$ of modular graph functions with integer exponents using the following instructions.
\begin{itemize}
\item Omit terms which are exponentially suppressed in $L$ (as announced in \autoref{foot:exp}) and terms which are powers of $L$ with non-zero exponents.
\sm
\item Retain $L$-independent terms and terms proportional to $\ln L$.
\end{itemize}
We introduce the following notation which renders these instructions explicit, 
\begin{align}
\label{eq:approx}
    \intML \cC
    \approx
    \cI_\cC
    + \mfc_\cC^{(1)} \ln L
\end{align}
In other words, the symbol $\approx$ denotes equality up to positive powers of $L$ and terms which vanish in the limit $L \to \infty$.

\sm

It will often be convenient to perform intermediate calculations using MGFs with complex non-integer exponents, such as the Eisenstein series $E_s^*$ with ${\Re(s) > 1}$ and $s$ not necessarily integer. MGFs with non-integer exponents do not have a Laurent polynomial near the cusp, so their integrals over $\cM_L$ will not be of the form~\eqref{eq:IC}. When writing these integrals, we shall retain all $L$-dependent terms except for those which are exponentially suppressed in~$L$ (as announced in \autoref{foot:exp}). The $\approx$ notation introduced above will not be used for the integrals of MGFs with non-integer exponents.

\subsection{Integrals of exact differentials}

Integrals of exact differentials over $\cM_L$ may be evaluated using Stokes's theorem and the fact that the boundary of $\cM_L$ is given by,
\begin{align}
    \p \cM_L
    =
    \{ \tau_1 \in \bbR/\bbZ, \, \tau_2 = L \}
\end{align}
This simple observation, used in conjunction with the differential relations between MGFs in~\eqref{eq:nablaMGF} and~\eqref{eq:DMGF}, provides a powerful tool to calculate the integrals of various MGFs, as shown in~\cite{DHoker:2019blr}. Moreover, we may systematically formulate the transcendental structure of the integrals of several exact differentials of MGFs.

\subsubsection{Integrals involving $\Delta$, $\nabla$, $\bar{\nabla}$}

The first-order Maass operators $\nabla$ and $\bar{\nabla}$ were defined in~\eqref{eq:nabla}. The second-order Laplace-Beltrami operator $\Delta$ was defined in~\eqref{eq:Delta}. Integrals of exact differentials involving these operators are given as follows.
\begin{prop}
\label{thm:IDC}
For arbitrary modular graph functions $\cC$, modular graph $(0,2)$-forms~$\cC^+$, and modular graph $(2,0)$-forms $\cC^-$,
\begin{align}
\label{eq:IDC}
    \intML \Delta \cC^{\phantom{+}}
    &=
    \int_0^1 d\tau_1 \,
    \p_{\tau_2} \cC
    \big|_{\tau_2 = L}
    \approx
    \mfc_\cC^{(1)}
\no \\
    \intML \nabla \cC^+
    &=
    \int_0^1 d\tau_1 \,\,
    \cC^+
    \big|_{\tau_2 = L}
    \phantom{\p_{\tau_2} \cC}
    \negphantom{\, \cC^+}
    \approx
    \mfc_{\cC^+}^{(0)}
\no \\
    \intML \bar{\nabla} \cC^-
    &=
    \int_0^1 d\tau_1 \,\,
    \cC^-
    \big|_{\tau_2 = L}
    \phantom{\p_{\tau_2} \cC}
    \negphantom{\, \cC^-}
    \approx
    \mfc_{\cC^-}^{(0)}
\end{align}
where \smash{$\mfc_\cC^{(1)}$} and \smash{$\mfc_{\cC^\pm}^{(0)}$} are the Laurent coefficients of $\cC$ and $\cC^\pm$, respectively. The $\approx$ notation applies only when $\cC$, $\nabla \cC^+$, and $\bar{\nabla} \cC^-$ are modular graph functions with integer exponents and weight $w$. In this case, the transcendental weight of each integral is also $w$.
\end{prop}

The proof of the equalities in~\eqref{eq:IDC} using Stokes's theorem is straightforward, and the transcendental weight assignments follow from the fact that \smash{$\mfc_\cC^{(1)}$} and \smash{$\mfc_{\cC^\pm}^{(0)}$} have transcendental weight $w$ when $\cC$, $\nabla \cC^+$, and $\bar{\nabla} \cC^-$ have weight $w$. In contrast with the integral~\eqref{eq:IC} of an arbitrary MGF with integer exponents, the integrals in~\eqref{eq:IDC} necessarily have vanishing $\ln L$ contributions since the asymptotic expansions of $\cC$ and $\cC^\pm$ are free of logarithms $\ln \tau_2$.

\subsubsection{The integral of $E_s^*$}

As an application of \autoref{thm:IDC}, we shall compute the integral of the Eisenstein series~$E^*_s$ with $\Re(s) > 1$ but $s$ not necessarily integer. At one loop, $E^*_s$ is the unique modular graph function with weight $s$. The Laplace equation~\eqref{eq:DE}, the asymptotic expansion~\eqref{eq:EAsy}, and the first equation of~\eqref{eq:IDC} yield,
\begin{align}
\label{eq:IE}
    \intML
    E^*_s
    &=
    \frac{1}{s(s-1)} \intML \, \Delta E^*_s
    =
    \zeta^*(2s) \,
    \frac{L^{s-1}}{s-1}
    +
    \zeta^*(2s-1) \,
    \frac{L^{-s}}{-s}
    \approx 0
\end{align}
where the $\approx$ notation applies for integer $s \geq 2$. In this case, the integral of $E^*_s$ vanishes up to non-zero powers of $L$.

\subsubsection{Integrals involving $\Delta_k$}

The differential operator $\Delta_k$ with integer $k \geq 1$ was used in~\cite{DHoker:2015sve, DHoker:2019txf} and may be equivalently defined in terms of the second-order operator $\Delta$ or in terms of the first-order operators $\nabla$ and $\bar{\nabla}$ as follows,
\begin{align}
\label{eq:Dk}
    \Delta_k
    =
    \prod_{\ell=1}^k
    \big( \Delta - \ell(\ell-1) \big)
    =
    \bar \nabla^k \tau_2^{-2k} \nabla^k
    =
    \nabla^k \tau_2^{-2k} \bar \nabla^k
\end{align}
Like the Laplace-Beltrami operator $\Delta=\Delta_1$, the operator $\Delta_k$ acts on modular graph functions and maps the space of MGFs into itself.

\sm

Both the monomial $\tau_2^s$ and the Eisenstein series $E_s^*$ are eigenfunctions of $\Delta_k$ with the eigenvalue $\lambda_k(s)$ given by,
\begin{align}
\label{eq:Dkeigen}
    \lambda_k(s)
    &=
    \prod_{\ell=1-k}^k (s-\ell)
    =
    \frac{\Gamma(s+k)}{\Gamma(s-k)}
    =
    \lambda_k(1-s)
\end{align}
The eigenvalue vanishes for integer $s$ with $1-k \leq s \leq k$ so that $\Delta_k$ annihilates an arbitrary Laurent polynomial in $\tau_2$ of degree $(k,1-k)$. As a result, for an arbitrary modular graph function~$\cC$ with integer exponents and weight $w$, $\Delta_w \cC$ is a cusp form with exponential decay at the cusp.

\sm

Additionally, the derivative of the eigenvalue, \smash{$\lambda_k'(s) = \frac{d}{ds} \lambda_k(s)$}, is itself an integer for integer values of $s$. This fact may be verified using the properties of the digamma function ${\psi(s) = \Gamma'(s) / \Gamma(s)}$ discussed in \autoref{apdx:zeta}.

\sm

The operator $\Delta_k$ may be used to construct a total derivative from two arbitrary MGFs. The integral of this total derivative is given as follows.
\begin{prop}
\label{thm:ICDC}
For arbitrary modular graph functions $\cC_1$ and $\cC_2$ with integer exponents and respective weights $w_1$ and $w_2$, $\cC_1 \Delta_k \cC_2 - \cC_2 \Delta_k \cC_1$ is a total derivative whose integral is,
\begin{align}
\label{eq:ICDC}
    \intML
    \big( \cC_1 \Delta_k \cC_2 - \cC_2 \Delta_k \cC_1 \big)
    & \approx
    \sum_{\ell_1=1-w_1}^{w_1}
    \sum_{\ell_2 = 1-w_2}^{w_2}
    \mfc_{\cC_1}^{(\ell_1)} \,
    \mfc_{\cC_2}^{(\ell_2)} \,
    \lambda_k'(\ell_2) \,
    \delta_{\ell_1+\ell_2,1}
\end{align}
where \smash{$\mfc_{\cC_1}^{(\ell_1)}$} and \smash{$\mfc_{\cC_2}^{(\ell_2)}$} are the Laurent coefficients of $\cC_1$ and $\cC_2$, respectively, and the sum is over $\ell_1$ and $\ell_2$ satisfying $\ell_1+\ell_2=1$. This integral has transcendental weight $w_1+w_2$.
\end{prop}

Before we prove this proposition, a few remarks are in order. First, the right-hand side of~\eqref{eq:ICDC} is anti-symmetric under the interchange of the labels $1$ and $2$, as required, since~${\lambda'(\ell_2) = - \lambda'(1-\ell_2) = - \lambda'(\ell_1)}$ for $\ell_1 + \ell_2 = 1$. Moreover, for $\cC_1 = 1$, $\cC_2 = \cC$, and $k=1$, the expression~\eqref{eq:ICDC} reproduces the integral of $\Delta \cC$ given in~\eqref{eq:IDC} upon using $\lambda_1'(1) = 1$.

\sm

To prove this proposition we first rewrite the combination $\cC_1 \Delta_k \cC_2 - \cC_2 \Delta_k \cC_1$ using the following relations,
\begin{align}
\label{eq:CDkC}
    \cC_1 \Delta_k \cC_2 
    &=
    \nabla
    \sum_{j=0}^{k-1} (-)^j
    \big( \nabla^j \cC_1 \big)
    \big( \nabla^{k-j-1} \tau_2^{-2k} \bar{\nabla}^k \cC_2 \big)
    +
    (-)^k \, \tau_2^{-2k}
    \big( \nabla^k \cC_1 \big)
    \big( \bar{\nabla}^k \cC_2 \big)
    \no \\
    \cC_2 \Delta_k \cC_1 
    &=
    \bar{\nabla}
    \sum_{j=0}^{k-1} (-)^j
    \big( \bar{\nabla}^j \cC_2 \big)
    \big( \bar{\nabla}^{k-j-1} \tau_2^{-2k} \nabla^k \cC_1 \big)
    +
    (-)^k \, \tau_2^{-2k}
    \big( \bar{\nabla}^k \cC_2 \big)
    \big( \nabla^k \cC_1 \big)
\end{align}
Subtracting the second relation from the first cancels the last term on the right-hand side of each line and expresses the integrand of~\eqref{eq:ICDC} in terms of total derivatives. Using the results on the last two lines of~\eqref{eq:IDC} and the asymptotic expansions~\eqref{eq:MGFAsy} for $\cC_1$ and $\cC_2$, the integral of~\eqref{eq:ICDC} evaluates as follows,
\begin{align}
    \sum_{\ell_1=1-w_1 \mathstrut}^{w_1}
    \sum_{\ell_2=1-w_2 \mathstrut}^{w_2}
    \sum_{j=0 \mathstrut}^{k-1} \,
    \mfc_{\cC_1}^{(\ell_1)}
    \mfc_{\cC_2}^{(\ell_2)} \,
    (-)^j \,
    \Big[
    &
    \big( \nabla^j \tau_2^{\ell_1} \big)
    \big( \nabla^{k-j-1} \tau_2^{-2k}
        \bar{\nabla}^k \tau_2^{\ell_2} \big)
\no \\[-1.5ex]
    {}-{}
    &
    \big( \bar{\nabla}^j \tau_2^{\ell_2} \big)
    \big( \bar{\nabla}^{k-j-1}
        \tau_2^{-2k} \nabla^k \tau_2^{\ell_1} \big)
    \Big]_{\tau_2 = L}
\end{align}
To proceed we use $\nabla^n \tau_2^s = \bar{\nabla}^n \tau_2^s = \Gamma(s+n) \, \tau_2^{s+n} / \, \Gamma{(s)} $ and carry out the sum over $j$ to obtain the right-hand side of~\eqref{eq:ICDC} after some straightforward simplifications. The transcendental weight assignment follows from the fact that \smash{$\mfc_{\cC_1}^{(\ell_1)}$} and \smash{$\mfc_{\cC_2}^{(\ell_2)}$} have transcendental weights $w_1$ and $w_2$, respectively. This completes our proof.

\sm

In the special case where one of the MGFs is $E^*_s$ with $\Re(s)>1$ but~$s$ not necessarily integer, the above construction does not directly apply since we assumed in \autoref{thm:ICDC} that the MGFs had integer exponents. The appropriate generalization is given as follows.

\begin{prop}
\label{thm:IEC}
For an arbitrary modular graph function $\cC$ with integer exponents and weight $w$, the combination \smash{$\cC \Delta_k E^*_s - E^*_s \Delta_k \cC$} with $\Re(s)>1$ is a total derivative whose integral is given by the analytic continuation in $s$ of the following expression,
\begin{align}
\label{eq:IEC}
    \intML
    \big(
    \cC \Delta_k E^*_s 
    -
    E^*_s \Delta_k \cC
    \big)
    &=
    \sum_{\ell=1-w}^{w}
    \mfc^{(\ell)}_\cC \,
    \big[
      \lambda_k(s)
    - \lambda_k(\ell)
    \big]
    \,
    \cG_{\ell}(s)
\end{align}
where \smash{$\mfc_{\cC}^{(\ell)}$} are the Laurent coefficients of $\cC$ and,
\begin{align}
\label{eq:F(s)}
    \cG_{\ell}(s)
    =
    \zeta^*(2s) \, 
    \frac{L^{\ell+s-1}}{\ell+s-1}
    + \zeta^*(2s-1) \, 
    \frac{L^{\ell-s}}{\ell-s}
\end{align}
For integer $s \geq 2$, this integral has transcendental weight $w+s$.
\end{prop}

To prove this proposition we write the combination \smash{$\cC \Delta_k E^*_s - E^*_s \Delta_k \cC$} in terms of total derivatives using the relations in~\eqref{eq:CDkC}. We then use the results on the last two lines of~\eqref{eq:IDC} and the asymptotic expansions~\eqref{eq:MGFAsy} and~\eqref{eq:EAsy} for $\cC$ and $E_s^*$, respectively. We find that the integral of~\eqref{eq:IEC} evaluates as follows,
\begin{align}
    \sum_{\ell=1-w \mathstrut}^{w}
    \sum_{j=0 \mathstrut}^{k-1} \,
    \mfc_{\cC}^{(\ell)} \,
    \zeta^*(2s) \,
    (-)^j \,
    \Big[
    &
    \big( \nabla^j \tau_2^{\ell} \big)
    \big( \nabla^{k-j-1} \tau_2^{-2k}
        \bar{\nabla}^k \tau_2^{s} \big)
\no \\[-1.5ex]
    {}-{}
    &
    \big( \bar{\nabla}^j \tau_2^{s} \big)
    \big( \bar{\nabla}^{k-j-1}
        \tau_2^{-2k} \nabla^k \tau_2^{\ell} \big)
    \Big]_{\tau_2 = L}
    + (s \to 1-s)
\end{align}
where the instruction $(s \to 1-s)$ refers to the entire previous term. We proceed as in the proof of \autoref{thm:ICDC} and obtain the right-hand side of~\eqref{eq:IEC} after some straightforward simplifications. The transcendental weight assignment follows from the fact that \smash{$\mfc_{\cC}^{(\ell)}$} has transcendental weight $w$ and the fact that both $\zeta^*(2s)$ and $\zeta^*(2s-1)$ have transcendental weight $s$ for integer $s \geq 2$, as discussed in \autoref{apdx:zeta}.

\subsubsection{The integral of $E_s^* E_t^*$}

As another application, we shall use the operator $\Delta_1 = \Delta$ to compute the integral of the product of two Eisenstein series. This integral is known as the Maass-Selberg relation~\cite{Zagier:1982}.

\begin{prop}
\label{thm:IEEMaass}
The integral of $E_s^* E_t^*$ with $\Re(s), \Re(t) > 1$ is given by the analytic continuation in $s$ and $t$ of the following expression,
\begin{align}
\label{eq:IEEMaass}
    \intML E^*_s E^*_t 
    &=
    \zeta^*(2s) \,
    \zeta^*(2t) \,
    \frac{L^{s+t-1}}{s+t-1}
    +
    \zeta^*(2s) \,
    \zeta^*(2t-1) \,
    \frac{L^{s-t}}{s-t}
\no \\
    & \quad
    +
    \zeta^*(2s-1) \,
    \zeta^*(2t) \,
    \frac{L^{t-s}}{t-s}
    +
    \zeta^*(2s-1) \,
    \zeta^*(2t-1) \,
    \frac{L^{1-s-t}}{1-s-t}
\end{align}
which vanishes up to non-zero powers of $L$ for integer $s, t \geq 2$ unless $s=t$.
\end{prop}

To prove this proposition, we consider $E^*_s \Delta E^*_t -  E^*_t \Delta E^*_s$ with $s \neq t$. This combination is proportional to $E^*_s E^*_t$ and may be written in terms of total derivatives using the relations in~\eqref{eq:CDkC}. We use the results on the last two lines of~\eqref{eq:IDC} and the asymptotic expansion~\eqref{eq:EAsy} to arrive at~\eqref{eq:IEC}, which may be analytically continued in $s$ and $t$.

\subsection{Integrals of Poincar\'e series}
\label{sec:unfold}

In addition to Stokes' theorem, we may use the method of unfolding a Poincar\'e series, also called \emph{the unfolding trick}, to evaluate the integrals of modular graph functions over $\cM_L$. The Poincar\'e series representation of a MGF was given in~\eqref{eq:Poincare}. A Poincar\'e seed function~$\Lambda$ for a modular graph function~$\cC$ is usually simpler than the function $\cC$ itself. The unfolding trick exploits this simplification.

\subsubsection{The standard unfolding trick}

We shall first consider a modular-invariant function $\hat \cC$ whose integral over $\cM$ is absolutely convergent and which has a Poincar\'e seed function $\hat \Lambda$. The standard unfolding trick replaces the integral of $\hat \cC$ over $\cM = \mathrm{PSL}(2,\bbZ) \backslash \cH$ with an integral of its seed function $\hat \Lambda$ over the upper half-strip~$\Gamma_\infty \backslash \cH$, which is described in~\eqref{eq:strip}. We have,
\begin{align}
\label{eq:unfold}
    \intM
    \hat \cC(\tau, \bar{\tau})
    = 
    \int\limits_{ \mathrm{PSL}(2,\bbZ) \backslash \cH }
    \!\!
    \frac{d^2\tau}{\tau_2^2}
    \!
    \mathop{\vphantom{\int}\sum}
        _{ \lambda \in \Gamma_\infty \backslash \mathrm{PSL}(2,\bbZ) }
    \!\!
    \hat \Lambda
        (\lambda \tau, \lambda \bar{\tau})
    =
    \int_0^\infty \frac{d\tau_2}{\tau_2^2}
    \int_0^1 d\tau_1 \,
    \hat \Lambda(\tau, \bar{\tau})
\end{align}
The absolute convergence of the integral and sum in the middle expression permits swapping their order and changing integration variables $\lambda \tau \to \tau$ to obtain the final expression.

\subsubsection{Unfolding modular graph functions}

We shall now consider an arbitrary modular graph function $\cC$ with integer exponents and weight~$w$. In general, $\cC$ has polynomial growth at the cusp and is not integrable on $\cM$.

\sm

To perform the unfolding trick, we construct from $\cC$ an associated function $\hat \cC$ which has a finite limit at the cusp by subtracting the positive powers of $\tau_2$ from the Laurent polynomial~\eqref{eq:MGFAsy} of $\cC$. Terms quadratic or higher in $\tau_2$ may be subtracted by Eisenstein series, and we use the two-loop modular graph function $C_{2,1,1}$ defined in~\eqref{eq:Cabc} to subtract any linear divergence in $\tau_2$, following~\cite{DHoker:2019mib}. For integers $\ell \geq 1$, we define,
\begin{align}
\label{eq:cE}
    \cE_\ell
    =
    \begin{cases}
    \frac{45}{\pi \zeta(3)}
    \big(
    C_{2,1,1}
    - \tfrac{2}{9} E_4^*
    \big)
    & \ell = 1
    \\[2ex]
    \frac{1}{ \zeta^*(2\ell) } \,
    E_\ell^*
    & \ell \geq 2
    \end{cases}
\end{align}
Our normalizations are chosen so that near the cusp \smash{$\cE_\ell = \tau_2^\ell + \cO(\tau_2^{-1})$}, and for $\ell \geq 2$ the Poincar\'e seeds are simply $\Lambda_{\cE_\ell} = \tau_2^\ell$. Given the positive-power Laurent coefficients \smash{$\mfc_\cC^{(1)}, \dots, \mfc_\cC^{(w)}$} of $\cC$, the modular-invariant function $\hat \cC$ defined by,
\begin{align}
\label{eq:CHat}
    \hat \cC
    =
    \cC
    - \sum_{\ell=1}^w
    \mfc_\cC^{(\ell)} \,
    \cE_\ell
    =
    \mfc_\cC^{(0)}
    + \cO(\tau_2^{-1})
\end{align}
is integrable over $\cM$. By unfolding the integral of $\hat \cC$ over $\cM$, we can calculate the integral of~$\cC$ over $\cM_L$. This procedure was carried out in~\cite{DHoker:2019mib} for the connected two-loop functions~$\cC_{u,v;w}$.

\sm

One could in principle continue to higher loops, but calculating the Laurent polynomial of higher-loop MGFs is prohibitively laborious. Only a few Laurent polynomials of higher-loop MGFs are explicitly known. For this reason, we present the following lemma, which describes a procedure for integrating MGFs which does not require a priori knowledge of the full Laurent polynomial.

\begin{lem}
\label{thm:Iunfold}
For an arbitrary modular graph function $\cC$ with integer exponents, weight $w$, and Poincar\'e seed $\Lambda$,
\begin{align}
\label{eq:Iunfold}
    \intML \cC
    &\approx
    \lim_{L \to \infty}
    \bigg(
    \int_0^L \frac{d\tau_2}{\tau_2^2}
    \int_0^1 d\tau_1 \,
    \Lambda
    - \mfc_\cC^{(1)} \ln(2L)
    - \sum_{\ell=2}^{w}
    \mfc_\cC^{(\ell)} \,
    \frac{L^{\ell-1}}{\ell-1}
    \bigg)
    + \mfc_\cC^{(1)} \ln(2L)
\end{align}
where \smash{$\mfc_{\cC}^{(\ell)}$} are the Laurent coefficients of $\cC$.
\end{lem}

Before we prove this lemma, we shall make a few remarks. First, we may identify the limit on the right-hand side of~\eqref{eq:Iunfold} with the constant term $\cI_\cC$ in~\eqref{eq:IC}. Second, the Laurent coefficients \smash{$\mfc_\cC^{(\ell)}$} are needed here only for $\ell \geq 1$, and they are usually easier to calculate from the Poincar\'e seed $\Lambda$ than from the full modular graph function $\cC$. In fact, the positive-power part of the Laurent polynomial of $\cC$ is equal to the positive-power part of the Laurent polynomial of $\Lambda$. Finally, the result of this lemma is reminiscent of the Rankin-Selberg-Zagier (RSZ) method of~\cite{Zagier:1982}. To integrate a function $\cC$ over the fundamental domain using RSZ, one regularizes the product $E_s^* \, \cC$ for sufficiently large $\Re(s)$, unfolds the Eisenstein series to its Poincar\'e seed $\zeta^*(2s) \, \tau_2^s$, integrates the resulting expression over the upper half-strip, and then finally computes the residue of this expression at $s = 1$. In comparison, the procedure described in \autoref{thm:Iunfold} unfolds $\cC$ to its Poincar\'e seed $\Lambda$ and then integrates this seed function over the upper half-strip.

\subsubsection{Proof of \autoref{thm:Iunfold}}

We shall use the two following propositions to prove our lemma.

\begin{prop}
\label{thm:Isum}
For $\tau \in \cH$ and $m, N \in \bbZ$ with $m \neq 0$, 
\begin{align}
\label{eq:Isum}
    \int_0^1 d\tau_1
    \sum_{n \in \bbZ}
    \frac{1}{|m\tau+n|^2 |m\tau+n+N|^2}
    =
    \frac{2 \pi }
        {|m| \tau_2 \(4m^2\tau_2^2 + N^2\)}
\end{align}
\end{prop}

\paragraph{Proof:}

We first separately decompose the holomorphic and anti-holomorphic factors in the summand using partial fractions as follows,
\begin{align}
   \frac{1}{(m\tau+n) (m\tau+n+N)}
   =
   \frac{1}{N}
   \(
   \frac{1}{m\tau+n}
   - \frac{1}{m\tau+n+N}
   \)
\end{align}
Since $N \in \bbZ$, we can shift the sums over $n$ and again use partial fraction decomposition on each term in the summand to write the left-hand side of~\eqref{eq:Isum} as,
\begin{align}
\label{eq:Isum2}
    \frac{1}{N^2}
    \(
    \frac{1}{m\tau_2}
    - \frac{1}{2m\tau_2+iN}
    - \frac{1}{2m\tau_2-iN}
    \)
    \int_0^1 d\tau_1
    \bigg(
    \sum_{n \in \bbZ}
    \frac{i}{m\tau+n}
    + c.c.
    \bigg)
\end{align}
where $c.c.$ stands for the complex conjugate of the previous term in parentheses. The sum over $n$ is given by,
\begin{align}
    \sum_{n \in \bbZ}
    \frac{i}{m\tau+n}
    =
    \pi \veps(m)
    \bigg(
    1 + 2
    \sum_{p > 0} e^{2\pi i |m| p \tau}
    \bigg)
\end{align}
where $\veps(m) = \pm 1$ is equal to the sign of $m$. Integrating over $\tau_1$ annihilates the sum over~$p$. The terms in the first set of parentheses in~\eqref{eq:Isum2} can then be combined with like denominators to yield the right-hand side of~\eqref{eq:Isum}.

\begin{prop}
\label{thm:IE1}
Let $\Lambda_{\cE_1}$ be a Poincar\'e seed for $\cE_1$ defined in~\eqref{eq:cE}. Then,
\begin{align}
\label{eq:IE1}
    \lim_{L \to \infty}
    \bigg(
    \int_0^L \frac{d\tau_2}{\tau_2^2}
    \int_0^1 d\tau_1 \,
    \Lambda_{\cE_1}
    - \ln(2L)
    \bigg)
    =
    \frac{\zeta'(4)}{\zeta(4)}
    - \frac{\zeta'(3)}{\zeta(3)}
\end{align}
\end{prop}

\paragraph{Proof:}

A Poincar\'e seed for $\cE_1$ is given by,
\begin{align}
\label{eq:LE1}
    \Lambda_{\cE_1}
    =
    \frac{45}{\pi\zeta(3)}
    \(
    \Lambda_{2,1,1}
    - \frac{2 \pi^4}{14175} \,
    \tau_2^4
    \)
\end{align}
where $\Lambda_{2,1,1}$ is a Poincar\'e seed for $C_{2,1,1}$. We obtain $\Lambda_{2,1,1}$ from the Kronecker-Eisenstein series representation~\eqref{eq:Cabc} of $C_{2,1,1}$ by rotating the integer pair ${(m_3,n_3) \neq (0,0)}$ to the pair~${(0,N) \neq (0,0)}$,
\begin{align}
    \Lambda_{2,1,1}
    =
    \frac{\tau_2^4}{\pi^4}
    \sum_{N \neq 0} \,
    \sum^{'}_{\substack{(m_r, n_r) \in \bbZ^2
        \\ r = 1,2}}
    \frac{\delta(m_1+m_2) \delta(n_1+n_2+N)}
        {|m_1\tau + n_1|^2 |m_2\tau + n_2|^2 N^4}
\end{align}
Next we split $\Lambda_{2,1,1}$ into contributions according to the number of non-vanishing summation variables $m_r$. The contribution with both $m_1=m_2=0$ cancels the $\tau_2^4$ term in~\eqref{eq:LE1}. The contribution with only one non-zero $m_r$ vanishes in view of the Kronecker delta function. Thus, the remaining contribution arises entirely from non-vanishing $m_1$ and $m_2$,
\begin{align}
    \Lambda_{\cE_1}
    &=
    \frac{45 \tau_2^4}{\pi^5 \zeta(3)}
    \sum_{m_1, m_2, N \neq 0} \,
    \sum_{n_1, n_2 \in \bbZ}
    \frac{\delta(m_1+m_2) \delta(n_1+n_2+N)}
        {|m_1\tau + n_1|^2 |m_2\tau + n_2|^2 N^4}
\end{align}
To evaluate this sum we use the delta symbols to solve $m=m_1=-m_2$ and ${n_2=-n_1-N}$. We then integrate over $\tau_1$ using~\eqref{eq:Isum} of \autoref{thm:Isum}. Restricting the sums to $m, N > 0$ and integrating over $\tau_2$, we find,
\begin{align}
    \int_0^L \frac{d\tau_2}{\tau_2^2}
    \int_0^1 d\tau_1 \,
    \Lambda_{\cE_1}
    &=
    \frac{1}{\zeta(3) \zeta(4)}
    \sum_{m, N > 0}
    \frac{1}{m^3 N^4}
    \times
    \tfrac{1}{2}
    \ln \Big( (2mL/N)^2 + 1 \Big)
\end{align}
The limit~\eqref{eq:IE1} simply follows.

\paragraph{Proof of \autoref{thm:Iunfold}:}

Returning to our lemma, we first construct the finite modular graph function~$\hat \cC$ as in~\eqref{eq:CHat} and use the standard unfolding trick~\eqref{eq:unfold} to find,
\begin{align}
\label{eq:ICproo1f}
    \intM
    \hat \cC
    =
    \int_0^\infty \frac{d\tau_2}{\tau_2^2}
    \int_0^1 d\tau_1
    \bigg(
    \Lambda
    - \mfc_\cC^{(1)}
    \Lambda_{\cE_1}
    - \sum_{\ell=2}^w
    \mfc_\cC^{(\ell)} \,
    \tau_2^\ell
    \bigg)
\end{align}
By construction, $\hat \cC$ is finite at the cusp, so the integrals of $\hat \cC$ over $\cM$ and $\cM_L$ differ by terms of order $\cO(L^{-1})$. Using the definition of $\hat \cC$ we obtain,
\begin{align}
    \intML
    \hat \cC
    = 
    \intML
    \cC
    - 
    \sum_{\ell=1}^w
    \mfc_\cC^{(\ell)}
    \intML
    \cE_\ell
\end{align}
The integrals of the Eisenstein series are given in~\eqref{eq:IE}. The integral of $\cE_1$ can be calculated using the differential equation $(\Delta-2) C_{2,1,1} = 9 E_4 - E_2^2$, the Laurent polynomial of $C_{2,1,1}$, and the integral of $E_2^2$. We simply quote the result from~\cite{DHoker:2019mib},
\begin{align}
    \intML \cE_1
    =
    \frac{\zeta'(4)}{\zeta(4)}
    - \frac{\zeta'(3)}{\zeta(3)}
    + \ln(2L)
    + \cO(L^{-1})
\end{align}
Now we return to~\eqref{eq:ICproo1f} and write the $\tau_2$ integral as the limit of a definite integral from $0$ to $L$. Combining the results above, we find,
\begin{align}
\label{eq:ICproof2}
    \intML \cC
    & \approx
    \lim_{L \to \infty}
    \bigg(
    \int_0^L \frac{d\tau_2}{\tau_2^2}
    \int_0^1 d\tau_1 \,
    \Lambda_\cC
    - \mfc_\cC^{(1)}
    \int_0^L \frac{d\tau_2}{\tau_2^2}
    \int_0^1 d\tau_1 \,
    \Lambda_{\cE_1}
    - \sum_{\ell=2}^{w}
    \mfc_\cC^{(\ell)} \,
    \frac{L^{\ell-1}}{\ell-1}
    \bigg)
\no \\
    & \quad 
    + \mfc_\cC^{(1)}
    \bigg(
    \frac{\zeta'(4)}{\zeta(4)}
    - \frac{\zeta'(3)}{\zeta(3)}
    + \ln(2L)
    \bigg)
\end{align}
Finally, we add and subtract \smash{$\mfc_\cC^{(1)} \ln(2L)$} within the limit in~\eqref{eq:ICproof2} and use~\eqref{eq:IE1} of~\autoref{thm:IE1} to arrive at~\eqref{eq:Iunfold}. This completes our proof of \autoref{thm:Iunfold}.

\newpage

\section{Integrating two-loop modular graph functions}
\label{sec:I2loops}

In this section, we shall evaluate the integrals of two-loop modular graph functions and discuss their transcendental structure. In \autoref{sec:2loopMGF}, we proved that the space of two-loop MGFs with non-negative integer exponents is spanned by the disconnected functions~\smash{$\cV_{s,t}^{(n)}$} defined in~\eqref{eq:V} and the connected functions~$\cC_{u,v;w}$ defined in~\eqref{eq:Cuvw}. We shall first evaluate the integral of \smash{$\cV_{s,t}^{(n)}$} using Stokes' theorem. Then we shall evaluate the integral of $\cC_{u,v;w}$ using the unfolding trick.

\subsection{The integral of \texorpdfstring{$\cV_{s,t}^{(n)}$}{}}

The infinite family of functions $\cV_{s,t}^{(n)}$ with $\Re(s), \Re(t) > 1$ and integer $n \geq 0$ spans the space of disconnected two-loop MGFs. Before specializing to integer $s, t \geq 2$, we shall consider complex non-integer values of $s,t$. In~\eqref{eq:IEEMaass}, we calculated the integral of the double product~$E^*_s E^*_t$ with $\Re(s), \Re(t) > 1$ and found that it vanishes up to non-zero powers of $L$ for integer~$s, t \geq 2$ unless $s=t$. This previous result is a special case of the following lemma.

\begin{lem}
\label{thm:IV}
The integral of $\cV_{s,t}^{(n)}$ with $\Re(s), \Re(t) > 1$ and integer $n \geq 0$ is given by the analytic continuation in $s$ and $t$ of the following expression,
\begin{align}
\label{eq:IV}
    \smash[b]{\intML \cV_{s,t}^{(n)}}
    &=
    4 \,
    \bigg\{
    \frac{ \zeta^*(2s) }{ \Gamma(s) }
    \frac{ \zeta^*(2t) }{ \Gamma(t) }
    \frac{ L^{s+t-1} }{ s+t-1 }
\no \\[1ex]
    \span
    +
    \frac{ \zeta^*(2s) }{ \Gamma(s) }
    \frac{ \zeta^*(2t-1) }{ \Gamma(t+n) }
    \frac{ \Gamma(1-t+n) }{ \Gamma(1-t) }
    \frac{ L^{s-t} }{ s-t }
    +
    \frac{ \zeta^*(2t) }{ \Gamma(t) }
    \frac{ \zeta^*(2s-1) }{ \Gamma(s+n) }
    \frac{ \Gamma(1-s+n) }{ \Gamma(1-s) }
    \frac{ L^{t-s} }{ t-s }
\no \\[1ex]
    & \quad
    +
    \frac{ \zeta^*(2s-1) }{ \Gamma(s+n) }
    \frac{ \zeta^*(2t-1) }{ \Gamma(t+n) }
    \frac{ \Gamma(1-s+n) }{ \Gamma(1-s) }
    \frac{ \Gamma(1-t+n) }{ \Gamma(1-t) }
    \frac{ L^{1-s-t} }{ 1-s-t }
    \bigg\}
\intertext{This integral has $L$-independent or $\ln L$ terms only when $s = t$. In this case,}
\label{eq:IVss}
    \smash[b]{\intML \cV_{s,s}^{(n)}}
    &=
    4 \,
    \frac{ \zeta^*(2s)^2 }{ \Gamma(s)^2 }
    \frac{ L^{2s-1} }{ 2s-1 }
    +
    4 \,
    \frac{ \zeta^*(2s-1)^2 }{ \Gamma(s+n)^2 }
    \frac{ \Gamma(1-s+n)^2 }{ \Gamma(1-s)^2 }
    \frac{ L^{1-2s} }{ 1-2s }
\\[1ex]
    & \quad
    + 8 \,
    \frac{ \zeta^*(2s) }{ \Gamma(s) }
    \frac{ \zeta^*(2s-1) }{ \Gamma(s+n) }
    \frac{ \Gamma(1-s+n) }{ \Gamma(1-s) }
\no \\
    & \qquad
    \times
    \[
    \tfrac{\zeta'(2s)}{\zeta(2s)}
    - \tfrac{\zeta'(2s-1)}{\zeta(2s-1)}
    - \psi(2s-1)
    + \half \psi(s+n)
    + \half \psi(s-n)
    + \ln(2L)
    \]
\no
\end{align}
where $\psi(z) = \Gamma'(z) / \Gamma(z)$ is the digamma function.
\end{lem}

To prove this lemma, we first consider $s \neq t$. We use~\eqref{eq:VE} and~\eqref{eq:CDkC} to write~\smash{$\cV_{s,t}^{(n)}$} in terms of $E_s^* E_t^*$ plus total derivative terms. The integral of $E_s^* E_t^*$ is given by~\eqref{eq:IEEMaass}. The total derivative terms may be integrated using the results on the last two lines of~\eqref{eq:IDC} and the asymptotic expansion~\eqref{eq:EAsy} of the Eisenstein series. After some straightforward simplifications, we arrive at~\eqref{eq:IV}, which agrees with~\eqref{eq:IEEMaass} when $n=0$. The result for~$s = t$ follows by analytic continuation. In this case, the singular terms on the second and third lines must cancel, generating constant and $\ln L$ contributions. We explicitly compute the limit~$t \to s$ using the properties of the starred zeta function and the digamma function given in \autoref{apdx:zeta}. After taking this limit, we use the duplication formula for the digamma function, $\psi(z) + \psi(z-\half) = 2 \, \psi(2z-1) - 2 \ln 2$, to find~\eqref{eq:IVss}. This completes our proof.

\sm

We shall now restrict to integer $s, t \geq 2$. In this case, we may write the integral in a form which makes its transcendental structure manifest.
\begin{thm}
\label{thm:IVint}
The integral of \smash{$\cV_{s,t}^{(n)}$} with integer $s,t \geq 2$ and $n \geq 0$ is given as follows.
\begin{itemize}
\item When $s \neq t$, the integral vanishes up to non-zero powers of $L$.
\item When $s = t \leq n$, the integral has transcendental weight $2s$ and is given by,
\begin{align}
\label{eq:IGG}
    \intML
    \cV_{s,s}^{(n)}
    & \approx
    8 \pi \, 
    \zeta(2s-1) \,
    \frac{ B_{2s} }{ (2s)! }
    \frac{ (2s-2)! (n-s)! }{ (n+s-1)! }
\intertext{\item When $s = t > n$, the integral has transcendental weight $2s+1$ and is given by,}
\label{eq:IVint}
    \intML
    \cV_{s,s}^{(n)}
    & \approx
    (-)^{s+n-1} \,
    16 \pi \, 
    \zeta(2s-1) \,
    \frac{ B_{2s} }{ (2s)! }
    \frac{ (2s-2)! }{ (s-n-1)! (s+n-1)! }
\\
    & \quad
    \times
    \[
    \tfrac{\zeta'(2s)}{\zeta(2s)}
    - \tfrac{\zeta'(2s-1)}{\zeta(2s-1)}
    - H_1(2s-1)
    + \half H_1(s+n)
    + \half H_1(s-n)
    + \ln(2L)
    \]
\no
\end{align}
where $H_1(m) = \sum_{k=1}^{m-1} \frac{1}{k}$ are finite harmonic sums. 
\end{itemize}
In each case, the integral of \smash{$\cV_{s,t}^{(n)}$} is of the form claimed in \autoref{thm:2looptrans}.
\end{thm}

To prove this theorem, we only need to consider the case $s=t$. In this case, we begin with~\eqref{eq:IVss} and use the properties in \autoref{apdx:zeta} to manipulate the starred zeta functions and digamma functions in the limit of integer~$s$. When $s = t \leq n$, we use,
\begin{align}
    \lim_{z \to s} \psi(z-n) / \Gamma(1-z)
    =
    (-)^{s-1} \, (s-1)!
\end{align}
and find~\eqref{eq:IGG}. When $s=t=n$, the integrand is given by \smash{$\cV_{n,n}^{(n)} = \tau_2^{2n} \, G_{2n} \, \bar{G}_{2n}$}, and our result agrees with Zagier's calculation of the same integral in~\cite{Zagier:1982}. When $s = t > n$, we find~\eqref{eq:IVint}.

\subsection{The integral of \texorpdfstring{$\cC_{u,v;w}$}{}}

We shall now evaluate the integral of the connected two-loop modular graph function $\cC_{u,v;w}$. This integral was calculated in~\cite{DHoker:2019mib} using the unfolding trick. This calculation made explicit use of the Laurent coefficients of $\cC_{u,v;w}$ which were calculated in~\cite{DHoker:2019txf}.

\sm

Our calculation will similarly use the unfolding trick but will not rely on a priori knowledge of the Laurent coefficients. Instead, we shall compute the integral of $\cC_{u,v;w}$ using \autoref{thm:Iunfold}. In the remainder of this subsection, we shall prove the following theorem.
\begin{thm}
\label{thm:ICuvw}
The integral of $\cC_{u,v;w}$ with integer $u,v,w$ satisfying ${1 \leq u,v \leq w-1}$ and ${u+v \geq 3}$ is given as follows. 
\begin{itemize}
\item For odd $w < u+v$, the integral vanishes up to non-zero powers of $L$.
\item For odd $w \geq u+v$, the integral has transcendental weight $w+1$ and is given by,
\begin{align}
\label{eq:ICuvwO}
    \intML
    \cC_{u,v;w}
    & \approx
    2 \pi \, \zeta(w) \,
    \frac{ B_{w-1} }{ (w-1)! }
    \frac{ (w-2)! }{ (u-1)! (v-1)! (w-u-v)! }
\intertext{\item For even $w < u+v$, the integral has transcendental weight $w$ and is given by,}
\label{eq:ICuvwE<}
    \intML
    \cC_{u,v;w}
    & \approx
    (-)^{u+v-1} \,
    8 \pi \, \zeta(w-1) \,
    \frac{ B_{w} }{ w! }
    \frac{ (w-2)! (u+v-w-1)! }{ (u-1)! (v-1)! }
\intertext{\item For even $w \geq u+v$, the integral has transcendental weight $w+1$ and is given by,}
\label{eq:ICuvwE>}
    \intML
    \cC_{u,v;w}
    & \approx
    - 8 \pi \, \zeta(w-1) \,
    \frac{ B_{w} }{ w! }
    \frac{ (w-2)! }{ (u-1)! (v-1)! (w-u-v)! }
\\
    & \quad
    \times
    \[
    \tfrac{\zeta'(w)}{\zeta(w)}
    - \tfrac{\zeta'(w-1)}{\zeta(w-1)}
    - H_1(w-1)
    + H_1(w-u-v+1)
    + \ln(2L)
    \]
\no
\end{align}
where $H_1(m) = \sum_{k=1}^{m-1} \frac{1}{k}$ are finite harmonic sums. 
\end{itemize}
In each case, the integral of $\cC_{u,v;w}$ is of the form claimed in \autoref{thm:2looptrans}.
\end{thm}

Before we prove this theorem, we note that the expressions for these integrals are compatible with the system of differential equations~\eqref{eq:DCuvw} obeyed by the functions $\cC_{u,v;w}$.

\subsubsection{The Poincar\'e seed $\Lambda_{u,v;w}$}

Following \autoref{thm:Iunfold}, we shall compute the integral of $\cC_{u,v;w}$ by integrating its Poincar\'e seed function $\Lambda_{u,v;w}$ over the truncated upper half-strip. This calculation is similar to the proof of \autoref{thm:IE1}.

\sm

We first obtain an expression for $\Lambda_{u,v;w}$ from the Kronecker-Eisenstein series representation~\eqref{eq:Cuvw} of $\cC_{u,v;w}$ by rotating the integer pair $(m_3,n_3) \neq (0,0)$ to $(0,N) \neq (0,0)$,
\begin{align}
    \Lambda_{u,v;w}
    =
    \frac{\tau_2^{w}}{\pi^{w}}
    \sum_{N \neq 0}
    \,
    \sum^{'}_{\substack{(m_r, n_r) \in \bbZ^2
        \\ r = 1,2}}
    \frac{\delta(m_1+m_2) \delta(n_1+n_2+N)}
        { (m_1\tau + n_1)^u (m_2\bar{\tau} + n_2)^v N^{2w-u-v}}
\end{align}
Splitting \smash{$\Lambda_{u,v;w}= \Lambda^{[0]}_{u,v;w} + \Lambda^{[1]}_{u,v;w} + \Lambda^{[2]}_{u,v;w}$}
into contributions according to the number of non-vanishing summation variables $m_r$, we see that \smash{$\Lambda^{[1]}_{u,v;w}$} vanishes thanks to the delta function constraint. The remaining contributions are given by,
\begin{align}
    \Lambda^{[0]}_{u,v;w}
    &=
    \frac{\tau_2^{w}}{\pi^{w}}
    \sum_{n_1, n_2, N \neq 0}
    \frac{\delta(n_1+n_2+N)}
        {n_1^u n_2^v N^{2w-u-v}}
\no \\
    \Lambda^{[2]}_{u,v;w}
    &=
    \frac{\tau_2^{w}}{\pi^{w}}
    \sum_{m_1, m_2, N \neq 0}
    \,
    \sum_{n_1, n_2 \in \bbZ}
    \frac{\delta(m_1+m_2) \delta(n_1+n_2+N)}
        {(m_1\tau + n_1)^u (m_2\bar{\tau} + n_2)^v N^{2w-u-v}}
\end{align}
The \smash{$\Lambda^{[0]}_{u,v;w}$} contribution yields a term proportional to $L^{w-1}$ in the integral of $\Lambda_{u,v;w}$ and may be ignored in light of \autoref{thm:Iunfold}. This leaves only \smash{$\Lambda^{[2]}_{u,v;w}$}. Hence,
\begin{align}
    \intML \cC_{u,v;w}
    \approx
    \int_0^L \frac{d\tau_2}{\tau_2^2}
    \int_0^1 d\tau_1 \,
    \Lambda^{[2]}_{u,v;w}
\end{align}
To evaluate \smash{$\Lambda^{[2]}_{u,v;w}$}, we first sum over the variables $m_2$ and $n_2$ using the two delta symbols. We then define~${m=m_1}$ and~$n=n_1$ and find,
\begin{align}
    \Lambda^{[2]}_{u,v;w}
    &=
    \frac{\tau_2^{w}}{\pi^{w}}
    \sum_{m, N \neq 0}
    \,
    \sum_{n \in \bbZ}
    \frac{ (-)^v }{ (m\tau+n)^u (m\bar{\tau}+n+N)^v N^{2w-u-v} }
\end{align}
The sum over $n$ may be carried out by partial fraction decomposition and the following standard summation formula,
\begin{align}
    \sum_{n \in \bbZ}
    \frac{1}{ (z+n)^{k+1} }
    =
    i \pi
    \frac{ (-)^k }{ k! }
    \frac{d^k}{dz^k}
    \(
    \frac{ 1 + e^{2\pi i z} }{ 1 - e^{2\pi i z} }
    \)
\end{align}
Integrating over $\tau_1$ projects onto the constant Fourier mode which contributes only when $k=0$. Thus we find,
\begin{align}
    \int_0^1 d\tau_1 \,
    \sum_{n \in \bbZ}
    \frac{1}{ (m\tau+n)^u (m\bar{\tau}+n+N)^v }
    =
    \tbinom{ u+v-2 }{ u-1 } \,
    \frac{ 2 \pi i \, \veps(m) \, (-)^{v-1} }
         { (2im\tau_2 - N)^{u+v-1} }
\end{align}
where $\veps(m) = \pm 1$ is equal to the sign of $m$. We now integrate over $\tau_2$ and find,
\begin{align}
    \int_0^L \frac{d\tau_2}{\tau_2^2}
    \int_0^1 d\tau_1 \,
    \Lambda^{[2]}_{u,v;w}
    & =
    - \frac{ 4i }{ \pi^{w-1} }
    \tbinom{ u+v-2 }{ u-1 }
    \sum_{m, N > 0}
    \frac{ 1 }{ N^{2w-u-v} }
\\
    & \quad
    \times
    \int_0^L
    d\tau_2^{\mathstrut} \,
    \tau_2^{w-2}
    \(
    \frac{ 1 }{ (2im\tau_2 - N)^{u+v-1} }
    +
    \frac{ (-)^{u+v} }{ (2im\tau_2 + N)^{u+v-1} }
    \)
\no
\end{align}
After some straightforward simplifications and a change of integration variables, we obtain the following expression,
\begin{align}
\label{eq:IL2uvw}
    \intML \cC_{u,v;w}
    & \approx
    \frac{ 8 }{ (2\pi)^{w-1} }
    \sum_{\ell=0}^{\floor{\frac{u+v-2}{2}}}
    \tbinom{ u+v-2 }{ u-1 }
    \tbinom{u+v-1}{2\ell+1} \,
    (-)^{u+v+\ell-1}
\no \\
    & \quad
    \times
    \sum_{m,N>0}
    \frac{1}{ m^{w-1} N^w }
    \int_0^{\frac{2mL}{N}}
    \frac{ dx \, x^{w+2\ell-1} }{ (x^2+1)^{u+v-1} }
\end{align}
This expression includes an infinite series of finite integrals which may be compactly expressed in terms of a more general family of functions.

\subsubsection{\texorpdfstring{$F_g$}{} functions}

Before we introduce the general family of functions $F_g$, we shall briefly review the aspects of arithmetic functions and Dirichlet series which are needed here.

\sm

An arithmetic function $f: \bbZ^+ \to \bbC$ is defined for positive integers $m$. For example, the constant (or trivial) function $\mathds{1} : m \mapsto 1$ and the finite harmonic series~\smash{${H_1 : m \mapsto \sum_{k=1}^{m-1} \frac{1}{k}}$} are both arithmetic functions.

\sm

For an arithmetic function $f$, the Dirichlet series $\zeta_f(s)$ and its derivative $\zeta_f'(s)$ are functions of $s \in \bbC$ defined by the following sums when absolutely convergent and elsewhere by analytic continuation in $s$,
\begin{align}
    \zeta_f(s)
    &=
    \sum_{m>0}
    \frac{f(m)}{m^s}
    &&
    \zeta_f'(s)
    =
    -
    \sum_{m>0}
    \frac{f(m)}{m^s}
    \ln(m) 
\end{align}
In particular, $\zeta_{\mathds{1}}(s) = \zeta(s)$ is the Riemann zeta function, and $\zeta_{H_1}(s) = \zeta(s,1)$ is a double zeta function. Both converge for $\Re(s) > 1$.

\sm

We shall now define the general family of functions $F_g(a,b;c,d;f;L)$ by the following infinite series of integrals,
\begin{align}
\label{eq:Fgdef}
    F_g(a,b;c,d;f;L)
    =
    (-)^{a+b+1}
    \sum_{m,n > 0}
    \frac{f(m)}{m^c n^d}
    \int_0^{\frac{2mL}{n}}
    \frac{dx \, x^{2a+g}}{(x^2+1)^b}
\end{align}
for $g=0$ or $1$, integers $a, b \geq 0$, real numbers $c,d > 1$, an arithmetic function $f$, and a finite cut-off $L > 1$. The sign prefactor is included for later convenience. The convergence properties and large-$L$ behavior of $F_g$ are given by the following lemma.

\begin{lem}
\label{thm:Fg}
For fixed $L > 1$, the sums in~\eqref{eq:Fgdef} which define $F_g$ converge if the following Dirichlet series is absolutely convergent,
\begin{align}
    \zeta_f \big( c - 2 (a-b) \Theta(a-b) - g - 1 \big)
\end{align}
where the step function is defined by $\Theta(x \geq 0) = 1$ and $\Theta(x < 0) = 0$.

\sm

When $F_g$ converges, its large-$L$ behavior up to powers of $L$ with non-zero exponents and terms which vanish in the limit $L \to \infty$ is given as follows,
\begin{align}
\label{eq:Fg}
    F_0(a,b;c,d;f;L)
    & \approx
    \pi \, \zeta_f(c) \, \zeta(d)
    \sum_{\ell=0}^{b-1}
    \tbinom{a}{b-1-\ell}
    \tbinom{2\ell}{\ell}
    \frac{ (-)^\ell }{ 2^{2\ell+1} }
\no \\
    F_1(a,b;c,d;f;L)
    & \approx
    \zeta_f(c) \, \zeta(d) \,
    \bigg\{
    \tbinom{a}{b-1}
    \Big[
    \tfrac{\zeta'_{\mathstrut}(d)}{\zeta_{\mathstrut}(d)}
    - \tfrac{\zeta_f'(c)}{\zeta_f(c)}
    + \ln(2L)
    \Big]
    +
    \half
    \displaystyle\sum_{\ell=1}^{b-1}
    \tbinom{a}{b-1-\ell}
    \frac{ (-)^\ell }{ \ell }
    \bigg\}
\end{align}
where the $\approx$ notation was introduced in~\eqref{eq:approx}.
\end{lem}

We shall begin our proof with the convergence condition on $F_g$. We may bound the integral in~\eqref{eq:Fgdef} as follows,
\begin{align}
    \int_0^{\frac{2mL}{n}}
    \frac{dx \, x^{2a+g}}{(x^2+1)^b}
    <
    \int_0^{\frac{2mL}{n}}
    dx \, x^{2(a-b) \Theta(a-b) + g}
    =
    \frac{(2mL/n)^{2(a-b) \Theta(a-b) + g + 1}}
        {2(a-b) \Theta(a-b) + g + 1}
\end{align}
Therefore, the sums in~\eqref{eq:Fgdef} converge if the following sums are convergent,
\begin{align}
    \sum_{m,n > 0}
    \frac{f(m)}{m^c n^d}
    \( \frac{m}{n} \)^{2(a-b) \Theta(a-b) + g + 1}
    &=
    \zeta_f \big( c - 2(a-b) \Theta(a-b) - g - 1 \big)
    \no \\[-2ex]
    & \quad \times
    \zeta \big( d + 2(a-b) \Theta(a-b) + g + 1 \big)
\end{align}
Since $d > 1$ by assumption, the sum over $n$ converges, leaving our claimed condition on the Dirichlet series. This convergence condition is sufficient but not strictly necessary. However, every instance of $F_g$ which appears in this paper will satisfy this condition.

\sm

To prove the remainder of \autoref{thm:Fg}, we decompose the integrand in the definition~\eqref{eq:Fgdef} into a sum of polynomials in $x$ plus terms of the form $x^g / (x^2+1)^\ell$ for positive integer~$\ell$ using a generalization of the finite geometric series formula. For integers $a, b \geq 0$,
\begin{align}
\label{eq:Fproof2}
    (-)^{a+b+1} \frac{x^{2a}}{(x^2+1)^b}
    =
    \sum_{\ell=0}^{b-1}
    \tbinom{a}{b-1-\ell}
    \frac{(-)^\ell}{(x^2+1)^{\ell+1}}
    + \sum_{\ell=0}^{a-b}
    \tbinom{a-1-\ell}{b-1}
    (-)^{\ell+1} x^{2\ell}
\end{align}
Using this identity, the definition of $F_g$ in~\eqref{eq:Fgdef} becomes,
\begin{align}
\label{eq:Fproof3}
F_g=    \sum_{m,n > 0}
    \frac{f(m)}{m^c n^d}
    \int_0^{\frac{2mL}{n}} dx
    \[
    \sum_{\ell=0}^{b-1}
    \tbinom{a}{b-1-\ell}
    \frac{(-)^\ell \, x^g}{(x^2+1)^{\ell+1}}
    + \sum_{\ell=0}^{a-b}
    \tbinom{a-1-\ell}{b-1}
    (-)^{\ell+1} \,
    x^{2\ell+g}
    \]
\end{align}
Integrating the polynomial terms in~\eqref{eq:Fproof3} at fixed $m$, $n$, and $L$ and then performing the sums over $m$ and $n$ produces terms which are proportional to positive powers of $L$. We define a function $\tilde F_g$ by subtracting these terms as follows,
\begin{align}
\label{eq:Fproof4}
    \tilde F_g
    &=
    F_g
    + \sum_{\ell=0}^{a-b}
    \tbinom{a-1-\ell}{b-1}
    (-)^{\ell} \,
    \frac{ (2L)^{2\ell+g+1}}{2\ell+g+1}
    \,
    \zeta_f(c-2\ell-g-1) \,
    \zeta(d+2\ell+g+1)
    \no \\
    &=
    \sum_{\ell=0}^{b-1}
    \tbinom{a}{b-1-\ell}
    (-)^\ell
    \sum_{m,n > 0}
    \frac{f(m)}{m^c n^d}
    \int_0^{\frac{2mL}{n}}
    \frac{dx \,x^g}{(x^2+1)^{\ell+1}}
\end{align}
The zeta functions in this expression converge if $F_g$ converges.

\sm

When $g=0$, the integrand in $\tilde F_0$ decays like $x^{-2}$ for large $x$, so $\tilde F_0$ converges as $L \to \infty$. In this case, the asymptotic expansion in~\eqref{eq:Fg} follows after using the integral,
\begin{align}
    \int_0^{\infty} \frac{dx}{(x^2+1)^{\ell+1}}
    =
    \frac{\pi}{2^{2\ell+1}}
    \binom{2\ell}{\ell}
\end{align}
When $g=1$, the integrand in $\tilde F_1$ decays like $x^{-1}$  for large $x$, so $\tilde F_1$ diverges like $\ln L$ as~$L \to \infty$. Isolating the $\ell=0$ term of~\eqref{eq:Fproof4} and evaluating the integral, we find,
\begin{align}
    \tilde F_1(a,b;c,d;f;L)
    \big|_{\ell=0}
    =
    \tbinom{a}{b-1}
    \sum_{m,n > 0}
    \frac{f(m)}{m^c n^d}
    \times
    \tfrac{1}{2}
    \ln \Big( (2mL/n)^2 + 1 \Big)
\end{align}
Subtracting the $\ln L$ divergence, taking the limit $L \to \infty$, and performing the sums over $m$ and $n$, we find,
\begin{align}
    \lim_{L \to \infty}
    \[
    \tilde F_1
    \big|_{\ell=0}
    - \tbinom{a}{b-1} \,
    \zeta_f(c) \,
    \zeta(d)
    \ln(2L)
    \]
    =
    \tbinom{a}{b-1} \,
    \zeta_f(c) \,
    \zeta(d)
    \[
    \tfrac{\zeta'_{\mathstrut}(d)}{\zeta_{\mathstrut}(d)}
    - \tfrac{\zeta_f'(c)}{\zeta_f(c)}
    \]
\end{align}
We return to~\eqref{eq:Fproof4}. The $\ell > 0$ terms are finite as $L \to \infty$ and may be evaluated using,
\begin{align}
    \int_0^{\infty} \frac{dx \, x}{(x^2+1)^{\ell+1}}
    =
    \frac{1}{2\ell}
\end{align}
Combining these results, we arrive at the asymptotic expansion in~\eqref{eq:Fg} for the case $g=1$. This completes our proof of \autoref{thm:Fg}.

\sm

With the general $F_g$ functions in hand, we shall now return to the integral of~$\cC_{u,v;w}$. We may clearly write the right-hand side of~\eqref{eq:IL2uvw} in terms of the $F_g$ functions. It will be convenient, however, to separately consider the cases of odd and even $w$.

\subsubsection{Odd $w$}

First we consider odd $w = 2\kappa+1$ with integer $\kappa \geq 1$. In this case, we may write~\eqref{eq:IL2uvw} in terms of the functions $F_0(\kappa+\ell, u+v-1; w-1; w; \mathds{1}; L)$. The large-$L$ behavior of $F_0$ is given by \autoref{thm:Fg}. After some simplifications, we find,
\begin{align}
    \intML \cC_{u,v;w}
    & \approx
    2 \pi \, \zeta(w) \,
    \frac{ B_{w-1} }{ (w-1)! } \,
    \cR_{u,v;w}^{(\text{odd})}
\end{align}
where we have written the even zeta-value $\zeta(w-1)$ in terms of a Bernoulli number and defined the rational number $\cR$ by,
\begin{align}
    \cR_{u,v;w}^{(\text{odd})}
    =
    \sum_{\ell=0}^{\floor{\frac{u+v-2}{2}}}
    \tbinom{ u+v-2 }{ u-1 }
    \tbinom{u+v-1}{2\ell+1}
    \sum_{j=0}^{u+v-2}
    \tbinom{\kappa+\ell}{u+v-2-j}
    \tbinom{2j}{j}
    \frac{ (-)^j }{ 2^{2j} }
\end{align}
To evaluate the sum over $j$, we use the following identity for integers $a, b \geq 0$,
\begin{align}
    \sum_{j=0}^{b-1}
    \tbinom{a}{b-1-j}
    \tbinom{2j}{j}
    \frac{ (-)^j }{ 2^{2j} }
    =
    \frac{ \Gamma(a+\half) }
         { \Gamma(b) \Gamma(a-b+\tfrac{3}{2}) }
\end{align}
We then find,
\begin{align}
    \cR_{u,v;w}^{(\text{odd})}
    =
    \frac{ (u+v-1)! }{ (u-1)! (v-1)! }
    \sum_{\ell=0}^{\floor{\frac{u+v-2}{2}}}
    \frac{ \Gamma(\kappa+\ell+\half) }{ \Gamma(2\ell+2) \Gamma(u+v-2\ell-1) \Gamma(\kappa+\ell-u-v+\tfrac{5}{2}) }
\end{align}
The sum over $\ell$ may be computed by separately considering the two cases of $u+v$ even or odd. In either case, we find,
\begin{align}
    \cR_{u,v;w}^{(\text{odd})}
    =
    \frac{ (w-2)! }{ (u-1)! (v-1)! (w-u-v)! }
\end{align}
which vanishes for $u+v > w$. This completes our proof of \autoref{thm:ICuvw} for odd~$w$.

\subsubsection{Even $w$}

Now we consider even $w = 2\kappa$ with integer $\kappa \geq 2$. In this case, we may write~\eqref{eq:IL2uvw} in terms of the functions $F_1(\kappa+\ell-1, u+v-1; w-1; w; \mathds{1}; L)$. The large-$L$ behavior is again given by \autoref{thm:Fg}. After some simplifications, we find,
\begin{align}
    \intML \cC_{u,v;w}
    & \approx
    - 8 \pi \, \zeta(w-1) \,
    \frac{ B_{w} }{ w! }
    \(
    \cR_{u,v;w}^{(\text{even})}
    \[
    \tfrac{\zeta'(w)}{\zeta(w)}
    - \tfrac{\zeta'(w-1)}{\zeta(w-1)}
    + \ln(2L)
    \]
    + \cH_{u,v;w}
    \)
\end{align}
where we have written the even zeta-value $\zeta(w)$ in terms of a Bernoulli number and defined the rational numbers $\cR$ and $\cH$ by,
\begin{align}
    \cR_{u,v;w}^{(\text{even})}
    &=
    \sum_{\ell=0}^{\floor{\frac{u+v-2}{2}}}
    \tbinom{ u+v-2 }{ u-1 }
    \tbinom{u+v-1}{2\ell+1}
    \tbinom{\kappa+\ell-1}{u+v-2}
\no \\
    \cH_{u,v;w}
    &=
    \half
    \sum_{\ell=0}^{\floor{\frac{u+v-2}{2}}}
    \tbinom{ u+v-2 }{ u-1 }
    \tbinom{u+v-1}{2\ell+1}
    \sum_{j=1}^{u+v-2}
    \tbinom{\kappa+\ell-1}{u+v-2-j}
    \frac{ (-)^j }{ j }
\end{align}
We may compute $\cR$ by separately considering the two cases of $u+v$ even or odd. In either case, we find,
\begin{align}
    \cR_{u,v;w}^{(\text{even})}
    =
    \frac{ (w-2)! }{ (u-1)! (v-1)! (w-u-v)! }
\end{align}
which vanishes for $u+v > w$. When $u + v \leq w$, $\cH$ is an integer times the difference of two finite harmonic sums,
\begin{align}
    \cH_{u,v;w}
    =
    - \frac{ (w-2)! }{ (u-1)! (v-1)! (w-u-v)! }
    \big[ 
    H_1(w-1) - H_1(w-u-v+1)
    \big]
\end{align}
When $u + v > w$, $\cH$ is the following rational number,
\begin{align}
    \cH_{u,v;w}
    =
    (-)^{u+v} \,
    \frac{ (w-2)! (u+v-w-1)! }{ (u-1)! (v-1)! }
\end{align}
This completes our proof of \autoref{thm:ICuvw} for even~$w$.

\newpage

\section{Integrating triple products of Eisenstein series}
\label{sec:Eisen}

In this section, we shall evaluate the integrals of some of the simplest three-loop modular graph functions, namely triple products of Eisenstein series and their derivatives, and shall obtain their the transcendental structure.

\sm

First we shall consider the infinite family of triple products of non-holomorphic Eisenstein series~$E_s^* E_t^* E_u^*$ with ${\Re(s), \Re(t), \Re(u) > 1}$. These triple products are disconnected three-loop modular graph functions with weight $s + t + u$ and the following decorated graph,
\begin{align}
    E_s^* E_t^* E_u^*
    ~=~
    \tfrac{1}{8} \,
    \Gamma(s) \,
    \Gamma(t) \,
    \Gamma(u) 
    \times~
    \begin{tikzpicture}[baseline=-0.5ex, scale=.7]
    \draw (-1,0) node[circle,fill,inner sep=1.5](a){};
    \draw (-1,0) arc(180:-180:1)
        node[style=draw,midway,fill=white,scale=.8]
        {$s,s$};
    \end{tikzpicture}
    ~~
    \begin{tikzpicture}[baseline=-0.5ex, scale=.7]
    \draw (-1,0) node[circle,fill,inner sep=1.5](a){};
    \draw (-1,0) arc(180:-180:1)
        node[style=draw,midway,fill=white,scale=.8]
        {$t,t$};
    \end{tikzpicture}
    ~~
    \begin{tikzpicture}[baseline=-0.5ex, scale=.7]
    \draw (-1,0) node[circle,fill,inner sep=1.5](a){};
    \draw (-1,0) arc(180:-180:1)
        node[style=draw,midway,fill=white,scale=.8]
        {$u,u$};
    \end{tikzpicture}
\end{align}
where the prefactor arises from the normalization of the starred Eisenstein series relative to the un-starred Eisenstein series.

\sm

We shall also consider the infinite family of disconnected three-loop modular graph functions~\smash{$\cW_s^{(m,n)}$} with $\Re(s) \geq 6$ and integers $m,n$ which satisfy \smash{${2 \leq m,n \leq \floor{\Re(s)/2}-1}$}. These functions are equal to a product of three one-loop modular graph forms and have the following decorated graph and matrix of exponents,
\begin{align}
\label{eq:W}
    \cW_s^{(m,n)}
    ~&=~
    \begin{tikzpicture}[baseline=-0.5ex, scale=.7]
    \draw (-1,0) node[circle,fill,inner sep=1.5](a){};
    \draw (-1,0) arc(180:-180:1)
        node[style=draw,midway,fill=white,scale=.6,xshift=-10]
        {$2m,0$};
    \end{tikzpicture}
    ~~
     \begin{tikzpicture}[baseline=-0.5ex, scale=.7]
    \draw (-1,0) node[circle,fill,inner sep=1.5](a){};
    \draw (-1,0) arc(180:-180:1)
        node[style=draw,midway,fill=white,scale=.6,xshift=-10]
        {$0,2n$};
    \end{tikzpicture}
    ~~
    \begin{tikzpicture}[baseline=-0.5ex, scale=.7]
    \draw (-1,0) node[circle,fill,inner sep=1.5](a){};
    \draw (-1,0) arc(180:-180:1)
        node[style=draw,midway,fill=white,scale=.6,xshift=-10]
        {$s-2m , s-2n$};
    \end{tikzpicture}
    ~=~
    \cC
    \! \begin{bmatrix}
    2m & 0 \\ 0 & 0
    \end{bmatrix}
    \cC
    \! \begin{bmatrix}
    0 & 0 \\ 2n & 0
    \end{bmatrix}
    \cC
    \! \begin{bmatrix}
    s-2m & 0 \\
    s-2n & 0
    \end{bmatrix}
\end{align}
The functions \smash{$\cW_s^{(m,n)}$} have weight $s$ and may be written in terms of derivatives acting on three Eisenstein series as follows,
\begin{align}
\label{eq:WE}
    \cW_s^{(m,n)}
    &=
    8 \,
    \Gamma(2m)^{-1} \,
    \Gamma(2n)^{-1} \,
    \nabla^{m} E_{m}^*
    \bar{\nabla}^{n} E_{n}^*
\no \\[1ex]
    & \quad
    \times
    \begin{cases}
    \Gamma(s-2m)^{-1} \,
    \tau_2^{ -2n } \,
    \nabla^{n-m} E_{s-m-n}^*
    & \quad
    m \leq n
    \\[2ex]
    \Gamma(s-2n)^{-1} \,
    \tau_2^{ -2m } \,
    \bar{\nabla}^{m-n} E_{s-m-n}^*
    & \quad
    m \geq n
    \end{cases}
\end{align}
The first two factors in~\eqref{eq:W} are respectively proportional to the holomorphic and anti-holomorphic Eisenstein series $G_{2m}$ and $\bar{G}_{2n}$, and we restrict to $m, n \geq 2$ so that their Kronecker-Eisenstein series are absolutely convergent. 

\subsection{The integral of \texorpdfstring{$E_s^* E_t^* E_u^*$}{}}

The integral of the triple product $E_s^* E_t^* E_u^*$ with ${\Re(s), \Re(t), \Re(u) > 1}$ was calculated by Zagier in~\cite{Zagier:1982} using the unfolding trick. This integral is given by the analytic continuation in $s,t,u$ of the following expression,
\begin{align}
\label{eq:IEEEZagier}
    \intML E^*_s E^*_t E^*_u
    &=
    \zeta^*(w-1) \,
    \zeta^*(w-2s) \,
    \zeta^*(w-2t) \,
    \zeta^*(w-2u)
\no \\[-1ex]
    & \quad
    +
    \sum_{x=s,1-s} \,
    \sum_{y=t,1-t} \,
    \sum_{z=u,1-u}
    \zeta^*(2x) \,
    \zeta^*(2y) \,
    \zeta^*(2z) \,
    \frac{ L^{x+y+z-1} }{ x+y+z-1 }
\end{align}
where $w=s+t+u$ and the finite sum over $x,y,z$ has eight total terms.

\sm

To analyze the transcendental structure of this integral, we restrict to integer ${s, t, u \geq 2}$ and thus $w \geq 6$. Without loss of generality, we shall choose the ordering ${u \geq t \geq s}$ so that we have ${w - 2s \geq w - 2t \geq 2}$. Thus~\eqref{eq:IEEEZagier} has two singular terms when $2u = w$ or when~${2u = w-1}$. In both cases, the two singular terms precisely cancel, generating $L$-independent and $\ln L$ terms. When $2u$ does not equal $w$ or $w-1$, there are no singular terms, and the integral simply equals the quadruple product of zeta-values on the first line of~\eqref{eq:IEEEZagier} up to non-zero powers of $L$. The following theorem makes these observations precise and writes the integral in a form with manifest transcendental weight.

\begin{thm}
\label{thm:IEEE}
The integral of $E_s^* E_t^* E_u^*$ with integer $u \geq t \geq s \geq 2$ and $w = s + t + u$ is given as follows. In each case, the top entry in the piecewise notation is for $w$ odd, and the bottom entry is for $w$ even.
\begin{itemize}
\item When $u < \floor{w/2}$, the integral has transcendental weight $w$ when $w$ is even and transcendental weight $w+1$ when $w$ is odd. It is given by,
\begin{align}
\label{eq:IEEE<}
    \intML E_s^* E_t^* E_u^*
    & \approx
    2 \pi \,
    (-)^{\floor{\frac{w}{2}}+1} \,
    (w-3)!! \, 
    (w - 2s - 2)!! \,
    (w - 2t - 2)!! \,
    (w - 2u - 2)!!
\no \\
    & \quad \times
    \begin{cases}
    \tfrac{ B_{w-1} }{ (w-1)! } \,
    \zeta(w-2s) \, \zeta(w-2t) \, \zeta(w-2u)
    \\[2ex]
    \tfrac{ B_{w-2s} }{ (w-2s)! }
    \tfrac{ B_{w-2t} }{ (w-2t)! }
    \tfrac{ B_{w-2u} }{ (w-2u)! } \,
    \zeta(w-1)
    \end{cases}
\intertext{\item When $u = \floor{w/2}$, the integral has transcendental weight $w+1$, except for a term proportional to $\ln(2\pi)$ that occurs when $w$ is even. It is given by,} 
\label{eq:IEEE=}
    \intML E_s^* E_t^* E_u^*
    & \approx
    2 \pi \,
    (-)^{\floor{\frac{w+1}{2}}} \,
    (w-3)!! \, 
    (w - 2s - 2)!! \,
    (w - 2t - 2)!!
\\
    & \quad \times
    \begin{cases}
    \tfrac{ B_{w-1} }{ (w-1)! } \,
    \zeta(w-2s) \, \zeta(w-2t)
    \\[1ex]
    \times
    \big[
      \cZ(w-1)
    - \cZ(w-2s)
    - \cZ(w-2t)
    + \gamma_E
    + \ln(2L)
    \big]
    \span
    \\[2ex]
    \tfrac{ B_{w-2s} }{ (w-2s)! }
    \tfrac{ B_{w-2t} }{ (w-2t)! } \,
    \zeta(w-1)
    \\[1ex]
    \times
    \big[
      \cZ(w-1)
    - \cZ(w-2s)
    - \cZ(w-2t)
    + \ln(2\pi)
    - \ln(2L)
    \big]
    \span
    \end{cases}
    \no
\intertext{where $\cZ$ is the combination of the logarithmic derivative of the Riemann zeta function and finite harmonic sums defined in~\eqref{eq:Zndef} and $\gamma_E$ is the Euler-Mascheroni constant.
\item When $u > \floor{w/2}$, the integral has transcendental weight $2u+1$ and is given by,}
\label{eq:IEEE>}
    \intML E_s^* E_t^* E_u^*
    & \approx
    2 \pi \,
    (-)^{u} \,
    (w-3)!! \, 
    (w - 2s - 2)!! \,
    (w - 2t - 2)!! \,
    (2u-w-1)!!
\no \\
    & \quad \times
    \begin{cases}
    \tfrac{ B_{w-1} }{ (w-1)! }
    \tfrac{ B_{2u-w+1} }{ (2u-w+1)! } \,
    \zeta(w-2s) \, \zeta(w-2t)
    \\[2ex]
    \tfrac{ B_{w-2s} }{ (w-2s)! }
    \tfrac{ B_{w-2t} }{ (w-2t)! } \,
    \zeta(w-1) \, \zeta(2u-w+1)
    \end{cases}
\end{align}
\end{itemize}
In each case, the integral of $E_s^* E_t^* E_u^*$ is trivially of the form claimed in \autoref{thm:3looptrans}.
\end{thm}

To prove this theorem, we begin with~\eqref{eq:IEEEZagier} and use the properties of the starred zeta functions in \autoref{apdx:zeta} to compute the appropriate limits. After some algebra we arrive at~\eqref{eq:IEEE<},~\eqref{eq:IEEE=}, and~\eqref{eq:IEEE>}.

\sm

In each case, the integral of $E_s^* E_t^* E_u^*$ is given by the product of a rational number, a single factor of $\pi$, and odd zeta-values. When $u = \floor{w/2}$ there is also a factor whose terms have unit transcendental weight (excluding the pesky $\ln(2\pi)$ term that occurs for $2u=w$). The transcendental structure of this integral may be summarized as follows.
\begin{center}
\begin{tabular}{|l|c|c|} 
    \hline
    \multicolumn{1}{|c|}{trans. weight} & $u$ & $w$
    \\ \hline \hline
    $w+1$ & $u < \floor{w/2}$ & odd
    \\
    $w$ & $u < \floor{w/2}$ & even
    \\
    $w+1$ & $u = \floor{w/2}$ & odd
    \\
    $w+1 \, ^*$ & $u = \floor{w/2}$ & even
    \\
    $2u+1$ & $u > \floor{w/2}$ & odd or even
    \\ \hline
\end{tabular}
\end{center}
\begin{itemize}
    \item When $2u \leq w-1$, the integral has transcendental weight $w$ or $w+1$ which is consistent with the integrals of total derivatives and the integrals of two-loop MGFs.
    \sm
    \item When $2u=w$, the asterisk in the table indicates the presence of $\ln(2\pi)$, which spoils the transcendental structure. The logarithm of a transcendental number cannot be consistently assigned a transcendental weight. Moreover, $\ln(2\pi)$ did not occur in the integrals of any two-loop MGFs and is new to these three-loop integrals.
    \sm
    \item When $2u > w$, the integral has transcendental weight ${2u+1 > w+1}$, a phenomenon which also did not occur at two loops.
\end{itemize}
Compared to the integrals of arbitrary two-loop MGFs, the integral of $E_s^* E_t^* E_u^*$ has novel transcendental structure when ${2u \geq w \geq 8}$. For example, at weight eight the integral of $E_2^* E_2^* E_4^*$ contains $\ln(2\pi)$, and at weight nine the integral of $E_2^* E_2^* E_5^*$ has transcendental weight eleven.

\subsection{The integral of \texorpdfstring{$\cW_s^{(m,n)}$}{}}

We shall now consider the infinite family of disconnected three-loop modular graph functions~\smash{$\cW_s^{(m,n)}$} defined in~\eqref{eq:W}. Before specializing to integer $s$, we shall consider complex~$s$.
\begin{lem}
\label{thm:IW}
The integral of \smash{$\cW_s^{(m,n)}$} with $\Re(s) \geq 6$ and integers ${2 \leq m,n \leq \floor{\Re(s)/2}-1}$ is given by the analytic continuation in $s$ of the following expression,
\begin{align}
\label{eq:IW}
    \intML
    \cW_s^{(m,n)}
    &=
    16 \, (-)^{m+n}
\no \\[-1ex]
    & \quad
    \times
    \bigg\{
    \frac{ \Gamma(s-1) }
         { \Gamma(2m) \Gamma(2n) }
    \frac{ \zeta(s-1) \,
           \zeta(s-2m) \,
           \zeta(s-2n) \,
           \zeta(s-2m-2n+1) }
         { (2\pi)^{ 2s-2m-2n-1 } }
\no \\[1ex]
    & \hspace{-2cm}
    +
    \frac{ \Gamma(2s-2m-2n-1) }
         { \Gamma(s-2m) \Gamma(s-2m) }
    \frac{ \zeta(2m) \,
           \zeta(2n) \, 
           \zeta(2s-2m-2n-1) }
         { (2\pi)^{s-1} }
    \frac{ (2L)^{2m+2n-s} }{ 2m+2n-s }
\no \\[1ex]
    & \hspace{-2cm}
    + (-)^{m+n} \,
    \frac{ \zeta(2m) \,
           \zeta(2n) \,
           \zeta(2s-2m-2n) }
         { (2\pi)^{s} }
    \frac{ (2L)^{s-1} }{ s-1 }
    \bigg\}
\end{align}
The first and second terms on the right-hand side each have simple poles at $s=2m+2n$ which cancel and generate constant and $\ln L$ terms.
\end{lem}

We shall prove this lemma using the unfolding trick. We begin with~\eqref{eq:WE} and the Laurent polynomial~\eqref{eq:EAsy} of the Eisenstein series to find the asymptotic expansion of \smash{$\cW_s^{(m,n)}$} near the cusp,
\begin{align}
\label{eq:WAsy}
    \cW_s^{(m,n)}
    &=
    8 \, 
    \frac{ \zeta^*(2m) }{ \Gamma(m) }
    \frac{ \zeta^*(2n) }{ \Gamma(n) }
    \bigg\{
    \frac{ \zeta^*(2s-2m-2n) }
         { \Gamma(s-m-n) }
    \,
    \tau_2^{s}
 \\
    & \qquad 
    +
    (-)^{m+n} \,
    \Gamma(s-m-n) \, 
    \frac{ \zeta^*(2s-2m-2n-1) }
         { \Gamma(s-2m) \Gamma(s-2n) }
    \,
    \tau_2^{2m+2n-s+1}
    \bigg\}
    +
    \cO( e^{-2\pi \tau_2} )
\no
\end{align}
When ${\Re(s) > 2m + 2n +1}$, the only positive-power part of this asymptotic expansion is proportional to $\tau_2^{s}$. When $\Re(s) \leq 2m + 2n + 1$, there is a second divergent term.

\sm

We shall work in the first range so that we may construct a regularized function which vanishes near the cusp by subtracting a single Eisenstein series from the original function. At the end of our calculation we shall then analytically continue our result to $\Re(s) \geq 6$. We define the regularized function as follows,
\begin{align}
    \tilde{\cW}^{(m,n)}_s
    =
    \cW^{(m,n)}_s
    - 8 \,
    \frac{ \zeta^*(2m) }{ \Gamma(m) }
    \frac{ \zeta^*(2n) }{ \Gamma(n) }
    \frac{ \zeta^*(2s-2m-2n) }{ \Gamma(s-m-n) }
    \frac{ 1 }{ \zeta^*(2s) }
    \, E_s^* 
\end{align}
A Poincar\'e seed for the regularized function is given by,
\begin{align}
    \tilde{\Lambda}^{(m,n)}_s
    =
    \Lambda^{(m,n)}_s
    - 8 \,
    \frac{ \zeta^*(2m) }{ \Gamma(m) }
    \frac{ \zeta^*(2n) }{ \Gamma(n) }
    \frac{ \zeta^*(2s-2m-2n) }{ \Gamma(s-m-n) }
    \, \tau_2^s
\end{align}
where \smash{$\Lambda_s^{(m,n)}$} is a Poincar\'e seed for the un-regularized function. We construct this seed function by unfolding the last factor in~\eqref{eq:W} and find,
{\begin{align}
    \Lambda_s^{(m,n)}
    &=
    2 \, 
    \frac{ \zeta^*(2s-2m-2n) }{ \Gamma(s-m-n) } \,
    \tau_2^{s} \,
    G_{2m} \,
    \bar{G}_{2n}
\end{align}
where $G_{2m}$ and $\bar{G}_{2n}$ are holomorphic and anti-holomorphic Eisenstein series, respectively. Using the Fourier expansion~\eqref{eq:GAsy} for these Eisenstein series, we find,
\begin{align}
    \int_0^1 d\tau_1 \, 
    \tilde{\Lambda}_s^{(m,n)}
    =
    8 \, (-4)^{m+n} \,
    \frac{ \zeta(2s-2m-2n) }
         { \Gamma(2m) \Gamma(2n) }
    \frac{ \tau_2^{s} }{ \pi^{s-2m-2n} }
    \sum_{k=1}^\infty
    \sigma_{2m-1}(k) \,
    \sigma_{2n-1}(k) \,
    (q \bar{q} )^k
\end{align}
where $q = e^{2\pi i \tau}$ and $\sigma_\ell(n)$ is the divisor sum defined below~\eqref{eq:GAsy}.

\sm

Using the standard unfolding trick~\eqref{eq:unfold}, we unfold the integral of the regularized function over the full fundamental domain and find,
\begin{align}
    \intM
    \tilde{\cW}_s^{(m,n)}
    =
    16 \, (-)^{m+n} \,
    \frac{ \Gamma(s-1)}{ \Gamma(2m) \Gamma(2n) }
    \frac{ \zeta(2s-2m-2n) }
         { (2\pi)^{2s-2m-2n-1 } }
    \sum_{k=1}^\infty
    \frac{ \sigma_{2m-1}(k) \,
           \sigma_{2m-1}(k) }
         { k^{s-1} }
\end{align}
To obtain this result, we have freely interchanged the sum over $k$ with the integral over $\tau_2$ using the positivity of the integrand and Fubini's theorem. The sum may be evaluated using the following formula of Ramanujan,
\begin{align}
    \sum_{k=1}^\infty
    \frac{ \sigma_{a}(k) \, \sigma_{b}(k) }{k^z}
    &=
    \frac{ \zeta(z) \, \zeta(z-a) \, \zeta(z-b) \, \zeta(z-a-b) }
         { \zeta(2z-a-b) }
\end{align}
which converges for $\Re(z), \Re(z-a), \Re(z-b), \Re(z-a-b) > 1$. We are working in the range ${\Re(s) > 2m + 2n +1}$, so the sum converges and yields,
\begin{align}
    \intM
    \tilde{\cW}_s^{(m,n)}
    &=
    16 \, (-)^{m+n} \,
    \frac{ \Gamma(s-1) }{ \Gamma(2m) \Gamma(2n) }
    \frac{ 1 }{ (2\pi)^{2s-2m-2n-1} }
\no \\
    & \quad
    \times
    \zeta(s-1) \,
    \zeta(s-2m) \,
    \zeta(s-2n) \,
    \zeta(s-2m-2n+1)
\end{align}
which is manifestly finite for ${\Re(s) > 2m + 2n +1}$.

\sm

At this point, we have integrated the regularized function over the full fundamental domain $\cM$. However, we set out to integrate the un-regularized function over the truncated fundamental domain~$\cM_L$. We may relate these two integrals as follows,
\begin{align}
    \intML
    \cW_s^{(m,n)}
    &=
    \intM
    \tilde{\cW} _s^{(m,n)}
    - \intMR \tilde{\cW}_s^{(m,n)}
 \\[1ex]
    & \quad
    + 8 \,
    \frac{ \zeta^*(2m) }{ \Gamma(m) }
    \frac{ \zeta^*(2n) }{ \Gamma(n) }
    \frac{ \zeta^*(2s-2m-2n) }{ \Gamma(s-m-n) }
    \frac{ 1 }{ \zeta^*(2s) }
    \intML
    E_s^* 
\no
\end{align}
where $\cM_R = \cM \setminus \cM_L$ is defined in~\eqref{eq:ML}. The second and third integrals depend on the cut-off $L$. The integral of the Eisenstein series is given in~\eqref{eq:IE}. The integral of \smash{$\tilde{\cW}_s^{(m,n)}$} over~$\cM_R$ may be calculated by replacing the integrand with its asymptotic expansion, which is given by~\eqref{eq:WAsy} and~\eqref{eq:EAsy}. Then using $\Re(s) > 2m + 2n + 1$,  we find,
\begin{align}
    \intMR \tilde{\cW}_s^{(m,n)}
    &=
    8 \,
    \frac{ \zeta^*(2m) }{ \Gamma(m) }
    \frac{ \zeta^*(2n) }{ \Gamma(n) }
    \bigg\{
    \frac{ \zeta^*(2s-2m-2n) }{ \Gamma(s-m-n) }
    \frac{ \zeta^*(2s-1) }
         { \zeta^*(2s) }
    \frac{ L^{-s} }{ -s }
\no \\
    & \quad
    - (-)^{m+n} \,
    \Gamma(s-m-n) \,
    \frac{ \zeta^*(2s-2m-2n-1) }{ \Gamma(s-2m) \Gamma(s-2n) }
    \frac{ L^{2m+2n-s} }{ 2m+2n-s }
    \bigg\}
\end{align}
Combining these results and rewriting the starred zeta functions in terms of Riemann zeta functions, we arrive at~\eqref{eq:IW}.

\sm

This final expression may be analytically continued in $s$ from $\Re(s) > 2n + 2m + 1$ to all $\Re(s) \geq 6$. When $s = 2m + 2n $, the two simple poles precisely cancel, and the integral is finite as required. Since $2 \leq m,n \leq \floor{\Re(s)/2}-1$, there are no other poles. This completes our proof of \autoref{thm:IW}.

\sm

To analyze the transcendental structure of this integral, we restrict to integer $s \geq 6$. In this case, the constant term in~\eqref{eq:IW} is proportional to the quadruple product of zeta-values,
\begin{align}
    \zeta(s-1) \, \zeta(s-2m) \, \zeta(s-2n) \, \zeta(s-2m-2n+1)
\end{align}
This quadruple product vanishes for odd $s$ satisfying ${6 \leq s < 2m+2n}$ since the Riemann zeta function vanishes at negative even integers. Moreover, this constant term is similar to the constant term in the integral~\eqref{eq:IEEEZagier} of the triple product $E_{m}^* E_{n}^* E_{s-m-n}^*$, which is given by the quadruple product of starred zeta-values,
\begin{align}
    \zeta^*(s-1) \, \zeta^*(s-2m) \, \zeta^*(s-2n) \, \zeta^*(s-2m-2n+1)
\end{align}
Both quadruple products have a simple pole at $s = 2m+2n$ which precisely cancels against an $L$-dependent pole in their respective integrals, generating constant and $\ln L$ terms. These two integrals are closely related and may be written in terms of each other.

\begin{thm}
\label{thm:IWint}
The integral of \smash{$\cW_s^{(m,n)}$} with integer $s \geq 6$ and ${2 \leq m,n \leq \floor{s/2}-1}$ may be written in terms of the integral $E_{m}^* E_{n}^* E_{s-m-n}^*$ as follows,
\begin{align}
\label{eq:IWint}
    \intML
    \(
    \cW_s^{(m,n)}
    -
    Q_{m,n,s-m-n}^{(\cW)}
    E_{m}^*
    E_{n}^*
    E_{s-m-n}^*
    \)
    & \approx
    \delta_{s, 2m+2n} \,
    \cI_{m,n}
\end{align}
where $Q_{m,n,s-m-n}^{(\cW)}$ is the following rational number,
\begin{align}
\label{eq:Q(W)}
    Q_{m,n,s-m-n}^{(\cW)}
    &=
    \frac{ 4 \, (-)^{m+n} }{ (2m-1)!  (2n-1)! }
    \frac{ (s-2)!! }{ (s-2m-2)!! (s-2n-2)!! }
\no \\
    & \qquad
    \times
    \begin{cases}
    \phantom{+} 0
    &
    \phantom{\textup{even }}
    \negphantom{\textup{odd }}
    \textup{odd }
    s < 2m+2n
    \\
    (-)^{m+n+s/2} \,
    (2m+2n-s-1)!!
    &
    \textup{even }
    s < 2m+2n
    \\
    -1
    &
    \phantom{\textup{even }}
    s = 2m+2n
    \\
    (s-2m-2n-1)!!^{-1}
    &
    \phantom{\textup{even }}
    s > 2m+2n
    \end{cases}
\end{align}
and $\cI_{m,n}$ has transcendental weight $2m+2n+1$,
\begin{align}
    \cI_{m,n}
    &=
    8 \pi \,
    \zeta(2m+2n-1) \,
    \frac{ B_{2m} }{ (2m)! }
    \frac{ B_{2n} }{ (2n)! }
    \frac{ (2m+2n-1)! }{ (2m-1)!  (2n-1)! }
\no \\[1ex]
    & \quad
    \times
    \big[
      H_1(2m)
    - \half H_1(m)
    + H_1(2n)
    - \half H_1(n)
    - \half H_1(m+n)
    \big]
\end{align}
The right-hand side of~\eqref{eq:IWint} is non-vanishing only when $s = 2m + 2n$, and in that case it has transcendental weight $s+1$. Thus, the integral of \smash{$\cW_s^{(m,n)}$} is of the form claimed in \autoref{thm:3looptrans}.
\end{thm}

To prove this theorem, we compute the analytic continuation of~\eqref{eq:IW} to integer~${s \geq 6}$ using the properties of the Riemann zeta function given in \autoref{apdx:zeta}. We then compare the result to the integrals of triple products of Eisenstein series presented in \autoref{thm:IEEE}.

\sm

The transcendental structure of the integral of \smash{$\cW_s^{(m,n)}$} is similar to that of the integral of $E_m^* E_n^* E_{s-m-n}^*$. When $s<2m+2n$, the integral of~\smash{$\cW_s^{(m,n)}$} vanishes or has transcendental weight $s$. When $s=2m+2n$, the presence of $\ln(2\pi)$ spoils the transcendental structure, as we described in detail below \autoref{thm:IEEE}. When~${s>2m+2n}$, the integral has transcendental weight $2s-2m-2n+1 > s+1$. Thus, compared to the integrals of arbitrary two-loop MGFs, the integral of \smash{$\cW_s^{(m,n)}$} has novel transcendental structure when $s \geq 2m+2n$. This structure first appears at weight eight with \smash{$\cW_8^{(2,2)}$} whose integral contains $\ln(2\pi)$. Then at weight nine the integral of \smash{$\cW_9^{(2,2)}$} has transcendental weight eleven. At any weight, we can remove the terms with novel transcendental structure in the integral of \smash{$\cW_s^{(m,n)}$} by subtracting a rational multiple of the integral of $E_m^* E_n^* E_{s-m-n}^*$, as shown in~\eqref{eq:IWint}.

\newpage

\section{Integrating \texorpdfstring{$E_k^* \, C_{a,b,c}$}{} and \texorpdfstring{$E_k^* \, \mfC_{w;m;p}$}{}}
\label{sec:ECabc}

In this section, we shall evaluate the integrals of the infinite family of disconnected three-loop modular graph functions $E_k^* C_{a,b,c}$ for arbitrary integers $a, b, c \geq 1$ and $k \geq 2$ as well as the infinite family of functions $E_k^* \mfC_{w;m;p}$. We shall also discuss the transcendental structure of these integrals.

\sm

The functions $E_k^* C_{a,b,c}$ are obtained by multiplying the Eisenstein series $E_k^*$ and the two-loop modular graph function $C_{a,b,c}$, which was defined in~\eqref{eq:Cabc}. These three-loop MGFs have weight~$a+b+c+k$ and the following decorated graph,
\begin{align}
\label{eq:EkCabc}
    E_k^* C_{a,b,c}
    ~&=~
    \half \, \Gamma(k)
    \times~
    \begin{tikzpicture}[baseline=-0.5ex, scale=.7]
    \draw (-1,0) node[circle,fill,inner sep=1.5](a){};
    \draw (-1,0) arc(180:-180:1)
        node[style=draw,midway,fill=white,scale=.8]
        {$k,k$};
    \end{tikzpicture}
    ~~
    \begin{tikzpicture}[baseline=-0.5ex, scale=.8, yscale=.8, xscale=.9]
    \draw (-1,0) node[circle,fill,inner sep=1.5](a){};
    \draw (1,0) node[circle,fill,inner sep=1.5](b){};
    \draw (-1,0) arc(180:0:1)
        node[style=draw,midway,fill=white,scale=.5]
        {$a,a$};
    \draw (-1,0) arc(-180:0:1)
        node[style=draw,midway,fill=white,scale=.5]
        {$b,b$};
    \draw (a) to [out=0,in=180] 
        node[style=draw,midway,fill=white,scale=.5] 
        {$c,c$} (b);
    \end{tikzpicture}
\end{align}
where the prefactor arises from the normalization of the starred Eisenstein series relative to the un-starred Eisenstein series. For example, $E_2^* \, C_{2,1,1}$ appears at weight six in the low-energy expansion of the genus-one four-graviton amplitude in Type~II superstring theory. Its integral over $\cM_L$ was calculated in~\cite{DHoker:2019blr}.

\sm

To tackle the general case, we recall our discussion in \autoref{sec:2loopMGF} of the system of differential equations~\eqref{eq:DCabc} obeyed by the functions $C_{a,b,c}$ and the corresponding eigenfunctions~$\mfC_{w;m;p}$. Each eigenfunction obeys the inhomogeneous Laplace eigenvalue equation~\eqref{eq:DCwmp}, which we repeat here for convenience,
\begin{align}
\label{eq:DCwmp2}
    \big(
    \Delta - (w-2m)(w-2m-1)
    \big)
    \mfC_{w;m;p}
    =
    \mfh^{(0)}_{w;m;p} \, E_w^*
    +
    \sum_{\ell = m+1}^{\floor{w/2}}
    \mfh^{(\ell)}_{w;m;p} \,
    E_\ell^* E_{w-\ell}^*
\end{align}
where we have used \autoref{thm:dhprop} to set \smash{$\mfh^{(\ell)}_{w;m;p} = 0$} for $2 \leq \ell \leq m$. The two bases of functions are related by~\eqref{eq:CabcCwmp}, which we also repeat here for convenience,
\begin{align}
\label{eq:CabcCwmp2}
    C_{a,b,c}
    &= 
    \!
    \sum_{m=1}^{\floor{\frac{w-1}{2}}} \!
    \sum_{p=0}^{\floor{\frac{w-2m-1}{3}}} \!
    d_{a,b,c}^{w;m;p} \,
    \mfC_{w;m;p}
    &
    \mfC_{w;m;p}
    &=
    \sum_{\substack{ a \geq b \geq c \geq 1 \\
                     a+b+c=w }}
    \mfd_{w;m;p}^{a,b,c} \,
    C_{a,b,c}
\end{align}
The $\mfh$ coefficients which appear in the inhomogeneous part of the Laplace equation and the expansion coefficients \smash{$d_{a,b,c}^{w;m;p}$} and \smash{$\mfd_{w;m;p}^{a,b,c}$} are all rational numbers. Explicit expressions for these coefficients are given in \autoref{apdx:CabcCwmp}. The Laurent coefficients of $C_{a,b,c}$ and $\mfC_{w;m;p}$ are  given in \autoref{apdx:Laurent}. These appendices contain novel results but are tangential to our discussions of integration and transcendentality. 

\sm

Although we are ultimately interested in the integral of $E_k^* C_{a,b,c}$, we shall first use the simple differential equation obeyed by $\mfC_{w;m;p}$ to calculate the integral of $E_k^* \mfC_{w;m;p}$. The integral of $E_k^* C_{a,b,c}$ will be given by a rational linear combination of the former integrals.

\subsection{The integral of \texorpdfstring{$E_k^* \, \mfC_{w;m;p}$}{} with \texorpdfstring{$k \neq w-2m$}{}}

We shall first consider integer $k \neq w-2m$. In this case, $E_k^*$ and $\mfC_{w;m;p}$ do not have the same eigenvalues under the action of the Laplacian, and we find the following result.

\begin{lem}
\label{thm:IEkCwmp}
The integral of $E_k^* \mfC_{w;m;p}$ with integer $k \geq 2$ and $k \neq w-2m$ is given by the following sum of integrals of double and triple products of Eisenstein series plus a boundary contribution with transcendental weight $k+w$,
\begin{align}
\label{eq:IEkCwmp}
    \intML
    E_k^* \mfC_{w;m;p}
    & \approx
    \frac{ \mfh^{(0)}_{w;m;p} }{ \mu_{k;w;m} }
    \intML
    E_k^* E_w^* 
    +
    \sum_{\ell=m+1}^{\floor{w/2}}
    \frac{ \mfh^{(\ell)}_{w;m;p} }{ \mu_{k;w;m} }
    \intML
    E_k^* E_\ell^* E_{w-\ell}^*
\no \\[1ex]
    & \quad
    +
    \frac{ (2k-1) }{ \mu_{k;w;m} }
    \Big(
    \zeta^*(2k) \,
    \mfc_{w;m;p}^{(1-k)}
    -
    \zeta^*(2k-1) \,
    \mfc_{w;m;p}^{(k)}
    \Big)
\end{align}
where $\mu_{k;w;m} = (k-w+2m)(k+w-2m-1)$ and \smash{$\mfc_{w;m;p}^{(\ell)}$} are the Laurent coefficients of $\mfC_{w;m;p}$. The boundary contribution on the second line vanishes when $k > w$.
\end{lem}

To prove this lemma, we use the Laplace equations for $E_k^*$ and $\mfC_{w;m;p}$ to write their product $E_k^* \mfC_{w;m;p}$ as follows,
\begin{align}
    \mu_{k;w;m} \,
    E_k^* \mfC_{w;m;p}
    &=
    \mfC_{w;m;p} \Delta E_k^*
    - E_k^* \Delta \mfC_{w;m;p}
    +
    \mfh^{(0)}_{w;m;p} \,
    E_k^* E_w^*
\no \\
    & \quad
    +
    \sum_{\ell = m+1}^{\floor{w/2}}
    \mfh^{(\ell)}_{w;m;p} \,
    E_k^* E_\ell^* E_{w-\ell}^*
\end{align}
where we have defined the difference of their eigenvalues,
\begin{align}
    \mu_{k;w;m}
    =
    k(k-1) - (w-2m)(w-2m-1)
    =
    (k-w+2m)(k+w-2m-1)
\end{align}
which is a non-zero integer for $k \neq w-2m$. Next we use \autoref{thm:ICDC} to evaluate the integral of the total derivative terms and find,
\begin{align}
    \intML
    \big(
    \mfC_{w;m;p} \Delta E_k^*
    - E_k^* \Delta \mfC_{w;m;p}
    \big)
    & \approx
    \lambda_1'(k) \,
    \zeta^*(2k) \,
    \mfc_{w;m;p}^{(1-k)}
\no \\
    & \quad
    +
    \lambda_1'(1-k) \,
    \zeta^*(2k-1) \,
    \mfc_{w;m;p}^{(k)}
\end{align}
where $\lambda_1(s) = \Gamma(s+1) / \Gamma(s-1) = s(s-1)$. This integral vanishes for $k > w$ since the Laurent coefficient~\smash{$\mfc^{(\ell)}_{w;m;p}$} is non-zero only for $1-w \leq \ell \leq w$. We now use,
\begin{align}
    \lambda'_1(k) = - \lambda'_1(1-k) = 2k-1
\end{align}
After dividing by $\mu_{k;w;m}$, we arrive at~\eqref{eq:IEkCwmp}.

\sm

The assignment of transcendental weight follows from the fact that~\smash{$\mfc^{(\ell)}_{w;m;p}$} has transcendental weight $w$ and the fact that the starred zeta-values~$\zeta^*(2k)$ and $\zeta^*(2k-1)$ each have transcendental weight $k$. This completes our proof of \autoref{thm:IEkCwmp}.

\sm

In each case, integral of $E_{k}^* \mfC_{w;m;p}$ with $k \neq w-2m$ is equal to a boundary contribution with transcendental weight $k+w$ plus an integral of the double product~$E_k^* E_w^*$ plus a sum of integrals of triple products of Eisenstein series. When $k=w$, the integral of $E_k^* E_w^*$ has transcendental weight $k+w+1$, and when $k \neq w$, it does not contribute. The transcendental structure of the integrals of the triple products is more complicated. As we described at length in \autoref{sec:Eisen}, the integrals of the triple products will have a novel transcendental structure if $2 \max(k,w-\ell,\ell) \geq k+w$, which first occurs at weight eight.

\sm

Thus, the integral of~$E_{k}^* \mfC_{w;m;p}$ will generally have a novel transcendental structure when its weight $k+w \geq 8$. For instance, at weight eight the integral of $E_{4}^* \mfC_{4;1;0}$ includes the integral of $E_2^* E_2^* E_4^*$ which contains $\ln(2\pi)$. Then at weight nine the integral of $E_{5}^* \mfC_{4;1;0}$ includes the integral of $E_2^* E_2^* E_5^*$ which has transcendental weight eleven. We have, however, clearly isolated these terms and identified their origin from the inhomogeneous Laplace equation for~$\mfC_{w;m;p}$. By subtracting these triple products, we may construct construct the following integral whose terms have transcendental weight $k+w$ or $k+w+1$,
\begin{align}
\label{eq:IQEkCwmp}
    \intML
    \bigg(
    E_{k}^* \mfC_{w;m;p}
    -
    \sum_{\ell = m+1}^{\floor{w/2}}
    \frac{ \mfh^{(\ell)}_{w;m;p}  }{\mu_{k;w;m}}
    \,
    E_{k}^* E_\ell^* E_{w-\ell}^* 
    \bigg)
\end{align}
Thus, the integral of \smash{$E_{k}^* \mfC_{w;m;p}$} with $k \neq w-2m$ is of the form claimed in \autoref{thm:3looptrans}.

\subsection{The integral of \texorpdfstring{$E_{s}^* \, \mfC_{w;m;p}$}{} with \texorpdfstring{$\Re(s) > 1$}{}}

So far we have evaluated the integral of \smash{$E_{k}^* \mfC_{w;m;p}$} for integer $k \neq w-2m$. Unfortunately, the case of $k=w-2m$ is more involved. Since the difference of eigenvalues $\mu_{w-2m;w-m}=0$, our previous proof fails. To tackle this case, we shall first compute the integral of $E_s^* \mfC_{w;m;p}$ for arbitrary complex~$s$ with~$\Re(s) > 1$.

\begin{lem}
\label{thm:IEsCwmp}
The integral of $E_s^* \mfC_{w;m;p}$ with $\Re(s) > 1$ is given by the analytic continuation in $s$ of the following expression,
\begin{align}
\label{eq:IEsCwmp}
    \intML
    E_s^* \mfC_{w;m;p}
    &=
    \sum_{\ell=m+1}^{\floor{w/2}}
    \mfh^{(\ell)}_{w;m;p} \,
    \frac{ \zeta^*(w+s-1) \, \zeta^*(w-s) \,
           \zeta^*(s+w-2\ell) \,
           \zeta^*(s-w+2\ell) }
         { (s-w+2m) (s+w-2m-1) }
\no \\
    & \quad
    +
    \sum_{\ell = 1-w}^{w}
    \mfc_{w;m;p}^{(\ell)} \,
    \cG_{\ell}(s)
\end{align}
where \smash{$\mfc_{w;m;p}^{(\ell)}$} are the Laurent coefficients of $\mfC_{w;m;p}$ and \smash{$\cG_{\ell}(s)$} is defined in~\eqref{eq:F(s)}.
\end{lem}

To prove this lemma, we shall first consider $\Re(s) > w$. The result for $\Re(s) > 1$ will then follow by analytic continuation. As in our previous proof, we use the Laplace equations for $E_s^*$ and $\mfC_{w;m;p}$ to write $E_s^* \mfC_{w;m;p}$ as follows,
\begin{align}
\label{eq:IEsCwmp2}
    E_s^* \mfC_{w;m;p} 
    &=
    \frac{1}{ s(s-1) - (w-2m)(w-2m-1) }
\\[-1ex]
    & \quad
    \times
    \bigg(
    \mfC_{w;m;p} \Delta E_s^*
    - E_s^* \Delta \mfC_{w;m;p}
    +
    \mfh^{(0)}_{w;m;p} \,
    E_s^* E_w^*
    +
    \sum_{\ell = m+1}^{\floor{w/2}}
    \mfh^{(\ell)}_{w;m;p} \,
    E_s^* E_\ell^* E_{w-\ell}^*
    \bigg)
\no
\end{align}
Unlike our previous proof, the weight $s$ of the Eisenstein series is not necessarily an integer, so the integral of the total derivative term is given by \autoref{thm:IEC} rather than by \autoref{thm:ICDC}. In this case, we have,
\begin{align}
\label{eq:IEsCwmp3}
    \intML
    \big(
    \mfC_{w;m;p} \Delta E_s^*
    - E_s^* \Delta \mfC_{w;m;p}
    \big)
    &=
    \sum_{\ell=1-w}^{w}
    \big(
    s(s-1) - \ell(\ell-1)
    \big) \,
    \mfc^{(\ell)}_{w;m;p} \,
    \cG_{\ell}(s)
\end{align}
where \smash{$\cG_{\ell}(s)$} was defined in~\eqref{eq:F(s)}. Since we are working in the range $\Re(s) > w$, the $\cG$ functions which appear here are all finite functions of~$s$.

\sm

Next we recall the expressions for the integrals of double and triple products of Eisenstein series in~\eqref{eq:IEEMaass} and~\eqref{eq:IEEEZagier}, respectively. We may rearrange these expressions so that they are written in terms of $\cG$ functions. For the integral of the double product, we find,
\begin{align}
\label{eq:IEsCwmp4}
    \intML
    E_s^* E_w^*
    &=
    \zeta^*(2w) \,
    \cG_{w}(s)
    +
    \zeta^*(2w-1) \,
    \cG_{1-w}(s)
\end{align}
and for the integral of the triple product,
\begin{align}
\label{eq:IEsCwmp5}
    \smash[b]{\intML}
    E_s^* E_\ell^* E_{w-\ell}^*
    &=
    \zeta^*(w+s-1) \, \zeta^*(w-s) \,
    \zeta^*(s+w-2\ell) \,
    \zeta^*(s-w+2\ell)
\no \\[1ex]
    & \quad
    +
    \zeta^*(2w-2\ell) \,
    \zeta^*(2\ell) \,
    \cG_{w}(s)
    +
    \zeta^*(2w-2\ell-1) \,
    \zeta^*(2\ell-1) \,
    \cG_{2-w}(s)
\no \\[1ex]
    & \quad
    +
    \zeta^*(2w-2\ell) \,
    \zeta^*(2\ell-1) \,
    \cG_{w-2\ell+1}(s)
\no \\[1ex]
    & \quad
    +
    \zeta^*(2w-2\ell-1) \,
    \zeta^*(2\ell) \,
    \cG_{1-w+2\ell}(s)
\end{align}
Since $\Re(s) > w$, the starred zeta functions and the $\cG$ functions which occur in these expressions are all finite functions of $s$.

\sm

It remains to combine~\eqref{eq:IEsCwmp2},~\eqref{eq:IEsCwmp3},~\eqref{eq:IEsCwmp4}, and~\eqref{eq:IEsCwmp5}. To simplify the result, we shall relate the Laurent coefficients of $\mfC_{w;m;p}$ and the $\mfh$ coefficients. Inserting the asymptotic expansions for the eigenfunctions $\mfC_{w;m;p}$ and the starred Eisenstein series into the Laplace equation~\eqref{eq:DCwmp2}, yields the following formula,
\begin{alignat}{2}
\label{eq:c(L)wmp}
    0
    &=
    \mfh^{(0)}_{w;m;p} \,
    \Big(
    \zeta^*(2w) \, \tau_2^w
    + \zeta^*(2w-1) \, \tau_2^{1-w}
    \Big)
    \span \span
\no \\[1ex]
    & \quad
    +
    \sum_{\ell = 1-w}^{w}
    \Big(
    (w-2m)(w-2m-1)
    -
    \ell(\ell-1)
    \Big)
    \,
    \mfc^{(\ell)}_{w;m;p} \,
    \tau_2^{\ell}
    \span \span
\no \\
    & \quad
    +
    \sum_{\ell = m+1}^{\floor{w/2}}
    \mfh^{(\ell)}_{w;m;p} \,
    &&
    \Big(
    \zeta^*(2w-2\ell) \, \tau_2^{w-\ell}
    + \zeta^*(2w-2\ell-1) \, \tau_2^{1-w+\ell}
    \Big)
\no \\
    & \span
    \times
    &
    \Big(
    \zeta^*(2\ell) \, \tau_2^{\ell}
    + \zeta^*(2\ell-1) \, \tau_2^{1-\ell}
    \Big)
\end{alignat}
which may be solved order-by-order in $\tau_2$ to derive a series of relations between the Laurent coefficients and the $\mfh$ coefficients. We shall use these relations to simplify the combination of equations~\eqref{eq:IEsCwmp2} through~\eqref{eq:IEsCwmp5}. After some straightforward but tedious algebra, we arrive at~\eqref{eq:IEsCwmp}. Thus far we have worked with $\Re(s) > w$, but~\eqref{eq:IEsCwmp} may be analytically continued in $s$ to $\Re(s) > 1$. This completes our proof of \autoref{thm:IEsCwmp}.

\sm

We note that individual terms in~\eqref{eq:IEsCwmp} have simple poles at integers $2 \leq s \leq w$. However, for all such $s$, the integral of $E^*_s \mfC_{w;m;p}$ over $\cM_L$ must be a finite function of $L$, so the poles necessarily cancel, generating $\ln L$ contributions to the integral.

\subsection{The integral of \texorpdfstring{$E_{w-2m}^* \, \mfC_{w;m;p}$}{}}

With the general result in hand, we shall now consider the integral of $E_{w-2m}^* \mfC_{w;m;p}$.

\begin{lem}
\label{thm:IEw2mCwmp}
The integral of $E_{w-2m}^* \mfC_{w;m;p}$ has transcendental weight $2w-2m+1$ and is given by,
\begin{align}
\label{eq:IEw2mCwmp}
    \intML
    E_{w-2m}^* \mfC_{w;m;p}
    & 
    \approx
    \frac{ 1 }{ (2w-4m-1) } \,
    \zeta^*(2w-2m-1) \, \zeta^*(2m)
\no \\
    & \quad
    \times    
    \sum_{\ell=m+1}^{\floor{w/2}}
    \mfh^{(\ell)}_{w;m;p} \,
    \zeta^*(2w-2\ell-2m) \,
    \zeta^*(2\ell-2m)
    \span
\no \\
    & \qquad
    \times
    \Big[
    \cZ(2w-2m-1)
    + \cZ(2w-2\ell-2m)
    - 2 \cZ(2w-4m)
    \span
\no \\
    & \hspace{0.7in}
    + \cZ(2\ell-2m)
    - \cZ(2m)
    - \ln(2L)
    \vphantom{\Big[}
\no \\
    & \hspace{0.7in}
    - H_1(2w-4m)
    + H_1(2w-4m-1)
    \Big]
    \span
\end{align}
where $\cZ$ is the combination of the logarithmic derivative of the Riemann zeta function and finite harmonic sums defined in~\eqref{eq:Zndef}.
\end{lem}

To prove this lemma, we simply compute the limit $s \to w-2m$ of~\eqref{eq:IEsCwmp} from \autoref{thm:IEsCwmp}. We first set $s=w-2m+\eps$ on the right-hand side of~\eqref{eq:IEsCwmp} and find,
\begin{multline}
\label{eq:IEw2mCwmp2}
    \frac{ \zeta^*(2w-2m-1+\eps) \, \zeta^*(2m-\eps) }
         { \eps \, (2w-4m-1+\eps) }
    \sum_{\ell=m+1}^{\floor{w/2}}
    \mfh^{(\ell)}_{w;m;p} \,
    \zeta^*(2w-2m-2\ell+\eps) \,
    \zeta^*(2m+2\ell+\eps)
\\[1ex]
    +
    \mfc_{w;m;p}^{(1-w+2m)} \,
    \zeta^*(2w-4m+2\eps) \, 
    \frac{L^{\eps}}{\eps}
    +
    \mfc_{w;m;p}^{(w-2m)} \,
    \zeta^*(2w-4m+2\eps-1) \, 
    \frac{L^{-\eps}}{-\eps}
    + \cO(L^\pm)
\end{multline}
where $\cO(L^\pm)$ denotes terms which are non-zero powers of $L$ in the limit $\eps \to 0$. We may ignore the third term in this expression because the Laurent coefficient \smash{$\mfc_{w;m;p}^{(w-2m)} = 0$}. This coefficient vanishes because the corresponding Laurent coefficients \smash{$\mfc_{a,b,c}^{(w-2m)}$} vanish, and the eigenfunctions $\mfC_{w;m;p}$ are linear combinations of the functions $C_{a,b,c}$. The Laurent polynomial of $C_{a,b,c}$ was computed in~\cite{DHoker:2017zhq} and is reviewed in \autoref{apdx:Laurent}.

\sm

The two remaining terms in~\eqref{eq:IEw2mCwmp2} are proportional to $1/\eps$ and must conspire in the limit~$\eps \to 0$ because the integral of $E_{w-2m}^* \mfC_{w;m;p}$ over $\cM_L$ is necessarily a finite function of the cut-off $L$. Thus, the remaining Laurent coefficient is given by,
\begin{align}
\label{eq:c(1w2m)wmp}
    \mfc_{w;m;p}^{(1-w+2m)}
    &=
    -
    \frac{ \zeta^*(2m) \, \zeta^*(2w-2m-1) }
         { (2w-4m-1) \, \zeta^*(2w-4m) }
\no \\
    & \qquad
    \times
    \sum_{\ell = m+1}^{\floor{w/2}}
    \mfh^{(\ell)}_{w;m;p} \,
    \zeta^*(2w-2\ell-2m) \,
    \zeta^*(2\ell-2m)
\end{align}
We now use this expression as well as the properties of the starred zeta function given in \autoref{apdx:zeta} to compute the limit $\eps \to 0$. Several of the resulting terms may be conveniently written in terms of the combinations $\cZ$ defined in~\eqref{eq:Zndef}. We also write the following fraction in terms of finite harmonic sums,
\begin{align}
    \frac{1}{2w-4m-1}
    =
    H_1(2w-4m)
    - H_1(2w-4m-1)
\end{align}
After some rearrangements, we arrive at~\eqref{eq:IEw2mCwmp}. The assignment of transcendental weight follows from the fact that the starred zeta-value~$\zeta^*(n)$ has transcendental weight~$\floor{(n+1)/2}$ for integer $n \geq 2$. This completes our proof of \autoref{thm:IEw2mCwmp}.

\subsection{The integral of \texorpdfstring{$E_k^* \, C_{a,b,c}$}{}}

Because the functions $C_{a,b,c}$ may be written as rational linear combinations of the eigenfunctions $\mfC_{w;m;p}$, the integral of $E_k^* C_{a,b,c}$ as well as its transcendental structure follow trivially from our results above. Our findings are summarized in the following theorem.

\begin{thm}
\label{thm:IECabc}
The integral of $E_k^* C_{a,b,c}$ for integer $k \geq 2$ and $a,b,c \geq 1$ is given by,
\begin{align}
\label{eq:IECabc}
    \intML
    E_k^* C_{a,b,c}
    &=
    \sum_{m=1}^{\floor{\frac{w-1}{2}}} \!
    \sum_{p=0}^{\floor{\frac{w-2m-1}{3}}} \!
    d_{a,b,c}^{w;m;p}
    \intML
    E_{k}^* \mfC_{w;m;p}
\end{align}
where $w=a+b+c$. The integral of $E_{k}^* \mfC_{w;m;p}$ is given in \autoref{thm:IEkCwmp} and \autoref{thm:IEw2mCwmp}. In each case, we may subtract the integral of a suitable linear combination of triple products of Eisenstein series to construct the following integral whose terms all have transcendental weight~$k+w$ or~$k+w+1$,
\begin{align}
\label{eq:IQECabc}
    \intML
    \bigg(
    E_k^* C_{a,b,c}
    -
    \sum_{\ell=2}^{\floor{w/2}}
    Q^{(EC)}_{k,\ell,w-\ell} \,
    E_k^* E_\ell^* E_{w-\ell}^*
    \bigg)
\end{align}
where \smash{$Q^{(EC)}_{k,\ell,w-\ell}$} is the following rational number,
\begin{align}
    Q^{(EC)}_{k,\ell,w-\ell}
    &=
    \sum_{\substack{ m=1 \\ m \neq (w-k)/2 }}^{\ell-1}
    \sum_{p=0}^{\floor{\frac{w-2m-1}{3}}}
    \frac{ d_{a,b,c}^{w;m;p} \, \mfh^{(\ell)}_{w;m;p} }
         { (k-w+2m)(k+w-2m-1) }
\end{align}
Thus, the integral of \smash{$E_k^* C_{a,b,c}$} is of the form claimed in \autoref{thm:3looptrans}.
\end{thm}

In general, the integrals of $E_k^* C_{a,b,c}$ and $E_{k}^* \mfC_{w;m;p}$ have terms with novel transcendental structure when the weight $k+w \geq 8$. For instance, at weight eight, the integral of $E_{4}^* C_{2,1,1}$ includes the integral of $E_2^* E_2^* E_4^*$ which contains $\ln(2\pi)$. Then at weight nine, the integral of~$E_{5}^* C_{2,1,1}$ includes the integral of $E_2^* E_2^* E_5^*$ which has transcendental weight eleven. We have, however, clearly isolated the terms with novel transcendental structure and identified their origin from the inhomogeneous Laplace equations obeyed by the functions $C_{a,b,c}$.

\newpage

\section{Integrating \texorpdfstring{$v_{k,3}$}{}}
\label{sec:vk3}

In this section, we shall evaluate the integrals of the infinite family of connected three-loop modular graph functions $v_{k,3}$ for integers $k \geq 2$ using their inhomogeneous Laplace equations. As a warm-up for this three-loop calculation, we shall first evaluate the integrals of the infinite family of connected two-loop modular graph functions $v_{k,2}$ which are the two-loop analogues of~$v_{k,3}$. We shall also discuss the transcendental structure of these integrals.

\sm

The connected two-loop functions $v_{k,2}$ were introduced in~\cite{DHoker:2016quv} and form an infinite subfamily of the functions $\cC_{u,v;w}$ which were defined in~\eqref{eq:Cuvw}. They satisfy $v_{k,2} = \cC_{k,k;k+1}$ and have the following decorated dihedral graph, matrix of exponents, and Kronecker-Eisenstein series representation,
\begin{align}
\label{eq:vk2}
    v_{k,2}
    ~&=~
    \begin{tikzpicture}[baseline=-0.5ex, scale=.8, yscale=.8, xscale=.9]
    \draw (-1,0) node[circle,fill,inner sep=1.5](a){};
    \draw (1,0) node[circle,fill,inner sep=1.5](b){};
    \draw (-1,0) arc(180:0:1)
        node[style=draw,midway,fill=white,scale=.5]
        {$k,0$};
    \draw (-1,0) arc(-180:0:1)
        node[style=draw,midway,fill=white,scale=.5]
        {$0,k$};
    \draw (a) to [out=0,in=180] 
        node[style=draw,midway,fill=white,scale=.5] 
        {$1,1$} (b);
    \end{tikzpicture}
    ~=~
    \cC
    \! \begin{bmatrix}
    k & 1 & 0 \\
    0 & 1 & k
    \end{bmatrix}
    =
    \frac{\tau_2^{k+1}}{\pi^{k+1}}
    \sum_{p_1, p_2, p_3 \in \Lambda'}
    \frac{ \delta(p_1+p_2+p_3)  }
         { p_1^{k} \, | p_2^{\mathstrut} |^{2}  \, \bar{p}_3^{k} }
\end{align}
The functions $v_{k,2}$ have weight $k+1$. For integers $k \geq 2$, the Kronecker-Eisenstein series converges and is real. The Laurent polynomial of $v_{k,2}$ was calculated in~\cite{DHoker:2019txf} and is reviewed in \autoref{apdx:Laurent}. Because they have only two exponents greater than one, these functions are one of the simplest families of connected two-loop MGFs.

\sm

The connected three-loop functions $v_{k,3}$ were also introduced in~\cite{DHoker:2016quv} and are the three-loop analogues of $v_{k,2}$. They have the following decorated dihedral graph, matrix of exponents, and Kronecker-Eisenstein series representation,
\begin{align}
\label{eq:vk3}
    v_{k,3}
    ~&=~
    \begin{tikzpicture}[baseline=-0.5ex, scale=.8, yscale=.8, xscale=.9]
    \draw (-1,0) node[circle,fill,inner sep=1.5](a){};
    \draw (1,0) node[circle,fill,inner sep=1.5](b){};
    \draw (-1,0) arc(180:0:1)
        node[style=draw,midway,fill=white,scale=.5]
        {$k,0$};
    \draw (-1,0) arc(-180:0:1)
        node[style=draw,midway,fill=white,scale=.5]
        {$0,k$};
    \draw (a) to [out=30,in=150]
        node[style=draw,midway,fill=white,scale=.5] 
        {$1,1$} (b);
    \draw (a) to [out=-30,in=-150] 
        node[style=draw,midway,fill=white,scale=.5] 
        {$1,1$} (b);
    \end{tikzpicture}
    ~=~
    \cC
    \! \begin{bmatrix}
    k & 1 & 1 & 0\\
    0 & 1 & 1 & k
    \end{bmatrix}
    =
    \frac{\tau_2^{k+2}}{\pi^{k+2}}
    \sum_{p_1,p_2,p_3,p_4 \in \Lambda '}
    \frac{ \delta(p_1+p_2+p_3+p_4) }
         { p_1^{k} \, 
           | p_2^{\mathstrut} |^{2} \,
           | p_3^{\mathstrut} |^{2} \, 
           p_4^{k} }
\end{align}
The functions $v_{k,3}$ have weight $k+2$. For integers $k \geq 2$, the Kronecker-Eisenstein series converges and is real. Because they have only two exponents greater than one, these functions are one of the simplest families of connected three-loop MGFs.

\subsection{The integral of \texorpdfstring{$v_{k,2}$}{}}

Because $v_{k,2} = \cC_{k,k;k+1}$, the integral of $v_{k,2}$ over $\cM_L$ is given by \autoref{thm:ICuvw}, but with the three-loop function~$v_{k,3}$ in mind, we shall present a new calculation for this integral using the inhomogeneous Laplace equation obeyed by $v_{k,2,}$. In the remainder of this subsection, we shall prove the following theorem.

\begin{thm}
\label{thm:Ivk2}
The integral of $v_{k,2}$ for integer $k \geq 2$ is given as follows.
\begin{itemize}
\item When $k$ is even, the integral vanishes up to non-zero powers of $L$.
\item When $k$ is odd, the integral has transcendental weight $k+1$ and is given by,
\begin{align}
\label{eq:Ivk2}
    \intML v_{k,2}
    &\approx
    - 8 \pi \,
    \zeta(k) \,
    \frac{B_{k+1}}{(k+1)!} \,
    \frac{1}{k-1}
\end{align}
\end{itemize}
In each case, the integral is of the form claimed in \autoref{thm:2looptrans}.
\end{thm}

\subsubsection{The inhomogeneous Laplace equation for $v_{k,2}$}

Like the two-loop modular graph functions $C_{a,b,c}$, the functions $v_{k,2}$ obey inhomogeneous Laplace eigenvalue equations. Unlike the case of $C_{a,b,c}$, the action of $\Delta$ on $v_{k,2}$ produces modular graph functions with two vanishing holomorphic or anti-holomorphic exponents. These inhomogeneous terms may then be simplified using holomorphic subgraph reduction~\eqref{eq:HSR}, yielding the following result.

\begin{lem}
\label{thm:Dvk2}
The functions $v_{k,2}$ with integer $k \geq 2$ obey the following Laplace equations.
\begin{itemize}
\item When $k$ is even,
\begin{align}
\label{eq:Dvk2E}
    \big(
    \Delta - k(k-1)
    \big) \,
    v_{k,2}
    &=
    k (k^2-2) \,
    E_{k+1}
    - k
    \sum_{\ell=2}^{\substack{ \hphantom{(k-1)/2} \\ k/2}}
    (2\ell-1)
    \(
    \cV^{(\ell)}_{\ell,k+1-\ell}
    + \cV^{(\ell)}_{k+1-\ell,\ell}
    \)
\intertext{\item When $k$ is odd,}
\label{eq:Dvk2O}
    \big(
    \Delta - k(k-1)
    \big) \,
    v_{k,2}
    &=
    - k (k^2-2) \,
    E_{k+1}
    + k
    \sum_{\ell=2}^{(k-1)/2}
    (2\ell-1)
    \(
    \cV^{(\ell)}_{\ell,k+1-\ell}
    + \cV^{(\ell)}_{k+1-\ell,\ell}
    \)
\no \\
    & \quad
    + k^2 \, \tau_2^{k+1} \,
    G_{k+1} \, \bar{G}_{k+1}
\end{align}
\end{itemize}
where $E_{k+1}$, $G_{k+1}$, and $\bar{G}_{k+1}$ are the non-holomorphic, holomorphic, and anti-holomorphic Eisenstein series, respectively, and $\cV$ is the disconnected two-loop modular graph function defined in~\eqref{eq:V}.
\end{lem}

To prove this lemma, we first use~\eqref{eq:DMGF} to evaluate the action of the Laplacian on~$v_{k,2}$. The result includes MGFs with exponents equal to $-1$. We use the momentum conservation identities~\eqref{eq:momcon} to remove these MGFs and find,
\begin{align}
    \big(
    \Delta - k(k-1)
    \big) \,
    v_{k,2}
    =
    k^2 \,
    \cC
    \!
    \[\begin{smallmatrix}
    k+1 & 0 & 0 \\
    0 & k+1 & 0
    \end{smallmatrix}\]
    &
    +
    k(k-1) \,
    \cC
    \!
    \[\begin{smallmatrix}
    k & 1 & 0 \\
    0 & 0 & k+1
    \end{smallmatrix}\]
    - k \,
    \cC
    \!
    \[\begin{smallmatrix}
    k-1 & 2 & 0 \\
    0 & 0 & k+1
    \end{smallmatrix}\]
\no \\[1ex]
    &
    +
    k(k-1) \,
    \cC
    \!
    \[\begin{smallmatrix}
    0 & 0 & k+1 \\
    k & 1 & 0
    \end{smallmatrix}\]
    - k \,
    \cC
    \!
    \[\begin{smallmatrix}
    0 & 0 & k+1 \\
    k-1 & 2 & 0
    \end{smallmatrix}\]
\end{align}
Since $k \geq 2$, each MGF in this expression is absolutely convergent. The first term on right-hand side may be simplified using the factorization identity~\eqref{eq:MGFfact}. The remaining terms may be simplified using holomorphic subgraph reduction~\eqref{eq:HSR}. After some straightforward rearrangements using the properties of MGFs discussed in \autoref{sec:MGFs}, we obtain~\eqref{eq:Dvk2E} and~\eqref{eq:Dvk2O}, which completes our proof of \autoref{thm:Dvk2}.

\sm

Alternatively, we could have proved this lemma using the fact that $v_{k,2} = \cC_{k,k;k+1}$. The action of the Laplacian on the functions~$\cC_{u,v;w}$ is given in~\eqref{eq:DCuvw}. For the case at hand, the right-hand side of this expression will include the functions $\cC_{u,v;w}$ with $u=w$ or $v=w$. Using the identities of \autoref{sec:MGFs}, we may rewrite these functions in terms of Eisenstein series and disconnected two-loop modular graph functions, yielding the inhomogeneous Laplace equations~\eqref{eq:Dvk2E} and~\eqref{eq:Dvk2O}.

\subsubsection{Proof of \autoref{thm:Ivk2}}

We shall now evaluate the integral of~$v_{k,2}$ over $\cM_L$ by considering the contributions from each term in its inhomogeneous Laplace equation.

\sm

Several of the terms do not contribute. The integral of $E_{k+1}$ is given in~\eqref{eq:IE} and vanishes up to powers of $L$ with non-zero exponents. Similarly, the integral of~\smash{$\cV^{(\ell)}_{\ell,k+1-\ell}$} with ${2 \leq \ell \leq \floor{k/2}}$ vanishes by \autoref{thm:IVint}. This leaves the contributions from the total derivative term $\Delta v_{k,2}$ and (for odd~$k$) from the double product $\tau_2^{k+1} \, G_{k+1} \, \bar{G}_{k+1}$.

\sm

By \autoref{thm:IDC}, the integral of $\Delta v_{k,2}$ is equal to the coefficient of the linear term in the Laurent polynomial of $v_{k,2}$. The Laurent polynomial of $v_{k,2}$ was calculated in~\cite{DHoker:2019txf} and is reviewed in \autoref{apdx:Laurent}. For all $k$, the linear coefficient vanishes. Thus, the integral of~$\Delta v_{k,2}$ vanishes, and for even $k$, the integral of $v_{k,2}$ vanishes.

\sm

For odd $k$, the contribution from $\tau_2^{k+1} \, G_{k+1} \, \bar{G}_{k+1}$ remains. This integral is given by~\eqref{eq:IGG} of \autoref{thm:IVint} with $n=s=t=(k+1)/2$. Combining these results with the Laplace equation~\eqref{eq:Dvk2O} yields~\eqref{eq:Ivk2}, which completes our proof of \autoref{thm:Ivk2}.

\subsection{The integral of \texorpdfstring{$v_{k,3}$}{}}

We now turn to the three-loop functions $v_{k,3}$. In the remainder of this subsection, we shall prove the following theorem using the inhomogeneous Laplace equations obeyed by~$v_{k,3}$.

\begin{thm}
\label{thm:Ivk3}
The integral of $v_{k,3}$ for integer $k \geq 2$ is given as follows.
\begin{itemize}
\item When $k$ is even, the integral of $v_{k,3}$ is equal to a sum of terms with transcendental weight~$k+2$ plus a sum of the integrals of the three-loop modular graph functions~\smash{$\cW_{k+2}^{(m,n)}$}, which are defined in~\eqref{eq:W} and whose integrals are is given in \autoref{thm:IWint},
\begin{align}
\label{eq:Ivk3E}
    \intML v_{k,3}
    &\approx
    16 \pi \, \zeta(k+1) \,
    \frac{1}{ k(k-1) } 
    \bigg\{
    \sum_{ \ell = 2 }^{k/2-1}
    \Big(
    \tbinom{k-2}{2\ell-2}
    + 4 \, (2\ell-1)
    \Big)
    \frac{B_{2\ell}}{(2\ell)!}
    \frac{B_{k+2-2\ell}}{(k+2-2\ell)!}
\no \\
    & \hspace{2.2in}
    - (k^2+k-1)
    \frac{B_{k+2}}{(k+2)!}
    \bigg\}
\no \\
    & \quad
    + 2
    \sum_{m,n=2}^{k/2}
    \frac{ (2m-1)(2n-1) }{ k(k-1) }
    \intML
    \cW_{k+2}^{(m,n)}
\intertext{\item When $k$ is odd, the integral of $v_{k,3}$ is equal to a sum of the integrals of \smash{$\cW_{k+2}^{(m,n)}$},}
\label{eq:Ivk3O}
    \intML v_{k,3}
    &\approx
    - 2
    \sum_{m,n=2}^{(k-1)/2}
    \frac{ (2m-1)(2n-1) }{ k(k-1) }
    \intML
    \cW_{k+2}^{(m,n)}
\end{align}
\end{itemize}
In each case, we may subtract the integral of a suitable linear combination of triple products of Eisenstein series to construct the following integral whose terms all have transcendental weight~$k+2$ or $k+3$,
\begin{align}
\label{eq:IQvk3}
    \intML
    \bigg(
    v_{k,3}
    -
    \sum_{ m,n = 2 }^{\floor{k/2}}
    Q_{m,n,k+2-m-n}^{(v)}
    \intML
    E_{m}^*
    E_{n}^*
    E_{k+2-m-n}^*
    \bigg)
\end{align}
where \smash{$Q_{m,n,k+2-m-n}^{(v)}$} is the following rational number,
\begin{align}
\label{eq:Q(vk3)}
    Q_{m,n,k+2-m-n}^{(v)}
    &=
    \frac{ 8 \, (-)^{k+m+n} \, (k-2)!! }
         { (k-1) (2m-1)! (2n-1)! (k-2m)!! (k-2n)!! }
\\
    & \qquad
    \times
    \begin{cases}
    \phantom{+} 0
    &
    \phantom{\textup{even }}
    \negphantom{\textup{odd }}
    \textup{odd }
    k+2 < 2m+2n
    \\
    (-)^{m+n+k/2+1} \,
    (2m+2n-k-3)!!
    &
    \textup{even }
    k+2 < 2m+2n
    \\
    -1
    &
    \phantom{\textup{even }}
    k+2 = 2m+2n
    \\
    (k+1-2m-2n)!!^{-1}
    &
    \phantom{\textup{even }}
    k+2 > 2m+2n
    \end{cases}
\no
\end{align}
Thus, the integral of $v_{k,3}$ is of the form claimed in \autoref{thm:3looptrans}.
\end{thm}

Before we prove this theorem, some remarks are in order. For all integers $k \geq 2$, we have written the integral of $v_{k,3}$ in terms of a piece with transcendental weight $k+2$ plus integrals of the disconnected three-loop modular graph functions \smash{$\cW_{k+2}^{(m,n)}$}. As described in \autoref{thm:IWint}, these latter integrals may have a novel transcendental structure when $k \geq 6$ (that is, for weight~${k+2 \geq 8}$). For instance, the integral of~\smash{$\cW_8^{(2,2)}$} contains $\ln(2\pi)$, and the integral of~\smash{$\cW_9^{(2,2)}$} has transcendental weight eleven. At any weight, we can remove the terms with novel transcendental structure by subtracting a rational multiple of the integral of the triple product $E_m^* E_n^* E_{k+2-m-n}^*$, as demonstrated in~\eqref{eq:IQvk3}.

\subsubsection{The inhomogeneous Laplace equation for $v_{k,3}$}

Just like their two-loop analogues $v_{k,2}$, the three-loop functions $v_{k,3}$ obey inhomogeneous Laplace eigenvalue equations. The action of $\Delta$ on $v_{k,3}$ produces MGFs with two vanishing holomorphic or anti-holomorphic exponents. These inhomogeneous terms may be simplified using holomorphic subgraph reduction~\eqref{eq:HSR}, yielding the following result.
\begin{lem}
\label{thm:Dvk3}
The functions $v_{k,3}$ with integer $k \geq 2$ obey the following Laplace equations.
\begin{itemize}
\item When $k$ is even,
\begin{align}
\label{eq:Dvk3E}
    \big(
    \Delta - k(k-1)
    \big) \,
    v_{k,3}
    &=
    \half (k+2)^2 (k-1)^2 \,
    E_{k+2}
    - 2k (k^2-2) \,
    v_{k+1,2}
\no \\
    & \quad
    - 2
    \sum_{m,n=2}^{k/2}
    (2m-1)(2n-1) \,
    \cW_{k+2}^{(m,n)}
\no \\
    & \quad
    - 2
    \sum_{\substack{ \ell=2 \\[-1ex] \hphantom{m,n=2} }}^{k/2}
    (2\ell-1)
    \(
    \nabla \,
    \cU_{k;\ell}^+
    +
    \bar{\nabla} \,
    \cU_{k;\ell}^-
    \)
\no \\
    & \quad    
    +
    2 (k+1)^2 \,
    \tau_2^{k+2} \,
    G_{k+2} \, \bar{G}_{k+2}
\intertext{\item When $k$ is odd,}
\label{eq:Dvk3O}
    \big(
    \Delta - k(k-1)
    \big) \,
    v_{k,3}
    &=
    - \half (k+2)^2 (k-1)^2 \,
    E_{k+2}
    - 2k (k^2-2) \,
    v_{k+1,2}
\no \\
    & \quad
    + 2
    \sum_{m,n=2}^{(k-1)/2}
    (2m-1)(2n-1) \,
    \cW_{k+2}^{(m,n)}
\no \\
    & \quad
    - 2
    \sum_{\ell=2}^{(k-1)/2}
    (2\ell-1)
    \(
    \nabla \,
    \cU_{k;\ell}^+
    +
    \bar{\nabla} \,
    \cU_{k;\ell}^-
    \)
\no \\
    & \quad
    +
    k (k+1) (k-2)
    \(
    \cV^{(\frac{k+1}{2})}_{\frac{k+1}{2},\frac{k+3}{2}}
    + \cV^{(\frac{k+1}{2})}_{\frac{k+3}{2},\frac{k+1}{2}}
    \)
\end{align}
\end{itemize}
where $E_{k+1}$, $G_{k+1}$, and $\bar{G}_{k+1}$ are the non-holomorphic, holomorphic, and anti-holomorphic Eisenstein series, respectively; $v_{k+1,2}$ is the connected two-loop modular graph function defined in~\eqref{eq:vk2}; $\cV$ is the disconnected two-loop modular graph function defined in~\eqref{eq:V}; $\cW$ is the disconnected three-loop modular graph function defined in~\eqref{eq:W}; and $\cU^\pm$ are disconnected three-loop modular graph forms with modular weights $(0,2)$ and $(2,0)$,
\begin{align}
\label{eq:Ukl}
    \cU_{k;\ell}^+
    &=
    \bar{G}_{2\ell} \,
    \cC^+
    \! \begin{bmatrix}
    k & 1 & 0 \\
    0 & 1 & k+2-2\ell 
    \end{bmatrix}
    &
    \cU_{k;\ell}^-
    &=
    \( \cU_{k;\ell}^+ \)^*
\end{align}
where $^*$ denotes complex conjugation.
\end{lem}

To prove this lemma, we first use~\eqref{eq:DMGF} to evaluate the action of the Laplacian on~$v_{k,3}$. We then use the momentum conservation identities~\eqref{eq:momcon} to remove MGFs with negative exponents and find,
\begin{align}
    \big(
    \Delta - k(k-1)
    \big) \,
    v_{k,3}
    &=
    - 2 \,
    \Big\{
    k \,
    \cC
    \!
    \[\begin{smallmatrix}
    k+1 & 1 & 0 & 0 \\
    0 & 0 & 2 & k
    \end{smallmatrix}\]
    -
    \cC
    \!
    \[\begin{smallmatrix}
    k & 2 & 0 & 0 \\
    0 & 0 & 2 & k
    \end{smallmatrix}\]
    \Big\}
\no \\
    & \quad
    - 2k \,
    \Big\{
    k \,
    \cC
    \!
    \[\begin{smallmatrix}
    k+1 & 1 & 1 & -1 \\
    0 & 0 & 1 & k+1
    \end{smallmatrix}\]
    -
    \cC
    \!
    \[\begin{smallmatrix}
    k & 2 & 1 & -1 \\
    0 & 0 & 1 & k+1
    \end{smallmatrix}\]
    \Big\}
\no \\
    & \quad
    + 2k \,
    \Big\{
    (k-1) \,
    \cC
    \!
    \[\begin{smallmatrix}
    0 & 0 & 1 & k+1 \\
    k & 1 & 1 & 0
    \end{smallmatrix}\]
    -
    \cC
    \!
    \[\begin{smallmatrix}
    0 & 0 & 1 & k+1 \\
    k-1 & 2 & 1 & 0
    \end{smallmatrix}\]
    \Big\}
\end{align}
where we have kept the two MGFs with negative exponents in the second line in order to not expose any divergent MGFs. Since $k \geq 2$, each MGF in this expression is absolutely convergent. Moreover, each term on the right-hand side may be simplified using holomorphic subgraph reduction~\eqref{eq:HSR}. Some of the resulting terms may then be simplified using a second application of~\eqref{eq:HSR}.

\sm

After some straightforward rearrangements using the properties of MGFs discussed in \autoref{sec:MGFs}, we obtain the following expression for the inhomogeneous part of the Laplace equation for $v_{k,3}$,
\begin{align}
    & \quad
    \half (k+2)^2 (k-1)^2 \,
    \cC
    \!
    \[\begin{smallmatrix}
    k+2 & 0 \\
    0 & k+2
    \end{smallmatrix}\]
    - 2k (k^2-2) \,
    \cC
    \!
    \[\begin{smallmatrix}
    k+1 & 1 & 0 \\
    0 & 1 & k+1
    \end{smallmatrix}\]
    \smash[t]{\vphantom{\sum_{\ell=4}^k}}
\no \\
    & \quad
    +
    2(k+1)^2 \,
    \cC
    \!
    \[\begin{smallmatrix}
    k+2 & 0 \\
    0 & 0 
    \end{smallmatrix}\]
    \cC
    \!
    \[\begin{smallmatrix}
    0 & 0 \\
    k+2 & 0
    \end{smallmatrix}\]
    + k(k+1)(k-2)
    \Big\{
    \cC
    \!
    \[\begin{smallmatrix}
    k+1 & 0 \\
    0 & 0 
    \end{smallmatrix}\]
    \cC
    \!
    \[\begin{smallmatrix}
    1 & 0 \\
    k+2 & 0
    \end{smallmatrix}\]
    + c.c.
    \Big\}
    \vphantom{\sum_{\ell=4}^k}
\no \\
    & \quad
    +
    \sum_{\ell=4}^k
    \, (\ell-1)
    \Big\{
    2k \,
    \cC
    \!
    \[\begin{smallmatrix}
    \ell & 0 \\
    0 & 0 
    \end{smallmatrix}\]
    \cC
    \!
    \[\begin{smallmatrix}
    k+1-\ell & 1 & 0 \\
    0 & 1 & k+1
    \end{smallmatrix}\]
    - (k+2)(k-1) \,
    \cC
    \!
    \[\begin{smallmatrix}
    \ell & 0 \\
    0 & 0 
    \end{smallmatrix}\]
    \cC
    \!
    \[\begin{smallmatrix}
    k+2-\ell & 0 \\
    0 & k+2
    \end{smallmatrix}\]
    + c.c.
    \Big\}
\no \\
    & \quad
    + 2 
    \sum_{\substack{ m,n=4 \\ \mathclap{ m+n \neq 2k+2 } }}^{k+1}
    (m-1)(n-1) \,
    \cC
    \!
    \[\begin{smallmatrix}
    m & 0 \\
    0 & 0 
    \end{smallmatrix}\]
    \cC
    \!
    \[\begin{smallmatrix}
    0 & 0 \\
    0 & n
    \end{smallmatrix}\]
    \cC
    \!
    \[\begin{smallmatrix}
    k+2-m & 0 \\
    0 & k+2-n
    \end{smallmatrix}\]
\end{align}
where $c.c.$ refers to the complex conjugate of the preceding term. We now use the following differential identity as well as its complex conjugate to simplify the last two lines,
\begin{align}
\label{eq:kGC-}
    \bar{\nabla}
    \Big\{
    \cC^-
    \!
    \[\begin{smallmatrix}
    \ell & 0 \\
    0 & 0
    \end{smallmatrix}\]
    \cC^-
    \!
    \[\begin{smallmatrix}
    k+2-\ell & 1 & 0 \\
    0 & 1 & k
    \end{smallmatrix}\]
    \Big\}
    &=
    \half
    (k+2)(k-1) \,
    \cC
    \!
    \[\begin{smallmatrix}
    \ell & 0 \\
    0 & 0
    \end{smallmatrix}\]
    \cC
    \!
    \[\begin{smallmatrix}
    k+2-\ell & 0 \\
    0 & k+2
    \end{smallmatrix}\]
    \smash[t]{\vphantom{\sum_{j=4}^{k+1}}}
\no \\
    & \quad
    - k \,
    \cC
    \!
    \[\begin{smallmatrix}
    \ell & 0 \\
    0 & 0
    \end{smallmatrix}\]
    \cC
    \!
    \[\begin{smallmatrix}
    k+1-\ell & 1 & 0 \\
    0 & 1 & k+1
    \end{smallmatrix}\]
    \vphantom{\sum_{j=4}^{k+1}}
\no \\
    & \quad
    -
    \sum_{j=4}^{k+1}
    \, (j-1) \,
    \cC
    \!
    \[\begin{smallmatrix}
    \ell & 0 \\
    0 & 0
    \end{smallmatrix}\]
    \cC
    \!
    \[\begin{smallmatrix}
    0 & 0 \\
    j & 0
    \end{smallmatrix}\]
    \cC
    \!
    \[\begin{smallmatrix}
    k+2-\ell & 0 \\
    0 & k+2-j
    \end{smallmatrix}\]
\end{align}
This identify follows from the action of the Maass operators~\eqref{eq:nabla} and holomorphic subgraph reduction~\eqref{eq:HSR}. After writing each inhomogeneous term using named modular graph functions, we arrive at~\eqref{eq:Dvk3E} and~\eqref{eq:Dvk3O}. This completes our proof of \autoref{thm:Dvk3}.

\subsubsection{Proof of \autoref{thm:Ivk3}}

We shall now evaluate the integral of~$v_{k,3}$ over $\cM_L$ by considering the contributions from each term in its inhomogeneous Laplace equation.

\sm

We have already calculated the integrals of many of these terms. The integral of $E_{k+2}$ is given in~\eqref{eq:IE} and vanishes up to powers of $L$ with non-zero exponents. The integral of~$v_{k+1,2}$ is given by \autoref{thm:Ivk2} and vanishes when $k$ is odd. The integral of~\smash{$\cW^{(m,n)}_{k+2}$} is given by \autoref{thm:IWint}. For even $k$, the integral of $\tau_2^{k+2} \, G_{k+2} \, \bar{G}_{k+2}$ is given by \autoref{thm:IVint}. For odd~$k$, the integrals of the $\cV$ terms vanish by \autoref{thm:IVint}. This leaves the contributions from the total derivative terms $\Delta v_{k,3}$, $\nabla \, \cU_{k;\ell}^+$, and $\bar{\nabla} \, \cU_{k;\ell}^-$. 

\sm

By \autoref{thm:IDC}, the integral of $\Delta v_{k,3}$ is equal to the coefficient of the linear term in the Laurent polynomial of $v_{k,3}$, and this linear term is fixed by the inhomogeneous Laplace equations~\eqref{eq:Dvk3E} and~\eqref{eq:Dvk3O}.

\sm

The Eisenstein series $E_{k+2}$, the two-loop function $v_{k+1,2}$, and the disconnected two-loop modular graph functions on the last lines of both~\eqref{eq:Dvk3E} and~\eqref{eq:Dvk3O} do not have linear terms in their Laurent polynomials. The asymptotic expansions of these functions are given in~\eqref{eq:EAsy},~\eqref{eq:vk2Asy}, and~\eqref{eq:VE}, respectively. Moreover, the total derivative terms cannot have linear terms in their Laurent polynomials. Thus, only the disconnected three-loop function~\smash{$\cW_{k+2}^{(m,n)}$} can contribute a linear term to the Laurent polynomial of $v_{k,3}$. These two Laurent coefficients are related by,
\begin{align}
    \mfc_{v_{k,3}}^{(1)}
    &=
    \frac{ 2 \, (-)^k }{ k(k-1) }
    \sum_{ m,n = 2 }^{\floor{k/2}}
    (2m-1)(2n-1) \,
    \mfc_{\cW_{k+2}^{(m,n)}}^{(1)}
\end{align}
The asymptotic expansion of~\smash{$\cW_{k+2}^{(m,n)}$} is given in~\eqref{eq:WAsy}, and its Laurent polynomial has a linear term if and only if $k+2=2m+2n$. Thus, the linear coefficient of the Laurent polynomial of $v_{k,3}$ is given by,
\begin{align}
\label{eq:vk3linear}
    \mfc_{v_{k,3}}^{(1)}
    &=
    16 \pi \,
    \zeta(k+1)
    \sum_{ m,n = 2 }^{\floor{k/2}}
    \frac{ (k-2)! }{ (2m-2)! (2n-2)! }
    \frac{B_{2m}}{(2m)!}
    \frac{B_{2n}}{(2n)!}
    \,
    \delta_{k+2,2m+2n}
\end{align}
which vanishes for all odd $k$ as well as for even $k \leq 4$.

\sm

By \autoref{thm:IDC}, the integrals of \smash{$\nabla \, \cU_{k;\ell}^+$} and \smash{$\bar{\nabla} \, \cU_{k;\ell}^-$} are equal to the coefficients of the constant terms in the Laurent polynomials of $\cU^+$ and $\cU^-$, respectively. To calculate these coefficients, we shall work with $\cU^+$. The case of $\cU^-$ is related by complex conjugation.

\sm

We first recall the definition~\eqref{eq:Ukl} of \smash{$\cU_{k;\ell}^+$}. Since $\bar{G}_{2\ell}$ has only a constant term in its Laurent polynomial~\eqref{eq:GAsy}, we may focus on the modular graph form \smash{$\cC^+ \! \[ \begin{smallmatrix} k & 1 & 0 \\ 0 & 1 & k+2-2\ell \end{smallmatrix} \]$}. This function obeys the following differential identity,
\begin{align}
    \cC^+
    \!
    \[\begin{smallmatrix}
    k & 1 & 0 \\
    0 & 1 & k+2-2\ell 
    \end{smallmatrix}\]
    &=
    (-)^{\ell-1} \,
    \tfrac{ (k-\ell)! }{ (k-1)! } \,
    \nabla^{\ell-1}
    \cC^+
    \!
    \[\begin{smallmatrix}
    k+1-\ell & 1 & 0 \\
    0 & 1 & k+1-\ell 
    \end{smallmatrix}\]
\\
    & \quad
    + (-)^{k-1}
    \sum_{j=1}^{\ell-1}
    \tfrac{ (k-1-j)! }{ (k-1)! } \,
    \nabla^{j-1}
    \bigg(
    \half
    (k+2-j) (k-1-j) \,
    \cC^+
    \!
    \[\begin{smallmatrix}
    k+2-j & 0 \\
    k+2-2\ell+j & 0
    \end{smallmatrix}\]
    \no \\[-2ex]
    & \hspace{2.1in}
    -
    \sum_{m=4}^{k+1-j}
    (m-1) \,
    \cC^+
    \!
    \[\begin{smallmatrix}
    m & 0 \\
    0 & 0
    \end{smallmatrix}\]
    \cC^+
    \!
    \[\begin{smallmatrix}
    k+2-j-m & 0 \\
    k+2-2\ell+j & 0
    \end{smallmatrix}\]
    \bigg)
\no
\end{align}
which may be proved through repeated use of holomorphic subgraph reduction~\eqref{eq:HSR}. The first term on the right-hand side of this expression is a derivative of the two-loop modular graph function $v_{k+1-\ell,2}$. The second and third lines can be written as derivatives of Eisenstein series and do not have constant terms in their Laurent polynomials. Thus,
\begin{align}
    \cC^+
    \!
    \[\begin{smallmatrix}
    k & 1 & 0 \\
    0 & 1 & k+2-2\ell 
    \end{smallmatrix}\]
    \big|_{ \text{const.} }
    &=
    (-)^{\ell-1} \,
    \tfrac{ (k-\ell)! }{ (k-1)! }
    \(
    \nabla^{\ell-1}
    v_{k+1-\ell,2}
    \)
    \big|_{ \text{const.} }
    =
    \tbinom{k-1}{\ell-1}^{-1} \,
    \mfc^{(1-\ell)}_{v_{k+1-\ell,2}}
\end{align}
The Laurent coefficients for $v_{k+1-\ell,2}$ are given below~\eqref{eq:vk2Asy}. After including the multiplicative contribution from the constant term in the Laurent polynomial of $\bar{G}_{2\ell}$, we obtain,
\begin{align}
\label{eq:Uklconst}
    \mfc^{(0)}_{\cU_{k;\ell}^\pm}
    &=
    ( 1 - \delta_{2\ell,k} ) \,
    16 \pi \,
    \zeta(k+1) \,
    \frac{B_{2\ell}}{(2\ell)!}
    \frac{B_{k+2-2\ell}}{(k+2-2\ell)!}
\end{align}
which vanishes for all odd $k$ as well as for even $k = 2\ell$.

\sm

Combining these results, we obtain the expressions for the integral of $v_{k,3}$ given in~\eqref{eq:Ivk3E} and~\eqref{eq:Ivk3O} of \autoref{thm:Ivk3}. The subtraction of triple products of Eisenstein series with rational coefficients described in~\eqref{eq:IQvk3} and~\eqref{eq:Q(vk3)} follows from the discussion of the transcendentality of the integral of $\cW_{k+2}^{(m,n)}$ in \autoref{thm:IWint}. This completes our proof of \autoref{thm:Ivk3}.

\newpage

\section{Integrating \texorpdfstring{$C_{k,1,1,1}$}{}}
\label{sec:Ck111}

In this section, we shall evaluate the integrals of the infinite family of connected three-loop modular graph functions $C_{k,1,1,1}$ for integer $k \geq 2$ using the unfolding trick.\footnote{The case $k=1$ has already been discussed in~\cite{DHoker:2019blr} and will not be reconsidered here as its treatment requires some extra technical modifications to the procedure used for $k \geq 2$.} As a warm-up for this three-loop calculation, we shall first evaluate the integrals of the infinite family of connected two-loop modular graph functions $C_{k,1,1}$ which are the two-loop analogues of~$C_{k,1,1,1}$. We shall also discuss the transcendental structure of these integrals.

\sm

The connected two-loop functions $C_{k,1,1}$ form an infinite subfamily of the two-loop functions $C_{a,b,c}$ defined in~\eqref{eq:Cabc} with $a=k$ and ${b=c=1}$. They have the following decorated dihedral graph, matrix of exponents, and Kronecker-Eisenstein series representation,
\begin{align}
\label{eq:Ck11}
    C_{k,1,1}
    ~&=~
    \begin{tikzpicture}[baseline=-0.5ex, scale=.8, yscale=.8, xscale=.9]
    \draw (-1,0) node[circle,fill,inner sep=1.5](a){};
    \draw (1,0) node[circle,fill,inner sep=1.5](b){};
    \draw (-1,0) arc(180:0:1)
        node[style=draw,midway,fill=white,scale=.5]
        {$k,k$};
    \draw (-1,0) arc(-180:0:1)
        node[style=draw,midway,fill=white,scale=.5]
        {$1,1$};
    \draw (a) to [out=0,in=180] 
        node[style=draw,midway,fill=white,scale=.5] 
        {$1,1$} (b);
    \end{tikzpicture}
    ~=~
    \cC
    \! \begin{bmatrix}
    k & 1 & 1 \\
    k & 1 & 1
    \end{bmatrix}
    =
    \frac{\tau_2^{k+2}}{\pi^{k+2}}
    \sum_{p_1, p_2, p_3 \in \Lambda'}
    \frac{ \delta(p_1+p_2+p_3)  }
         { |p_1|^{2} \,
           |p_2|^{2} \,
           |p_3|^{k} }
\end{align}
The functions $C_{k,1,1}$ have weight $k+2$. For integer $k \geq 1$, the Kronecker-Eisenstein series converges and is real.

\sm

The connected three-loop functions $C_{k,1,1,1}$ are the three-loop analogues of $C_{k,1,1}$. They have the following decorated dihedral graph, matrix of exponents, and Kronecker-Eisenstein series representation,
\begin{align}
\label{eq:Ck111}
    C_{k,1,1,1}
    ~&=~
    \begin{tikzpicture}[baseline=-0.5ex, scale=.8, yscale=.8, xscale=.9]
    \draw (-1,0) node[circle,fill,inner sep=1.5](a){};
    \draw (1,0) node[circle,fill,inner sep=1.5](b){};
    \draw (-1,0) arc(180:0:1)
        node[style=draw,midway,fill=white,scale=.5]
        {$k,k$};
    \draw (-1,0) arc(-180:0:1)
        node[style=draw,midway,fill=white,scale=.5]
        {$1,1$};
    \draw (a) to [out=30,in=150]
        node[style=draw,midway,fill=white,scale=.5] 
        {$1,1$} (b);
    \draw (a) to [out=-30,in=-150] 
        node[style=draw,midway,fill=white,scale=.5] 
        {$1,1$} (b);
    \end{tikzpicture}
    ~=~
    \cC
    \! \begin{bmatrix}
    k & 1 & 1 & 1\\
    k & 1 & 1 & 1
    \end{bmatrix}
    =
    \frac{\tau_2^{k+3}}{\pi^{k+3}} \,
    \sum_{p_1, p_2, p_3, p_4 \in \Lambda '}
    \frac{\delta(p_1+p_2+p_3+p_4) }
         { |p_1|^2 \, |p_2|^2 \,
           |p_3|^2 \, |p_4|^{2k} }
\end{align}
The functions $C_{k,1,1,1}$ have weight $k+3$. For integer $k \geq 1$, the Kronecker-Eisenstein series converges and is real. Unlike the functions of previous two sections, the Laplacian does not have a nice action on the functions $C_{k,1,1,1}$. The cases $C_{1,1,1,1}$, $C_{2,1,1,1}$, and $C_{3,1,1,1}$ appear in the low-energy expansion of the genus-one four-graviton amplitude in Type~II superstring theory, and their integrals over $\cM_L$ were calculated in~\cite{DHoker:2019blr}. The function $C_{1,1,1,1}$ is a so-called melon modular graph function and is often denoted by $D_4$~\cite{DHoker:2019xef}.

\subsection{The integral of \texorpdfstring{$C_{k,1,1}$}{}}

The two-loop functions $C_{k,1,1}$ may be written as a linear combination of the functions~$\cC_{u,v;w}$ using \autoref{thm:CuvwProp}. Thus, the integral of $C_{k,1,1}$ may be computed from the expressions for the integrals of $\cC_{u,v;w}$ in \autoref{thm:ICuvw}. However, with the three-loop function~$C_{k,1,1,1}$ in mind, we shall present a new calculation using the unfolding trick and \autoref{thm:Iunfold} which is more readily generalized to three loops. In the remainder of this subsection, we shall prove the following theorem.

\begin{thm}
\label{thm:ICk11}
The integral of $C_{k,1,1}$ for integer $k \geq 1$ is given as follows.
\begin{itemize}
\item When $k$ is odd, the integral has transcendental weight $k+3$ and is given by,
\begin{align}
\label{eq:ICk11O}
    \intML C_{k,1,1}
    &\approx
    4 \pi \,
    \zeta(k+2) \,
    \dfrac{B_{k+1}}{(k+1)!}
\intertext{ \item When $k$ is even, the integral has transcendental weight $k+3$ and is given by,}
\label{eq:ICk11E}
    \intML C_{k,1,1}
    &\approx
    - 16 \pi \,
    \zeta(k+1) \,
    \dfrac{B_{k+2}}{(k+2)!}
    \[
    \dfrac{\zeta'(k+2)}{\zeta(k+2)}
    - \dfrac{\zeta'(k+1)}{\zeta(k+1)}
    + \ln(2L)
    \]
\end{align}
\end{itemize}
In each case, the integral is of the form claimed in \autoref{thm:2looptrans}.
\end{thm}

\subsubsection{Proof of \autoref{thm:ICk11}}

To prove this theorem, we shall integrate the Poincar\'e seed function $\Lambda_{k,1,1}$ for $C_{k,1,1}$ over the truncated upper half-strip. This calculation is similar to the proofs of both \autoref{thm:IE1} and \autoref{thm:ICuvw}.

\sm

We first obtain an expression for $\Lambda_{k,1,1}$ from the Kronecker-Eisenstein series representation~\eqref{eq:Ck11} of $C_{k,1,1}$ by rotating the integer pair $(m_3,n_3) \neq (0,0)$ to $(0,N) \neq (0,0)$,
\begin{align}
    \Lambda_{k,1,1}
    =
    \frac{\tau_2^{k+2}}{\pi^{k+2}}
    \sum_{N \neq 0}
    \,
    \sum^{'}_{\substack{(m_r, n_r) \in \bbZ^2
        \\ r = 1,2}}
    \frac{\delta(m_1+m_2) \delta(n_1+n_2+N)}
        {|m_1\tau + n_1|^2 |m_2\tau + n_2|^2 N^{2k}}
\end{align}
Splitting \smash{$\Lambda_{k,1,1} = \Lambda^{[0]}_{k,1,1} + \Lambda^{[1]}_{k,1,1} + \Lambda^{[2]}_{k,1,1}$} into contributions according to the number of non-vanishing summation variables $m_r$, we see that \smash{$\Lambda^{[1]}_{k,1,1}$} vanishes thanks to the delta function constraint. The remaining contributions are given by,
\begin{align}
    \Lambda^{[0]}_{k,1,1}
    &=
    \frac{\tau_2^{k+2}}{\pi^{k+2}}
    \sum_{n_1, n_2, N \neq 0}
    \frac{\delta(n_1+n_2+N)}
        {n_1^2 n_2^2 N^{2k}}
\no \\
    \Lambda^{[2]}_{k,1,1}
    &=
    \frac{\tau_2^{k+2}}{\pi^{k+2}}
    \sum_{m_1, m_2, N \neq 0}
    \,
    \sum_{n_1, n_2 \in \bbZ}
    \frac{\delta(m_1+m_2) \delta(n_1+n_2+N)}
        {|m_1\tau + n_1|^2 |m_2\tau + n_2|^2 N^{2k}}
\end{align}
The \smash{$\Lambda^{[0]}_{k,1,1}$} contribution yields a term proportional to $L^{k+1}$ in the integral of \smash{$\Lambda_{k,1,1}$} and may be ignored in light of \autoref{thm:Iunfold}, leaving only \smash{$\Lambda^{[2]}_{k,1,1}$}.

\sm

To evaluate \smash{$\Lambda^{[2]}_{k,1,1}$}, we first sum over the variables $m_2$ and $n_2$ using the two delta symbols. We then define $m=m_1$ and $n=n_1$, integrate over $\tau_1$ using~\eqref{eq:Isum} of \autoref{thm:Isum}, and finally integrate over $\tau_2$. After some straightforward simplifications and a change of integration variables, we obtain the following infinite series of finite integrals,
\begin{align}
\label{eq:ILk11[2]}
    \int_0^L \frac{d\tau_2}{\tau_2^2}
    \int_0^1 d\tau_1 \,
    \Lambda^{[2]}_{k,1,1}
    =
    \frac{16}{(2\pi)^{k+1}}
    \sum_{m, N > 0}
    \frac{1}{m^{k+1} N^{k+2}}
    \int_0^{\frac{2mL}{N}}
    \frac{dx \, x^{k-1}}{x^2 + 1}
\end{align}
This expression may be compactly expressed in terms of the $F_g$ functions defined in~\eqref{eq:Fgdef}. Defining $\kappa = \lfloor k/2 \rfloor$ and $\delta = k - 2\kappa$ so that $\delta = 0$ for even $k$ and $\delta = 1$ for odd $k$, we write,
\begin{align}
    \int_0^L \frac{d\tau_2}{\tau_2^2}
    \int_0^1 d\tau_1 \,
    \Lambda^{[2]}_{k,1,1}
    =
    16 \,
    \frac{(-)^{\kappa+\delta+1}}
        {(2\pi)^{k+1}} \,
    F_{1-\delta}
        (\kappa+\delta-1, 1; k+1, k+2; \mathds{1}; L)
\end{align}
For odd $k$, the integral is proportional to $F_0$ and has only polynomial divergences in $L$. For even~$k$, it is proportional to $F_1$ and has both polynomial and $\ln L$ divergences. In either case, \autoref{thm:Fg} details its large $L$ behavior. Combining these results with \autoref{thm:Iunfold} to calculate the integral of $C_{k,1,1}$, we obtain~\eqref{eq:ICk11O} and~\eqref{eq:ICk11E}. This completes our proof of \autoref{thm:ICk11}.

\subsection{The integral of \texorpdfstring{$C_{k,1,1,1}$}{}}

We now turn to the connected three-loop functions $C_{k,1,1,1}$. In the remainder of this subsection, we shall prove the following theorem using a Poincar\'e seed function~$\Lambda_{k,1,1,1}$ for~$C_{k,1,1,1}$ and the unfolding trick. This calculation is significantly more involved than the two-loop case, and several technical details are contained in \autoref{apdx:ILk111} and \autoref{apdx:Ik111exp}.

\begin{thm}
\label{thm:ICk111}
The integral over $\cM_L$ of $C_{k,1,1,1}$ for integer $k \geq 2$ is given as follows.
\begin{itemize}
\item When $k$ is odd, the integral is equal to a sum of terms with transcendental weight $k+3$ or $k+4$ plus a term proportional to $\ln(2\pi)$ and terms with odd transcendental weight greater than $k+4$,
\begin{align}
\label{eq:ICk111O}
    &
    \intML C_{k,1,1,1}
    \approx
    48 \pi \, \zeta(k+2)
    \Bigg\{
    \sum_{\ell=2}^{(k-1)/2}
    \tfrac{ B_{2\ell\vphantom{k+1}} }{ (2\ell)! }
    \tfrac{ B_{k+3-2\ell} }{ (k+3-2\ell)! }
    \Big[
    (k+4-6\ell) \,
    \tfrac{ \zeta'(2\ell) }{ \zeta(2\ell) }
    - \ln(2\pi)
    \Big]
\no \\[2ex]
    & \qquad
    -
    \tfrac{ B_{2\vphantom{k+1}} }{ 2! \mathstrut }
    \tfrac{ B_{k+1} }{ (k+1)! }
    \Big[
    \tfrac{ \zeta'(k+2) }{ \zeta(k+2) }
    + 2 \, (k+2) \, \tfrac{ \zeta'(k+1) }{ \zeta(k+1) }
    - (k+2) \, \tfrac{ \zeta'(2) }{ \zeta(2) }
    - \ln(2L)
    - 2
    \Big]
\no \\
    & \qquad
    -
    \tfrac{ B_{k+3} }{ (k+3)! }
    \Big[
    5 \, (k+1) \, \tfrac{ \zeta'(k+3) }{ \zeta(k+3) }
    + \half (k-1)(k-2) \, \tfrac{ \zeta'(k+2) }{ \zeta(k+2) }
    - \half (k-1)(k-2) \ln(2L)
    - 6
    \Big]
    \vphantom{\sum_{\ell=2}^{(k-1)/2}}
\no \\
    & \qquad
    +
    \sum_{\substack{ \ell_{1,2}=2 \\ \mathclap{\ell_1+\ell_2 < (k+3)/2} }}
         ^{(k-1)/2}
    \tfrac{ B_{k+3-2\ell_1} }{ (k+3-2\ell_1)! }
    \tfrac{ B_{k+3-2\ell_2} }{ (k+3-2\ell_2)! }
    (k+3-2\ell_1-2\ell_2)! \,
    \zeta(k+4-2\ell_1-2\ell_2)
    \Bigg\}
\end{align}
\item When $k$ is even, the integral is equal to a sum of terms with transcendental weight $k+4$ plus terms with odd transcendental weight greater than $k+4$,
\begin{align}
\label{eq:ICk111E}
    &
    \intML C_{k,1,1,1}
    \approx
    4 \pi \,
    \tfrac{ B_{k+2} }{ (k+2)! }
    \Bigg\{
    -(2k+1)(k+1) \, \zeta(k+3)
\no \\[1ex]
    & \qquad
    + 12
    \sum_{\ell=1}^{k/2-1}
    \zeta(2\ell+1) \,
    \zeta(k+1-2\ell)
    \Big[
    \tfrac{ \zeta'(k+2) }{ \zeta(k+2) }
    - \tfrac{ \zeta'(k+1-2\ell) }{ \zeta(k+1-2\ell) }
    - \tfrac{ \zeta'(2\ell) }{ \zeta(2\ell) }
    + \gamma_E
    + \ln(2L)
    \Big]
\no \\
    & \qquad
    + 4
    \phantom{12}\negphantom{4}
    \sum_{\substack{ \ell_{1,2}=1 \\ \mathclap{\ell_1+\ell_2 < k/2} }}^{k/2-1}
    \zeta(2\ell_1+1) \,
    \zeta(2\ell_2+1) \, 
    \zeta(k+1-2\ell_1-2\ell_2)
\no \\
    & \qquad
    - 12
    \sum_{\substack{ \ell_{1,2}=2 \\ \mathclap{\ell_1+\ell_2 < (k+4)/2} }}
         ^{ k/2 }
    \tfrac{ B_{k+4-2\ell_1-2\ell_2} }{ (k+4-2\ell_1-2\ell_2)! } \,
    \zeta(k+3-2\ell_1) \,
    \zeta(k+3-2\ell_2)
    \Bigg\}
\end{align}
\end{itemize}
In each case, we may subtract the integral of a suitable linear combination of triple products of Eisenstein series to construct the following integral whose terms all have transcendental weight~$k+3$ or~$k+4$,
\begin{align}
\label{eq:IQCk111}
    \intML
    \bigg(
    C_{k,1,1,1}
    -
    \sum_{\substack{ \ell_{1,2}=2 \\ \mathclap{\ell_1+\ell_2 < (k+4)/2} }}
        ^{\floor{k/2}}
    Q_{\ell_1,\ell_2,k+3-\ell_1-\ell_2}^{(C)} \,
    E_{\ell_1}^*
    E_{\ell_2}^*
    E_{k+3-\ell_1-\ell_2}^*
    \bigg)
\end{align}
where \smash{$Q_{\ell_1,\ell_2,k+3-\ell_1-\ell_2}^{(C)}$} is the following rational number,
\begin{align}
\label{eq:Q(Ck111)}
    Q_{\ell_1,\ell_2,k+3-\ell_1-\ell_2}^{(C)}
    &=
    24 \,
    (-)^{\ell_1+\ell_2+\delta_{k+3,2\ell_1+2\ell_2}} \,
    \frac{ (k+3-2\ell_1-2\ell_2)!! }
         { k!! (k+1-2\ell_1)!! (k+1-2\ell_2)!! }
\end{align}
Thus, the integral of $C_{k,1,1,1}$ is of the form claimed in \autoref{thm:3looptrans}.
\end{thm}

Before we prove this theorem, some remarks are in order. For all $k \geq 2$, we have written the integral of $C_{k,1,1,1}$ as a piece with transcendental weight $k+3$ or $k+4$ plus terms with novel transcendental structure. These novel terms include $\ln(2\pi)$ which occurs for odd~${k \geq 5}$. There are also terms with odd transcendental weight greater than~${k+4}$ for all~${k \geq 6}$. For instance, at weight eight, the integral of $C_{5,1,1,1}$ contains $\ln(2\pi)$. Then at weight nine, the integral of $C_{6,1,1,1}$ includes a term with transcendental weight eleven. This is precisely the transcendental structure of the integrals of triple products of Eisenstein series which we described at length in \autoref{sec:Eisen}. In any case, the terms with novel transcendental structure may be removed by subtracting the integral of a suitable linear combination of triple products of Eisenstein series as detailed in~\eqref{eq:IQCk111}. We speculate that this structure originates from triple products of Eisenstein series (or their derivatives) which appear in the inhomogeneous Laplace equation obeyed by a larger family of three-loop MGFs. 

\subsubsection{Proof of \autoref{thm:ICk111}}

Following \autoref{thm:Iunfold} and our previous two-loop calculation, we shall integrate the Poincar\'e seed function $\Lambda_{k,1,1,1}$ over the truncated upper half-strip. 

\sm

We first obtain an expression for $\Lambda_{k,1,1,1}$ from the Kronecker-Eisenstein series representation~\eqref{eq:Ck111} of $C_{k,1,1,1}$ by rotating the integer pair $(m_4,n_4) \neq (0,0)$ to $(0,N) \neq (0,0)$,
\begin{align}
    \Lambda_{k,1,1,1}
    =
    \frac{\tau_2^{k+3}}{\pi^{k+3}}
    \sum_{N \neq 0}
    \,
    \sum^{'}_{\substack{(m_r, n_r) \in \bbZ^2
        \\ r = 1,2,3}}
    \frac{\delta(m_1+m_2+m_3) \delta(n_1+n_2+n_3+N)}
        {|m_1\tau + n_1|^2 |m_2\tau + n_2|^2
        |m_3\tau + n_3|^2 N^{2k}}
\end{align}
Splitting \smash{$\Lambda_{k,1,1,1} = \Lambda^{[0]}_{k,1,1,1} + \Lambda^{[1]}_{k,1,1,1} + \Lambda^{[2]}_{k,1,1,1} + \Lambda^{[3]}_{k,1,1,1}$} into contributions according to the number of non-vanishing summation variables $m_r$, we see that \smash{$\Lambda^{[1]}_{k,1,1,1}$} vanishes thanks to the delta function constraint. The remaining contributions are given by,
\begin{align}
    \label{eq:Lk111}
    \Lambda^{[0]}_{k,1,1,1}
    &=
    \frac{\tau_2^{k+3}}{\pi^{k+3}}
    \sum_{N \neq 0}
    \,
    \sum_{\substack{n_r \neq 0 \\
        r = 1,2,3}}
    \frac{\delta(n_1+n_2+n_3+N)}
        {n_1^2 n_2^2 n_3^2 N^{2k}}
\no \\
    \Lambda^{[2]}_{k,1,1,1}
    &=
    3 \,
    \frac{\tau_2^{k+3}}{\pi^{k+3}}
    \sum_{n_3, N \neq 0}
    \,
    \sum_{\substack{m_r \neq 0 \\
        n_r \in \bbZ \\
        r = 1,2}}
    \frac{\delta(m_1+m_2) \delta(n_1+n_2+n_3+N)}
        {|m_1\tau + n_1|^2 |m_2\tau + n_2|^2
        n_3^2 N^{2k}}
\no \\
    \Lambda^{[3]}_{k,1,1,1}
    &=
    \frac{\tau_2^{k+3}}{\pi^{k+3}}
    \sum_{N \neq 0}
    \,
    \sum_{\substack{m_r \neq 0 \\
        n_r \in \bbZ \\
        r = 1,2,3}}
    \frac{\delta(m_1+m_2+m_3) \delta(n_1+n_2+n_3+N)}
        {|m_1\tau + n_1|^2 |m_2\tau + n_2|^2
         |m_3\tau + n_3|^2 N^{2k}}
\end{align}
The factor of $3$ in \smash{$\Lambda^{[2]}_{k,1,1,1}$} arises from the three different $m_r$ which can vanish. The \smash{$\Lambda^{[0]}_{k,1,1,1}$} contribution is proportional to $\tau_2^{k+3}$ and will produce a contribution proportional to $L^{k+2}$ in the integral of $\Lambda_{k,1,1,1}$ which may be ignored in light of \autoref{thm:Iunfold}.

\sm

Hence, only \smash{$\Lambda^{[2]}_{k,1,1,1}$} and \smash{$\Lambda^{[3]}_{k,1,1,1}$} will contribute to the integral of $\Lambda_{k,1,1,1}$ (and thus to the integral of $C_{k,1,1,1}$). Their integrals may be massaged into the following form.

\begin{lem}
\label{thm:ILk111}
Up to exponentially suppressed corrections, the integral of \smash{$\Lambda^{[2]}_{k,1,1,1} + \Lambda^{[3]}_{k,1,1,1}$} over the truncated upper half-strip may be written as follows,
\begin{align}
\label{eq:ILk111}
    \int_0^L \frac{d\tau_2}{\tau_2^2}
    \int_0^1 d\tau_1 \,
    \Big(
    \Lambda^{[2]}_{k,1,1,1}
    + \Lambda^{[3]}_{k,1,1,1}
    \Big)
    &=
    \cI_{k,1,1,1}^{\mathrm{(exp)}}
    +
    \cI_{k,1,1,1}^{\mathrm{(pow)}}(L)
\end{align}
where $\cI_{k,1,1,1}^{\mathrm{(exp)}}$ and $\cI_{k,1,1,1}^{\mathrm{(pow)}}(L)$ are the following infinite series of integrals,
\begin{align}
\label{eq:Ik111}
    \cI_{k,1,1,1}^{\mathrm{(exp)}}
    &=
    \frac{96 \, \zeta(k+2) }{(2\pi)^{k+1}}
    \sum_{N_1, N_2 > 0}
    \frac{1}{N_1^{k-1} N_2^{k-1}}
    \int_0^\infty
    \frac{dx \, x^{k-1}}
        {(x^2 + N_1^2)(x^2 + N_2^2)}
    \frac{e^{-2\pi x}}{1-e^{-2\pi x}}
\no \\[2ex]
    \cI_{k,1,1,1}^{\mathrm{(pow)}}(L)
    &=
    \frac{48}{(2\pi)^{k+2}} \,
    \sum_{m, N > 0}
    \Bigg\{
    \frac{2 \, \zeta(2) }{m^{k+2} N^{k+1}}
    \int_0^{\frac{2mL}{N}}
    \frac{dx \, x^k}{x^2 + 1}
    + \frac{1}{m^{k+2} N^{k+3}}
    \int_0^{\frac{2mL}{N}}
    \frac{dx \, x^k}{(x^2 + 1)^2}
\no \\
    & \hspace{0.7in}
    - \frac{4}{m^{k+2} N^{k+3}}
    \int_0^{\frac{2mL}{N}}
    \frac{dx \, x^k}{(x^2 + 1)^3}
    - \frac{\pi}{m^{k+2} N^{k+2}}
    \int_0^{\frac{2mL}{N}}
    \frac{dx \, x^{k-1}}{x^2 + 1}
\no \\[1ex]
    & \hspace{0.7in}
    + \frac{2\pi}{m^{k+2} N^{k+2}}
    \int_0^{\frac{2mL}{N}}
    \frac{dx \, x^{k-1}}{(x^2 + 1)^2}
    + \frac{4\pi \, H_1(m)}{m^{k+1} N^{k+2}}
    \int_0^\frac{2mL}{N} 
    \frac{dx \, x^{k-1}}{x^2 + 1}
    \Bigg\}
\end{align}
These two contributions are so named because their integrands decay exponentially or are power-behaved for large $x$.
\end{lem}

In \autoref{apdx:ILk111}, we prove this lemma by manipulating the Kronecker-Eisenstein series for \smash{$\Lambda^{[2]}_{k,1,1,1}$} and \smash{$\Lambda^{[3]}_{k,1,1,1}$}. It now remains to explicitly evaluate \smash{$\cI_{k,1,1,1}^{\mathrm{(pow)}}(L)$} and \smash{$\cI_{k,1,1,1}^{\mathrm{(exp)}}$}.

\sm

We shall begin with the series of power-behaved integrals. The six terms in \smash{$\cI_{k,1,1,1}^{\mathrm{(pow)}}(L)$} are each similar to the integral of \smash{$\Lambda_{k,1,1}^{[2]}$} in~\eqref{eq:ILk11[2]} which we encountered in our two-loop warm-up. Similarly, these integrals may be compactly expressed in terms of the general family $F_g$ functions defined in~\eqref{eq:Fgdef}. The large-$L$ behavior of \smash{$\cI_{k,1,1,1}^{\mathrm{(pow)}}(L)$} is then given by \autoref{thm:Fg}, and we find the following lemma.

\begin{lem}
\label{thm:Ik111pow}
The power-behaved contribution \smash{$\cI_{k,1,1,1}^{\mathrm{(pow)}}(L)$} for integer $k \geq 2$ is given by,
\begin{align}
\label{eq:Ik111pow}
    \cI_{k,1,1,1}^{\mathrm{(pow)}}(L)
    & \approx
    24 \,
    (-)^{\lfloor k/2 \rfloor} \,
    \frac{ \zeta(k+2) }{(2\pi)^{k+1}} \,
    \cJ_{k,1,1,1}^{\mathrm{(pow)}}(L)
\end{align}
where $\cJ_{k,1,1,1}^{\mathrm{(pow)}}(L)$ is given as follows.
\begin{itemize}
\item When $k$ is odd,
\begin{align}
\label{eq:Ik111powO}
    \pi \,
    \cJ_{k,1,1,1}^{\mathrm{(pow)}}(L)
    & =
    \pi^2
    \Big(
    2 \, \zeta(k+1,1)
    - \half (k-1) \, \zeta(k+2)
    \Big)
\\
    & \quad
    + 2 \, \zeta(2) \, \zeta'(k+1)
    - \half (k-1) (k-2) \, \zeta'(k+3)
    + ( k-\tfrac{3}{2} ) \, \zeta(k+3)
    \phantom{\Big)}
\no \\
    & \quad
    +
    \Big(
    \half (k-1) (k-2) \, \zeta(k+3)
    - 2 \, \zeta(2) \, \zeta(k+1)
    \Big)
    \Big(
    \tfrac{ \zeta'(k+2) }{ \zeta(k+2) }
    - \ln(2L)
    \Big)
\no
\intertext{\item When $k$ is even,}
\label{eq:Ik111powE}
    \cJ_{k,1,1,1}^{\mathrm{(pow)}}(L)
    & =
    \zeta(k+1) \, \zeta(2)
    - \zeta(k+2)
    - (k-1) \, \zeta'(k+2)
    + 4 \, \tfrac{d}{dk} \zeta(k+1,1)
\no \\
    & \quad
    - \tfrac{1}{4} (k-1)(k-2) \, \zeta(k+3)
    \phantom{\Big)}
\no \\
    & \quad
    +
    \Big(
    (k-1) \, \zeta(k+2)
    - 4 \, \zeta(k+1,1)
    \Big)
    \Big(
    \tfrac{ \zeta'(k+2) }{ \zeta(k+2) }
    + \ln(2L)
    \Big)
\end{align}
\end{itemize}
We have arranged these expressions in a specific form for later convenience.
\end{lem}

To prove this lemma, we first define $\kappa = \lfloor k/2 \rfloor$ and $\delta = k - 2\kappa$. We then write the expression for~\smash{$\cI_{k,1,1,1}^{\mathrm{(pow)}}(L)$} in~\eqref{eq:Ik111} in terms of the $F_g$ functions as follows,
\begin{align}
    \cI_{k,1,1,1}^{\mathrm{(pow)}}(L)
    &=
    48 \, (-)^{\kappa} \, / \, (2\pi)^{k+2} \,
    \Big\{
    2 \, \zeta(2) \,
    F_\delta(\kappa, 1; k+2, k+1; \mathds{1}; L)
    \vphantom{\Big]}
\no \\
    & \qquad
    - F_\delta(\kappa, 2; k+2, k+3; \mathds{1}; L)
    \vphantom{\Big]}
\no \\
    & \qquad
    - 4 \,
    F_\delta(\kappa, 3; k+2, k+3; \mathds{1}; L)
    \vphantom{\Big]}
\no \\
    & \qquad
    + \pi \, (-)^\delta \,
    F_{1-\delta}(\kappa+\delta-1, 1; k+2, k+2; \mathds{1}; L)
    \vphantom{\Big]}
\no \\
    & \qquad
    + 2\pi \, (-)^\delta \,
    F_{1-\delta}(\kappa+\delta-1, 2; k+2, k+2; \mathds{1}; L)
    \vphantom{\Big]}
\no \\
    & \qquad
    - 4\pi \, (-)^\delta \,
    F_{1-\delta}(\kappa+\delta-1, 1; k+1, k+2; H_1; L)
    \Big\}
\end{align}
We note that here both $F_0$ and $F_1$ appear regardless of whether $k$ is odd or even. In the two-loop case, only $F_0$ or $F_1$ respectively appeared for odd or even $k$. We now use \autoref{thm:Fg} to extract the constant and $\ln L$ terms from this expression. The arithmetic functions $\mathds{1}$ and~$H_1(m)$ appear in the second-to-last arguments of the $F_g$ functions, so both the Riemann zeta function $\zeta(s)$ and the double zeta function $\zeta(s,1)$ will appear. Some straightforward algebra then yields~\eqref{eq:Ik111pow},~\eqref{eq:Ik111powE}, and~\eqref{eq:Ik111powO}. This completes our proof of \autoref{thm:Ik111pow}.

\sm

The remaining contribution \smash{$\cI_{k,1,1,1}^{\mathrm{(exp)}}$} is more difficult to evaluate but may be written in terms of zeta-values and related functions. In \autoref{apdx:Ik111exp}, we prove the following lemma using contour integral methods and analytic continuation.

\begin{lem}
\label{thm:Ik111exp}
The exponential contribution \smash{$\cI_{k,1,1,1}^{\mathrm{(exp)}}$} for integer $k \geq 2$ is given by,
\begin{align}
\label{eq:Ik111exp}
    \cI_{k,1,1,1}^{\mathrm{(exp)}}
    &=
    24 \,
    (-)^{\lfloor k/2 \rfloor} \,
    \frac{ \zeta(k+2) }{(2\pi)^{k+1}} \,
    \cJ_{k,1,1,1}^{\mathrm{(exp)}}
\end{align}
where $\cJ_{k,1,1,1}^{\mathrm{(pow)}}(L)$ is given as follows.
\begin{itemize}
\item When $k$ is odd,
\begin{align}
\label{eq:Ik111expO}
    \pi \,
    \cJ_{k,1,1,1}^{\mathrm{(exp)}}
    &=
    - \pi^2
    \Big(
    2 \, \zeta(k+1,1)
    - \half (k-1) \, \zeta(k+2)
    \Big)
\no \\
    & \quad
    - ( k + \tfrac{9}{2} ) \, \zeta(k+3)
    + 4 \, \zeta(2) \, \zeta(k+1)
    + \half (k+3)(k+4) \, \zeta'(k+3)
    \vphantom{\sum_{\ell=1}^{(k-1)/2}}
\no \\
    & \quad
    - 8
    \sum_{\ell=1}^{(k-1)/2}
    (\ell+1) \,
    \zeta(k+1-2\ell) \, \zeta'(2\ell+2)
    - 6 \, \zeta(2) \, \zeta'(k+1)
\no \\
    & \quad
    + 4
    \sum_{\ell_{1,2}=1}^{(k-1)/2}
    \,
    \zeta(2\ell_1) \,
    \zeta(2\ell_2) \,
    \zeta'(k+3-2\ell_1-2\ell_2)
\intertext{\item When $k$ is even,}
\label{eq:Ik111expE}
    \cJ_{k,1,1,1}^{\mathrm{(exp)}}
    &=
    - \zeta(k+1) \, \zeta(2)
    + \zeta(k+2)
    + (k-1) \, \zeta'(k+2)
    - 4 \, \tfrac{d}{dk} \zeta(k+1,1)
\no \\
    & \quad
    + \tfrac{1}{4} (k^2-k-4) \, \zeta(k+3)
    - 2k \, \zeta(k+2,1)
    + 4 \, \zeta(k+1,1,1)
    \vphantom{\sum_{\ell=1}^{k/2-1}}
\no \\
    & \quad
    + 4
    \sum_{\ell=1}^{k/2-1}
    \zeta(k+1-2\ell)
    \Big(
    \zeta(2\ell+1,1)
    - \ell \, \zeta(2\ell+2)
\no \\
    & \hspace{2.3in}
    - \zeta'(2\ell+1)
    + \half \gamma_E \, \zeta(2\ell+1)
    \Big)
\no \\
    & \quad
    + 2
    \sum_{\substack{ \ell_{1,2}=1 \\ \mathclap{\ell_1+\ell_2 \neq k/2} }}^{k/2-1}
    \,
    \zeta(2\ell_1+1) \,
    \zeta(2\ell_2+1) \,
    \zeta(k+1-2\ell_1-2\ell_2)
\end{align}
\end{itemize}
We have arranged these expressions in a specific form for later convenience.
\end{lem}

By \autoref{thm:Iunfold} and \autoref{thm:ILk111}, the integral of $C_{k,1,1,1}$ is equal to the sum of~\smash{$\cI_{k,1,1,1}^{\mathrm{(exp)}}$} and~\smash{$\cI_{k,1,1,1}^{\mathrm{(pow)}}(L)$}. Thus, to complete our proof we must simply combine these two terms. At this point we shall separately consider the cases of odd and even $k$.

\sm

We first consider odd $k$. In this case, we have the sum of~\eqref{eq:Ik111powO} and~\eqref{eq:Ik111expO} which is messy but simplifies considerably. We have arranged these expressions so that their first lines precisely cancel. We then use~\eqref{eq:evenZetaSum} to perform the finite sums over even zeta-values and~\eqref{eq:zeta(-2n)} to rewrite the derivatives of the Riemann zeta function at even negative integers. Finally, we write the remaining even zeta-values in terms of Bernoulli numbers. After some rearrangements, we arrive at~\eqref{eq:ICk111O}.

\sm

Next we consider even $k$. Once more, the sum of~\eqref{eq:Ik111powE} and~\eqref{eq:Ik111expE} is quite messy, but we have arranged these expressions so that their first lines precisely cancel. We use~\eqref{eq:evenZetaSum} to perform the finite sums over even zeta-values and~\eqref{eq:zeta(n,1)} to rewrite the double zeta-values of the form $\zeta(N,1)$ in terms of Riemann zeta-values. We also use~\eqref{eq:zeta(2n+1,1,1)} to write the triple zeta-value $\zeta(k+1,1,1)$ in terms of Riemann zeta-values, and we write the remaining even zeta-values in terms of Bernoulli numbers. After some rearrangements, we arrive at~\eqref{eq:ICk111E}.

\sm

With~\eqref{eq:ICk111O} and~\eqref{eq:ICk111E} in hand, it remains to prove that the integral~\eqref{eq:IQCk111} of $C_{k,1,1,1}$ minus a particular linear combination of triple products of Eisenstein series contains only terms with transcendental weight $k+3$ or $k+4$. This follows directly from the expressions for the integrals of triple products in \autoref{thm:IEEE}. This completes our proof of \autoref{thm:ICk111}.

\newpage

\section{Conclusion}
\label{sec:conc}

In this paper, we have evaluated the integrals over $\cM_L$ of several special infinite families of three-loop modular graph functions using a variety of methods. In \autoref{sec:Eisen}, we reviewed the integrals of triple products of Eisenstein series $E_s^* E_t^* E_u^*$, and we used the unfolding trick to integrate the disconnected three-loop functions \smash{$\cW_s^{(m,n)}$}. In \autoref{sec:ECabc}, we integrated the disconnected three-loop functions $E_k^* C_{a,b,c}$ using the system of inhomogeneous Laplace equations obeyed by the functions $C_{a,b,c}$. In \autoref{sec:vk3}, we integrated the connected three-loop functions $v_{k,3}$ using their inhomogeneous Laplace equations. Finally, in \autoref{sec:Ck111}, we integrated the connected three-loop functions $C_{k,1,1,1}$ using their Poincar\'e series and the unfolding trick. 

\sm

The integrals of each of these infinite families of three-loop MGFs contain the same mathematical ingredients and exhibit similar transcendental structures. Each integral is composed of Riemann zeta-values, their first derivatives, and reducible multiple zeta-values. No irreducible multiple zeta-values nor derivatives of multiple zeta-values appear in our final expressions.

\sm

For weight $w \leq 7$, the transcendental structure of the integrals of each individual three-loop MGF is consistent with the uniform transcendentality of superstring amplitudes. Specifically, each integral is the sums of terms with transcendental weight~$w+1$ and terms with transcendental weight~$w$ multiplied by rational numbers which may be interpreted as arising from finite harmonic sums.

\sm

For weight $w \geq 8$, the transcendental structure of the integrals of certain three-loop MGFs is not consistent with the uniform transcendentality of superstring amplitudes. Moreover, the violations of uniform transcendentality occurring in the integrals of individual three-loop~MGFs are all of the same form as the violations endemic to the integrals of triple products of Eisenstein series. Thus, the violations in the integrals of each three-loop~MGF may be precisely removed by subtracting the integral of a suitable linear combination of these triple products. These subtractions are generally described in \autoref{thm:3looptrans} and are specifically demonstrated in~\eqref{eq:IWint},~\eqref{eq:IQEkCwmp},~\eqref{eq:IQECabc},~\eqref{eq:IQvk3}, and~\eqref{eq:IQCk111}.

\sm

Our results have physical implications for the conjectured uniform transcendentality of the genus-one four-graviton amplitude in Type~II superstring theory. The integrals of certain three-loop MGFs violate uniform transcendentality in a controlled way, but these violations may be consistently subtracted with triple products of Eisenstein series. Thus, if the full genus-one superstring amplitude is to exhibit uniform transcendentality at weight eight and higher, special cancellations between the integrals of individual three-loop MGFs must occur. In other words, if physical amplitudes exhibit uniform transcendentality, then the violations described in this paper constrain the combinations of MGFs which may appear in the integrand of any superstring amplitude.

\sm

Future work on this front may calculate the contributions of order $D^{14} \cR^4$ and $D^{16} \cR^4$ in the low-energy expansion to see if any violations occur at weight eight. It may also be fruitful to study the transcendental structure of the genus-one five-point amplitude in Type~II superstring theory~\cite{Green:2013bza} since considerable progress has been recently made with the genus-two five-point amplitude~\cite{DHoker:2020prr, DHoker:2020tcq, DHoker:2021kks}.

\sm

Moreover, transcendentality may offer a significant guide to discovering a higher-genus generalization of the Kawai-Lewellen-Tye relations~\cite{Kawai:1985xq} between open and closed string amplitudes (and thus between gauge theory and gravity) at genus zero. The single-valued map, which acts on motivic multiple zeta-values~\cite{Brown:2013gia}, relates open and closed string amplitudes at genus zero~\cite{Schlotterer:2012ny, Stieberger:2013wea, Stieberger:2014hba, Schlotterer:2018zce, Vanhove:2018elu, Brown:2019wna}. A genus-one generalization, the elliptic single-valued map, was studied in~\cite{Zerbini:2015rss, Zerbini:2018hgs, Broedel:2018izr, Zagier:2019eus, Gerken:2020xfv, Vanhove:2020qtt}. The precise map between open and closed strings at genus one is under active study.

\sm

Future work may also attempt to evaluate the integrals of other infinite families of three-loop~MGFs using the technology developed in this paper. For instance, the family of functions $C_{k,1,1,1}$ forms a subset of the larger family of dihedral three-loop functions~$C_{a,b,c,d}$ which are analogous to the two-loop functions~$C_{a,b,c}$. Unfortunately, integrating this infinite family using the Laplace operator seems intractable since the action of the Laplacian does not close on this space. In fact, it was shown in~\cite{Basu:2019idd, Kleinschmidt:2017ege} that the Laplacian mixes the spaces of dihedral, trihedral, and tetrahedral three-loop~MGFs. It may be fruitful to apply a generating function approach, as in \autoref{apdx:CabcCwmp}, to the larger space of three-loop~MGFs with mixed topologies. It may also be possible to evaluate the integral of $C_{a,b,c,d}$ using its Poincar\'e series and the unfolding trick by generalizing the computation of \autoref{sec:Ck111}.

\sm

The integrals of four-loop MGFs are completely unknown beyond a few simple examples at small weights. Not even the integral of a quadruple product of Eisenstein series has been explicitly evaluated. Perhaps our methods may be adapted to integrate some simple infinite families of four-loop MGFs.

\sm

Orthogonally, our methods might also be used in other physical contexts to evaluate the integrals over $\cM$ or $\cM_L$ of modular-invariant functions outside the space of MGFs. Such integrals have recently appeared in string phenomenology~\cite{Abel:2021tyt} and in the context of holography between three-dimensional theories of gravity and ensembles of two-dimensional conformal field theories~\cite{Maloney:2020nni, Afkhami-Jeddi:2020ezh, Benjamin:2021ygh}.

\sm

Perhaps the most pressing open question is whether the transcendental structure of the integrals of modular graph functions may be obtained without having to explicitly evaluate the integrals. Understanding the source(s) of the violations of uniform transcendentality in some way other than by explicit calculations would significantly help progress in this direction. The fact that violations occur only for special ranges or arrangements of the exponents of the modular graph functions studied in this paper provides one piece of circumstantial evidence that such a characterization may well exist.

\newpage


\appendix
\addtocontents{toc}{\protect\setcounter{tocdepth}{1}}

\section{Zeta functions, zeta-values, and multiple zeta-values} 
\label{apdx:zeta}
Zeta functions, zeta-values, and multiple zeta-values play an important role in this paper. A review of these concepts may be found in~\cite{Zagier1994}. In this appendix, we shall review their salient properties.

\subsection{The Riemann zeta function}

The Riemann zeta function is defined for $\Re(z) > 1$ by the absolutely convergent series,
\begin{align}
    \zeta(z)
    =
    \sum_{m>0}
    \frac{1}{m^z}
\end{align}
and may be analytically continued to a meromorphic function of $z \in \bbC$ with a single simple pole at $z=1$. Near this pole,
\begin{align}
    \zeta(z)
    =
    \frac{1}{z-1}
    +
    \gamma_E
    +
    \cO(z-1)
\end{align}
where $\gamma_E$ is the Euler-Mascheroni constant.

\subsubsection{Riemann zeta-values}

For integer arguments $n \neq 1$, $\zeta(n)$ is referred to as a zeta-value. The zeta-value $\zeta(n)$ with~$n \geq 2$ is assigned transcendental weight $n$. Zeta-values for even positive integers and for all negative integers may be expressed in terms of the Bernoulli numbers~$B_n$,
\begin{align}
    \zeta(2n)
    &=
    \half (-)^{n+1} \,
    (2\pi)^{2n} \,
    \frac{ B_{2n} }{ (2n)! }
    &
    \zeta(-n)
    &=
    (-)^n \,
    \frac{B_{n+1}}{ n+1}
\end{align}
The Bernoulli numbers are rational, and the negative even zeta-values vanish since the odd Bernoulli numbers, other than $B_1$, vanish. The expression for $\zeta(2n)$ implies that $\pi$ has transcendental weight one. Additionally, $\zeta(0) = - \half$.

\sm

Finite sums of even zeta-values satisfy several identities which correspond to properties of the Bernoulli numbers. For instance, for integer $n \geq 2$,
\begin{align}
\label{eq:evenZetaSum}
    \sum_{\ell=1}^{n-1}
    \zeta(2\ell) \,
    \zeta(2n-2\ell)
    &=
    ( n + \tfrac{1}{2} ) \,
    \zeta(2n)
\no \\
    \sum_{\ell=1}^{n-1}
    \ell \,
    \zeta(2\ell) \,
    \zeta(2n-2\ell)
    &=
    \tfrac{1}{2} n
    ( n + \tfrac{1}{2} ) \,
    \zeta(2n)
\end{align}
The first identity is proved in~\cite{Zagier1994}, and the second may be reduced to the first by changing summation variables $\ell \to n-\ell$.

\subsubsection{The starred zeta function}

It is often convenient to perform intermediate calculations in terms of the starred zeta function $\zeta^*(z)$ rather than the Riemann zeta function $\zeta(z)$. The starred function is defined~by,
\begin{align}
\label{eq:zstar}
    \zeta^*(z)
    & =
    \pi^{-z/2} \, \Gamma(z/2) \,
    \zeta(z)
\end{align}
and obeys the functional equation $\zeta^*(1-z) = \zeta^*(z)$ which provides an analytic continuation for $\zeta^*(z)$, and thus $\zeta(z)$, to the complex $z$-plane. The starred zeta function has simple poles at $z=0$ and $z=1$.

\sm

For positive integer arguments, the starred zeta function is given by,
\begin{align}
    \zeta^*(2n)
    &=
    (-)^{n+1} \,
    (2n-2)!! \,
    (2\pi)^n \,
    \frac{B_{2n}}{(2n)!}
\no \\[1ex]
    \zeta^*(2n+1)
    &=
    (2n-1)!! \,
    (2\pi)^{-n} \,
    \zeta(2n+1)
\end{align}
where $n \geq 1$ and the double factorial is defined by,
\begin{align}
    n!! &= n \, (n-2)!!
    &
    1!! &= 0!! = (-1)!! = 1
\end{align}
The starred zeta-values $\zeta^*(n)$ have transcendental weight $\floor{(n+1)/2}$ for integer $n \geq 2$. Thus, the use of starred zeta-values obscures an expression's transcendental weight. For this reason, we shall always write our final expressions in terms of explicit factors of $\pi$, odd zeta-values, rational numbers such as the Bernoulli numbers $B_n$, and other objects with manifest transcendental weight.

\subsubsection{Derivatives, harmonic sums, and the digamma function}

The derivative of the Riemann zeta function obeys several identities which stem from its functional equation. For instance, the derivative at even negative integers is given by,
\begin{align}
\label{eq:zeta(-2n)}
    \zeta'(-2n)
    &=
    \half
    (-)^n \,
    (2\pi)^{-2n} \, (2n)! \,
    \zeta(2n+1)
\end{align}
which has transcendental weight one. Additionally, $\zeta'(0) = - \half \ln(2\pi)$.

\sm

The logarithmic derivative of the Riemann zeta function $\zeta'(z) / \zeta(z)$ and the digamma function ${\psi(z) = \Gamma'(z)/\Gamma(z)}$ often arise together because the logarithmic derivative of the starred zeta function is given by,
\begin{align}
    \frac{{\zeta^*}'(z)}{\zeta^*(z)}
    =
    \frac{\zeta'(z)}{\zeta(z)}
    + \half \psi ( z/2 )
    - \half \ln \pi
\end{align}
The digamma function obeys the reflection identity,
\begin{align}
    \psi(1-z)
    =
    \psi(z)
    + \pi \cot(\pi z)
\end{align}
and the duplication formula,
\begin{align}
    \psi(z)
    + \psi(z-\half)
    =
    2 \, \psi(2z-1)
    - 2 \ln 2
\end{align}
which may be derived from the respective properties of the gamma function.

\sm

At positive integer and half-integer arguments, the digamma function may be written in terms of finite harmonic sums and the Euler-Mascheroni constant $\gamma_E$. The finite harmonic sums $H_1(m)$ with integer $m \geq 1$ are defined as follows,
\begin{align}
    H_1(m) = \sum_{k=1}^{m-1} \frac{1}{k}
\end{align}
with $H_1(1) = 0$. Definitions of $H_1(m)$ with a different upper limit are also common. For integer~$n \geq 1$, we then have,
\begin{align}
    \psi(n)
    &=
    H_1(n) - \gamma_E
\no \\[1ex]
    \psi ( n+\half )
    &=
    2 \, H_1(2n+1)
    - H_1(n+1)
    - 2 \ln 2
    - \gamma_E
\end{align}
As discussed in \autoref{sec:trans}, the terms $H_1(m)$, $\ln 2$, and $\gamma_E$ each have transcendental weight one. Thus, $\psi(n)$ and $\psi(n+\half)$ also have unit transcendental weight. 

\sm

These objects will often occur in combinations such that the Euler-Mascheroni constants cancel, leaving only logarithmic derivative of the Riemann zeta function and finite harmonic sums. For this reason, it will be convenient to define the following combination,
\begin{align}
\label{eq:Zndef}
    \cZ(n)
    =
    \frac{\zeta'(n)}{\zeta(n)}
    +
    \begin{cases}
    \half H_1 \big( \frac{n}{2} \big)
    &
    n \text{ even}
    \\[1ex]
    H_1(n) - \half H_1 \big( \frac{n+1}{2} \big)
    &
    n \text{ odd}
    \end{cases}
\end{align}
for integer $n \geq 2$. The combination $\cZ(n)$ has transcendental weight one.

\subsection{Multiple zeta-values}

The Riemann zeta function may be generalized to the multiple zeta function of depth~$\ell$, which is defined by the following $\ell$-fold infinite sum,
\begin{align}
    \zeta(z_1, \dots, z_\ell)
    =
    \sum_{m_1>\dots>m_\ell>0}
    \frac{1}{m_1^{z_1} \cdots m_\ell^{z_\ell}}
\end{align}
The series definition of the multiple zeta function converges for $\Re(z_1) > 1$ and ${\Re(z_{j \geq 2}) \geq 1}$ and may be analytically continued in $\bbC^\ell$~\cite{Zhao:2000, Akiyama:2002}.

\sm

The multiple zeta function with positive integer arguments, is called a multiple zeta-value (MZV). The MZV~$\zeta(n_1, \dots, n_\ell)$ with integer $n_1 \geq 2$ and ${n_{j \geq 2} \geq 1}$ is assigned transcendental weight $n_1 + \dots + n_\ell$. All known identities between zeta-values and MZVs respect their (conjectural) grading by transcendental weight~\cite{Blumlein:2009cf}.

\sm

Multiple zeta-values which can be written in terms of zeta-values with rational coefficients are said to be reducible. The MZVs which occur in this paper are all reducible. For example, the double zeta-value $\zeta(N,1)$ with integer $N \geq 2$ obeys,
\begin{align}
    \zeta(N,1)
    &=
    \half N \, \zeta(N+1)
    - \half
    \sum_{\ell=1}^{N-2}
    \zeta(\ell+1) \, \zeta(N-\ell)
\end{align}
which may be further simplified (using~\eqref{eq:evenZetaSum} to obtain the second equation) as follows,
\begin{align}
\label{eq:zeta(n,1)}
    \zeta(2n,1)
    &=
    n \, \zeta(2n+1)
    -
    \sum_{\ell=1}^{n-1}
    \zeta(2\ell+1) \, \zeta(2n-2\ell)
    \no \\
    \zeta(2n+1,1)
    &=
    \tfrac{1}{4} (2n-1) \, \zeta(2n+2)
    - \half
    \sum_{\ell=1}^{n-1}
    \zeta(2\ell+1) \, \zeta(2n+1-2\ell)
\end{align}
for integer $n \geq 1$. Similarly, the triple zeta-value $\zeta(2n+1,1,1)$ is reducible  and obeys~\cite{Borwein:2006},
\begin{align}
\label{eq:zeta(2n+1,1,1)}
    \zeta(2n+1,1,1)
    &=
    \tfrac{1}{6}(2n+1)(2n+2) \,
    \zeta(2n+3)
    - n \, \zeta(2) \, \zeta(2n+1)
    \vphantom{ \sum_{\ell=1}^{n-1} }
    \no \\
    & \quad
    - \tfrac{1}{2}
    \sum_{\ell=1}^{n-1}
    \(n+\ell-\tfrac{1}{2}\)
    \zeta(2\ell+1) \,
    \zeta(2n+2-2\ell)
    \no \\
    & \quad
    + \tfrac{1}{6}
    \sum_{\ell_1=1}^{n-1}
    \sum_{\ell_2=1}^{n-1-\ell_1}
    \zeta(2\ell_1+1) \,
    \zeta(2\ell_2+1) \,
    \zeta(2n-2\ell_1-2\ell_2+1)
\end{align}
for integer $n \geq 1$. We note that these expression explicitly respect transcendental weight.

\sm

Multiple zeta-values may also be written in terms of generalized finite harmonic sums. We shall define the generalized finite harmonic sums $H_k(m)$ for integer $k, m \geq 1$ as follows,
\begin{align}
    H_k(m) = \sum_{M=1}^{m-1} \frac{1}{M^k}
\end{align}
with $H_k(1) = 0$. Definitions of $H_k(m)$ with a different upper limit are also common. In analogy with our discussion of $H_1(m)$ in \autoref{sec:trans}, $H_k(m)$ is assigned transcendental weight $k$. Again, one should think of $H_k(m)$ not as its value for a single~$m$ (which would give a rational number whose natural transcendental weight assignment is zero) but instead as a function of $m$ to be inserted into an infinite series in $m$. For instance,~$H_k(m)$ occurs in this manner in the double zeta-value $\zeta(n,k)$,
\begin{align}
    \zeta(n,k)
    =
    \sum_{m_1 > m_2 > 0}
    \frac{1}{m_1^n \, m_2^k}
    =
    \sum_{m = 2}^\infty
    \frac{H_k(m)}{m^n}
\end{align}
The standard transcendental weight assignments of $\zeta(n)$ and $\zeta(n,k)$ are $n$ and $n+k$, respectively, which justifies assigning transcendental weight $k$ to $H_k(m)$.

\newpage

\section{Relating \texorpdfstring{$C_{a,b,c}$}{} and \texorpdfstring{$\mfC_{w;m;p}$}{}}
\label{apdx:CabcCwmp}

In this appendix, we shall explicitly relate the two-loop modular graph functions~$C_{a,b,c}$ and~$\mfC_{w;m,p}$ which were introduced in \autoref{sec:2loopMGF}. The functions $\mfC_{w;m,p}$ are linear combinations of the functions~$C_{a,b,c}$ which diagonalize the homogeneous part of their system of inhomogeneous Laplace eigenvalue equations.

\sm

The linear relations~\eqref{eq:CabcCwmp} between these functions define the expansion coefficients~$d$ and~$\mfd$, which we repeat here to render this appendix reasonably self-contained, 
\begin{align}
\label{eq:CabcCwmp3} 
    C_{a,b,c}
    &=
    \sum_{m=1}^{\floor{\frac{w-1}{2}}} \!
    \sum_{p=0}^{\floor{\frac{w-2m-1}{3}}}
    d_{a,b,c}^{w;m;p} \,
    \mfC_{w;m;p}
    &
    \mfC_{w;m;p}
    &=
    \sum_{\substack{ a \geq b \geq c \geq 1 \\
                     a+b+c=w }}
    \mfd_{w;m;p}^{a,b,c} \,
    C_{a,b,c}
    \quad
\end{align}
The coefficients $\mfh$ were defined in~\eqref{eq:DCwmp3}, which we also repeat here for convenience,
\begin{align}
\label{eq:DCwmp3}
    \big(
    \Delta - (w-2m)(w-2m-1)
    \big)
    \mfC_{w;m;p}
    =
    \mfh^{(0)}_{w;m;p} \, E_w^*
    +
    \sum_{\ell=2}^{\floor{w/2}}
    \mfh^{(\ell)}_{w;m;p} \,
    E_\ell^* E_{w-\ell}^*
\end{align}
Throughout this appendix, the ranges of the variables $m$ and $p$ are those indicated in the first sum of~\eqref{eq:CabcCwmp3},
\begin{align}
    {1 \leq m \leq \floor{\tfrac{w-1}{2}}}
    &&
    {0 \leq p \leq \floor{\tfrac{w-2m-1}{3}}}
\end{align}
where $m$ determines the eigenvalue of the Laplacian and $p$ labels the degeneracy of the corresponding eigenspace.

\sm

In this appendix, we shall obtain explicit formulas for the coefficients $d$, $\mfd$, and $\mfh$. Our results are packaged in the following three theorems.

\begin{thm}
\label{thm:dabc}
The expansion coefficients $d$ are rational numbers given by,
\begin{align}
\label{eq:dabc}
    d^{w;m;p}_{a,b,c}
    &=
    \sum_{k=m \mathstrut }^{m_+ \mathstrut}
    \sum_{\kappa=0 \mathstrut }^{k-1 \mathstrut}
    \sum_{ \a_j \geq 0 }
    \sum_{ \b_j \geq 0 }
    \sum_{ \g_j \geq 0 }
    \tfrac{ (w-3p-2m-1)! }
          { (w-3p-2k-1)! }
    \tfrac{ (2w-2m-2k-3)!! }
          { (2k-2m)!! }
    \tbinom{k-1}{\kappa}
\no \\
    & \qquad
    \times
    \tfrac{ (\a_1+\a_2+\a_3)! }{ \a_1! \a_2! \a_3!}
    \tfrac{ (\b_1+\b_2+\b_3)! }{ \b_1! \b_2! \b_3!} 
    \tfrac{ (\g_1+\g_2+\g_3)! }{ \g_1! \g_2! \g_3!}
    \,
    2^{w-2m-1} \,
    3^{k-1} \,
    (-4)^\kappa \,
    (-)^{m-1}
    \vphantom{\Big[}
\no \\
    & \qquad \times
    \delta_{\a_1+\a_2+\a_3,3p} \,
    \delta_{\b_1+\b_2+\b_3,\kappa} \,
    \delta_{\g_1+\g_2+\g_3, w-3p-2\kappa-3}
    \vphantom{\Big[}
\no \\
    & \qquad \times
    \psi(\a_1,\a_2,\a_3) \,
    \delta_{a, \a_1+\b_2+\b_3+\g_1+1} \,
    \delta_{b, \a_2+\b_3+\b_1+\g_2+1} \,
    \delta_{c, \a_3+\b_1+\b_2+\g_3+1}
    \smash[b]{\vphantom{\Big[}}
\end{align}
where $m_+ = \floor{(w-3p-1)/2}$ and the function $\psi(\a,\b,\g)$ is given by, 
\begin{align}
\label{eq:psi(a,b,c)}
    \psi(\a,\b,\g)
    &=
    \tfrac{2}{3}
    \big\{
      \cos \big( \tfrac{2 \pi}{3} (\a-\b) \big)
    + \cos \big( \tfrac{2 \pi}{3} (\b-\g) \big)
    + \cos \big( \tfrac{2 \pi}{3} (\g-\a) \big)
    \big\}
\end{align}
The function $\psi(\a,\b,\g)$ depends on its integer arguments modulo $3$ and is invariant under permutations, simultaneous shifts, and simultaneous sign reversal of its arguments. It takes the values $\psi(0,0,0)=2$, $\psi(0,0,1)=0$, and $\psi(0,1,-1)=-1$ on its inequivalent orbits which implies that $\psi(\a,\b,\g)=0$ when $\a+\b+\g \not \equiv 0 \pmod{3}$.
\end{thm}

\begin{thm}
\label{thm:dwmp}
The  expansion coefficients $\mfd$ are rational numbers given by,
\begin{align}
\label{eq:dwmp}
    ( \sigma_{a,b,c} )^{-1} \,
    \mfd^{a,b,c}_{w;m;p}
    &=
    \sum_{s=1 \mathstrut}^{m \mathstrut}
    \sum_{\ell=s \mathstrut}^{m_+}
    \tfrac{ (w-3p-2s-1)! }{ (w-3p-2m-1)! }
    \tfrac{ (2w-4m-1) }{ (2w-2s-2m-1)!! }
    \tfrac{ 1 }{ (2m-2s)!! }
    \tbinom{\ell-1}{s-1} \,
    (-)^{s-1} \,
\no \\
    & \qquad
    \times
    2^{w-3p-2m-2\ell-\delta_{p,0}+1} \,
    \big( \tfrac{1}{3} \big)^{w-3} \,
    \cM_{a,b,c}(\ell-1,3p+\ell-1) 
    \end{align}
The normalization factor $\sigma_{a,b,c}$ provides proper combinatorial counting and is given by,
\begin{align}
\label{eq:sabc}
    (\sigma_{a,b,c})^{-1}
    =
    1 + \delta_{a,b} + \delta_{b,c} + \delta_{c,a}
    + 2 \, \delta_{a,b} \, \delta_{b,c}
\end{align}
The value of $(\sigma_{a,b,c})^{-1}$ is equal to $n!$ when $n$ of its three arguments $a,b,c$ are equal to one another. The function $\cM_{a,b,c}(\ell_1,\ell_2) $ is given by, 
\begin{align}
\label{eq:Mabc}
    \cM_{a,b,c}(\ell_1,\ell_2)
    &=
    \sum_{\a_j=0}^{a-1}
    \sum_{\b_j=0}^{b-1}
    \sum_{\g_j=0}^{c-1}
    \tfrac{ (a-1)! }{ \a_1! \a_2! (a-1-\a_1-\a_2)! }
    \tfrac{ (b-1)! }{ \b_1! \b_2! (b-1-\b_1-\b_2)! }
    \tfrac{ (c-1)! }{ \g_1! \g_2! (c-1-\g_1-\g_2)! }
\no \\
    & \qquad \times 
    3 \,
    \psi(\a_1-\a_2,\b_1-\b_2,\g_1-\g_2) \,
    \delta_{\ell_1, \a_1+\b_1+\g_1} \,
    \delta_{\ell_2, \a_2+\b_2+\g_2}
\end{align}
where $\psi(\a,\b,\g)$ was defined in~\eqref{eq:psi(a,b,c)}. The function $\cM_{a,b,c}(\ell_1,\ell_2)$ is invariant under permutations of $a,b,c$ and under swapping $\ell_1, \ell_2$. Also, $\cM_{a,b,c}(\ell_1,\ell_2)=0$ when  $\ell_1 \not \equiv \ell_2 \pmod{3}$ as well as when $\ell_1+\ell_2 > w-3$. 
\end{thm}

\begin{thm}
\label{thm:hwmp}
The coefficients $\mfh^{(\ell)}$ with $2 \leq \ell \leq m$ are rational numbers given in terms of the coefficients~$\mfd$ by the following relation,
\begin{alignat}{4}
\label{eq:hlwmp}
    \mfh^{(\ell)}_{w;m;p}
    &=
    -
    \frac{4}{ (w-\ell-2)! (\ell-2)! } \,
    \Big\{
    &&
    \big(
    1 + \delta_{\ell,2} - \delta_{\ell,2} \, \delta_{w,4}
    \big)
    &&
    \mfd_{w;m;p}^{w-\ell,\ell-1,1}
\no \\
    & \span
    {} + \Theta(\ell-3)
    &
    \big(
    1 + \delta_{2\ell,w-1} - \delta_{2\ell,w}
    \big)
    &&
    \mfd_{w;m;p}^{w-\ell-1,\ell,1}
    \vphantom{\Big\}}
\no \\
    & \span
    {} - \Theta(\ell-3)
    &
    \big(
    1 + \delta_{\ell,3} + \delta_{\ell,3} \, \delta_{w,6}
    \big)
    &&
    \mfd_{w;m;p}^{w-\ell-1,\ell-1,2}
    \,
    \Big\}
\intertext{where the step function is defined by $\Theta(x \geq 0) = 1$ and $\Theta(x < 0) = 0$. The coefficients~$\mfh^{(0)}$ are also rational numbers given in terms of the coefficients~$\mfd$ and~$\mfh^{(\ell)}$ by,}
\label{eq:h0wmp}
    \mfh^{(0)}_{w;m;p}
    &=
    \frac{2}{ (w-1)! } \,
    \bigg\{
    \sum_{\ell=1}^{\floor{(w-1)/2}}
    (w-\ell-1)
    \big(
    2\ell
    + 2 \delta_{\ell,1} 
    + \delta_{\ell,1} \, \delta_{w,3}
    - \delta_{\ell,2}
    \big) \,
    \mfd_{w;m;p}^{w-\ell-1,\ell,1}
    \span \span \span \span
\no \\[-1ex]
    & \hspace{1.4in}
    - \tfrac{1}{4}
    \sum_{\ell=2}^{ \floor{w/2} }
    (w-\ell-1)! (\ell-1)! \,
    \mfh^{(\ell)}_{w;m;p}
    \bigg\}
    \span \span \span \span
\end{alignat}
The coefficients $\mfh^{(\ell)}$ vanish for $2 \leq \ell \leq m$.
\end{thm}

\subsection{The generating function \texorpdfstring{$\cW$}{}}

To prove these theorems, we shall appeal to the generating function~$\cW$ for the modular graph functions $C_{a,b,c}$ which was introduced in~\cite{DHoker:2015gmr} and is defined by,
\begin{align}
\label{eq:Wgen1}
    \cW(t_1,t_2,t_3 | \tau, \bar{\tau} )
    &=
    \sum_{a,b,c=1}^\infty
    t_1^{a-1} t_2^{b-1} t_3^{c-1} \,
    C_{a,b,c}( \tau, \bar{\tau} )
\end{align}
Since each $C_{a,b,c}$ is invariant under permutations of its indices, the generating function is itself invariant under permutations of the auxiliary variables $t_1,t_2,t_3$. To make this symmetry more manifest we recast $\cW$ as a sum over the monomial \smash{$t_1^{a-1} t_2^{b-1}t_3^{c-1}$} plus its five permutations, irrespective of whether some of the exponents are equal to one another or not. Restricting the sum in~\eqref{eq:Wgen1} to ordered $a \geq b \geq c \geq 1$, we obtain,
\begin{align}
\label{eq:Wgen2}
    \cW(t_1,t_2,t_3 | \tau, \bar{\tau} )
    &=
    \sum_{a \geq b \geq c \geq 1}
    \sigma_{a,b,c} \,
    \cL_{a,b,c}(t_1,t_2,t_3) \,
    C_{a,b,c}(\tau, \bar{\tau} )
\end{align}
where the symmetry factor $\sigma_{a,b,c}$ was defined in~\eqref{eq:sabc} and $\cL_{a,b,c}$ is given by,
\begin{align}
    \cL_{a,b,c}(t_1,t_2,t_3)
    =
    t_1^{a-1} t_2^{b-1} t_3 ^{c-1}
    + 5 \text{ permutations of } t_1, t_2, t_3 
\end{align}
By a slight abuse of terminology we shall refer to $\cL_{a,b,c}$ as symmetrized monomials.

\sm

The modular graph functions $\mfC_{w;m;p}$ emerge by expanding the generating function~$\cW$ in a basis of the simultaneous eigenfunctions $\mfW_{w;m;p}$ of the following mutually commuting differential operators,
\begin{align}
\label{eq:diffops}
    \mfD
    &=
    t_1 \p_{t_1} + t_2 \p_{t_2} + t_3 \p_{t_3}
    \vphantom{\big)^2}
\no \\
    \mfL^2
    &=
    \mfD^2 + \mfD
    + ( t_1^2 + t_2^2 + t_3^2 - 2 t_1 t_2 - 2 t_2 t_3 - 2 t_3 t_1 )
        ( \p_{t_1} \p_{t_2} + \p_{t_2} \p_{t_3} + \p_{t_3} \p_{t_1} )
    \vphantom{\big)^2}
\no \\
    \mfL_0^2
    &=
    \tfrac{1}{3}
    \big(
    (t_1-t_2) \, \p_{t_3} + (t_2-t_3) \, \p_{t_1} + (t_3-t_1) \, \p_{t_2}
    \big)^2
\end{align}
with the following eigenvalues,
\begin{align}
    \mfD \, \mfW_{w;m;p}
    &=
    (w-3) \, \mfW_{w;m;p}
\no \\
    \mfL^2 \, \mfW_{w;m;p}
    &=
    (w-2m)(w-2m-1) \, \mfW_{w;m;p}
\no \\
    \mfL_0^2 \, \mfW_{w;m;p}
    &=
    -9 p^2 \, \mfW_{w;m;p}
\end{align}
Eigenfunctions of the scaling operator $\mfD$ are homogeneous polynomials in~${t_1,t_2,t_3}$. Moreover, all three differential operators are manifestly invariant under permutations of these variables. The construction of these operators relies on a certain $\mathrm{SO}(2,1)$ algebra with the generators~${\mfL_0, \mfL_\pm}$ obeying the structure relations $[\mfL_0, \mfL_\pm] = \pm i \mfL_\pm$ and ${[\mfL_+, \mfL_-]= -i \mfL_0}$. The quadratic Casimir operator of this algebra is given by ${\mfL^2=\mfL_+ \mfL_- + \mfL_- \mfL_+ - \mfL_0^2}$. For more details, see~\cite{DHoker:2015gmr} or Appendix A of~\cite{Kleinschmidt:2017ege}.

\sm

Explicit expressions for $\mfW_{w;m;p}$ may be conveniently obtained in terms of a new set of auxiliary variables $u,z,\bar{z}$ which are well-adapted to the $\mathrm{SO}(2,1)$ structure of the problem. These new variables are related to the original variables $t_1, t_2, t_3$ by the following relations,\footnote{The variables $v,\bar{v}$ used in~\cite{DHoker:2015gmr} are related to the variables $z,\bar{z}$ used here by $\sqrt{2} \, v = u z$ and $\sqrt{2} \, \bar{v} = u \bar{z}$.}
\begin{align}
\label{eq:uzz}
    \sqrt{3} \, t_1
    &=
    u \( 1 + \half z + \half \bar{z} \)
    &
    \sqrt{3} \, u \phantom{z}
    &=
    t_1 + t_2 + t_3 
\no \\
    \sqrt{3} \, t_2
    &=
    u \( 1 + \half \ep^2 z + \half \ep \bar{z} \)
    &
    \half \sqrt{3} \, u z
    &= 
    t_1 + \ep t_2 + \ep^2 t_3 
\no \\
    \sqrt{3} \, t_3
    &=
    u \( 1 + \half \ep z + \half \ep^2 \bar{z} \)
    &
    \half \sqrt{3} \, u \bar{z}
    &=
    t_1 + \ep^2 t_2 + \ep t_3 
\end{align}
where $\ep = e^{2\pi i /3}$. Assuming that the variables $t_1,t_2,t_3$ are real-valued, then $u$ is real-valued while~${z,\bar{z}}$ are complex-valued and mutual complex conjugates. In terms of these new variables, the eigenfunctions $\mfW_{w;m;p}$ take the following form,\footnote{Our eigenfunctions $\mfW_{w;m;p}$ are proportional to the eigenfunctions $\cW_{w;s;p}$ with $s=w-2m$ in~\cite{DHoker:2015gmr}.}
\begin{align}
\label{eq:Wwmp}
    \mfW_{w;m;p}(u,z,\bar{z})
    &=
    \big( \sqrt{3} \, u \big)^{w-3}
    \big( z^{3p} + \bar{z}^{3p} \big)
\no \\
    & \quad \times
    \sum_{k=m}^{m_+}
    \tfrac{ (w-3p-2m-1)! }{ (w-3p-2k-1)! }
    \tfrac{ (2w-2m-2k-2)! }{ (w-m-k-1)! }
    \tfrac{ 1 }{ (k-m)! }
    \,
    (-)^{k-m} \,
    ( 1 - z \bar{z} )^{k-1} 
\end{align}
where $m_+ = \floor{(w-3p-1)/2}$. We have chosen an overall normalization for $\mfW_{w;m;p}$ which will be convenient in the sequel. The dependence of $\mfW_{w;m;p}$ on the variables $u$ and $z / \bar z$ is completely determined by the quantum numbers $w$ and $p$, respectively, and factors out of the finite sum over~$k$.

\sm

We now return to the generating function $\cW$. Using the Laplace equation~\eqref{eq:DCabc} for~$C_{a,b,c}$ and the definition of the generating function in~\eqref{eq:Wgen1}, we readily verify that the action of the differential operator $\mfL^2$ reproduces the action of the Laplacian~$\Delta$ on $\cW$,
\begin{align}
    \( \Delta - \mfL^2 \)
    \cW
    =
    \text{inhomogeneous terms}
\end{align}
where the inhomogeneous terms are non-holomorphic Eisenstein series and their double products, as described in \autoref{sec:2loopMGF}. Thus, the eigenfunctions~$\mfC_{w;m;p}$ of $\Delta$ must multiply the eigenfunctions~$\mfW_{w;m;p}$ of $\mfL^2$ when $\cW$ is expanded in this basis. In fact, we shall define the eigenfunctions~$\mfC_{w;m;p}$ by this expansion,
\begin{align}
\label{eq:Wgen3}
    \cW(t_1,t_2,t_3 | \tau, \bar{\tau} )
    &=
    \sum_{w=3}^\infty
    \sum_{m=1}^{\floor{\frac{w-1}{2}}} \!
    \sum_{p=0}^{\floor{\frac{w-2m-1}{3}}}
    \mfW_{w;m;p}(u,z,\bar{z}) \,
    \mfC_{w;m;p}( \tau, \bar{\tau} )
\end{align}
where the variables $u,z,\bar{z}$ implicitly depend on $t_1,t_2,t_3$. 
\sm

We may now use the two expressions~\eqref{eq:Wgen2} and~\eqref{eq:Wgen3} for the generating function~$\cW$ to relate the two bases of modular graph functions, $C_{a,b,c}$ and $\mfC_{w;m;p}$. The linear relations~\eqref{eq:CabcCwmp3} between these functions define the expansion coefficients~$d$ and~$\mfd$. Equating~\eqref{eq:Wgen2} and~\eqref{eq:Wgen3} and expanding either basis of modular graph functions using~\eqref{eq:CabcCwmp3}, we obtain,
\begin{align}
\label{eq:LW}
    \sigma_{a,b,c} \,
    \cL_{a,b,c}
    &= 
    \sum_{m=1}^{\floor{\frac{w-1}{2}}} \!
    \sum_{p=0}^{\floor{\frac{w-2m-1}{3}}} \!
    \mfd_{w;m;p}^{a,b,c} \,
    \mfW_{w;m;p}
\\[1ex]
\label{eq:WL}
    \mfW_{w;m;p}
    &=
    \sum_{\substack{ a \geq b \geq c \geq 1 \\ a+b+c=w }}
    d_{a,b,c}^{w;m;p} \,
    \sigma_{a,b,c} \,
    \cL_{a,b,c}
\end{align}
In other words, we may compute the expansion coefficients~$d$ and~$\mfd$ by converting the symmetrized monomials~$\cL_{a,b,c}$ and the eigenfunctions~$\mfW_{w;m;p}$ into one another. It is these formulas which we shall use to compute $d$ and $\mfd$.

\subsection{Proof of \autoref{thm:dabc}: expressing \texorpdfstring{$\mfW_{w;m;p}$}{} in terms of \texorpdfstring{$\cL_{a,b,c}$}{}}

We shall first express the eigenfunctions $\mfW_{w;m;p}$ of~\eqref{eq:Wwmp} in terms of the symmetrized monomials $\cL_{a,b,c}$ as in~\eqref{eq:WL} to extract the coefficients $d$. To carry out this expansion, we shall write $\mfW_{w;m;p}$ in terms of the variables $u, uz, u \bar{z}$ since these variables are each linear in~$t_1, t_2, t_3$. Therefore, our starting point is the following expression,
\begin{align}
    \mfW_{w;m;p}
    &=
    \big( \sqrt{3} \, u \big)^{w-3}
    \big( (uz)^{3p} + (u \bar z)^{3p} \big)
\\
    & \quad 
    \times
    \sum_{k=m}^{m_+}
    \tfrac{ (w-3p-2m-1)! }{ (w-3p-2k-1)! }
    \tfrac{ (2w-2m-2k-2)! }{ (w-m-k-1)! }
    \tfrac{ 1 }{ (k-m)! } \,
    (-)^{k-m} \,
    u^{w-3p-1-2k} \,
    \big( u^2 - (uz)(u\bar{z}) \big)^{k-1} 
\no
\end{align}
We shall decompose each factor in this expression into powers of $t_1, t_2, t_3$ while maintaining manifest permutation symmetry.

\sm

We begin with the factor \smash{$( (uz)^{3p} + (u \bar z)^{3p} )$}. Both $uz$ and $u\bar{z}$ are trinomials in $t_1,t_2,t_3$ which may be expanded while maintaining manifest permutation symmetry by using three summation variables constrained by a Kronecker delta function,
\begin{align}
    (uz)^{3p} + (u\bar{z})^{3p}
    &=
    \big( \tfrac{1}{\sqrt{3}} \big)^{3p}
    \sum_{ \a_j \geq 0 }
    \tfrac{ (\a_1+\a_2+\a_3)! }{ \a_1! \a_2! \a_3! } \,
    \psi(\a_1,\a_2,\a_3) \,
    \delta_{\a_1+\a_2+\a_3,3p} \,
    t_1^{\a_1} t_2 ^{\a_2} t_3 ^{\a_3}
\end{align}
where the function $\psi(\a_1,\a_2,\a_3)$ was defined in~\eqref{eq:psi(a,b,c)}. This function arises from the explicit symmetrization of~$\ep^{2\a_2+\a_3} + \ep^{\a_2+2\a_3}$, where we recall the notation $\ep = e^{2\pi i /3}$. For the allowed values of its arguments $\a_j$, namely those which satisfy $\a_1+\a_2+\a_3 \equiv 0 \pmod{3}$, the function~$\psi(\a_1,\a_2,\a_3)$ can take the values $-1$ or $2$.

\sm

Next, we shall expand the factor $( u^2 - (uz)(u\bar{z}) )^{k-1}$ in the summand using the relation, 
\begin{align}
    u^2 - (uz)(u\bar{z})
    =
    4 \, (t_1 t_2 + t_2 t_3 + t_3 t_1)
    - (t_1+t_2+t_3)^2
\end{align}
It will be convenient to first use a binomial expansion for the $k-1$ power of the two terms on the right side of the above relation. We will then use a trinomial expansion only for the term $( t_1 t_2 + t_2 t_3 + t_3 t_1 )$ while leaving the powers of $(t_1+t_2+t_3)$ un-expanded. These steps produce the following expansion, 
\begin{align}
    \big( u^2 - (uz)(u\bar{z}) \big)^{k-1}
    &=
    (-)^{k-1} 
    \sum_{\kappa=0}^{k-1}
    \tbinom{k-1}{\kappa} \,
    (-4)^\kappa
    \sum_{ \b_j \geq 0 }
    \tfrac{ (\b_1+\b_2+\b_3)! }{ \b_1! \b_2! \b_3!} \, 
    \delta_{\b_1+\b_2+\b_3,\kappa} \,
\no \\
    & \qquad
    \times 
    t_1^{\b_2+\b_3} t_2^{\b_3+\b_1} t_3^{\b_1+\b_2} \,
    (t_1+t_2+t_3)^{2k-2-2\kappa}
\end{align}
where we have again expanded the trinomial using three summation variables constrained by a Kronecker delta function in order to maintain manifest permutation symmetry. At this stage we could eliminate the sum over $\kappa$ by solving the delta function constraint to get~${\kappa = \b_1+\b_2+\b_3}$, but it saves space to keep the sum over $\kappa$ as it stands.

\sm

Thus far, we have refrained from expanding the powers of $(t_1 + t_2 + t_3)$ as they will profitably combine with the remaining powers of $u$. The combined exponent of $(t_1 + t_2 + t_3)$ is $w-3p-2\kappa-3$, and we have the following final trinomial expansion, 
\begin{align}
    (t_1+t_2+t_3)^{w-3p-2\kappa-3}
    =
    \sum_{ \g_j \geq 0 }
    \tfrac{ (\g_1+\g_2+\g_3)! }{ \g_1! \g_2! \g_3!} \,
    \delta_{\g_1+\g_2+\g_3, w-3p-2\kappa-3} \,
    t_1^{\g_1} t_2^{\g_2} t_3^{\g_3}
\end{align}
Putting all the contributions together, we obtain,
\begin{align}
    \mfW_{w;m;p}
    &=
    \sum_{k=m \mathstrut}^{m_+ \mathstrut}
    \sum_{\kappa=0 \mathstrut}^{k-1 \mathstrut}
    \sum_{ \a_j \geq 0 }
    \sum_{ \b_j \geq 0 }
    \sum_{ \g_j \geq 0 }
    \tfrac{ (w-3p-2m-1)! }{ (w-3p-2k-1)! }
    \tfrac{ (2w-2m-2k-2)! }{ (w-m-k-1)! }
    \tfrac{ 1 }{ (k-m)! } \,
    \tbinom{k-1}{\kappa} \,
    3^{k-1} \,
    (-4)^\kappa \,
    (-)^{m-1}
\no \\[1ex]
    & \qquad
    \times
    \tfrac{ (\a_1+\a_2+\a_3)! }{ \a_1! \a_2! \a_3!}
    \tfrac{ (\b_1+\b_2+\b_3)! }{ \b_1! \b_2! \b_3!} 
    \tfrac{ (\g_1+\g_2+\g_3)! }{ \g_1! \g_2! \g_3!}
\no \\[1ex]
    & \qquad \times
    \delta_{\a_1+\a_2+\a_3,3p} \,
    \delta_{\b_1+\b_2+\b_3,\kappa} \,
    \delta_{\g_1+\g_2+\g_3, w-3p-2\kappa-3}
\no \\[1ex]
    & \qquad \times
    \psi(\a_1,\a_2,\a_3) \,
    t_1^{\a_1+\b_2+\b_3+\g_1} \,
    t_2^{\a_2+\b_3+\b_1+\g_2} \,
    t_3^{\a_3+\b_1+\b_2+\g_3}
\end{align}
This expression is manifestly invariant under permutations of $t_1,t_2,t_3$. One may also readily verify that the sum of the exponents of $t_1,t_2,t_3$ equals $w-3$ as expected.

\sm

To extract the expansion coefficient \smash{$d_{a,b,c}^{w;m;p}$} from this expression, we use the fact that the normalized symmetrized monomial $\sigma_{a,b,c} \, \cL_{a,b,c}$ includes as a term the monomial \smash{$t_1^{a-1} t_2^{b-1} t_3^{c-1}$} with unit coefficient. Hence, the expansion coefficient \smash{$d_{a,b,c}^{w;m;p}$} is equal to the coefficient of this monomial in $\mfW_{w;m;p}$. An explicit expression is given in~\eqref{eq:dabc} of \autoref{thm:dabc}. This expression is manifestly symmetric under permutations of $a,b,c$. The finite sum is over eleven variables constrained by six Kronecker delta functions, leaving effectively a sum over five variables. Manifestly, we have \smash{$d_{a,b,c}^{w;m;p} \in \bbQ$}. This completes our proof of \autoref{thm:dabc}.

\subsection{Proof of \autoref{thm:dwmp}: expressing \texorpdfstring{$\cL_{a,b,c}$}{} in terms of \texorpdfstring{$\mfW_{w;m;p}$}{}}

To calculate the coefficients $\mfd$, we use~\eqref{eq:LW}. Unfortunately, obtaining the decomposition of the symmetrized monomial $\cL_{a,b,c}$ in terms of the eigenfunctions $\mfW_{w;m;p}$ is considerably more complicated than the inverse problem solved in the previous subsection.

\sm

We begin by expanding each factor of the monomial \smash{$t_1^{a-1} t_2^{b-1} t_3^{c-1}$} in terms of the variables~$u,z,\bar{z}$ using the relations of~\eqref{eq:uzz} and the trinomial expansion,
\begin{align}
    t_1^{a-1}
    &=
    \Big(
    \tfrac{ \displaystyle u \mathstrut }{\sqrt{3}}
    \Big)^{a-1} \,
    \sum_{\a_j=0}^{a-1}
    \tfrac{ (a-1)! }{ \a_1! \a_2! (a-1-\a_1-\a_2)! }
    \( \half z \)^{\a_1}
    \( \half \bar{z} \)^{\a_2}
\no \\
    t_2^{b-1}
    &=
    \Big(
    \tfrac{ \displaystyle u \mathstrut }{\sqrt{3}}
    \Big)^{b-1} \,
    \sum_{\b_j=0}^{b-1}
    \tfrac{ (b-1)! }{ \b_1! \b_2! (b-1-\b_1-\b_2)! }
    \( \half \ep^2 z \)^{\b_1}
    \( \half \ep \bar{z} \)^{\b_2}
\no \\
    t_1^{c-1}
    &=
    \Big(
    \tfrac{ \displaystyle u \mathstrut }{\sqrt{3}}
    \Big)^{c-1} \,
    \sum_{\g_j=0}^{c-1}
    \tfrac{ (c-1)! }{ \g_1! \g_2! (c-1-\g_1-\g_2)! }
    \( \half \ep z \)^{\g_1}
    \( \half \ep^2 \bar{z} \)^{\g_2}
\end{align}
where we recall the notation $\ep = e^{2\pi i / 3}$. Taking the product of these three expressions and adding the five permutations of $t_1,t_2,t_3$, we obtain the following expression for $\cL_{a,b,c}$ in terms of the variables~$u,z,\bar{z}$,
\begin{align}
\label{eq:Labc1}
    \cL_{a,b,c}
    &=
    \Big(
    \tfrac{ \displaystyle u \mathstrut }{\sqrt{3}}
    \Big)^{w-3} \,
    \sum_{\a_j=0}^{a-1}
    \sum_{\b_j=0}^{b-1}
    \sum_{\g_j=0}^{c-1}
    \tfrac{ (a-1)! }{ \a_1! \a_2! (a-1-\a_1-\a_2)! }
    \tfrac{ (b-1)! }{ \b_1! \b_2! (b-1-\b_1-\b_2)! }
    \tfrac{ (c-1)! }{ \g_1! \g_2! (c-1-\g_1-\g_2)! }
\no \\
    & \qquad \times
    3 \,
    \psi(\a_1-\a_2,\b_1-\b_2,\g_1-\g_2) \,
\no \\
    & \qquad \times
    \( \half \)^{ \a_1+\a_2+\b_1+\b_2+\g_1+\g_2 }
    z^{\a_1+\b_1+\g_1} \, \bar{z}^{\a_2+\b_2+\g_2}
\end{align}
The function $\psi(\a,\b,\g)$, which was defined in~\eqref{eq:psi(a,b,c)}, results from summing the various powers of~$\ep$ over the six permutations of $t_1, t_2, t_3$ and using the identity, 
\begin{align}
    \psi(\a,\b,\g)
    =
    \tfrac{1}{3} 
    \big| \ep^{\a}+ \ep^{\b}+ \ep ^{\g}\big|^2
    - 1
\end{align}
The properties of the function $\psi(\a,\b,\g)$ given at the end of \autoref{thm:dabc} will be crucial to the decomposition of~\eqref{eq:Labc1} into the eigenfunctions~$\mfW_{w;m;p}$. 

\subsubsection{Extracting the $(z^{3p}+\bar{z}^{3p})$-dependence}

We now return to the expression~\eqref{eq:Labc1} for $\cL_{a,b,c}$ and introduce two Kronecker delta functions to parametrize the exponents of $z$ and $\bar{z}$ as follows,
\begin{align}
    \cL_{a,b,c}
    &=
    \Big(
    \tfrac{ \displaystyle u \mathstrut }{\sqrt{3}}
    \Big)^{w-3}
    \sum_{\substack{ \ell_1,\ell_2 \geq 0 \\ \mathclap{ \ell_1+\ell_2 \leq w-3} }}
    \cM_{a,b,c}(\ell_1,\ell_2) \,
    \big( \half \big)^{\ell_1+\ell_2} \,
    z^{\ell_1} \bar{z}^{\ell_2}
\end{align}
where $\cM_{a,b,c}(\ell_1,\ell_2)$ was defined in~\eqref{eq:Mabc}.  The function $\cM_{a,b,c}(\ell_1,\ell_2)$ inherits the following properties from the properties of the function $\psi(\a,\b,\g)$. First, $\cM_{a,b,c}(\ell_1,\ell_2)$ is invariant under permutations of $a,b,c$ and under swapping $\ell_1,\ell_2$. Second, it satisfies ${\cM_{a,b,c}(\ell_1,\ell_2)=0}$ when ${\ell_1 \not \equiv \ell_2 \pmod{3}}$ or when $\ell_1+\ell_2 > w-3$, as stated at the end of \autoref{thm:dwmp}. 

\sm

The fact that $\cM_{a,b,c}(\ell_1,\ell_2)$ vanishes unless $\ell_1 \equiv \ell_2 \pmod{3}$ allows us to change summation variables from $\ell_1$ and $\ell_2$ to the new variables $\ell$ and $p$ defined by~${\min{(\ell_{1,2})} +1 = \ell}$ and~${\max{(\ell_{1,2})} +1 = \ell+3p}$. Carrying out this change of summation variables, we obtain,
\begin{align}
\label{eq:Labc2}
    \cL_{a,b,c}
    &=
    \Big(
    \tfrac{ \displaystyle u \mathstrut }{\sqrt{3}}
    \Big)^{w-3}
    \,
    \sum_{p=0 \mathstrut}^{\floor{\frac{w-3}{3}}}
    \sum_{\ell=1 \mathstrut}^{m_+}
    \big( \half \big)^{2\ell-2+3p+\delta_{p,0}} \,
    \big( z^{3p} + \bar{z}^{3p} \big)
    (z \bar{z})^{\ell-1}
\no \\
    & \qquad
    \times
    \cM_{a,b,c}(\ell-1,\ell+3p-1)
\end{align}
where we recall that $m_+=\floor{(w-3p-1)/2}$. Comparing this expression with the formula~\eqref{eq:Wwmp} for the eigenfunction $\mfW_{w;m;p}$, we see that the factor~$u^{w-3} \, (z^{3p} + \bar z^{3p})$ matches precisely and in both cases multiplies a function of the single variable~${z \bar{z}}$. 

\subsubsection{Extracting the $z \bar{z}$-dependence}

To compare the $z \bar{z}$-dependence of $\cL_{a,b,c}$ and~$\mfW_{w;m;p}$, we shall instead work with the variable ${r^2 = 1-z \bar{z}}$. Binomial expanding the factor $(z\bar{z})^{\ell-1}$ in~\eqref{eq:Labc2}, we obtain,
\begin{align}
\label{eq:Labc3}
    \cL_{a,b,c}
    &=
    \Big(
    \tfrac{ \displaystyle u \mathstrut }{\sqrt{3}}
    \Big)^{w-3}
    \,
    \sum_{p=0 \mathstrut}^{\floor{\frac{w-3}{3}}}
    \sum_{s=1 \mathstrut}^{m_+}
    \sum_{\ell=s \mathstrut}^{m_+}
    \tbinom{\ell-1}{s-1}
    \big( \half \big)^{2\ell-2+3p+\delta_{p,0}} \,
    (-)^{s-1} \,
    \big( z^{3p} + \bar{z}^{3p} \big)
    \,
    r^{2s-2} 
\no \\
    & \qquad \times 
    \cM_{a,b,c}(\ell-1,3p+\ell-1)
\end{align}
Next, we recall the expression~\eqref{eq:Wwmp} for the eigenfunctions $\mfW_{w;m;p}$ and re-express them in terms of the variable $r$,
\begin{align}
    \mfW_{w;m;p}(u,z,\bar{z})
    &=
    \big( \sqrt{3} \, u \big)^{w-3}
    \big( z^{3p} + \bar{z}^{3p} \big) \,
    \tfrac{  (2w-4m-2)!   } {  (w-2m-1)! } \,
    \cR_{w;m;p} (r) 
\end{align}
where the function $\cR_{w;m;p}$ is given by,
\begin{align}
\label{eq:Rwmp}
    \cR_{w;m;p}(r) 
    &=
    \sum_{k=0}^{m_+-m}
    \tfrac{ (w-3p-2m-1)! }{ (w-3p-2m-2k-1)! }
    \tfrac{ (2w-4m-2k-3)!! }{ (2w-4m-3)!! }
    \tfrac{ 1 }{ (2k)!! } \,
    (-)^{k} \,
    r^{2m+2k-2}
\end{align}
for $m = 1, \dots, m_+$. The normalization of $\cR_{w;m;p}$ has been chosen so that the~${k=0}$ term is simply $r^{2m-2}$ with unit coefficient.

\sm

For fixed $w$ and $p$, we may view the definition of $\cR_{w;m;p}$ as a system of linear equations expressing the $m_+$ number of functions $\cR_{w;1;p}, \dots, \cR_{w;m_+;p}$ in terms of the $m_+$ number of monomials $1, r^2, \dots, r^{2m_+-2}$. To express $\cL_{a,b,c}$ in terms of the eigenfunctions $\mfW_{w;m;p}$ we need to invert this system and obtain the various powers of $r^2$ as linear combinations of the functions $\cR_{w;1;p}, \dots, \cR_{w;m_+;p}$. This inversion is given by the following lemma.

\begin{lem}
\label{thm:r(R)}
For fixed $w$ and $p$, the system of equations~\eqref{eq:Rwmp} may be inverted as follows,
\begin{align}
\label{eq:r(R)}
    r^{2s-2}
    &=
    \sum_{n=0}^{m_+-s}
    \frac{ (w-3p-2s-1)! }{ (w-3p-2s-2n-1)! }
    \frac{ (2w-4s-4n-1)!! }{ (2w-4s-2n-1)!! }
    \frac{ 1 }{ (2n)!! } \,
    \cR_{w;s+n;p}(r) 
\end{align}
for $s = 1, \dots, m_+ = \floor{(w-3p-1)/2}$.
\end{lem}

We shall prove this lemma after we complete our decomposition of $\cL_{a,b,c}$. Combining~\eqref{eq:Labc3} and~\eqref{eq:r(R)} yields,
\begin{align}
    \cL_{a,b,c}
    &= 
    \sum_{p=0 \mathstrut}^{\floor{\frac{w-3}{3}}}
    \!
    \sum_{m=1 \mathstrut}^{m_+ \mathstrut}
    \sum_{s=1 \mathstrut}^{m \mathstrut}
    \sum_{\ell=s \mathstrut}^{m_+ \mathstrut}
    \tfrac{ (w-3p-2s-1)! }{ (w-3p-2m-1)! }
    \tfrac{ (2w-4m-1)!! }{ (2w-2s-2m-1)!! }
    \tfrac{ (w-2m-1)! }{ (2w-4m-2)! }
    \tfrac{ 1 }{ (2m-2s)!! }
    \tbinom{\ell-1}{s-1} \,
    (-)^{s-1}
\no \\
    & \qquad
    \times
    \big( \tfrac{1}{3} \big)^{w-3} 
    \big( \half \big)^{3p+2\ell-2+\delta_{p,0}} \,
    \cM_{a,b,c}(\ell-1,3p+\ell-1) \,
    \mfW_{w;m;p}
\end{align}
To obtain the expansion coefficient \smash{$\mfd^{a,b,c}_{w;m;p}$} from this expression, we simply multiply by~$\sigma_{a,b,c}$ and extract the coefficient of $\mfW_{w;m;p}$. The explicit expression is given in~\eqref{eq:dwmp} of \autoref{thm:dwmp}. To complete the proof of this theorem, it remains only to provide a proof of \autoref{thm:r(R)}.

\subsubsection{Proof of \autoref{thm:r(R)}}

Before we analytically prove~\eqref{eq:r(R)}, we shall first provide numerical evidence for this formula. We define $\eps= w-3p-1-2m_+$ so that $\eps=0$ when $w-3p-1$ is even and $\eps=1$ when~$w-3p-1$ is odd. We also define the parameter $k=m_+-m$ and introduce the following abbreviations, 
\begin{align}
    R_k &= \cR_{w;m_+-k;p}(r)
    &
    T_k &= r^{2m_+-2k-2}
\end{align}
In terms of $R_k$ and $T_k$, the system of linear relations~\eqref{eq:Rwmp} which define~$\cR_{w;m;p}$ becomes,
\begin{align}
    R_k
    &=
    \sum_{n=0}^k
    \frac{ \Gamma(w- 2m_+ +2k -n-\half) }
         { \Gamma( w-2m_+ +2k -\half ) }
    \frac{ (2k+\eps)! }{ (2k -2n+\eps)! \, 2^{2n} \, n! } \,
    (-)^n \,
    T_{k-n}
\end{align}
This linear system may be described by a triangular matrix with unit diagonal elements and may be solved by forward substitution.

\sm

To carry out this process, we parametrize the relations between $R_k$ and $T_k$, including all dependence on $w$ and $m_+$, in terms of the following functions,
\begin{align}
    f_k
    =
    \frac{1}{ 2w - 4 m_+ + k }
\end{align}
for integer $k \geq 0$. In terms of these functions,
\begin{align}
\label{eq:Rk}
    R_k
    &=
    T_k
    +
    \sum_{n=1}^k
    \frac{ (2k+\eps)! }{ (2k -2n+\eps)! \, 2^{2n} \, n! } \,
    (-)^n \,
    f_{4k-3} f_{4k-5} \cdots f_{4k-2n-1} \,
    T_{k-n}
\end{align}
For $k \not= \ell$, these functions obey the partial fraction decomposition $f_k f_\ell = \tfrac{1}{\ell-k} ( f_k - f_\ell )$ which may be used to express each coefficient in~\eqref{eq:Rk} as a linear combination of $f_k$ functions.

\sm

At this point we shall invert the system of relations between $R_k$ and $T_k$ for low values of~$k$ using MAPLE. For the case $\eps=0$, we find the following relations for $k \leq 5$,
\begin{align}
T_0 & = R_0
 \\
T_1 & = R_1 + f_1 R_0
\no \\
T_2 & = R_2 + 6 f_5 R_1 + \tfrac{3}{2} (f_1-f_3) R_0
\no \\
T_3 & = R_3 + 15 f_9 R_2 + \tfrac{45}{2} (f_5-f_7) R_1 + \tfrac{15}{8} (f_1-2f_3+f_5) R_0
\no \\
T_4 & = R_4 + 28 f_{13} R_3 +105(f_9-f_{11} ) R_2 + \tfrac{105}{2} (f_5 -2 f_7 + f_9) R_1
\no \\ & \quad
+ \tfrac{ 35}{16} (f_1-3f_3+3f_5-f_7) R_0
\no \\
T_5 & = R_5 + 45 f_{17} R_4 +315(f_{13}-f_{15}) R_3 
+ \tfrac{1575}{4}(f_9-2f_{11}+f_{13}) R_2
\no \\ & \quad
+ \tfrac{1575}{16}(f_5-3f_7+3f_9-f_{11})R_1 
+ \tfrac{315}{128} (f_1-4f_3+6f_5-4f_7+f_9) R_0
\no
\end{align}
For the case $\ep=1$, we similarly find,
\begin{align}
T_0 & = R_0
 \\
T_1 & = R_1 + 3 f_1 R_0
\no \\
T_2 & = R_2 + 10 f_5 R_1 + \tfrac{15}{2} (f_1-f_3) R_0
\no \\
T_3 & = R_3 + 21 f_9 R_2 + \tfrac{105}{2} (f_5-f_7) R_1 + \tfrac{105}{8} (f_1-2f_3+f_5) R_0
\no \\
T_4 & = R_4 + 36 f_{13} R_3 +189(f_9-f_{11} ) R_2 + \tfrac{315}{2} (f_5 -2 f_7 + f_9) R_1
\no \\& \quad
+ \tfrac{ 315}{16} (f_1-3f_3+3f_5-f_7) R_0
\no \\
T_5 & = R_5 + 55 f_{17} R_4 + 495(f_{13}-f_{15} R_3 + \tfrac{3465}{4}(f_9-2f_{11}+f_{13}) R_2
\no \\ & \quad
+ \tfrac{5775}{16}(f_5-3f_7
+3f_9-f_{11})R_1 
+ \tfrac{3465}{128} (f_1-4f_3+6f_5-4f_7+f_9) R_0
\no
\end{align}
Based on these low order expressions, we readily observe the following properties which have been verified by MAPLE to order $k \leq 22$, 
\begin{itemize}
\item The coefficient of $R_{k-n}$ in $T_k$ is proportional to the alternating binomial sum,
\begin{align}
    \sum_{m=0}^{n-1}
    \tbinom{n-1}{m} \,
    (-)^m \,
    f_{4k+1-4n+2m}
    &=
    2^{n-1} \,
    (n-1)!
    \prod_{m=0}^{n-1}
    f_{4k+1-4n+2m}
\end{align}
\item The coefficient multiplying this alternating binomial sum of $f_k$ functions depends on the two numbers $k$ and $n$ and is given by the binomial \smash{$\binom{2k+\eps}{2n}$} multiplied by the following function which is independent of $k$ and $\eps$,
\begin{align}
    \frac{ (2n)! }{ 2^{2n} \, n! \, (n-1)!}
\end{align}
\end{itemize}
Combining these ingredients, we obtain the following conjectural formula for $T_k$,
\begin{align}
    T_k
    &=
    R_k
    +
    \sum_{n=1}^k
    \frac{ (2k+\ep)! }{ (2k-2n+\ep)! \, 2^n \, n! }
    \(
    \prod_{m=0}^{n-1}
    f_{4k+1-4n+2m}
    \)
    R_{k-n}
\end{align}
which becomes~\eqref{eq:r(R)} with $s = m_+-k$ when written in terms of $\cR_{w;m;p}$ and $r$.

\sm

To prove this formula, we shall simply show that the right-hand side equals $r^{2s-2}$. Substituting the definition~\eqref{eq:Rwmp} of $\cR_{w;m;p}$ into the right-hand side of~\eqref{eq:r(R)}, we obtain,
\begin{align}
    \sum_{n=0}^{m_+-s}
    \sum_{k=0}^{m_+-s-n}
    \tfrac{ (w-3p-2s-1)! }{ (w-3p-2s-2n-2k-1)! }
    \tfrac{ (2w-4s-4n-2k-3)!! }{ (2w-4s-2n-1)!! }
    \tfrac{ (2w-4s-4n-1) }{ (2n)!! (2k)!! } \,
    (-)^{k} \,
    r^{2s+2n+2k-2}
\end{align}
We change summation variables from $k$ to $N=n+k$ and interchange the order of the two finite sums to find,
\begin{align}
    \sum_{N=0}^{m_+-s}
    \tfrac{ (w-3p-2s-1)! }{ (w-3p-2s-2N-1)! } \,
    (-)^{N} \,
    r^{2s+2N-2} \,
    \sum_{n=0}^{N}
    \tfrac{ (2w-4s-2n-2N-3)!! }{ (2w-4s-2n-1)!! }
    \tfrac{ (2w-4s-4n-1) }{ (2n)!! (2N-2n)!! } \,
    (-)^{n} 
\end{align}
The $N=0$ term of this sum is just $r^{2s-2}$. For $N \geq 1$, the sum over $n$ may be written as a difference of two hypergeometric functions by writing the factorials and double factorials in terms of gamma functions. After some careful algebra, we find,
\begin{align}
\label{eq:r(R)proof}
    &
    \sum_{n=0}^{N}
    \tfrac{ (2w-4s-2n-2N-3)!! }{ (2w-4s-2n-1)!! }
    \tfrac{ (2w-4s-4n-1) }{ (2n)!! (2N-2n)!! } \,
    (-)^{n}
\\
    &=
    \sum_{n=0}^{N}
    \tfrac{ (2w-4s-2n-2N-3)!! }{ (2w-4s-2n-3)!! }
    \tfrac{ 1 }{ (2n)!! (2N-2n)!! } \,
    (-)^{n}
    +
    \sum_{n=0}^{N-1}
    \tfrac{ (2w-4s-2n-2N-5)!! }{ (2w-4s-2n-3)!! }
    \tfrac{ 1 }{ (2n)!! (2N-2-2n)!! } \,
    (-)^{n}
\no \\[1ex]
    &=
    ( -\tfrac{1}{4} )^N \,
    \tfrac{ \Gamma( 1/2-w+2s+1) }
          { N! \, \Gamma( 1/2-w+2s+N+1) } \,
    {}_2F_1(-N, 1/2-w+2s+1; 1/2-w+2s+N+1; 1) 
\no \\[1ex]
    & \quad
    -
    ( -\tfrac{1}{4} )^N \,
    \tfrac{ \Gamma( 1/2-w+2s+1) }
          { (N-1)! \, \Gamma( 1/2-w+2s+N+2) } \,
    {}_2F_1(-N, 1/2-w+2s+1; 1/2-w+2s+N+2; 1)
\no
\end{align}
where the hypergeometric function is defined by,
\begin{align}
    {}_2F_1(a,b;c;z)
    &=
    \sum_{n=0}^\infty
    \frac{ \Gamma(a+n) }{ \Gamma(a) }
    \frac{ \Gamma(b+n) }{ \Gamma(b) }
    \frac{ \Gamma(c) }{ \Gamma(c+n) }
    \frac{ z^n }{ n! }
\end{align}
Using Gauss's evaluation of the hypergeometric function at unit argument,
\begin{align}
    {}_2F_1(a,b;c;1)
    &=
    \frac{ \Gamma(c) \Gamma(c-a-b) }
         { \Gamma(c-a) \Gamma(b-c) }
\end{align}
we see that~\eqref{eq:r(R)proof} vanishes. Thus, the right-hand side of~\eqref{eq:r(R)} equals $r^{2s-2}$. This completes our proof or \autoref{thm:r(R)} and thus also our proof of \autoref{thm:dwmp}.

\subsection{The \texorpdfstring{$\mfd$}{} coefficients at different weights}

Although~\eqref{eq:dwmp} provides an explicit analytic expression for the $\mfd$ coefficients, it will be useful to have a formula which relates these coefficients at weight $w$ to those at the lower weight $w-2$. For fixed $w$ and~${m \geq 2}$, this relationship provides an embedding of the space of eigenfunctions $\mfC_{w-2;m-1;p}$ into the space of $\mfC_{w;m;p}$ as discussed in~\cite{DHoker:2015gmr}.

\begin{lem}
\label{thm:dabcdabc}
The coefficient \smash{$\mfd_{w;m;p}^{a,b,c}$} with $m \geq 2$ may be written as follows,
\begin{align}
\label{eq:dabcdabc}
    \mfd_{w;m;p}^{a,b,c}
    &=
    \tfrac{ \sigma_{a,b,c} }{ 3 \, (2m-2) (2w-2m-3) \,}
    \Big\{
    \,
    \tfrac{ (a-1)(b-1) }{ \sigma_{a,b,c+1} } \,
    \mfd_{w-2;m-1;p}
        ^{\max(a-1,c), \mathrm{med}(a-1,b-1,c), \min(b-1,c)}
\no \\[0.5ex]
    & \hspace{1.33in}
    + \tfrac{ (b-1)(c-1) }{ \sigma_{a+1,b,c} } \,
    \mfd_{w-2;m-1;p}^{a,b-1,c-1}
\no \\[0.5ex]
    & \hspace{1.33in}
    + \tfrac{ (c-1)(a-1) }{ \sigma_{a,b+1,c} } \,
    \mfd_{w-2;m-1;p}^{\max(a-1,b), \min(a-1,b), c-1}
    \,
    \Big\}
\end{align}
where $\sigma_{a,b,c}$ is defined in~\eqref{eq:sabc}. The $\min$, $\mathrm{med}$, and $\max$ functions on the right-hand side of this expression ensure that the coefficients \smash{$\mfd_{w-2;m-1;p}^{a',b',c'}$} have ordered indices $a' \geq b' \geq c'$.
\end{lem}

We shall prove this lemma using the differential operator $\mfP = ( \p_{t_1} \p_{t_2} + \p_{t_2} \p_{t_3} + \p_{t_3} \p_{t_1} )$. The action of $\mfP$ on the symmetrized monomial $\cL_{a,b,c}$ is given by,
\begin{align}
    \mfP \,
    \cL_{a,b,c}
    =
    (a-1)(b-1) \, \cL_{a-1,b-1,c}
    &
    + (b-1)(c-1) \, \cL_{a,b-1,c-1}
\no \\
    &
    + (c-1)(a-1) \, \cL_{a-1,b,c-1}
\end{align}
To calculate the action of $\mfP$ on $\mfW_{w;m;p}$, we first write $\mfP$ in terms of the operators $\mfD$ and~$\mfL^2$ using~\eqref{eq:diffops} and the following relation between the variables $t_1,t_2,t_3$ and $u,z,\bar{z}$,
\begin{align}
    - u^2 \, (1-z\bar{z})
    =
    t_1^2 + t_2^2 + t_3^2
    - 2t_1t_2 - 2t_2t_3 - 2t_3t_1
\end{align}
We thus obtain $\mfP = ( u^2 \, (1-z\bar{z}) )^{-1} ( \mfD^2 + \mfD - \mfL^2 )$. Now, $\mfW_{w;m;p}$ is an eigenfunction with eigenvalue $(2m-2)(2w-2m-3)$ under the action of $( \mfD^2 + \mfD - \mfL^2 )$. The other factor in~$\mfP$ produces the following action,
\begin{align}
    ( u^2 \, (1-z\bar{z}) )^{-1} \,
    \mfW_{w;m;p}
    &=
    3 \, \mfW_{w-2;m-1;p}
\end{align}
which follows directly from the definition~\eqref{eq:Wwmp} of $\mfW_{w;m;p}$. Thus,
\begin{align}
    \mfP \,
    \mfW_{w;m;p}
    &=
    3 \,
    (2m-2)(2w-2m-3) \,
    \mfW_{w-2;m-1;p}
\end{align}
We now act with $\mfP$ on the expansion~\eqref{eq:LW} of $\cL_{a,b,c}$ in terms the eigenfunctions $\mfW_{w;m;p}$ with coefficients \smash{$\mfd_{w;m;p}^{a,b,c}$}. After some straightforward rearrangements we find~\eqref{eq:dabcdabc}.

\subsection{Proof of \autoref{thm:hwmp}: the \texorpdfstring{$\mfh$}{} coefficients}

We shall now consider the $\mfh$ coefficients which multiply the inhomogeneous terms in the Laplace equation for $\mfC_{w;m;p}$. These coefficients are written in terms of the $\mfd$~coefficients in \autoref{thm:hwmp}.

\sm

To begin our proof this theorem, we recall the expansion~\eqref{eq:CabcCwmp3} of the eigenfunction~$\mfC_{w;m;p}$ in terms of the functions $C_{a,b,c}$ with ordered indices $a \geq b \geq c \geq 1$. We shall use the Laplace equation~\eqref{eq:DCabc} for $C_{a,b,c}$ and the Laplace equation~\eqref{eq:DCwmp} for~$\mfC_{w;m;p}$ to isolate the inhomogeneous terms, i.e. the Eisenstein series, which arise from the functions $C_{a,b,c}$ with any index~$a,b,c = 0,-1$. The functions $C_{w-\ell,\ell,0}$ and $C_{w-\ell+1,\ell,-1}$ are written in terms of Eisenstein series in~\eqref{eq:inhom}. Since the action of $\Delta$ lowers an index by at most two, only the functions~$C_{a,b,c}$ with $c=1,2$ in the expansion~\eqref{eq:CabcCwmp3} will contribute inhomogeneous terms. 

\sm

Using these results, we may collect together all the Eisenstein series $E^*_w$ and the double products~$E^*_{w-\ell} E^*_\ell$ with $2 \leq \ell \leq \floor{w/2}$ which result from the action of $\Delta$ on $\sum_{a,b,c} \mfd \, C_{a,b,c}$. The coefficient of $E^*_w$ is equal to \smash{$\mfh^{(0)}$}, and the coefficients of the double products are equal to \smash{$\mfh^{(\ell)}$}. Some careful algebra then yields~\eqref{eq:hlwmp} and~\eqref{eq:h0wmp} of \autoref{thm:hwmp}. Since the $\mfd$ coefficients are rational, each $\mfh \in \bbQ$ manifestly.

\sm

We shall now combine~\eqref{eq:hlwmp}, which writes the \smash{$\mfh^{(\ell)}$} coefficients in terms of the $\mfd$ coefficients, and~\eqref{eq:dabcdabc}, which relates the $\mfd$ coefficients at different weights, to find a relation between the \smash{$\mfh^{(\ell)}$} coefficients at weights $w$ and $w-2$. After some tedious algebra, we find,
\begin{align}
    \mfh^{(\ell)}_{w;m;p}
    &=
    \Theta(\ell-3) \,
    \tfrac{ 1 }{ 3 \, (2m-2)(2w-2m-3) } \,
    \mfh^{(\ell-1)}_{w-2;m-1;p}
\end{align}
for $m \geq 2$, where the step function is defined by $\Theta(x \geq 0) = 1$ and $\Theta(x < 0) = 0$. This expression may then be iterated $m-1$ times to yield,
\begin{align}
\label{eq:hlhl}
    \mfh^{(\ell)}_{w;m;p}
    &=
    \Theta(\ell-m-1) \,
    \( \tfrac{1}{3} \)^{m-1}
    \tfrac{ (2w-4m-1)!! }{ (2m-2)!! \, (2w-2m-3)!! } \,
    \mfh^{(\ell-m+1)}_{w-2(m-1);1;p}
\end{align}
In other words, the coefficient \smash{$\mfh^{(\ell)}_{w;m;p}$} is proportional to $\mfh^{(\ell-m+1)}_{w-2(m-1);1;p}$ for $m+1 \leq \ell \leq \floor{w/2}$ and vanishes for $2 \leq \ell \leq m$. This completes our proof of \autoref{thm:hwmp}.

\newpage

\section{The Laurent polynomials of \texorpdfstring{$C_{a,b,c}$}{}, \texorpdfstring{$\mfC_{w;m;p}$}{}, and \texorpdfstring{$v_{k,2}$}{}}
\label{apdx:Laurent}

In this appendix, we shall discuss the Laurent polynomials of the infinite families of two-loop modular graph functions $C_{a,b,c}$, $\mfC_{w;m;p}$, and $v_{k,2}$.

\subsection{The Laurent polynomial of \texorpdfstring{$C_{a,b,c}$}{}}

The two-loop modular graph function $C_{a,b,c}$ was defined in~\eqref{eq:Cabc}. The Laurent polynomial of~$C_{a,b,c}$ was computed in~\cite{DHoker:2017zhq}. Near the cusp, $C_{a,b,c}$ has the following asymptotic expansion,
\begin{align}
\label{eq:CabcAsy}
    C_{a,b,c}
    &=
    \mfc^{(w)}_{a,b,c} \, \tau_2^w
    + \sum_{\ell=1}^{w-1}
      \mfc^{(w-2\ell-1)}_{a,b,c} \,
      \tau_2^{w-2\ell-1}
    + \mfc^{(2-w)}_{a,b,c} \,
      \tau_2^{2-w}
    + \cO(e^{-2\pi \tau_2})
\end{align}
The Laurent coefficients each have transcendental weight $w=a+b+c$ and are given by,\footnote{The expression for \smash{$\mfc^{(2-w)}_{a,b,c}$} was obtained in~\cite{DHoker:2017zhq} assuming the validity of a conjectured identity between rational numbers which was numerically verified to high order in MAPLE. Our expression for \smash{$\mfc^{(2-w)}_{a,b,c}$} corrects a factor of $2$ error in (1.26), an error in the upper limit of the finite sum in (1.26), and a minus sign error in (1.28) of the second version of~\cite{DHoker:2017zhq}.}
\begin{align}
    \mfc^{(w)}_{a,b,c}
    &=
    (-4\pi)^w \!\!
    \sum_{k=0}^{\max(b,c)}
    \frac{ B_{2k } }{ (2k)! }
    \frac{ B_{2w-2k} }{ (2w-2k)! }
    \Big[
    \tbinom{2b+2c-2k-1}{2b-1}
    +
    \tbinom{2b+2c-2k-1}{2c-1}
    \Big]
\no \\[1ex]
    \mfc^{(w-2\ell-1)}_{a,b,c}
    &=
    (4\pi)^{w-2\ell-1} \,
    \mfq^{(w-2\ell-1)}_{a,b,c} \,
    \zeta(2\ell+1)
\no \\[1ex]
    \mfc^{(2-w)}_{a,b,c}
    &=
    (4\pi)^{2-w}
    \sum_{k=1}^{w-3}
    \lambda_{a,b,c}(k) \,
    \zeta(2k+1) \,
    \zeta(2w-2k-3)
\end{align}
where \smash{$\mfq^{(w-2\ell-1)}_{a,b,c}$} and \smash{$\lambda_{a,b,c}(k)$} are rational numbers defined as follows,
\begin{align}
    \mfq^{(w-2\ell-1)}_{a,b,c}
    &=
    \frac{ 2 \, B_{2w-2\ell-2} }{ (2w-2\ell-2)! }
    \sum_{\alpha=0 \mathstrut}^{a-1}
    \sum_{\beta=0 \mathstrut}^{a-1-\alpha}
    (-)^{\beta+c+1} \,
    \Theta\!\( c + \floor{\tfrac{b+\beta}{2}} - w + \ell + 1 \)
    \no \\[-1.8ex]
    & \hspace{2in}
    \times
    \tbinom{2a-2-\alpha-\beta}{a-1}
    \tbinom{b+\alpha-1}{b-1}
    \tbinom{b+\beta-1}{b-1}
    \tbinom{2\ell-2a+\alpha+\beta+1}{b+\alpha-1}
    \no \\
    & \quad
    + 5 \text{ permutations of } a,b,c
    \no \\[2ex]
    \lambda_{a,b,c}(k)
    &=
    - Z_0(a,b,c)
    + 2 \, Z_k(a,b,c) \, \Theta(a-k-1)
    \no \\
    & \quad
    -
    \sum_{\alpha=1 \mathstrut}^{a-1}
    \sum_{\beta=0 \mathstrut}^{2\alpha-1}
    Z_\alpha(a,b,c) \,
    \mathrm{E}_\beta(0)
    \tbinom{2k}{2\alpha-\beta}
    \tbinom{2w-2\alpha+\beta-4}{\beta}
    \no \\
    & \quad
    + 5 \text{ permutations of } a,b,c
\end{align}
The step function is defined by $\Theta(x \geq 0) = 1$ and $\Theta(x < 0) = 0$. Here $\mathrm{E}_n(x)$ are the Euler polynomials (not to be confused with the Eisenstein series $E_s$) which are defined by the following generating function,
\begin{align}
    \frac{2 \, e^{xt}}{e^x + 1}
    =
    \sum_{n=1}^\infty
    \mathrm{E}_n(t) \frac{x^n}{n!}
\end{align}
Finally, the functions $Z_\alpha(a,b,c)$ are integers defined by,
\begin{align}
    Z_\alpha(a,b,c)
    &=
    \sum\nolimits_{k=\max(1,2\alpha+2-a)}^{\min(a,2\alpha+1)}
    \tbinom{a+c-k-1}{c-1}
    \tbinom{a+c-2\alpha+k-3}{c-1}
    \tbinom{2\alpha}{k-1}
    \tbinom{2w-2\alpha-4}{w-k-1}
\end{align}
We will not use these expressions in this paper, but we have included them for completeness.

\sm

We note that these Laurent coefficients have a particular transcendental structure. The leading coefficient is a rational multiple of $\pi^w$. The coefficient of $\tau_2^{w-2\ell-1}$ for $1 \leq \ell \leq w-1$ is a rational multiple of $\pi^{w-2\ell-1} \zeta(2\ell+1)$. The coefficient of $\tau_2^{2-w}$ is equal to $\pi^{2-w}$ times a sum of double products of odd zeta-values with transcendental weight $2w-2$.

\subsection{The Laurent polynomial of \texorpdfstring{$\mfC_{w;m;p}$}{}}

The two-loop modular graph functions $\mfC_{w;m;p}$ were introduced in \autoref{sec:2loopMGF}. Because the functions $\mfC_{w;m;p}$ are rational linear combinations of the functions $C_{a,b,c}$, their Laurent polynomials must have the same transcendental structure. However, for fixed~$m$, several of the Laurent coefficients \smash{$\mfc^{(\ell)}_{w;m;p}$} may vanish while the corresponding \smash{$\mfc^{(\ell)}_{a,b,c}$} are generally non-zero. Moreover, the Laurent coefficients of $\mfC_{w;m;p}$ may be written in terms of the rational coefficients $\mfh$ which appear in the inhomogeneous Laplace equation~\eqref{eq:DCwmp} for $\mfC_{w;m;p}$. Explicit expressions for these coefficients are given in \autoref{apdx:CabcCwmp}.

\begin{thm}
\label{thm:CwmpAsy}
Near the cusp, $\mfC_{w;m;p}$ has the following asymptotic expansion,
\begin{align}
\label{eq:CwmpAsy}
    \mfC_{w;m;p}
    & =
    \mfc_{w;m;p}^{(w)} \,
    \tau_2^{w}
    +
    \sum_{k = m}^{w-m-1}
    \mfc_{w;m;p}^{(w-2k-1)} \,
    \tau_2^{w-2k-1}
\no \\
    & \quad
    +
    \mfc_{w;m;p}^{(2-w)} \,
    \tau_2^{2-w}
    +
    \mfc_{w;m;p}^{(1-w)} \,
    \tau_2^{1-w}
    +
    \cO(e^{-2 \pi \tau_2})
\end{align}
The Laurent coefficients each have transcendental weight $w$ and are given by,
\begin{align}
\label{eq:c(L)wmp2}
    \mfc_{w;m;p}^{(w)}
    &=
    \frac{ 1 }{ (2m)(2w-2m-1) }
    \Big(
    \mfh^{(0)}_{w;m;p} \,
    \zeta^*(2w)
    +
    \sum_{\ell = m+1}^{\floor{w/2}}
    \mfh^{(\ell)}_{w;m;p} \,
    \zeta^*(2w-2\ell) \,
    \zeta^*(2\ell)
    \Big)
    \smash[t]{\vphantom{\Bigg\{}}
\no \\
    \mfc_{w;m;p}^{(w-2k-1)}
    &=
    -
    \frac{ \( 1 + \delta_{2k,w-2} \) \, 
            \mfh^{ \( \min(k+1,w-k-1) \) }_{w;m;p} }
         { (2k-2m+1) (2w-2k-2m-2) } \,
    \zeta^*(2w-2k-2) \,
    \zeta^*(2k+1)
    \vphantom{\Bigg\{}
\no \\
    \mfc_{w;m;p}^{(1-w+2m)}
    &=
    -
    \frac{ \zeta^*(2m) \, \zeta^*(2w-2m-1) }
         { (2w-4m-1) \, \zeta^*(2w-4m) }
    \sum_{\ell = m+1}^{\floor{w/2}}
    \mfh^{(\ell)}_{w;m;p} \,
    \zeta^*(2w-2\ell-2m) \,
    \zeta^*(2\ell-2m)
    \vphantom{\Bigg\{}
\no \\
    \mfc_{w;m;p}^{(2-w)}
    &=
    \frac{ 1 }{ (2m-1)(2w-2m-2) }
    \sum_{\ell = m+1}^{\floor{w/2}}
    \mfh^{(\ell)}_{w;m;p} \,
    \zeta^*(2w-2\ell-1) \,
    \zeta^*(2\ell-1)
    \vphantom{\Bigg\{}
\no \\
    \mfc_{w;m;p}^{(1-w)}
    &=
    \frac{ 1 }{ (2m)(2w-2m-1) } \,
    \mfh^{(0)}_{w;m;p} \,
    \zeta^*(2w-1)
    \smash[b]{\vphantom{\Bigg\{}}
\end{align}
In the second line, the integer $k$ runs over the range $m \leq k \leq w-m-2$.
\end{thm}

The form of the Laurent polynomial in~\eqref{eq:CwmpAsy} follows from the asymptotic expansion of the function~$C_{a,b,c}$ in~\eqref{eq:CabcAsy} and from the relations~\eqref{eq:c(L)wmp} implied by the inhomogeneous Laplace equation for $\mfC_{w;m;p}$. In fact, these relations determine every Laurent coefficient except for~\smash{$\mfc_{w;m;p}^{(1-w+2m)}$} in terms of starred zeta-values and the coefficients $\mfh$. The remaining Laurent coefficient was determined in~\eqref{eq:c(1w2m)wmp} by demanding that the integral of $E_{s}^* \mfC_{w;m;p}$ over~$\cM_L$ in~\eqref{eq:IEsCwmp} was finite in the limit $s \to w-2m$.

\subsection{The Laurent polynomial of \texorpdfstring{$v_{k,2}$}{}}

The two-loop modular graph functions $v_{k,2}$ were defined in~\eqref{eq:vk2}. The Laurent polynomial of $v_{k,2}$ was computed in~\cite{DHoker:2019txf}. Near the cusp, $v_{k,2}$ has the following asymptotic expansion,
\begin{align}
\label{eq:vk2Asy}
    v_{k,2}
    &=
    \mfc^{(k+1)}_{k,2} \, \tau_2^{k+1}
    + \sum_{\ell=2}^{\floor{k/2}}
      \mfc^{(-k+2\ell)}_{k,2} \,
      \tau_2^{-k+2\ell}
    + \mfc^{(1-k)}_{k,2} \,
      \tau_2^{1-k}
    + \mfc^{(-k)}_{k,2} \,
      \tau_2^{-k}
    + \cO(e^{-2\pi \tau_2})
\end{align}
The only positive-power term in this Laurent polynomial is proportional to $\tau_2^{k+1}$. We also note that there is no $\tau_2^{2-k}$ term. The coefficients have transcendental weight $k+1$ and are given as follows,
\begin{align}
    \mfc^{(k+1)}_{k,2}
    &=
    2 \, (-4\pi)^{k+1}
    \sum_{\ell=0}^{\floor{k/2}}
    \frac{ B_{2\ell} }{ (2\ell)! }
    \frac{ B_{2k-2\ell+2} }{ (2k-2\ell+2)! }
    \tbinom{2k-2\ell-1}{k-1}
    \no \\[1ex]
    \mfc^{(-k+2\ell)}_{k,2}
    &=
    2 \, (-4\pi)^{-k+2\ell} \,
    \zeta(2k-2\ell+1) \,
    \frac{ B_{2\ell} }{ (2\ell)! }
    \Big[
    \Theta( 1- \ell) \,
    \tbinom{2k-2\ell}{2-2\ell}
    \tbinom{2k-2}{k-1}
    -
    2 \, 
    \tbinom{2k-2\ell-1}{k-2\ell}
    \Big]
    \no \\[1ex]
    \mfc^{(1-k)}_{k,2}
    &=
    2 \, (-4\pi)^{1-k} \,
    \tbinom{2k-2}{k-1}
    \sum_{\ell=1}^{k-2}
    \zeta(2\ell+1) \,
    \zeta(2k-2\ell-1)
\end{align}
where the step function is defined by $\Theta(x \geq 0) = 1$ and $\Theta(x < 0) = 0$. These Laurent coefficients have the same transcendental structure as those of $C_{a,b,c}$ and~$\mfC_{w;m;p}$.

\newpage

\section{Proof of \autoref{thm:ILk111}}
\label{apdx:ILk111}

In this appendix, we shall prove \autoref{thm:ILk111} by explicitly manipulating the Kronecker-Eisenstein series for \smash{$\Lambda^{[2]}_{k,1,1,1}$} and \smash{$\Lambda^{[3]}_{k,1,1,1}$} in~\eqref{eq:Lk111}.

\subsection{The integral of \texorpdfstring{$\Lambda^{[2]}_{k,1,1,1}$}{}}

We begin with~\eqref{eq:Lk111} and sum over $m_2$ and $n_2$ using the two delta symbols. We rename $m_1$ to $m$, define $\nu = n_3 + N$, and restrict the sum over $m$ to $m > 0$. Performing the integral over $\tau_1$ using~\eqref{eq:Isum} of \autoref{thm:Isum}, we find,
\begin{align}
    \frac{1}{\tau_2^2}
    \int_0^1 d\tau_1 \,
    \Lambda^{[2]}_{k,1,1,1}
    =
    \frac{12 \, \tau_2^{k}}{\pi^{k+2}}
    \sum_{m > 0 \mathstrut}
    \sum_{N \neq 0 \mathstrut}
    \sum_{\nu \neq N \mathstrut}
    \frac{1}{m N^{2k} (N-\nu)^2 (4 m^2 \tau_2^2 + \nu^2)}
\end{align}
To perform the sum over $\nu$, we use the following partial fraction decomposition,
\begin{align}
    \frac{1}{(N-\nu)^2 (y^2 + \nu^2)}
    =
    \frac{1}{(N-\nu)^2 (y^2 + N^2)}
    &
    + \frac{2 N}{(N-\nu) (y^2+N^2)^2}
\no \\
    &
    + \frac{N^2 + 2\nu N - y^2}
           {(y^2 + \nu^2) (y^2 + N^2)^2}
\end{align}
where $y = 2m\tau_2$. The sum over $\nu$ may then be performed in terms of the Riemann zeta function and the series,
\begin{align}
    \frac{1}{2\pi}
    \sum_{\nu \in \bbZ}
    \frac{y}{y^2 + \nu^2}
    =
    \frac{1}{2}
    + \frac{e^{-2\pi y}}{1-e^{-2\pi y}}
\end{align}
which gives the following result,
\begin{align}
    \sum_{\nu \neq N}
    \frac{1}{(N-\nu)^2 (y^2 + \nu^2)}
    &=
    \frac{2 \, \zeta(2) }{y^2 + N^2}
    + \frac{1}{(y^2 + N^2)^2}
    - \frac{4 N^2}{(y^2 + N^2)^3}
\no \\
    & \quad
    -
    \[
    \frac{2\pi}{y(y^2 + N^2)}
    - \frac{4\pi N^2}{y(y^2 + N^2)^2}
    \]
    \[
    \frac{1}{2}
    + \frac{e^{-2\pi y}}{1-e^{-2\pi y}}
    \]
\end{align}
At this point we begin to see a distinction between power-behaved and exponential terms. We shall separate the integral over $\tau_2$ along these lines. Restricting the $N$ sum to ${N > 0}$, integrating over $\tau_2$, and changing variables, we find,
\begin{align}
    \int_0^L \frac{d\tau_2}{\tau_2^2}
    \int_0^1 d\tau_1 \,
    \Lambda^{[2]}_{k,1,1,1}
    =
    \cI_{k,1,1,1}^{[2] \mathrm{(exp)}}(L)
    +
    \cI_{k,1,1,1}^{[2] \mathrm{(pow)}}(L)
\end{align}
where the two contributions are given by,
\begin{align}
\label{eq:I2k111}
    \cI_{k,1,1,1}^{[2] \mathrm{(exp)}}(L)
    &=
    \frac{48}{(2\pi)^{k+1}}
    \sum_{m, N > 0} \,
    \int_0^{2mL}
    \frac{dx \, x^{k-1}}{m^{k+2} N^{2k}} \,
    \frac{N^2-x^2}{(x^2 + N^2)^2} \,
    \frac{e^{-2\pi x}}{1-e^{-2\pi x}}
\no \\[1ex]
    \cI_{k,1,1,1}^{[2] \mathrm{(pow)}}(L)
    &=
    \frac{48}{(2\pi)^{k+2}}
    \sum_{m, N > 0} \,
    \int_0^{\frac{2mL}{N}}
    \frac{dx \, x^{k-1} }{m^{k+2} N^{k+3}}
\no \\
    & \quad
    \times
    \[
    \frac{2 \, \zeta(2) \, N^2 x}{x^2 + 1}
    + \frac{x}{(x^2 + 1)^2}
    - \frac{4 x}{(x^2 + 1)^3}
    - \frac{\pi N}{x^2 + 1}
    + \frac{2\pi N}{(x^2 + 1)^2}
    \]
\end{align}

\subsection{The integral of \texorpdfstring{$\Lambda^{[3]}_{k,1,1,1}$}{}}

We begin with~\eqref{eq:Lk111} and perform partial fraction decomposition in $n_r$ on each factor of~$|p_r|^{-2}$ for $r = 1$, $2$, $3$ to find,\footnote{Recall that $p_r = m_r \tau + n_r$.}
\begin{align}
    \Lambda^{[3]}_{k,1,1,1}
    &=
    \frac{i \tau_2^k}{8\pi^{k+3}}
    \sum_{N \neq 0} \,
    \sum_{\substack{m_r \neq 0 \\
        n_r \in \bbZ \\
        r = 1,2,3}}
    \frac{\delta(m) \delta(n+N)}
        {m_1 m_2 m_3 N^{2k}}
    \( \frac{1}{\bar p_1} - \frac{1}{p_1} \)
    \( \frac{1}{\bar p_2} - \frac{1}{p_2} \)
    \( \frac{1}{\bar p_3} - \frac{1}{p_3} \)
\end{align}
where $m = \sum_{r=1}^3 m_r$ and $n = \sum_{r=1}^3 n_r$. The summand is invariant under permutations of the labels $(1,2,3)$. Using this symmetry, we make the following replacement within the summand,
\begin{align}
    \( \frac{1}{\bar p_1} - \frac{1}{p_1} \)
    \( \frac{1}{\bar p_2} - \frac{1}{p_2} \)
    \( \frac{1}{\bar p_3} - \frac{1}{p_3} \)
    \to
    \(
    \frac{3}{\bar p_1 p_2 p_3}
    - \frac{1}{p_1 p_2 p_3}
    \)
    - c.c.
\end{align}
where $c.c.$ stands for the complex conjugate of the preceding term. We then use the following partial fraction decomposition,
\begin{align}
    \frac{1}{ABC}
    =
    \frac{1}{A+B+C}
    \( \frac{1}{A} + \frac{1}{B+C} \)
    \( \frac{1}{B} + \frac{1}{C} \)
\end{align}
along with the delta function constraint $p_1 + p_2 + p_3 = - N$ and the remaining permutation symmetry of the labels $(2,3)$ to make the following replacement within the summand,
\begin{align}
    \frac{3}{\bar p_1 p_2 p_3}
    - \frac{1}{p_1 p_2 p_3}
    \to
    \frac{2}{p_2}
    \[
    \frac{3}{\bar p_1 - p_1 - N}
    \( \frac{1}{\bar p_1} - \frac{1}{p_1 + N} \)
    + \frac{1}{N}
    \( \frac{1}{p_1} - \frac{1}{p_1 + N} \)
    \]
\end{align}
The term with a factor of $1/N$ vanishes in the full sum.

\sm

We now return to \smash{$\Lambda^{[3]}_{k,1,1,1}$} with the above replacements, sum over $m_3$ and $n_3$ using the delta symbols, and restrict the sum over $N$ to $N > 0$. The sums over $n_1$ and $n_2$ factorize and may each be expressed in terms of the following function,
\begin{align}
    \cB(\tau, m)
    =
    \frac{1}{2\pi}
    \sum_{n \in \bbZ}
    \(
    \frac{i}{m\tau+n}
    -
    \frac{i}{m\bar\tau+n}
    \)
\end{align}
which satisfies $B(\tau,m)= \ep (m) B(\tau, |m|)$ where $\veps(m) = \pm 1$ is equal to the sign of $m$. After some algebra, we find,
\begin{align}
    \Lambda^{[3]}_{k,1,1,1}
    =
    \frac{12 \, \tau_2^{k+1}}{\pi^{k+1}} \,
    \sum_{N > 0} \,
    \sum_{\substack{m_1, m_2 \neq 0 \\ m_1+m_2 \neq 0}}
    \frac{ \veps(m_1) \veps(m_1) \cB(\tau, |m_1|) \cB(\tau, |m_2|) }
         { m_2 (m_1+m_2) N^{2k} (4 m_1^2 \tau_2^2 + N^2) }
\end{align}
Only the $\cB$ functions depend on $\tau_1$. Their Fourier expansion is given by,
\begin{align}
    \cB(\tau, |m|)
    =
    1 + \sum_{p > 0}
    \( q^{|m| \, p} + \bar q^{|m| \, p} \)
\end{align}
with $q = e^{2\pi i \tau}$. The constant Fourier mode of their product is thus,
\begin{align}
    \int_0^1 d\tau_1 \,
    \cB(\tau, |m_1|) \, \cB(\tau, |m_2|)
    &=
    1 + \sum_{p_1, p_2 > 0}
    \( q^{|m_1| \, p_1} \, \bar q^{|m_2| \, p_2}
    + c.c. \)
    \big|_{|m_1| \, p_1 = |m_2| \, p_2}
    \no \\
    &=
    1 + 2 \sum_{p > 0}
    (q \bar q)^{\{|m_1|,|m_2|\} \, p}
    \no \\
    &=
    1 +
    \frac{2 \, e^{-4\pi \{|m_1|,|m_2|\} \tau_2}}
         {1 - e^{-4\pi \{|m_1|,|m_2|\} \tau_2} }
\end{align}
where we have defined the arithmetic operation $\{m_1,m_2\} = m_1 m_2 / \gcd(m_1,m_2)$. To go from the first to the second line, we solved the constraint $|m_1| \, p_1 = |m_2| \, p_2$ by,
\begin{align}
    p_1
    &=
    \frac{|m_2| \, p}{\gcd\big(|m_1|, |m_2|\big)}
    &
    p_2
    &=
    \frac{|m_1| \, p}{\gcd\big(|m_1|, |m_2|\big)}
\end{align}
for positive integers $p$. Thus, the sums over $p_1$ and $p_2$ in the first line collapse into a single sum over $p > 0$. In the final line we again see a split between power-behaved and exponential terms. We separate the integral over $\tau_2$ along these lines and find,
\begin{align}
    \int_0^L \frac{d\tau_2}{\tau_2^2}
    \int_0^1 d\tau_1 \,
    \Lambda^{[3]}_{k,1,1,1}
    =
    \cI_{k,1,1,1}^{[3] \mathrm{(exp)}}(L)
    +
    \cI_{k,1,1,1}^{[3] \mathrm{(pow)}}(L)
\end{align}
where the two contributions are given by,
\begin{align}
\label{eq:I3k111}
    \cI_{k,1,1,1}^{[3] \mathrm{(exp)}}(L)
    &=
    \frac{48}{(2\pi)^{k+1}}
    \sum_{\substack{m_1, m_2 \neq 0 \\ m_1+m_2 \neq 0}}
    \sum_{N > 0} \,
    \frac{\veps(m_1) \veps(m_2)}
        {m_2 (m_1+m_2) N^{2k}}
\no \\
    & \quad
    \times
    \int_0^{2L} 
    \frac{dx \, x^{k-1}}
        {m_1^2 x^2 + N^2}
    \frac{e^{-2\pi \{|m_1|,|m_2|\} x}}
        { 1 - e^{-2\pi \{|m_1|,|m_2|\} x} }
\no \\
    \cI_{k,1,1,1}^{[3] \mathrm{(pow)}}(L)
    &=
    \frac{24}{(2\pi)^{k+1}}
    \sum_{\substack{m_1, m_2 \neq 0 \\ m_1+m_2 \neq 0}}
    \sum_{N > 0} \,
    \frac{\veps(m_1) \veps(m_2)}
        {|m_1|^k m_2 (m_1+m_2) N^{k+2}}
    \int_0^\frac{2 |m_1| L}{N} 
    \frac{dx \, x^{k-1}}{x^2 + 1}
\end{align}

\subsection{The exponential integrals}

We shall now combine the two exponential integrals. Symmetrizing the summand of \smash{$\cI_{k,1,1,1}^{[3] \mathrm{(exp)}}$} in $m_1$ and $m_2$ eliminates the factor of $1/(m_1 + m_2)$ and allows us to restore the ${m_1+m_2=0}$ contribution to the sum. This contribution precisely equals the expression for \smash{$\cI_{k,1,1,1}^{[2] \mathrm{(exp)}}$} in~\eqref{eq:I2k111}. After some algebra, we find,
\begin{align}
    \cI_{k,1,1,1}^{[2] \mathrm{(exp)}}(L)
    +
    \cI_{k,1,1,1}^{[3] \mathrm{(exp)}}(L)
    &=
    \frac{96}{(2\pi)^{k+1}}
    \sum_{m_1, m_2, N > 0} \,
    \frac{1}{ \tilde m_1^{k-1} \tilde m_2^{k-1}
              \gcd(m_1,m_2)^{k+2} N^{2k-2}}
\no \\[1ex]
    & \quad \times
    \int_0^{2 \{ m_1, m_2 \} L}
    \hspace{-1cm}
    \frac{dx \, x^{k-1}}
        {(x^2 + \tilde m_1^2 N^2)
        (x^2 + \tilde m_2^2 N^2)}
    \frac{e^{-2\pi x}}{1-e^{-2\pi x}}
\end{align}
where $\tilde m_{1,2} = m_{1,2}/\gcd(m_1,m_2)$ so that the integers $\tilde m_1$ and $\tilde m_2$ are coprime. 

\sm

To proceed, we raise the upper the limit of the integral to infinity which removes the pesky appearance $\{ m_1, m_2 \}$ and introduces corrections of order \smash{$\cO(e^{-2\pi L})$}. We then parametrize the sums over $m_1$ and $m_2$ by splitting them into the independent sums over their greatest common divisor and over pairs of coprime integers ${\tilde m_1, \tilde m_2 > 0}$. That is,
\begin{align}
\label{eq:msums}
    \sum_{m_1, m_2 > 0}
    &=
    \sum_{\gcd(m_1,m_2) > 0}
    \quad
    \sum_{\substack{
        \tilde m_1, \tilde m_2 > 0 \\
        \text{coprime}}}
\end{align}
The sum over $\gcd(m_1,m_2)$ yields the Riemann zeta-value $\zeta(k+2)$. The sum over~$N$ may then be carried out by defining the independent variables $N_1 = N \, \tilde m_1$ and $N_2 = N \, \tilde m_2$ and using~\eqref{eq:msums} in reverse. In total, we find \smash{$\cI_{k,1,1,1}^{[2] \mathrm{(exp)}} + \cI_{k,1,1,1}^{[3] \mathrm{(exp)}} = \cI_{k,1,1,1}^{\mathrm{(exp)}}$} as in~\eqref{eq:Ik111}.

\subsection{The power-behaved integrals}

We shall now combine the two power-behaved integrals. The summand of \smash{$\cI_{k,1,1,1}^{[3] \mathrm{(pow)}}$} is invariant under ${(m_1, m_2) \to - (m_1, m_2)}$, so we fix $m_1 > 0$. We then evaluate the sum over $m_2$ for fixed~$m_1$ by decomposing its summand in partial fractions and splitting the sum as,
\begin{align}
    \sum_{m_2 \neq 0, -m_1}
    \frac{\veps(m_2)}{ m_2 (m_1+m_2) }
    &=
    \frac{1}{m_1}
    \sum_{m_2 > 0} \!
    \( \frac{1}{m_2}
    - \frac{1}{m_1+m_2} \)
    + \frac{1}{m_1}
    \sum_{\substack{m_2 > 0 \\ m_2 \neq m_1}} \!\!
    \( \frac{1}{m_2}
    + \frac{1}{m_1-m_2} \)
\no \\
    &=
    \frac{2 H_1(m_1)}{m_1}
\end{align}
where $H_1(m) = \sum_{k=1}^{m-1} \frac{1}{k}$ are finite harmonic sums.

\sm

After some straightforward rearrangements, we find \smash{$\cI_{k,1,1,1}^{[2] \mathrm{(pow)}} + \cI_{k,1,1,1}^{[3] \mathrm{(pow)}} = \cI_{k,1,1,1}^{\mathrm{(pow)}}$} as in~\eqref{eq:Ik111}. This completes our proof of \autoref{thm:ILk111}.

\newpage

\section{Proof of \autoref{thm:Ik111exp}}
\label{apdx:Ik111exp}

In this appendix, we shall prove \autoref{thm:Ik111exp} using contour integral methods and analytic continuation.

\subsection{The function \texorpdfstring{$\cJ_{k}(\eps)$}{}}

For integer $k \geq 2$ and $\eps \in \bbC$, we define the following function,
\begin{align}
\label{eq:Jkdef}
    \cJ_{k}(\eps)
    &=
    4 \,
    (-)^{\lfloor k/2 \rfloor}
    \sum_{M, N > 0}
    \frac{1}{M^{k-1} N^{k-1}}
    \int_0^\infty
    \frac{dx \, x^{\eps}}
         {(x^2 + M^2)(x^2 + N^2)}
    \frac{e^{-2\pi x}}{1-e^{-2\pi x}}
\end{align}
which converges for $\Re(\eps) > 0$ and obeys \smash{$\cJ_{k}(k-1) = \cJ_{k,1,1,1}^{\mathrm{(exp)}}$} with \smash{$\cJ_{k,1,1,1}^{\mathrm{(exp)}}$} defined in~\eqref{eq:Ik111exp}. We shall proceed with $\eps$ in the range $1 < \Re(\eps) < 2$. In this range, we may decompose $\cJ_{k}(\eps)$ into a sum of elementary integrals and perform the infinite sums in terms of convergent zeta functions. The resulting expression will then admit an analytic continuation to $\eps = k-1$.

\subsection{Splitting \texorpdfstring{$\cJ_{k}(\eps)$}{}}

To begin, we write the exponential factor in the integrand of $\cJ_{k}(\eps)$ in terms of an infinite sum over positive integers using the following identity,
\begin{align}
    \frac{e^{-2\pi x}}{1-e^{-2\pi x}}
    &=
    - \frac{1}{2}
    + \frac{1}{2\pi x}
    + \frac{1}{\pi}
    \sum_{P>0} \frac{x}{x^2+P^2}
\end{align}
For $1 < \Re(\eps) < 2$, we may interchange the order of the integral over~$x$ and the sum over~$P$ in $\cJ_{k}(\eps)$. We then split the unrestricted sums over $M$, $N$, and $P$ into sums over distinct positive integers and write,
\begin{align}
    \cJ_{k}(\eps)
    &=
    (-)^{\lfloor k/2 \rfloor}
    \sum_{\ell=1}^8
    \cJ_k^{(\ell)}(\eps)
\end{align}
where the eight contributions \smash{$\cJ_k^{(\ell)}(\eps)$} are given by,
\begin{align}
\label{eq:JkSplit}
    \cJ_k^{(1)}(\eps)
    &=
    - 2
    \sum_{M>0}
    \frac{1}{M^{2k-2}}
    \int_0^\infty
    \frac{dx \, x^{\eps}}
        {(x^2+M^2)^2}
    \no \\
    \cJ_k^{(2)}(\eps)
    &=
    - 2
    \sum_{\substack{M,N>0 \\ \text{dist}}}
    \frac{1}{M^{k-1} N^{k-1}}
    \int_0^\infty
    \frac{dx \, x^{\eps}}
        {(x^2+M^2)(x^2+N^2)}
    \no \\
    \cJ_k^{(3)}(\eps)
    &=
    \frac{2}{\pi}
    \sum_{M>0}
    \frac{1}{M^{2k-2}}
    \int_0^\infty
    \frac{dx \, x^{\eps-1}}
        {(x^2+M^2)^2}
    \no \\
    \cJ_k^{(4)}(\eps)
    &=
    \frac{2}{\pi}
    \sum_{\substack{M,N>0 \\ \text{dist}}}
    \frac{1}{M^{k-1} N^{k-1}}
    \int_0^\infty
    \frac{dx \, x^{\eps-1}}
        {(x^2+M^2)(x^2+N^2)}
    \no \\
    \cJ_k^{(5)}(\eps)
    &=
    \frac{4}{\pi}
    \sum_{M>0}
    \frac{1}{M^{2k-2}}
    \int_0^\infty
    \frac{dx \, x^{\eps+1}}
        {(x^2+M^2)^3}
    \no \\
    \cJ_k^{(6)}(\eps)
    &=
    \frac{8}{\pi}
    \sum_{\substack{M,N>0 \\ \text{dist}}}
    \frac{1}{M^{k-1} N^{k-1}}
    \int_0^\infty
    \frac{dx \, x^{\eps+1}}
        {(x^2+M^2)^2(x^2+N^2)}
    \no \\
    \cJ_k^{(7)}(\eps)
    &=
    \frac{4}{\pi}
    \sum_{\substack{M,P>0 \\ \text{dist}}}
    \frac{1}{M^{2k-2}}
    \int_0^\infty
    \frac{dx \, x^{\eps+1}}
        {(x^2+M^2)^2(x^2+P^2)}
    \no \\
    \cJ_k^{(8)}(\eps)
    &=
    \frac{4}{\pi}
    \sum_{\substack{M,N,P>0 \\ \text{dist}}}
    \frac{1}{M^{k-1} N^{k-1}}
    \int_0^\infty
    \frac{dx \, x^{\eps+1}}
        {(x^2+M^2)(x^2+N^2)(x^2+P^2)}
\end{align}
Here ``$\mathrm{dist}$" denotes that the summation variables must be distinct. The additional factor of~$2$ in \smash{$\cJ_k^{(6)}(\eps)$} counts the two equivalent contributions of ${M = P \neq N}$ and ${M \neq N = P}$.

\subsection{Computing the integrals}

The integrals in each contribution converge for $1 < \Re(\eps) < 2$ and may be calculated using a keyhole contour that begins at the origin, runs up the positive real axis from above, circles counter-clockwise around the complex plane at infinity, and then runs down the positive real axis from below.

\sm

The eight integrals evaluate to rational expression in the relevant summation variables multiplied by $\sec( \eps \tfrac{\pi}{2} )$ or $\csc( \eps \tfrac{\pi}{2} )$. After integration, the single sums in \smash{$\cJ_k^{(1)}(\eps)$}, \smash{$\cJ_k^{(3)}(\eps)$}, and~\smash{$\cJ_k^{(5)}(\eps)$} may be performed in terms of the Riemann zeta function,
\begin{align}
    \cos( \eps \tfrac{\pi}{2} )
    \times
    \cJ_k^{(1)}(\eps)
    &=
    \half (\eps-1) \,
    \pi  \,
    \zeta(2k+1-\eps)
    \no \\
    \sin( \eps \tfrac{\pi}{2} )
    \times
    \cJ_k^{(3)}(\eps)
    &=
    - \half (\eps-2) \,
    \zeta(2k+2-\eps)
    \no \\
    \sin( \eps \tfrac{\pi}{2} )
    \times
    \cJ_k^{(5)}(\eps)
    &=
    - \tfrac{1}{4} \eps (\eps-2) \,
    \zeta(2k+2-\eps)
\end{align}
The contributions \smash{$\cJ_k^{(2)}(\eps)$}, \smash{$\cJ_k^{(4)}(\eps)$}, and \smash{$\cJ_k^{(6)}(\eps)$} may be written as,
\begin{align}
    \cos( \eps \tfrac{\pi}{2} )
    \times
    \cJ_k^{(2)}(\eps)
    &=
    \sum_{\substack{M,N>0 \\ \text{dist}}}
    \frac{ -2 \pi }{ M^{k-\eps} N^{k-1} (N^2 - M^2) }
    \no \\
    \sin( \eps \tfrac{\pi}{2} )
    \times
    \cJ_k^{(4)}(\eps)
    &=
    \sum_{\substack{M,N>0 \\ \text{dist}}}
    \frac{ 2 }{ M^{k+1-\eps} N^{k-1} (N^2 - M^2) }
    \no \\
    \sin( \eps \tfrac{\pi}{2} )
    \times
    \cJ_k^{(6)}(\eps)
    &=
    \sum_{\substack{M,N>0 \\ \text{dist}}}
    \frac{ 2 \eps }{ M^{k+1-\eps} N^{k-1} (N^2 - M^2) }
\end{align}
The final contributions, \smash{$\cJ_k^{(7)}(\eps)$} and \smash{$\cJ_k^{(8)}(\eps)$}, may be simplified using the following identities,
\begin{align}
    \sum_{\substack{N>0 \\ N \neq M}}
    \frac{1}{N^2 - M^2}
    &=
    \frac{3}{4 M^2}
    &
    \sum_{\substack{N>0 \\ N \neq M}}
    \frac{1}{(N^2 - M^2)^2}
    &=
    \frac{ \zeta(2) }{2 M^2}
    - \frac{11}{16 M^4}
\end{align}
which may be derived by decomposing the denominator of each summand in partial fractions. Using these sums we find,
\begin{align}
    \sin( \eps \tfrac{\pi}{2} )
    \times
    \cJ_k^{(7)}(\eps)
    &=
    \zeta(2) \, \zeta(2k-\eps)
    +
    ( \tfrac{3}{4} \eps - \tfrac{11}{8} ) \,
    \zeta(2k+2-\eps)
\no \\[1ex]
    & \quad
    +
    \sum_{\substack{M,N>0 \\ \text{dist}}}
    \frac{ - 2 \, N^\eps }{ M^{2k-2} (N^2 - M^2)^2 }
\no \\
    \sin( \eps \tfrac{\pi}{2} )
    \times
    \cJ_k^{(8)}(\eps)
    &=
    \sum_{\substack{M,N>0 \\ \text{dist}}}
    \frac{ - 3 }{ M^{k+1-\eps} N^{k-1} (N^2 - M^2) }
\no \\
    & \quad
    +
    \sum_{\substack{M,N>0 \\ \text{dist}}}
    \frac{ 4 }{ M^{k-1-\eps} N^{k-1} (N^2 - M^2)^2 }
\no \\
    & \quad
    +
    \sum_{\substack{M,N,P>0 \\ \text{dist}}}
    \frac{ - 2 \, P^\eps }{ M^{k-1} N^{k-1} (M^2 - P^2) (N^2 - P^2) }
\end{align}
We now reassemble the eight contributions and find,
\begin{align}
\label{eq:JkRe}
    (-)^{\lfloor k/2 \rfloor} 
    \times
    \cJ_{k}(\eps)
    &=
    \half (\eps-1)
    \sec( \eps \tfrac{\pi}{2} ) \,
    \pi \, \zeta(2k+1-\eps)
\no \\[1ex]
    & \quad
    +
    \csc( \eps \tfrac{\pi}{2} )
    \Big[
    -\tfrac{1}{4}
    ( \eps^2 - 3 \eps + \tfrac{3}{2} ) \,
    \zeta(2k+2-\eps)
    +
    \zeta(2) \, \zeta(2k-\eps)
    \Big]
\no \\[1ex]
    & \quad
    +
    \sum_{M>0}
    \frac{1}{ M^{k+1-\eps} }
    \Big[
    (2\eps-1) \csc( \eps \tfrac{\pi}{2} )
    -
    2 \pi M \sec( \eps \tfrac{\pi}{2} )
    \Big]
\no \\
    & \qquad
    \times
    \sum_{\substack{N>0 \\ N \neq M}}
    \frac{1}{ N^{k-1} (N^2 - M^2) }
\no \\
    & \quad
    +
    4 \csc( \eps \tfrac{\pi}{2} )
    \sum_{M>0}
    \frac{ 1 }{ M^{k-1-\eps} }
    \sum_{\substack{N>0 \\ N \neq M}}
    \frac{1}{ N^{k-1} (N^2 - M^2)^2 }
\no \\
    & \quad
    - 2
    \csc( \eps \tfrac{\pi}{2} )
    \sum_{M>0}
    \frac{ 1 }{ M^{-\eps} }
    \Bigg\{
    \sum_{\substack{N>0 \\ N \neq M}}
    \frac{1}{ N^{k-1} (N^2 - M^2) }
    \Bigg\}^2
\end{align}
The squared sum on the last line arises from the combination of the final terms in the previous expressions for \smash{$\cJ_k^{(7)}(\eps)$} and \smash{$\cJ_k^{(8)}(\eps)$}.

\subsection{Summing over \texorpdfstring{$N$}{}}

There are two distinct infinite sums over the variable $N$ in~\eqref{eq:JkRe}. Both summands may be decomposed in partial fractions. The sums may then be explicitly evaluated as functions of the integers $M \geq 1$ and $k \geq 2$. The result depends on whether $k$ is even or odd. With some algebra, we find,
\begin{align}
\label{eq:Nsums}
    \sum_{\substack{N>0 \\ N \neq M}}
    \frac{1}{ N^{k-1} (N^2 - M^2) }
    &=
    \frac{ \half (k+\half) }{ M^{k+1} }
    -
    \sum_{\ell=1}^{\floor{(k-1)/2}}
    \frac{ \zeta(k+1-2\ell) }{ M^{2\ell} }
\no \\
    & \quad
    -
    \begin{cases}
    \dfrac{ H_1(M+1) }{ M^{k} }
    & 
    k \text{ even}
    \\[2ex]
    \hfill
    0
    \hfill
    &
    k 
    \text{ odd}
    \end{cases}
\no \\[2ex]
    \sum_{\substack{N>0 \\ N \neq M}}
    \frac{1}{ N^{k-1} (N^2 - M^2)^2 }
    &=
    \frac{ - \tfrac{1}{8} ( k^2 + 3k + \tfrac{3}{2} ) }
         { M^{k+3} }
    +
    \sum_{\ell=1}^{\floor{(k-1)/2}}
    \frac{ \ell \, \zeta(k+1-2\ell) }{ M^{2\ell+2} }
\no \\
    & \quad
    + 
    \begin{cases}
    \dfrac{ \half k \, H_1(M+1) }{ M^{k+2} }
    +
    \dfrac{ \half \, H_2(M+1) }{ M^{k+1} }
    &
    k \text{ even}
    \\[1ex]
    \hfill
    \dfrac{ \half \, \zeta(2) }{ M^{k+1} }
    \hfill
    &
    k \text{ odd}
    \end{cases}
\end{align}
where $H_2(n) = \sum_{N=1}^{n-1} \frac{1}{N^2}$ is a generalized finite harmonic series. It remains to insert these expressions into~\eqref{eq:JkRe}, perform the sums over $M$ in terms of convergent zeta functions, and then analytically continue the result to $\eps = k-1$. At this point, we shall discuss the cases of odd and even $k$ separately.

\subsection{Odd \texorpdfstring{$k$}{}}

For odd $k \geq 3$, we insert~\eqref{eq:Nsums} into~\eqref{eq:JkRe}, perform the sums over $M$, and find,
\begin{align}
    \cJ_{k}(\eps)
    &=
    (-)^{(k-1)/2} \,
    \pi 
    \sec( \eps \tfrac{\pi}{2} )
    \bigg\{
    - ( k + 1 - \tfrac{\eps}{2} ) \,
    \zeta(2k+1-\eps)
\no \\
    & \qquad
    + 2
    \sum_{\ell=1}^{(k-1)/2}
    \zeta(k+1-2\ell) \,
    \zeta(k+2\ell-\eps)
    \bigg\}
\no \\
    & \quad
    +
    (-)^{(k-1)/2} \,
    \csc( \eps \tfrac{\pi}{2} )
    \bigg\{
    - \tfrac{1}{4}
    (2k+2-\eps) (2k+3-\eps) \,
    \zeta(2k+2-\eps)
\no \\
    & \qquad
    +
    3 \,
    \zeta(2) \, \zeta(2k-\eps)
    \vphantom{\bigg\{}
\no \\
    & \qquad
    + 2
    \sum_{\ell=1}^{(k-1)/2}
    ( k+2\ell+1-\eps) \,
    \zeta(k+1-2\ell) \,
    \zeta(k+1+2\ell-\eps)
\no \\
    & \qquad
    - 2
    \sum_{\mathclap{\ell_{1,2}=1}}^{(k-1)/2}
    \,
    \zeta(k+1-2\ell_1) \,
    \zeta(k+1-2\ell_2) \,
    \zeta(2\ell_1+2\ell_2-\eps)
    \bigg\}
\end{align}
This expression is an analytic function of $\eps$ valid for $\Re(\eps) > 0$. Near $\eps = k-1$ with $k$ odd, the zeta functions are finite, \smash{$\sec( \eps \tfrac{\pi}{2} ) \sim (-)^{\frac{k-1}{2}} $}, and \smash{$\csc( \eps \tfrac{\pi}{2} ) \sim (-)^{\frac{k+1}{2}} \, \frac{2}{\pi} \frac{1}{k-1-\eps} $}. Thus, the terms in brackets which multiply $\csc( \eps \tfrac{\pi}{2} )$ must vanish at $\eps = k-1$ to cancel the simple pole. Indeed, we may show that they vanish using the identities obeyed by finite sums of even zeta-values in~\eqref{eq:evenZetaSum}. Hence, $\cJ_k(k-1)$ for odd $k$ is finite as required and given by~\eqref{eq:Ik111expO}.

\sm

In the first line of~\eqref{eq:Ik111expO} we have used~\eqref{eq:zeta(n,1)} to write a finite sum of odd zeta-values with transcendental weight $k+2$ in terms of the double zeta-value $\zeta(k+1,1)$ for later convenience. This completes our proof of \autoref{thm:Ik111exp} for odd $k$.

\subsection{Even \texorpdfstring{$k$}{}}

For even $k \geq 2$, we again insert~\eqref{eq:Nsums} into~\eqref{eq:JkRe}. The square of the finite harmonic sum~$H_1(M+1)$ is given by,
\begin{align}
    H_1(M+1)^2
    &=
    \sum_{m_1, m_2 = 1}^M
    \frac{1}{m_1 m_2}
\no \\
    &=
    \frac{1}{M^2}
    +
    \sum_{m=1}^{M-1} 
    \frac{1}{m^2}
    +
    \frac{2}{M}
    \sum_{m=1}^{M-1}
    \frac{1}{m}
    +
    \sum_{m_1=1}^{M-1} \,
    \sum_{m_2=1}^{m_1-1}
    \frac{2}{m_1 m_2}
\end{align}
The second term is equal to the generalized harmonic sum $H_2(M)$ and cancels. The last term yields a triple zeta-value $\zeta(s,1,1)$ upon summing over $M$. Performing these sums, we find the following lengthy expression,
\begin{align}
    \cJ_{k}(\eps)
    &=
    (-)^{k/2} \,
    \pi 
    \sec( \eps \tfrac{\pi}{2} )
    \bigg\{
    - ( k - 1 - \tfrac{\eps}{2} ) \,
    \zeta(2k+1-\eps)
    + 2 \, \zeta(2k-\eps,1)
\no \\
    & \qquad
    + 2
    \sum_{\ell=1}^{k/2-1}
    \zeta(k+1-2\ell) \,
    \zeta(k+2\ell-\eps)
    \bigg\}
\no \\
    & \quad
    +
    (-)^{k/2}
    \csc( \eps \tfrac{\pi}{2} )
    \bigg\{
    - \tfrac{1}{4}
    ( 4k^2 - 6k - 4\eps k + \eps^2 +3\eps - 2 ) \,
    \zeta(2k+2-\eps)
\no \\
    & \qquad
    + \zeta(2) \, \zeta(2k-\eps)
    + 2 \, (2k-1-\eps) \, \zeta(2k+1-\eps,1)
    - 4 \, \zeta(2k-\eps,1,1)
    \vphantom{\bigg\{}
\no \\
    & \qquad
    + 2
    \sum_{\ell=1}^{k/2-1}
    ( k+2\ell-1-\eps) \,
    \zeta(k+1-2\ell) \,
    \zeta(k+1+2\ell-\eps)
\no \\
    & \qquad
    - 4
    \sum_{\ell=1}^{k/2-1}
    \,
    \zeta(k+1-2\ell) \,
    \zeta(k+2\ell-\eps,1)
\no \\
    & \qquad
    - 2
    \sum_{\mathclap{\ell_{1,2}=1}}^{k/2-1}
    \,
    \zeta(k+1-2\ell_1) \,
    \zeta(k+1-2\ell_2) \,
    \zeta(2\ell_1+2\ell_2-\eps)
    \bigg\}
\end{align}
This is again an analytic function of $\eps$ valid for $\Re(\eps) > 0$. Near $\eps = k-1$ with $k$ even, we have~\smash{$\csc( \eps \tfrac{\pi}{2} ) \sim (-)^{k/2+1} $} and \smash{$\sec( \eps \tfrac{\pi}{2} ) \sim (-)^{k/2+1} \, \frac{2}{\pi} \frac{1}{k-1-\eps} $}. The zeta functions, however, are not all finite. The terms in the last line with $2\ell_1+2\ell_2 = k$ have a simple pole since $\zeta(k-\eps) \sim \frac{1}{k-1-\eps}$. These poles must combine with the $\sec( \eps \tfrac{\pi}{2} )$ poles to produce a finite result at $\eps = k-1$. We must then have,
\begin{align}
     0
    &=
    (k-1) \,
    \zeta(k+2)
    - 4 \, \zeta(k+1,1)
    - 2
    \sum_{\ell=1}^{k/2-1}
    \zeta(k+1-2\ell) \,
    \zeta(2\ell+1)
\end{align}
Using the multiple zeta-value identity~\eqref{eq:zeta(n,1)}, we see that this expression indeed vanishes. Hence, $\cJ_k(k-1)$ for even $k$ is finite as required and given by~\eqref{eq:Ik111expE}. This completes our proof of \autoref{thm:Ik111exp} for even $k$.

\newpage

\bibliography{3LoopMGFs}

\end{document}

%% file: preambledefs.tex
\newcommand{\cB}{\mathcal{B}}
\newcommand{\cC}{\mathcal{C}}

\newcommand{\cE}{\mathcal{E}}

\newcommand{\cG}{\mathcal{G}}
\newcommand{\cH}{\mathcal{H}}
\newcommand{\cI}{\mathcal{I}}
\newcommand{\cJ}{\mathcal{J}}
\newcommand{\cK}{\mathcal{K}}
\newcommand{\cL}{\mathcal{L}}
\newcommand{\cM}{\mathcal{M}}
\newcommand{\cN}{\mathcal{N}}
\newcommand{\cO}{\mathcal{O}}

\newcommand{\cR}{\mathcal{R}}

\newcommand{\cU}{\mathcal{U}}
\newcommand{\cV}{\mathcal{V}}
\newcommand{\cW}{\mathcal{W}}

\newcommand{\cZ}{\mathcal{Z}}

\newcommand{\mfC}{\mathfrak{C}}
\newcommand{\mfD}{\mathfrak{D}}

\newcommand{\mfL}{\mathfrak{L}}

\newcommand{\mfP}{\mathfrak{P}}

\newcommand{\mfW}{\mathfrak{W}}

\newcommand{\mfc}{\mathfrak{c}}
\newcommand{\mfd}{\mathfrak{d}}

\newcommand{\mfh}{\mathfrak{h}}

\newcommand{\mfq}{\mathfrak{q}}

\newcommand{\bbC}{\mathbb{C}}

\newcommand{\bbQ}{\mathbb{Q}}
\newcommand{\bbR}{\mathbb{R}}

\newcommand{\bbZ}{\mathbb{Z}}

\newcommand{\eps}{\epsilon}
\newcommand{\veps}{\varepsilon}

\renewcommand{\(}{\left(}
\renewcommand{\)}{\right)}
\renewcommand{\[}{\left[}
\renewcommand{\]}{\right]}

\newcommand{\p}{\partial}

\renewcommand{\Re}{\operatorname{Re}}
\renewcommand{\Im}{\operatorname{Im}}

\newcommand{\floor}[1]{\cramped{\left\lfloor #1 \right\rfloor}}

\newcommand{\half}{\tfrac{1}{2}}

\newcommand{\negphantom}[1]{
    \ifmmode\settowidth{\dimen0}{$#1$}
    \else\settowidth{\dimen0}{#1}
    \fi
    \hspace*{-\dimen0}}

%% file: 3LoopMGFs.bbl
\providecommand{\href}[2]{#2}\begingroup\raggedright\begin{thebibliography}{10}

\bibitem{Green:1999pv}
M.B.~Green and P.~Vanhove, \emph{{The low-energy expansion of the one-loop Type
  II superstring amplitude}},
  \href{https://doi.org/10.1103/PhysRevD.61.104011}{\emph{Phys. Rev. D}
  {\bfseries 61} (2000) 104011}
  [\href{https://arxiv.org/abs/hep-th/9910056}{{\ttfamily hep-th/9910056}}].

\bibitem{Green:2008uj}
M.B.~Green, J.G.~Russo and P.~Vanhove, \emph{{Low-energy expansion of the
  four-particle genus-one amplitude in Type II superstring theory}},
  \href{https://doi.org/10.1088/1126-6708/2008/02/020}{\emph{JHEP} {\bfseries
  02} (2008) 20} [\href{https://arxiv.org/abs/0801.0322}{{\ttfamily
  0801.0322}}].

\bibitem{DHoker:2015gmr}
E.~D'Hoker, M.B.~Green and P.~Vanhove, \emph{{On the modular structure of the
  genus-one Type II superstring low-energy expansion}},
  \href{https://doi.org/10.1007/JHEP08(2015)041}{\emph{JHEP} {\bfseries 08}
  (2015) 41} [\href{https://arxiv.org/abs/1502.06698}{{\ttfamily 1502.06698}}].

\bibitem{Zerbini:2015rss}
F.~Zerbini, \emph{{Single-valued multiple zeta values in genus-one superstring
  amplitudes}},
  \href{https://doi.org/10.4310/CNTP.2016.v10.n4.a2}{\emph{Commun. Num. Theor.
  Phys.} {\bfseries 10} (2016) 703}
  [\href{https://arxiv.org/abs/1512.05689}{{\ttfamily 1512.05689}}].

\bibitem{DHoker:2015wxz}
E.~D'Hoker, M.B.~Green, O.~G\"urdogan and P.~Vanhove, \emph{{Modular graph
  functions}}, \href{https://doi.org/10.4310/CNTP.2017.v11.n1.a4}{\emph{Commun.
  Num. Theor. Phys.} {\bfseries 11} (2017) 165}
  [\href{https://arxiv.org/abs/1512.06779}{{\ttfamily 1512.06779}}].

\bibitem{DHoker:2016mwo}
E.~D'Hoker and M.B.~Green, \emph{{Identities between modular graph forms}},
  \href{https://doi.org/10.1016/j.jnt.2017.11.015}{\emph{J. Number Theory}
  {\bfseries 189} (2018) 25}
  [\href{https://arxiv.org/abs/1603.00839}{{\ttfamily 1603.00839}}].

\bibitem{DHoker:2016quv}
E.~D'Hoker and J.~Kaidi, \emph{{Hierarchy of modular graph identities}},
  \href{https://doi.org/10.1007/JHEP11(2016)051}{\emph{JHEP} {\bfseries 11}
  (2016) 51} [\href{https://arxiv.org/abs/1608.04393}{{\ttfamily 1608.04393}}].

\bibitem{Gerken:2018zcy}
J.E.~Gerken and J.~Kaidi, \emph{{Holomorphic subgraph reduction of higher-point
  modular graph forms}},
  \href{https://doi.org/10.1007/JHEP01(2019)131}{\emph{JHEP} {\bfseries 01}
  (2019) 131} [\href{https://arxiv.org/abs/1809.05122}{{\ttfamily
  1809.05122}}].

\bibitem{Gerken:2019cxz}
J.E.~Gerken, A.~Kleinschmidt and O.~Schlotterer, \emph{{All-order differential
  equations for one-loop closed-string integrals and modular graph forms}},
  \href{https://doi.org/10.1007/JHEP01(2020)064}{\emph{JHEP} {\bfseries 01}
  (2020) 064} [\href{https://arxiv.org/abs/1911.03476}{{\ttfamily
  1911.03476}}].

\bibitem{Gerken:2020aju}
J.E.~Gerken, \emph{{Basis decompositions and a Mathematica package for modular
  graph forms}},  \href{https://arxiv.org/abs/2007.05476}{{\ttfamily
  2007.05476}}.

\bibitem{Gerken:2020xte}
J.E.~Gerken, \emph{{Modular graph forms and scattering amplitudes in string
  theory}}, Ph.D. thesis, Humboldt U., Berlin, 2020.
\newblock \href{https://arxiv.org/abs/2011.08647}{{\ttfamily 2011.08647}}.

\bibitem{DHoker:2015sve}
E.~D'Hoker, M.B.~Green and P.~Vanhove, \emph{{Proof of a modular relation
  between 1-, 2-, and 3-loop Feynman diagrams on a torus}},
  \href{https://doi.org/10.1016/j.jnt.2017.07.022}{\emph{J. Number Theory}
  {\bfseries 196} (2019) 381}
  [\href{https://arxiv.org/abs/1509.00363}{{\ttfamily 1509.00363}}].

\bibitem{Basu:2015ayg}
A.~Basu, \emph{{Poisson equation for the Mercedes diagram in string theory at
  genus one}},
  \href{https://doi.org/10.1088/0264-9381/33/5/055005}{\emph{Class. Quant.
  Grav.} {\bfseries 33} (2016) 055005}
  [\href{https://arxiv.org/abs/1511.07455}{{\ttfamily 1511.07455}}].

\bibitem{Basu:2016xrt}
A.~Basu, \emph{{Poisson equation for the three-loop ladder diagram in string
  theory at genus one}},
  \href{https://doi.org/10.1142/S0217751X16501694}{\emph{Int. J. Mod. Phys. A}
  {\bfseries 31} (2016) 1650169}
  [\href{https://arxiv.org/abs/1606.02203}{{\ttfamily 1606.02203}}].

\bibitem{Basu:2016kli}
A.~Basu, \emph{{Proving relations between modular graph functions}},
  \href{https://doi.org/10.1088/0264-9381/33/23/235011}{\emph{Class. Quant.
  Grav.} {\bfseries 33} (2016) 235011}
  [\href{https://arxiv.org/abs/1606.07084}{{\ttfamily 1606.07084}}].

\bibitem{Basu:2019idd}
A.~Basu, \emph{{Eigenvalue equation for the modular graph $C_{a,b,c,d}$}},
  \href{https://doi.org/10.1007/JHEP07(2019)126}{\emph{JHEP} {\bfseries 07}
  (2019) 126} [\href{https://arxiv.org/abs/1906.02674}{{\ttfamily
  1906.02674}}].

\bibitem{Kleinschmidt:2017ege}
A.~Kleinschmidt and V.~Verschinin, \emph{{Tetrahedral modular graph
  functions}}, \href{https://doi.org/10.1007/JHEP09(2017)155}{\emph{JHEP}
  {\bfseries 09} (2017) 155}
  [\href{https://arxiv.org/abs/1706.01889}{{\ttfamily 1706.01889}}].

\bibitem{Brown:2013gia}
F.~Brown, \emph{{Single-valued motivic periods and multiple zeta values}},
  \href{https://doi.org/10.1017/fms.2014.18}{\emph{SIGMA} {\bfseries 2} (2014)
  e25} [\href{https://arxiv.org/abs/1309.5309}{{\ttfamily 1309.5309}}].

\bibitem{Blumlein:2009cf}
J.~Blumlein, D.~Broadhurst and J.~Vermaseren, \emph{{The multiple zeta value
  data mine}}, \href{https://doi.org/10.1016/j.cpc.2009.11.007}{\emph{Comput.
  Phys. Commun.} {\bfseries 181} (2010) 582}
  [\href{https://arxiv.org/abs/0907.2557}{{\ttfamily 0907.2557}}].

\bibitem{Zerbini:2018hgs}
F.~Zerbini, \emph{{Modular and holomorphic graph functions from superstring
  amplitudes}},  in \emph{{KMPB Conference}: {Elliptic Integrals, Elliptic
  Functions and Modular Forms in Quantum Field Theory}}, pp.~459--484, 2019,
  \href{https://doi.org/10.1007/978-3-030-04480-0_18}{DOI}
  [\href{https://arxiv.org/abs/1807.04506}{{\ttfamily 1807.04506}}].

\bibitem{Zerbini:2018sox}
F.~Zerbini, \emph{{Elliptic multiple zeta values, modular graph functions, and
  genus-one superstring scattering amplitudes}}, Ph.D. thesis, Bonn U., 2017.
\newblock \href{https://arxiv.org/abs/1804.07989}{{\ttfamily 1804.07989}}.

\bibitem{Broedel:2018izr}
J.~Broedel, O.~Schlotterer and F.~Zerbini, \emph{{From elliptic multiple zeta
  values to modular graph functions: open and closed strings at one loop}},
  \href{https://doi.org/10.1007/JHEP01(2019)155}{\emph{JHEP} {\bfseries 01}
  (2019) 155} [\href{https://arxiv.org/abs/1803.00527}{{\ttfamily
  1803.00527}}].

\bibitem{Broedel:2019vjc}
J.~Broedel and O.~Schlotterer, \emph{{One-loop string scattering amplitudes as
  iterated Eisenstein integrals}},  in \emph{{KMPB Conference}: {Elliptic
  Integrals, Elliptic Functions and Modular Forms in Quantum Field Theory}},
  pp.~133--159, 2019, \href{https://doi.org/10.1007/978-3-030-04480-0_7}{DOI}.

\bibitem{Gerken:2020yii}
J.E.~Gerken, A.~Kleinschmidt and O.~Schlotterer, \emph{{Generating series of
  all modular graph forms from iterated Eisenstein integrals}},
  \href{https://doi.org/10.1007/JHEP07(2020)190}{\emph{JHEP} {\bfseries 07}
  (2020) 190} [\href{https://arxiv.org/abs/2004.05156}{{\ttfamily
  2004.05156}}].

\bibitem{DHoker:2017zhq}
E.~D'Hoker and W.~Duke, \emph{{Fourier series of modular graph functions}},
  \href{https://doi.org/10.1016/j.jnt.2018.04.012}{\emph{J. Number Theory}
  {\bfseries 192} (2018) 1} [\href{https://arxiv.org/abs/1708.07998}{{\ttfamily
  1708.07998}}].

\bibitem{Ahlen:2018wng}
O.~Ahl\'en and A.~Kleinschmidt, \emph{{$D^{6} \mathcal{R}^{4}$ curvature
  corrections, modular graph functions, and Poincar\'e series}},
  \href{https://doi.org/10.1007/JHEP05(2018)194}{\emph{JHEP} {\bfseries 05}
  (2018) 194} [\href{https://arxiv.org/abs/1803.10250}{{\ttfamily
  1803.10250}}].

\bibitem{DHoker:2019txf}
E.~D'Hoker and J.~Kaidi, \emph{{Modular graph functions and odd cuspidal
  functions: Fourier and Poincar\'e series}},
  \href{https://doi.org/10.1007/JHEP04(2019)136}{\emph{JHEP} {\bfseries 04}
  (2019) 136} [\href{https://arxiv.org/abs/1902.04180}{{\ttfamily
  1902.04180}}].

\bibitem{Dorigoni:2019yoq}
D.~Dorigoni and A.~Kleinschmidt, \emph{{Modular graph functions and asymptotic
  expansions of Poincar\'e series}}, {\emph{Commun. Num. Theor. Phys.}
  {\bfseries 13} (2019) 569}
  [\href{https://arxiv.org/abs/1903.09250}{{\ttfamily 1903.09250}}].

\bibitem{Basu:2020kka}
A.~Basu, \emph{{Zero mode of the Fourier series of some modular graphs from
  Poincar\'e series}},
  \href{https://doi.org/10.1016/j.physletb.2020.135715}{\emph{Phys. Lett. B}
  {\bfseries 809} (2020) 135715}
  [\href{https://arxiv.org/abs/2005.07793}{{\ttfamily 2005.07793}}].

\bibitem{DHoker:2019mib}
E.~D'Hoker, \emph{{Integral of two-loop modular graph functions}},
  \href{https://doi.org/10.1007/JHEP06(2019)092}{\emph{JHEP} {\bfseries 06}
  (2019) 092} [\href{https://arxiv.org/abs/1905.06217}{{\ttfamily
  1905.06217}}].

\bibitem{Rankin:1939}
R.~Rankin, \emph{{Contributions to the theory of Ramanujan's function $\tau(n)$
  and similar arithmetical functions}},
  \href{https://doi.org/10.1017/S0305004100021095}{\emph{Proc. of the Cambridge
  Philosophical Society} {\bfseries 35} (1939) 351}.

\bibitem{Selberg:1940}
A.~Selberg, \emph{{Bemerkungen über eine Dirichletsche Reihe, die mit der
  Theorie der Modulformen nahe verbunden ist}}, {\emph{Arch. Math. Naturvid.}
  {\bfseries 43} (1940) 47}.

\bibitem{Zagier:1982}
D.~Zagier, \emph{{The Rankin-Selberg method for automorphic functions which are
  not of rapid decay}}, \href{https://doi.org/10.15083/00039589}{\emph{J. of
  the Faculty of Science, the University of Tokyo. Sect. 1 A, Mathematics}
  {\bfseries 28} (1982) 415}.

\bibitem{DHoker:2019blr}
E.~D'Hoker and M.B.~Green, \emph{{Exploring transcendentality in superstring
  amplitudes}}, \href{https://doi.org/10.1007/JHEP07(2019)149}{\emph{JHEP}
  {\bfseries 07} (2019) 149}
  [\href{https://arxiv.org/abs/1906.01652}{{\ttfamily 1906.01652}}].

\bibitem{Kotikov:2002ab}
A.V.~Kotikov and L.N.~Lipatov, \emph{{DGLAP and BFKL equations in the
  $\mathcal{N} = 4$ supersymmetric gauge theory}},
  \href{https://doi.org/10.1016/S0550-3213(03)00264-5}{\emph{Nucl. Phys. B}
  {\bfseries 661} (2003) 19}
  [\href{https://arxiv.org/abs/hep-ph/0208220}{{\ttfamily hep-ph/0208220}}].

\bibitem{Beccaria:2009vt}
M.~Beccaria and V.~Forini, \emph{{Four-loop reciprocity of twist-two operators
  in $\mathcal{N} = 4$ SYM}},
  \href{https://doi.org/10.1088/1126-6708/2009/03/111}{\emph{JHEP} {\bfseries
  03} (2009) 111} [\href{https://arxiv.org/abs/0901.1256}{{\ttfamily
  0901.1256}}].

\bibitem{Arkani-Hamed:2010pyv}
N.~Arkani-Hamed, J.L.~Bourjaily, F.~Cachazo and J.~Trnka, \emph{{Local
  integrals for planar scattering amplitudes}},
  \href{https://doi.org/10.1007/JHEP06(2012)125}{\emph{JHEP} {\bfseries 06}
  (2012) 125} [\href{https://arxiv.org/abs/1012.6032}{{\ttfamily 1012.6032}}].

\bibitem{Broedel:2018qkq}
J.~Broedel, C.~Duhr, F.~Dulat, B.~Penante and L.~Tancredi, \emph{{Elliptic
  Feynman integrals and pure functions}},
  \href{https://doi.org/10.1007/JHEP01(2019)023}{\emph{JHEP} {\bfseries 01}
  (2019) 023} [\href{https://arxiv.org/abs/1809.10698}{{\ttfamily
  1809.10698}}].

\bibitem{DHoker:1994gnm}
E.~D'Hoker and D.H.~Phong, \emph{{The box graph in superstring theory}},
  \href{https://doi.org/10.1016/0550-3213(94)00526-K}{\emph{Nucl. Phys. B}
  {\bfseries 440} (1995) 24}
  [\href{https://arxiv.org/abs/hep-th/9410152}{{\ttfamily hep-th/9410152}}].

\bibitem{Dorigoni:2021jfr}
D.~Dorigoni, A.~Kleinschmidt and O.~Schlotterer, \emph{{Poincar\'e series for
  modular graph forms at depth two I: seeds and Laplace systems}},
  \href{https://arxiv.org/abs/2109.05017}{{\ttfamily 2109.05017}}.

\bibitem{Dorigoni:2021ngn}
D.~Dorigoni, A.~Kleinschmidt and O.~Schlotterer, \emph{{Poincar\'e series for
  modular graph forms at depth two II: iterated integrals of cusp forms}},
  \href{https://arxiv.org/abs/2109.05018}{{\ttfamily 2109.05018}}.

\bibitem{DHoker:2019xef}
E.~D'Hoker and M.~Green, \emph{{Absence of irreducible multiple zeta values in
  melon modular graph functions}},
  \href{https://doi.org/10.4310/CNTP.2020.v14.n2.a2}{\emph{Commun. Num. Theor.
  Phys.} {\bfseries 14} (2020) 315}
  [\href{https://arxiv.org/abs/1904.06603}{{\ttfamily 1904.06603}}].

\bibitem{Green:2013bza}
M.B.~Green, C.R.~Mafra and O.~Schlotterer, \emph{{Multiparticle one-loop
  amplitudes and S-duality in closed superstring theory}},
  \href{https://doi.org/10.1007/JHEP10(2013)188}{\emph{JHEP} {\bfseries 10}
  (2013) 188} [\href{https://arxiv.org/abs/1307.3534}{{\ttfamily 1307.3534}}].

\bibitem{DHoker:2020prr}
E.~D'Hoker, C.R.~Mafra, B.~Pioline and O.~Schlotterer, \emph{{Two-loop
  superstring five-point amplitudes I: construction via chiral splitting and
  pure spinors}}, \href{https://doi.org/10.1007/JHEP08(2020)135}{\emph{JHEP}
  {\bfseries 08} (2020) 135}
  [\href{https://arxiv.org/abs/2006.05270}{{\ttfamily 2006.05270}}].

\bibitem{DHoker:2020tcq}
E.~D'Hoker, C.R.~Mafra, B.~Pioline and O.~Schlotterer, \emph{{Two-loop
  superstring five-point amplitudes II: low-energy expansion and S-duality}},
  \href{https://doi.org/10.1007/JHEP02(2021)139}{\emph{JHEP} {\bfseries 02}
  (2021) 139} [\href{https://arxiv.org/abs/2008.08687}{{\ttfamily
  2008.08687}}].

\bibitem{DHoker:2021kks}
E.~D'Hoker and O.~Schlotterer, \emph{{Two-loop superstring five-point
  amplitudes III: construction via the RNS formulation: even spin structures}},
   \href{https://arxiv.org/abs/2108.01104}{{\ttfamily 2108.01104}}.

\bibitem{Kawai:1985xq}
H.~Kawai, D.~Lewellen and S.~Tye, \emph{{A relation between tree amplitudes of
  closed and open strings}},
  \href{https://doi.org/10.1016/0550-3213(86)90362-7}{\emph{Nucl. Phys. B}
  {\bfseries 269} (1986) 1}.

\bibitem{Schlotterer:2012ny}
O.~Schlotterer and S.~Stieberger, \emph{{Motivic multiple zeta values and
  superstring amplitudes}},
  \href{https://doi.org/10.1088/1751-8113/46/47/475401}{\emph{J. Phys. A}
  {\bfseries 46} (2013) 475401}
  [\href{https://arxiv.org/abs/1205.1516}{{\ttfamily 1205.1516}}].

\bibitem{Stieberger:2013wea}
S.~Stieberger, \emph{{Closed superstring amplitudes, single-valued multiple
  zeta values, and the Deligne associator}},
  \href{https://doi.org/10.1088/1751-8113/47/15/155401}{\emph{J. Phys. A}
  {\bfseries 47} (2014) 155401}
  [\href{https://arxiv.org/abs/1310.3259}{{\ttfamily 1310.3259}}].

\bibitem{Stieberger:2014hba}
S.~Stieberger and T.R.~Taylor, \emph{{Closed string amplitudes as single-valued
  open string amplitudes}},
  \href{https://doi.org/10.1016/j.nuclphysb.2014.02.005}{\emph{Nucl. Phys. B}
  {\bfseries 881} (2014) 269}
  [\href{https://arxiv.org/abs/1401.1218}{{\ttfamily 1401.1218}}].

\bibitem{Schlotterer:2018zce}
O.~Schlotterer and O.~Schnetz, \emph{{Closed strings as single-valued open
  strings: a genus-zero derivation}},
  \href{https://doi.org/10.1088/1751-8121/aaea14}{\emph{J. Phys. A} {\bfseries
  52} (2019) 045401} [\href{https://arxiv.org/abs/1808.00713}{{\ttfamily
  1808.00713}}].

\bibitem{Vanhove:2018elu}
P.~Vanhove and F.~Zerbini, \emph{{Closed string amplitudes from single-valued
  correlation functions}},  \href{https://arxiv.org/abs/1812.03018}{{\ttfamily
  1812.03018}}.

\bibitem{Brown:2019wna}
F.~Brown and C.~Dupont, \emph{{Single-valued integration and superstring
  amplitudes in genus zero}},
  \href{https://doi.org/10.1007/s00220-021-03969-4}{\emph{Commun. Math. Phys.}
  {\bfseries 382} (2021) 815}
  [\href{https://arxiv.org/abs/1910.01107}{{\ttfamily 1910.01107}}].

\bibitem{Zagier:2019eus}
D.~Zagier and F.~Zerbini, \emph{{Genus-zero and genus-one string amplitudes and
  special multiple zeta values}},
  \href{https://doi.org/10.4310/CNTP.2020.v14.n2.a4}{\emph{Commun. Num. Theor.
  Phys.} {\bfseries 14} (2020) 413}
  [\href{https://arxiv.org/abs/1906.12339}{{\ttfamily 1906.12339}}].

\bibitem{Gerken:2020xfv}
J.E.~Gerken, A.~Kleinschmidt, C.R.~Mafra, O.~Schlotterer and B.~Verbeek,
  \emph{{Towards closed strings as single-valued open strings at genus one}},
  \href{https://arxiv.org/abs/2010.10558}{{\ttfamily 2010.10558}}.

\bibitem{Vanhove:2020qtt}
P.~Vanhove and F.~Zerbini, \emph{{Building blocks of closed and open string
  amplitudes}},  in \emph{{MathemAmplitudes 2019: Intersection Theory and
  Feynman Integrals}}, 7, 2020
  [\href{https://arxiv.org/abs/2007.08981}{{\ttfamily 2007.08981}}].

\bibitem{Abel:2021tyt}
S.~Abel and K.R.~Dienes, \emph{{Calculating the Higgs mass in string theory}},
  \href{https://arxiv.org/abs/2106.04622}{{\ttfamily 2106.04622}}.

\bibitem{Maloney:2020nni}
A.~Maloney and E.~Witten, \emph{{Averaging over Narain moduli space}},
  \href{https://doi.org/10.1007/JHEP10(2020)187}{\emph{JHEP} {\bfseries 10}
  (2020) 187} [\href{https://arxiv.org/abs/2006.04855}{{\ttfamily
  2006.04855}}].

\bibitem{Afkhami-Jeddi:2020ezh}
N.~Afkhami-Jeddi, H.~Cohn, T.~Hartman and A.~Tajdini, \emph{{Free partition
  functions and an averaged holographic duality}},
  \href{https://doi.org/10.1007/JHEP01(2021)130}{\emph{JHEP} {\bfseries 01}
  (2021) 130} [\href{https://arxiv.org/abs/2006.04839}{{\ttfamily
  2006.04839}}].

\bibitem{Benjamin:2021ygh}
N.~Benjamin, S.~Collier, A.L.~Fitzpatrick, A.~Maloney and E.~Perlmutter,
  \emph{{Harmonic analysis of 2d CFT partition functions}},
  \href{https://arxiv.org/abs/2107.10744}{{\ttfamily 2107.10744}}.

\bibitem{Zagier1994}
D.~Zagier, \emph{Values of zeta functions and their applications},  in
  \emph{First European Congress of Mathematics Paris, July 6--10, 1992: Vol.
  II: Invited Lectures (Part 2)}, A.~Joseph, F.~Mignot, F.~Murat, B.~Prum and
  R.~Rentschler, eds., (Basel), pp.~497--512, Birkh{\"a}user Basel (1994),
  \href{https://doi.org/10.1007/978-3-0348-9112-7_23}{DOI}.

\bibitem{Zhao:2000}
J.~Zhao, \emph{Analytic continuation of multiple zeta functions},
  \href{https://doi.org/10.1090/S0002-9939-99-05398-8}{\emph{Proceedings of the
  American Mathematical Society} {\bfseries 128} (2000) 1275}.

\bibitem{Akiyama:2002}
S.~Akiyama and H.~Ishikawa, \emph{{On analytic continuation of multiple
  $L$-functions and related zeta-functions}},  in \emph{Analytic Number
  Theory}, C.~Jia and K.~Matsumoto, eds., (Boston, MA), pp.~1--16, Springer US
  (2002), \href{https://doi.org/10.1007/978-1-4757-3621-2_1}{DOI}.

\bibitem{Borwein:2006}
D.~Borwein, J.M.~Borwein and D.M.~Bradley, \emph{{Parametric Euler sum
  identities}},
  \href{https://doi.org/{10.1016/j.jmaa.2005.04.040}}{\emph{Journal of
  Mathematical Analysis and Applications} {\bfseries 316} (2006) 328–338}
  [\href{https://arxiv.org/abs/505058}{{\ttfamily 505058}}].

\end{thebibliography}\endgroup
